\documentclass[12pt]{article}
\usepackage{youngtab}
\makeatletter \@addtoreset{equation}{section} \makeatother
\renewcommand{\theequation}{\thesection.\arabic{equation}}
\newcommand{\ba}{\begin{array}}
\newcommand{\ea}{\end{array}}
\newcommand{\beq}{\begin{equation}}
\newcommand{\eeq}{\end{equation}}
\newcommand{\bea}{\begin{eqnarray}}
\newcommand{\eea}{\end{eqnarray}}

\def\bce{\begin{center}}
\def\ece{\end{center}}

\def\nonu{\nonumber}

\def\pa{\partial}

\def\be{\beta}

\def\de{\delta}

\def\la{\lambda}
\def\La{\Lambda}

\def\si{\sigma}

\def\eps6{{\displaystyle \mathop{\epsilon}^{6}}{}}
\def\g6{{\displaystyle \mathop{g}^{6}}{}}

\def\nab6{{\displaystyle \mathop{\nabla}^{6}}{}}

\def\0{{\sst{(0)}}}
\def\1{{\sst{(1)}}}
\def\2{{\sst{(2)}}}
\def\3{{\sst{(3)}}}
\def\4{{\sst{(4)}}}
\def\5{{\sst{(5)}}}
\def\6{{\sst{(6)}}}
\def\7{{\sst{(7)}}}
\def\8{{\sst{(8)}}}

\def\ba{\begin{array}}
\def\ea{\end{array}}
\def\beq{\begin{equation}}
\def\eeq{\end{equation}}
\def\be{\begin{equation}}
\def\ee{\end{equation}}

\def\la{\lambda}
\def\eps{\epsilon}

\def\ba{\begin{array}}
\def\ea{\end{array}}
\def\beq{\begin{equation}}
\def\eeq{\end{equation}}
\def\be{\begin{equation}}
\def\ee{\end{equation}}

\def\la{\lambda}
\def\eps{\epsilon}

\def\eps6{{\displaystyle \mathop{\epsilon}^{6}}{}}

\def\nab6{{\displaystyle \mathop{\nabla}^{6}}{}}

\newcommand{\bean}{\begin{eqnarray*}}
\newcommand{\eean}{\end{eqnarray*}}

\begin{document}
\thispagestyle{empty} \addtocounter{page}{-1}
   \begin{flushright}
\end{flushright}

\vspace*{1.3cm}
  
\centerline{ \Large \bf   
Higher Spin Currents
with Manifest $SO(4)$ Symmetry } 
\vspace*{0.3cm}
\centerline{ \Large \bf 
 in the Large ${\cal N}=4$ Holography} 
\vspace*{1.5cm}
\centerline{{\bf Changhyun Ahn}
\footnote{On leave from the Department of Physics, Kyungpook National University, Taegu
  41566, Korea and 
address until Aug. 31, 2018:
C.N. Yang Institute for Theoretical Physics,
Stony Brook University,
Stony Brook, NY 11794-3840, USA
}
} 
\vspace*{1.0cm} 
\centerline{\it
C.N. Yang Institute for Theoretical Physics,
Stony Brook University,
Stony Brook, NY 11794-3840, USA}
 \centerline{\it 
Department of Physics, Kyungpook National University, Taegu
41566, Korea} 
\vspace*{0.8cm} 
\centerline{\tt ahn@knu.ac.kr
} 
\vskip1cm

\centerline{\bf Abstract}
\vspace*{0.5cm}

The large ${\cal N}=4$ nonlinear superconformal algebra
is generated by six spin-$1$ currents, four spin-$\frac{3}{2}$ currents and
one spin-$2$ current.
The simplest extension of these $11$ currents
is described by  
the $16$ higher spin currents of spins $(1,\frac{3}{2},\frac{3}{2},
\frac{3}{2}, \frac{3}{2}, 2,2,2,2,2,2,
\frac{5}{2}, \frac{5}{2},\frac{5}{2}, \frac{5}{2}, 3)$.
In this paper,
by using the defining operator product expansions (OPEs) between
the $11$ currents and $16$ higher spin currents, we determine the
$16$ higher spin currents
(the higher spin-$1, \frac{3}{2}$ currents were found previously)
in terms of affine Kac-Moody spin-$\frac{1}{2}, 1$
currents in the Wolf space coset model completely.
An antisymmetric second rank tensor, three antisymmetric
almost complex structures or the structure constant are contracted with
the multiple product of spin-$\frac{1}{2}, 1$ currents.
The Wolf space coset contains the group $SU(N+2)$
and the level $k$ is characterized by the affine Kac-Moody spin-$1$
currents.
After calculating the eigenvalues of the zeromode of the
higher spin-$3$ current acting on the higher representations up to
three (or four) boxes of Young tableaux in $SU(N+2)$ in the
Wolf space coset, we obtain the corresponding
three-point functions with two scalar operators at finite $(N,k)$.
Furthermore,
under the large $(N,k)$ 't Hooft like limit, the eigenvalues  
associated with any boxes of Young tableaux are obtained
and the corresponding three-point functions
are written in terms of the 't Hooft coupling constant in simple form
in addition to the two-point functions of scalars and the number of
boxes.

\baselineskip=18pt
\newpage
\renewcommand{\theequation}
{\arabic{section}\mbox{.}\arabic{equation}}


\section{Introduction}

The duality \cite{GG1011,GG1205} between the higher spin gauge theory on
$AdS_3$ space \cite{PV9806,PV9812}
and the large $N$ 't Hooft-like limit
of the $W_N$ minimal models \cite{GKO1,GKO2,BBSS1,BBSS2,FL}  
implies that it is necessary to obtain the higher spin currents
in the boundary theory in order to see some consistency
with the bulk theory \cite{GG1207}.
For example, the three-point functions of the higher spin currents
with two scalar operators can be determined from the explicit forms
for the
higher spin currents.

The Goddard-Kent-Olive (GKO)
coset construction \cite{GKO1,GKO2} for
the higher spin-$3$ Casmir operator
has been found in \cite{BBSS1,BBSS2} some time ago.
The role of completely symmetric $SU(N)$ invariant tesor of rank-$3$
is crucial in this construction.
In \cite{GH}, the eigenvalues for the higher spin-$3$ current
were studied in the context of \cite{GG1011,GG1205,GG1207}.
The same coset construction for the higher spin-$4$ Casimir operator
has been studied in \cite{Ahn1111} where
the completely symmetric $SU(N)$ invariant tensor of rank $4$
can be written in terms of the above rank-$3$ tensor and rank-$2$ Kronecker
delta tensor.
The higher spin-$4$ Casimir operator occurs 
in the  operator product expansion (OPE) between the
higher spin-$3$ Casimir operator and itself.
The rank-$4$ tensor is given by the quadratic rank-$3$ tensor
and quadratic rank-$2$ tensor.
The eigenvalues for the higher spin-$4$ current
are simply the ones for the higher spin-$3$ current
multiplied by a factor which depends on the 't Hooft-like coupling
constant
under the large $N$ 't Hooft-like limit.

For  the higher spin-$5$ Casimir operator,
the above coset
construction has been described in \cite{AK1308} by focusing on the
particular pole in the OPE between
the known higher spin-$3$ Casmir operator and
the known higher spin-$4$ Casmir operator.
The triple product of rank-$3$ tensor appears in this construction:
one factor from the former and the two factors from the latter.
As before,
the eigenvalues for the higher spin-$5$ current
can be written in terms of the ones
for the higher spin-$4$ current with additional simple factor
under the large $N$ 't Hooft-like limit. 
The GKO
coset construction for
the higher spin-$4$ Casmir operator in the orthogonal coset model
has been discussed in \cite{Ahn1701}.
The completely symmetric $SO(2N)$ invariant tensor of rank-$4$
played an important role. See also the relevant papers in
\cite{Ahn1202,AP1301}.

The ${\cal N}=1$ supersymmetric (higher spin currents) extension
of the GKO coset construction in \cite{BBSS1,BBSS2}
has been studied in \cite{Ahn1305,Ahn1211}
by considering
some condition for one of the levels of the coset model,
which allows us to construct the fermionic spin-$\frac{3}{2}$
current, the superpartner of spin-$2$ stress energy tensor current
(and the superpartners of bosonic
higher spin currents)
using a fermion field, along the line of \cite{ASS,HR}.
Still one can take the large $N$ 't Hooft-like limit.
Similarly, the  ${\cal N}=1$ supersymmetric higher spin extension
of the GKO coset construction in
\cite{Ahn1701} (in the orthogonal coset model)
was obtained in the same paper
\cite{Ahn1701}
by applying the above level condition to the bosonic coset theory. 
The other type of ${\cal N}=1$ supersymmetric 
GKO
coset construction for
the higher spin Casmir operators in the orthogonal coset model
was found in \cite{AP1310} by looking at the paper
\cite{CHR1209} on the ${\cal N}=1$ higher spin holography closely.

The ${\cal N}=2$ supersymmetric (higher spin currents) extension
of the GKO coset construction in \cite{BBSS1,BBSS2} (or in
\cite{Ahn1305,Ahn1211})
has been studied in \cite{Ahn1604}
by restricting
further condition on the the remaining level in the ${\cal N}=1$
supersymmetric coset model,
which allows us to construct the additional fermionic spin-$\frac{3}{2}$
current and bosonic spin-$1$ current (and the superpartners of ${\cal N}=1$
higher spin currents)
using an additional fermion field, along the line of \cite{GHKSS}.
There are two kinds of adjoint fermions.
The other type of ${\cal N}=2$ supersymmetric 
GKO
coset construction for
the higher spin Casmir operators 
was found in \cite{Ahn1208,Ahn1206} by considering the ${\cal N}=2$
Kazama-Suzuki model \cite{KS1,KS2}
in the context of
the ${\cal N}=2$ higher spin holography \cite{CHR1111,CG,HP}.

The ${\cal N}=3$ supersymmetric (higher spin currents) extension
of the GKO coset construction in \cite{Ahn1208,Ahn1206}
has been studied in \cite{AK1607}
by considering
some condition for the level of the coset model,
which allows us to construct the additional spin-$\frac{3}{2}$
current, two bosonic spin-$1$ current and fermionic spin-$\frac{1}{2}$
current (and the ${\cal N}=3$ superpartners of ${\cal N}=2$
higher spin currents)
using the various fermion fields, along the line of
the ${\cal N}=3$ higher spin holography
\cite{CHR1406}.

Note that compared to the bosonic cases studied in
\cite{Ahn1111,AK1308,Ahn1701,Ahn1202,AP1301},
one can use the supersymmetry (or the fermionic spin-$\frac{3}{2}$ current)
to
determine the superpartners corresponding to the (higher spin) currents
(inside of given supermultiplet)
for the supersymmetric cases in
\cite{Ahn1305,Ahn1211,AP1310,Ahn1604,Ahn1208,Ahn1206,AK1607}.
In other words, for example, once the lowest component higher spin current
in the given supermultiplet is known explicitly,
then in principle all the other components in the same 
supermultiplet can be determined successively
by starting with the OPE between the fermionic
spin-$\frac{3}{2}$ current and the above lowest higher spin current.   

It is natural to ask whether one can construct
the GKO coset construction for the higher spin currents
in the ${\cal N}=4$ superconformal Wolf space coset model
in the context of
the ${\cal N}=4$ higher spin holography \cite{GG1305}.
See also later works in \cite{GG1406,GG1501,GG1512}.
In \cite{AK1411}, the explicit higher spin-$1$ current
was determined in terms of adjoint spin-$1, \frac{1}{2}$ currents
(See also \cite{Ahn1311}).
The antisymmetric rank-$2$ tensor which has the coset indices
appears in the overall factor in the higher spin-$1$ current.
The structure constant tensor also appears in the linear term of
adjoint spin-$1$ current. Furthermore, the four higher spin-$\frac{3}{2}$ 
currents was obtained. One of them has the above
antisymmetric rank-$2$ tensor and each three of them contains
the symmetric (and traceless) rank-$2$ tensor. These three
symmetric tensors are given by the contraction between the three
almost complex structures (which are antisymmetric), above
antisymmetric rank-$2$ tensor and the metric.
Two of the six higher spin-$2$ currents
were obtained explicitly and four of them were determined but contain
some composite field between the known (higher spin) currents.
Similarly, four higher spin-$\frac{5}{2}$ currents are given by
the combination of the above adjoint spin-$1, \frac{1}{2}$ currents
and the composite fields between the known (higher spin) currents.
Finally, the higher spin-$3$ current is written in mixed form. 
For orthogonal Wolf space coset, see also \cite{AKP1510,AP1410}. 

In this paper, one would like to continue to the previous work \cite{AK1411}
but in different basis. Although the
six spin-$1$ currents of the large ${\cal N}=4$ `nonlinear'
superconformal algebra \cite{GS,VanProeyen,GPTV,GK} are manifest in $SU(2) \times SU(2)$ basis
(their linear combination provides the six independent spin-$1$ currents
transforming as $SO(4)$ vector representation),
the spin-$\frac{3}{2}$ currents are manifest in $SO(4)$ basis.
One can use the ${\cal N}=4$ primary condition for the $16$ higher spin
current as done in previous supersymmetric examples.
This holds for the extension of the large ${\cal N}=4$ `linear'
superconformal algebra \cite{STV,STVS,Schoutens,ST,Saulina} generated by $16$ currents.
It is known that the ${\cal N}=4$ higher spin multiplet
transforms as linearly under the ${\cal N}=4$ stress energy tensor
in ${\cal N}=4$ superspace.
In other words, the right hand side of the ${\cal N}=4$ super OPE
between them contains only the linear term in
the  ${\cal N}=4$ higher spin multiplet \cite{AK1509}. 
Moreover, one can write down this ${\cal N}=4$ super OPE in
component approach. By the work of  Goddard-Schwimmer \cite{GS},  
one can write down the $11$ currents in the nonlinear version
using $16$ currents in the linear version.
Similarly,
the $16$ higher spin currents in the nonlinear version
can be written in terms of
$16$ higher spin currents in the linear version \cite{BCG}.
Therefore, the OPEs between the $11$ currents
and the $16$ higher spin currents in nonlinear version
can be written explicitly in \cite{AKK1703}.

What is the usefulness of the description in the $SO(4)$ basis rather than
$SU(2) \times SU(2)$ basis?
For the higher spin-$3$ case, according to the observation of \cite{AK1411}
in $SU(2) \times SU(2)$ basis,
one should consider both the first and the second order poles
of the OPE between
the spin-$\frac{3}{2}$ current and the higher spin-$\frac{5}{2}$ current. 
The second order pole of this OPE contains many
composite fields.
One should also subtract some composite fields appearing in the
first order pole of this OPE.
On the other hands, what one should calculate in the $SO(4)$ basis is  
the first order pole
of the OPE between
the spin-$\frac{3}{2}$ current
and the higher spin-$\frac{5}{2}$ current.
Moreover, what one subtracts in the first order pole
is simply the composite field between
the higher spin-$1$ current and the spin-$2$ current.

In \cite{Ahn1711},
the conformal dimension,
two $SU(2)$ quantum numbers, and $U(1)$ charge
in the minimal (and higher) representations
 up to three boxes
(of Young tableaux)
in the Wolf space coset have been studied.
The eigenvalues associated with the higher spin currents in the
(minimal and higher) representations up to two boxes
are also obtained.
This implies that
the three-point functions \cite{CY} of the higher spin currents
with two scalar operators are determined.
Although the explicit (closed)
expressions for the higher spin currents in terms of
adjoint spin-$1, \frac{1}{2}$ currents 
are known for the higher spin-$1,\frac{3}{2}$ currents
and the remaining higher spin-$2, \frac{5}{2}, 3$ currents
are known for several $N$ values, one could calculate
the above eigenvalues explicitly.
If the explicit realization for the higher spin currents
with the help of adjoint spin-$1,\frac{1}{2}$ currents is known completely,
then
one can analyze the various eigenvalues for the higher spin currents
(and three-point functions) further because one can calculate
each eigenvalue for each term.
There are two $SO(4)$ singlets, the higher spin-$1$ current and
the higher spin-$3$ current. Maybe one can express the three-point
functions for the higher spin-$3$ current with two scalar operators
(associated with {\it any} boxes of Young tableaux) in terms of those for
the higher spin-$1$ current with the same
two scalar operators at least under the large $(N,k)$ 't Hooft like
limit as in
\cite{Ahn1701,AK1308}. 


In section $2$,
the $11$ currents, for the large ${\cal N}=4$
nonlinear superconformal algebra, in terms of adjoint spin-$1,\frac{1}{2}$
currents are reviewed. 
In section $3$,
the higher spin-$2, \frac{5}{2}, 3$ currents
in terms of  the adjoint spin-$1,\frac{1}{2}$
currents are determined explicitly. Together with the known
higher spin-$1, \frac{3}{2}$ currents, all these $16$ higher spin currents 
are obtained explicitly.
In section $4$, the eigenvalues and three-point functions
are analyzed based on the results in section $3$.
In section $5$, the summary of this paper is given and some open problems
are presented.
In Appendices, some details, which are related to the contents in
previous sections, are given \footnote{
The Thielemans package in \cite{Thielemans} with a mathematica
\cite{mathematica}
is useful to check
the OPEs.}.

In \cite{HU1708} (See also the works of \cite{HU1708-1,HU1801}),
using the decomposition of the scalar four-point functions
by Virasoro conformal blocks, the three-point functions including
$\frac{1}{N}$ corrections in the two dimensional (bosonic)
$W_N$ minimal model
were obtained using the result of
\cite{PR} (see also the works of \cite{CY1112,CY1302}).
As observed in \cite{HU1708}, it is an open problem to
obtain the three-point functions from the decomposition of
four-point functions in the large ${
\cal N }=4$ holography.

In \cite{EGGL}, by analyzing the BPS spectrum of string theory
and supergravity theory on $AdS_3 \times {\bf S^3} \times {\bf S^3}
\times {\bf S^1}$, it has been found that the BPS spectra of both
descriptions agree (where the world sheet approach is used).
The appearance of an infinite stringy tower of massless higher spin
fields at the critical level of WZW model (in the bosonic string on $AdS_3$)
was found in \cite{GGH}. 
The similar analysis for the superstring on $AdS_3$ has been found
in \cite{FGJ}. See also the relevant works in \cite{EGL,EGR}.
Very recently, some signal for the transition
from the black holes to long strings in the superstring on $AdS_3$
at the critical level is interpreted as
the infinite tower of modes that become massless
in \cite{GHKPR}.
See also the work of \cite{GG1803}. It would
be interesting to see how the large ${\cal N}=4$ superconformal higher spin
and CFT duality arises in the context of these world sheet approaches.

\section{The $11$ currents which generate the large
${\cal N}=4$ nonlinear superconformal algebra}

Let us consider the Wolf space coset in the ``supersymmetric'' version
with groups $G=SU(N+2)$
and $H=SU(N) \times SU(2) \times U(1)$.
The operator product expansion between the spin-$1$ current $V^a(z)$ and
the spin-$\frac{1}{2}$ current $Q^a(z)$
is described as \cite{KT}
\bea
V^a(z) \, V^b(w) & = & \frac{1}{(z-w)^2} \, k \, g^{ab}
-\frac{1}{(z-w)} \, f^{ab}_{\,\,\,\,\,\,c} \, V^c(w) 
+\cdots,
\nonu \\
Q^a(z) \, Q^b(w) & = & -\frac{1}{(z-w)} \, (k+N+2) \, g^{ab} + \cdots,
\qquad
V^a(z) \, Q^b(w)  =  + \cdots.
\label{opevq}
\eea
The level $k$ is the positive integer. 

The metric can be obtained from $g_{ab} =
\frac{1}{2c_G} \, f_{a c}^{\,\,\, d}
\, f_{b d}^{\,\,\,c}$ where $c_G$ is the dual Coxeter number of
the group $G=SU(N+2)$.
The metric $g_{ab}$ is given by the generators
of $SU(N+2)$ in the complex basis \cite{AK1506} as follows
\bea
g_{ab} = \mbox{Tr} (T_a T_b)=
\left(
\begin{array}{cc}
0 & 1 \\
1 & 0 \\
\end{array}
\right) = g^{ab}, \qquad
a,b=1,2, \cdots, (N+2)^2-1.
\label{metric}
 \eea
 The commutation relation for the $SU(N+2)$ generators
 is given by $[T_a, T_b] = f_{ab}^{\,\,\,c} \, T_c$.
 Due to the regular behavior between the spin-$1$ current
 and the spin-$\frac{1}{2}$ current, one can move any current
 in the composite
 fields between them freely. 
 Any $SU(N+2)$ group indices, $a, b, \cdots$ can be raised or lowered
 by using the above metric $g_{ab}$.

 The four supersymmetry currents of spin-$\frac{3}{2}$,
 $\hat{G}^0(z)$ and
 $\hat{G}^i(z)$, the six spin-$1$ currents of $SU(2)_k \times SU(2)_N$,
 $A^{\pm i}(z)$ and
 the spin-$2$ stress energy tensor $\hat{T}(z)$ can be
 described as follows \cite{VanProeyen,GK,AK1411}  
\bea
\hat{G}^{0}(z) &  = &   \frac{i}{(k+N+2)}  \, g_{\bar{a} \bar{b}} \,
Q^{\bar{a}} \, V^{\bar{b}}(z),
\qquad
\hat{G}^{i}(z)  =  \frac{i}{(k+N+2)} 
\, h^{i}_{\bar{a} \bar{b}} \, Q^{\bar{a}} \, V^{\bar{b}}(z),
\nonu \\
A^{+i}(z) &  = & 
-\frac{1}{4N} \, f^{\bar{a} \bar{b}}_{\,\,\,\,\,\, c} \, h^i_{\bar{a} \bar{b}} \, V^c(z), 
\qquad
A^{-i}(z)  =  
-\frac{1}{4(k+N+2)} \, h^i_{\bar{a} \bar{b}} \, Q^{\bar{a}} \, Q^{\bar{b}}(z),
\nonu \\
\hat{T}(z)  & = & 
\frac{1}{2(k+N+2)^2} \Bigg[ (k+N+2) \, g_{\bar{a} \bar{b}}
\,  V^{\bar{a}} \, V^{\bar{b}} 
+k \, g_{\bar{a} \bar{b}} \, Q^{\bar{a}} \, \pa \, Q^{\bar{b}} 
+f^{\bar{a} \bar{b}}_{ \,\,\,\,\,\, c} \, g_{\bar{a} \bar{c}} \, g_{\bar{b} \bar{d}} \,
Q^{\bar{c}} \, Q^{\bar{d}} \, V^c  \Bigg] (z)
\nonu \\
&&- \frac{1}{(k+N+2)} ( A^{+i}+A^{-i} )^2 (z), \qquad i=1,2,3.
\label{11currents}
\eea
The Wolf coset indices, $\bar{a}, \bar{b}, \cdots$,
run over  $\bar{a}, \bar{b}, \cdots = 1, 2, \cdots, 4N$. 
The quantity $4N$ can be obtained from the number of generators
$(N+2)^2-1$ in $G=SU(N+2)$ and the number of generators
$(N^2-1) +4$ in $H=SU(N) \times SU(2) \times U(1)$.
In the spin-$2$ stress energy tensor, the terms $A^{-i} \, A^{-i}$
which contain  $ (Q^{\bar{a}} \, Q^{\bar{b}})( Q^{\bar{c}} \, Q^{\bar{d}})(z)$
can be further simplified by using the rearrangement lemmas
\cite{BS,Fuchs}. 

The three almost complex structures are given by the following
$4N \times 4N$ matrices \cite{AK1506}
\bea
h^1_{\bar{a} \bar{b}} = 
\left(
\begin{array}{cccc}
0 & 0  & 0 & -i \\
0 & 0 & -i & 0 \\
0 & i & 0 & 0 \\
i & 0 & 0 & 0 \\
\end{array}
\right), \quad
h^2_{\bar{a} \bar{b}} = 
\left(
\begin{array}{cccc}
0 & 0  & 0 & 1 \\
0 & 0 & -1 & 0 \\
0 & 1 & 0 & 0 \\
-1 & 0 & 0 & 0 \\
\end{array}
\right), \qquad
h^{3}_{\bar{a} \bar{b}}
=
\left(
\begin{array}{cccc}
0 & 0  & i & 0 \\
0 & 0 & 0 & -i \\
-i & 0 & 0 & 0 \\
0 & i & 0 & 0 \\
\end{array}
\right).
\label{himatrix}
\eea
Each entry is $N \times N$ matrix. 
The last almost complex structure
can be obtained from the first two $h^3_{\bar{a} \bar{b}} =
h^1_{\bar{a} \bar{c}} \,
g^{\bar{c} \bar{d}} \, h^2_{\bar{d} \bar{b}}$.
They are antisymmetric and satisfy the quaternionic algebra
\cite{Saulina} where one sees the presence of the Wolf space coset metric
$g_{\bar{a} \bar{b}} \equiv h^0_{\bar{a} \bar{b}}$.
See also (\ref{hhrel}).
Note that this coset metric and three almost complex structures
appear in the above four supersymmetry currents of spin-$\frac{3}{2}$.

Then the large ${\cal N}=4$ nonlinear superconformal algebra \cite{GS}
can be obtained explicitly using the above $11$ currents.
The $10$ currents of spin $\frac{3}{2}$ and $1$ are primary under the
stress energy tensor. The nonlinear structure appears in the OPE
between the spin-$\frac{3}{2}$ currents. The two levels of $SU(2)$'s are
given by $k$ and $N$ respectively.

One can rewrite the large ${\cal N}=4$ nonlinear superconformal
algebra in $SO(4)$ manifest way. 
With the $SO(4)$ singlet
\bea
T(z) = \hat{T}(z),
\label{Spin2}
\eea
the spin-$\frac{3}{2}$ currents, transforming as the $SO(4)$ vector
representation,
are given by \cite{AKK1703}
\bea
G^1(z) &=& \hat{G}^3(z), \qquad
G^2(z) = \hat{G}^0(z),
\nonu \\
G^3(z) &=& \hat{G}^1(z), \qquad
G^4(z) = -\hat{G}^2(z).
\label{spin3halfrelation}
\eea
Furthermore, the six spin-$1$ currents, $T^{\mu \nu}(z)$
transforming as the $SO(4)$ adjoint representation,
can be obtained from
the corresponding two adjoint spin-$1$ currents $A^{\pm i}(z)$ as follows
\cite{AK1509,AKK1703}
\bea
T^{12}(z) & = & -i (A^{+3}-A^{-3})(z), \qquad
T^{13}(z)  =  -i (A^{+2}+A^{-2})(z), \nonu \\
T^{14}(z) & = &  -i (A^{+1}+A^{-1})(z), \qquad
T^{23}(z)  =  i (A^{+1}-A^{-1})(z), \nonu \\
T^{24}(z) & = &  -i (A^{+2}-A^{-2})(z), \qquad
T^{34}(z)  =  i (A^{+3}+A^{-3})(z). 
\label{TA}
\eea
The $SO(4)$ invariant Kronecker delta $\delta^{\mu \nu}$ and epsilon
tensors $\varepsilon^{\mu \nu \rho \si}$ appear in the 
corresponding large ${\cal N}=4$ nonlinear superconformal algebra
\cite{GS,VanProeyen,GPTV,GK}.
The six $4 \times 4$ matrices $\alpha_{\mu \nu}^{\pm i}$ \cite{STVS}
relate the spin-$1$ currents
$A^{\pm i}(z)$ to the spin-$1$ currents $T^{\mu \nu}(z)$
\footnote{Because
  the $SU(N+2)$ generators one uses in this paper are given by
  the ones in \cite{AK1506} rather than the ones in \cite{AK1411}, one
  cannot use some identities appeared in \cite{AK1411} directly.}.

\section{Construction of the higher spin currents}

In this section, we would like to obtain  the higher spin-$3$ current
together with the higher spin-$2$ currents and the higher spin-$\frac{5}{2}$
currents in terms of the Kac-Moody currents, $V^a(z)$ and $Q^a(z)$.
Some of the expressions for these higher spin currents are given in
\cite{AK1411} but they are mixed with the currents presented in the
previous section from the large
${\cal N}=4$ nonlinear superconformal algebra.
The main things in this section are to write them in terms of
$V^a(z)$ and $Q^a(z)$ completely.

\subsection{Known higher spin-$1$ current}

The higher spin-$1$ current, which transforms as
the $SO(4)$ singlet representation,
can be written in terms of
the spin-$1$ currents $V^a(z)$ and the spin-$\frac{1}{2}$
currents $Q^a(z)$. The tensorial structure with exact coefficients
are determined by \cite{AK1411}
\bea
\Phi_0^{(1)} (z) &=&
-\frac{1}{2(k+N+2)} \, d^0_{\bar{a} \bar{b}} \,
f^{\bar{a} \bar{b}}_{\,\,\,\,\,\, c}  V^c  (z)
+ \frac{k}{2(k+N+2)^2} \, d^0_{\bar{a} \bar{b}} \, Q^{\bar{a}} \, Q^{\bar{b}} (z).
\label{spinone}
\eea
The overall tensor is antisymmetric and is given by
the following $4N \times 4N$ matrix
\bea
d^0_{\bar{a} \bar{b}}  = 
\left(
\begin{array}{cccc}
0 & 0 & -1 & 0 \\
0 & 0 & 0 & -1 \\
1 & 0 & 0 & 0  \\
0 & 1 & 0 & 0 \\
\end{array}
\right).
\label{dzero}
\eea
Each entry is $N \times N$ matrix.
Note that there exists no coset spin-$1$ current $V^{\bar{c}}(z)$ term
in the first term of (\ref{spinone})
because of the fact that $f^{\bar{a}
  \bar{b}}_{\,\,\,\,\,\, \bar{c}}=0$. On the other hand, the quadratic
spin-$\frac{1}{2}$ current terms
contain only the coset spin-$\frac{1}{2}$
current.
Because this higher spin-$1$ current belongs to the
lowest component of the ${\cal N}=4$ higher spin multiplet,
the other $15$ higher spin currents can be
determined by the help of the four spin-$\frac{3}{2}$ supersymmetry
currents presented in the previous section.

\subsection{Known higher spin-$\frac{3}{2}$ currents}

From the defining equation for the OPE
between the spin-$\frac{3}{2}$ current presented in previous section
and the higher spin-$1$
current described in the previous subsection,
one can obtain the higher spin-$\frac{3}{2}$ current.
From Appendix (\ref{1116}), one has  the OPE \footnote{We use the
  hat notation for the (higher spin) currents in the $SU(2) \times SU(2)$
  basis. The OPEs with hat notation are the same as the ones in Appendix
  $A$.  One has $\hat{\Phi}_{0}^{(1)}(z) =\Phi_{0}^{(1)}(z)$. }
\bea
\hat{G}^{\mu}(z) \, \Phi_{0}^{(1)}(w) & = & 
-\frac{1}{(z-w)} \, \hat{\Phi}_{\frac{1}{2}}^{(1),\mu}(w)
+\cdots.
\label{higherspin3half}
\eea
By using the explicit expressions in (\ref{11currents})
and (\ref{spinone}),
one can calculate the OPE
$\hat{G}^{\mu}(z) \, \Phi_{0}^{(1)}(w)$.
On the one hand, the following relation holds by using (\ref{opevq})
with the help of the description in \cite{BS}
\bea
Q^{\bar{a}} V^{\bar{b}}(z) \, V^c(w) =
\frac{1}{(z-w)^2} \, k \, g^{c\bar{b}} \, Q^{\bar{a}}(w)
+\frac{1}{(z-w)} \, \Bigg[ k
\, g^{c \bar{b}} \, \pa Q^{\bar{a}} + f^{c \bar{b}}_{\,\,\,\,\,\, d}
\, Q^{\bar{a}} \, V^{d}
\Bigg](w) +\cdots.
\label{firstcont}
\eea
On the other hand, there exist other contributions from
\bea
Q^{\bar{a}} V^{\bar{b}}(z) \, Q^{\bar{c}} Q^{\bar{d}}(w) =
\frac{1}{(z-w)} \, (k+N+2) \, \Bigg[
-g^{\bar{c} \bar{a}} V^{\bar{b}} \, Q^{\bar{d}} +
g^{\bar{d} \bar{a}} Q^{\bar{c}} \, V^{\bar{b}}
\Bigg](w) + \cdots.
\label{secondcont}
\eea
Then one obtains the final OPE by combining
the two equations (\ref{firstcont}) and (\ref{secondcont})
with the appropriate metric, almost complex
structures, $d$ tensors and the structure constants.

\subsubsection{The second order pole}

The second order pole of $
\hat{G}^{\mu}(z) \, \Phi_{0}^{(1)}(w)$ (and the similar term in the
first order pole) vanishes 
due to the fact that
\bea
f^{\bar{c} \bar{d}}_{\,\,\,\,\,\, e} \, g^{e \bar{b}} =0.
\label{fgrelation}
\eea
Note that the structure constant $f^{\bar{c} \bar{d}}_{\,\,\,\,\,\, \bar{e}}=0$
and the nonzero metric components are given by the coset indices
or the subgroup indices (\ref{metric}) separately.

\subsubsection{The first order pole}

The first order pole of
$
\hat{G}^{\mu}(z) \, \Phi_{0}^{(1)}(w)$
contains
three terms.
One can use the following identity
\bea
d^0_{\bar{c}\bar{d}} \,
f^{\bar{c} \bar{d}}_{\,\,\,\,\,\, e} \, f^{e \bar{b}}_{\,\,\,\,\,\, g}
= -2(N+2) \, d^{0}_{\bar{h}\bar{g}} \, g^{\bar{h}\bar{b}},
\label{dffrelation-1}
\eea
to simplify the last term in (\ref{firstcont}). 
The $N$ dependence can be seen from the several $N$ values,
$N=3,5,7, \cdots$.
By multiplying the almost complex structure originating from the
spin-$\frac{3}{2}$ currents into (\ref{dffrelation-1}),
it is obvious that
\bea
h^{\mu}_{\bar{a}\bar{b}} \, d^0_{\bar{c}\bar{d}} \,
f^{\bar{c} \bar{d}}_{\,\,\,\,\,\, e} \, f^{e \bar{b}}_{\,\,\,\,\,\, g}
= -2(N+2) \, d^{\mu}_{\bar{a}\bar{g}},
\label{Right}
\eea
where we, in the right hand side of (\ref{Right}),
introduce the product of $d^0_{\bar{a} \bar{b}}$
tensor (\ref{dzero}),
which is antisymmetric, 
and complex structures (\ref{himatrix})
together with the metric tensor as follows \cite{AK1411}
\bea
d^0_{\bar{a} \bar{b}} \, g^{\bar{b} \bar{c}} \, h^{\mu}_{\bar{c} \bar{d}} \equiv
d^{\mu}_{\bar{a} \bar{d}}, \qquad \mu=0,1,2,3.
\label{dmu}
\eea
The $d_{\bar{a} \bar{b}}^i$ tensor  where $i=1,2,3$ is symmetric,
compared to the antisymmetric $d_{\bar{a} \bar{b}}^0$ tensor.
Note that we introduce the (symmetric) metric tensor 
\bea
g_{\bar{a} \bar{b}} \equiv h^{0}_{\bar{a} \bar{b}},
\label{hzero}
\eea
compared to the antisymmetric  almost complex structures
$h_{\bar{a} \bar{b}}^i$. 

Therefore, the higher spin-$\frac{3}{2}$ currents
from (\ref{higherspin3half}) in $SU(2) \times SU(2)$ basis are given by
\bea
\hat{\Phi}_{\frac{1}{2}}^{(1),\mu}(z) = 
-\frac{i}{(k+N+2)} \, d^{\mu}_{\bar{a} \bar{b}} \, Q^{\bar{a}} \,
V^{\bar{b}} (z), \qquad \mu =0,1,2,3,
\label{3halfhat}
\eea
which corresponds to the higher spin-$\frac{3}{2}$ currents
$G^{\prime \mu}(z)$   
with minus sign in \cite{AK1411}.

As done in (\ref{spin3halfrelation})(the last one has different sign),
one obtains 
the higher spin-$\frac{3}{2}$ currents
which are in the $SO(4)$ vector representation as follows
\cite{AK1411}
\bea
\Phi_{\frac{1}{2}}^{(1),1}(z) & = &  
-\frac{i}{(k+N+2)} \, d^{3}_{\bar{a} \bar{b}} \, Q^{\bar{a}} \,
V^{\bar{b}} (z), \qquad
\Phi_{\frac{1}{2}}^{(1),2}(z) = 
-\frac{i}{(k+N+2)} \, d^{0}_{\bar{a} \bar{b}} \, Q^{\bar{a}} \,
V^{\bar{b}} (z),
\nonu \\
\Phi_{\frac{1}{2}}^{(1),3}(z) & = &  
-\frac{i}{(k+N+2)} \, d^{1}_{\bar{a} \bar{b}} \, Q^{\bar{a}} \,
V^{\bar{b}} (z), \qquad
\Phi_{\frac{1}{2}}^{(1),4}(z) = 
\frac{i}{(k+N+2)} \, d^{2}_{\bar{a} \bar{b}} \, Q^{\bar{a}} \,
V^{\bar{b}} (z).
\label{spin3half}
\eea
The $d^{\mu}_{\bar{a} \bar{b}}$ tensor appearing in these
expressions, which are characteristic of the higher spin-$\frac{3}{2}$
currents, is given by (\ref{dmu}).
In other words, one also has the similar OPE relation
given in (\ref{higherspin3half}), for given the spin-$\frac{3}{2}$
current $G^{\mu}(z)$ and the higher spin-$1$ current $\Phi_0^{(1)}(w)$
in Appendix $A$.

\subsection{Higher spin-$2$ currents}

Let us consider the higher spin-$2$ currents.
It is known that the defining OPEs between
the spin-$\frac{3}{2}$ currents and the higher spin-$\frac{3}{2}$
currents are given by (from Appendix (\ref{1116})) 
{\small
  \bea
\hat{G}^{\mu}(z)\,\hat{\Phi}_{\frac{1}{2}}^{(1),\nu}(w)  = 
-\frac{1}{(z-w)^{2}} 2 \delta^{\mu\nu}\,\Phi_{0}^{(1)}(w)
+  
\frac{1}{(z-w)}
\Bigg[-\delta^{\mu \nu} \partial\Phi_{0}^{(1)}+ \frac{1}{2}
  \varepsilon^{\mu \nu \rho \si}  \hat{\Phi}_{1}^{(1),\rho \si}
  \Bigg](w)+\cdots.
\label{3half3half}
\eea}
Let us analyze each pole in (\ref{3half3half}). In particular,
the first order pole provides the higher spin-$2$ currents.

\subsubsection{The third order pole}

The third order pole of $\hat{G}^{\mu}(z)\,
\hat{\Phi}_{\frac{1}{2}}^{(1),\nu}(w)$
contains
\bea
h^{\mu}_{\bar{a} \bar{b}} \, d^{\nu}_{\bar{c} \bar{d}} \, g^{\bar{c} \bar{a}}
\, g^{\bar{b} \bar{d}}.
\label{hdgg}
\eea
One can check that the $16$ cases of this expression ($\mu, \nu=0,1,2,3$)
vanish
by using (\ref{himatrix}), (\ref{hzero}), (\ref{dzero}), (\ref{dmu})
and (\ref{metric}). This is consistent with the structure of the OPE
in (\ref{3half3half}).

\subsubsection{The second order pole}

The second order pole is given by
\bea
\frac{1}{(k+N+2)} \, h^{\mu}_{\bar{a} \bar{b}} \, g^{\bar{c} \bar{a}} \,
d^{\nu}_{\bar{c} \bar{d}} \, f^{\bar{b} \bar{d}}_{\,\,\,\,\,\, e} \, V^e(w)
-\frac{k}{(k+N+2)^2} \,  h^{\mu}_{\bar{a} \bar{b}} \, g^{\bar{d} \bar{b}} \,
d^{\nu}_{\bar{c} \bar{d}} \, Q^{\bar{c}} \, Q^{\bar{a}}(w).
\label{secondterm}
\eea
One can check the above OPE by considering five cases
\footnote{We will treat the simplification on each pole of  OPE
  for general indices without fixing them explicitly. But it is rather
  difficult to use some identities between the tensors when there are
  too many indices one should simplify. Later, we will fix them and
describe each pole for fixed indices.},
$(\mu,\nu)=(0,0),
(i,i), (0,i), (i,0)$ and $(i,j)$.
One expects that the first two will contribute to the OPE
while the last three will do not contribute at all
due to the delta tensor $\delta^{\mu \nu}$ in the second order pole of
(\ref{3half3half}). 

$\bullet$
$(\mu, \nu)=(0,i)$ case

One obtains
$h^{0}_{\bar{a} \bar{b}} \, g^{\bar{c} \bar{a}} \,
d^{i}_{\bar{c} \bar{d}}= d^{i}_{\bar{b} \bar{d}}$ for the first term of
(\ref{secondterm}) from  the relation (\ref{hzero}).
Note that this is symmetric in the indices.
Then the first term of (\ref{secondterm}) vanishes
because the structure constant is antisymmetric in the indices
of $\bar{b}$ and ${\bar{d}}$.
Similarly, one obtains $h^{0}_{\bar{a} \bar{b}} \, g^{\bar{d} \bar{b}} \,
d^{i}_{\bar{c} \bar{d}}= d^{i}_{\bar{c} \bar{a}}$, which is also symmetric,
for the second term of (\ref{secondterm}).
Then due to the fermionic property of spin-$\frac{1}{2}$ current,
the quantity $Q^{\bar{c}} \, Q^{\bar{a}}(w)$ is also antisymmetric by
interchanging the indices. There is no nontrivial contribution from the
second term. The total contribution is zero.

$\bullet$ $(\mu, \nu)=(i, 0)$ case

One obtains
$-h^{i}_{\bar{b} \bar{a}} \, g^{\bar{a} \bar{c}} \,
d^{0}_{\bar{c} \bar{d}}= -d^{i}_{\bar{b} \bar{d}}$ (for the first term) where
the following identiy is used
\bea
h^{\mu}_{\bar{a} \bar{b}} \, g^{\bar{b} \bar{c}} \, d^0_{\bar{c} \bar{d}} =
d^{\mu}_{\bar{a} \bar{d}}.
\label{hgd-1}
\eea
This can be obtained from (\ref{dmu}).
Similarly, one obtains
$-h^{i}_{\bar{a} \bar{b}} \, g^{\bar{b} \bar{d}} \,
d^{0}_{\bar{d} \bar{c}}= -d^{i}_{\bar{a} \bar{c}}$ (for the second term)
which is symmetric in the indices. 
Therefore, the final contribution from (\ref{secondterm}) vanishes
from the same analysis before.

$\bullet$ $(\mu,\nu)=(i,j)$ case where $i \neq j$

What happens in this case?
One can use the identity which will appear in
(\ref{higdi}) and show that there is no
contribution.

$\bullet$ $(\mu,\nu)=(0,0)$ case

For the first term of (\ref{secondterm}), one has
$h^{0}_{\bar{b} \bar{a}} \, g^{\bar{a} \bar{c}} \,
d^{0}_{\bar{c} \bar{d}}= d^{0}_{\bar{b} \bar{d}}$ by using
the relation (\ref{hgd-1})
and for the second term, one has
$h^{0}_{\bar{a} \bar{b}} \, g^{\bar{d} \bar{b}} \,
d^{0}_{\bar{c} \bar{d}}= d^{0}_{\bar{c} \bar{a}}$
\footnote{For simplicity, we will concentrate  on the main part of
  the coefficient in front of composite field. Sometimes, we do not
  care
  about the signs, the numerical values, $k$ factors or
  $(k+N+2)$ factors in their descriptions.
  We will consider them in the final results.}.
Then one can check the second order pole of (\ref{3half3half})
together with (\ref{spinone})
by combining the factors in (\ref{secondterm}). 

$\bullet$ $(\mu,\nu)=(i,i)$ case

In this case,
one should describe some identity from the relation between
the three almost complex structures.
By multiplying $d^0_{\bar{e} \bar{f}} \, g^{\bar{f} \bar{b}}$ into
the following relation between the three almost complex structures
\bea
h^i_{\bar{a} \bar{c}} \, g^{\bar{c} \bar{d}} \, h^j_{\bar{d} \bar{b}} =
\varepsilon^{ijk} \, h^k_{\bar{a} \bar{b}} -
\delta^{ij} \, h^0_{\bar{a} \bar{b}}, \qquad i,j, \cdots = 1,2,3,
\label{hhrel}
\eea
one obtains the following identity, by using (\ref{dmu}) together with the 
antisymmetric properties of $h^i_{\bar{a} \bar{b}}$ and $d^0_{\bar{a} \bar{b}}$
(symmetric properties of $h^0_{\bar{a} \bar{b}}$ and $d^i_{\bar{a} \bar{b}}$),
\bea
h^i_{\bar{a} \bar{c}} \, g^{\bar{c} \bar{d}} \, d^j_{\bar{d} \bar{e}} =
\varepsilon^{ijk} \, d^k_{\bar{a} \bar{e}} - \delta^{ij} \, d^0_{\bar{a} \bar{e}},
\qquad i,j, \cdots = 1,2,3.
\label{higdi}
\eea
For the $(\mu,\nu)=(i,i)$ case (there is no sum in the index $i$),
one obtains
$-h^{i}_{\bar{b} \bar{a}} \, g^{\bar{a} \bar{c}} \,
d^{i}_{\bar{c} \bar{d}}= d^{0}_{\bar{b} \bar{d}}$ by using (\ref{higdi})
for the first term of (\ref{secondterm}).
Similarly, one has 
$h^i_{\bar{a} \bar{b}} \, g^{\bar{b} \bar{d}} \,
d^{i}_{\bar{d} \bar{c}}= -d^0_{\bar{a} \bar{c}} =d^{0}_{\bar{c} \bar{a}}$
for the second term. Again, the identity (\ref{higdi})
is used.
It is straightforward to see the second order pole of (\ref{3half3half})
for this case also \footnote{One can check the relation (\ref{hdgg})
  using the identity (\ref{higdi}). Note that $d^i$ tensor is traceless.}.

\subsubsection{The first order pole}

Let us consider the first order pole.
The explicit form is as follows
\bea
-\frac{1}{(k+N+2)^2} \, h^{\mu}_{\bar{a} \bar{b}} \, d^{\nu}_{\bar{c} \bar{d}}
\Bigg[
 (k+N+2) \, g^{\bar{c} \bar{a}} \, V^{\bar{b}} \, V^{\bar{d}}
  +k \, g^{\bar{d} \bar{b}} \, Q^{\bar{c}} \, \pa Q^{\bar{a}} +
  f^{\bar{d} \bar{b}}_{\,\,\,\,\,\, e} \, Q^{\bar{c}} \, Q^{\bar{a}} \, V^{e}
  \Bigg](w).
\label{poleone}
\eea

$\bullet$ $\mu=\nu$ case

When one differentiates the second term of (\ref{secondterm})
with respect to the coordinate $w$, then
one obtains the second term of (\ref{poleone}) with
an extra coefficient $2$.
Similarly, one obtains the first term of (\ref{poleone})
with an extra coefficient $2$ after one differentiates
the first term of (\ref{secondterm})
with respect to the coordinate $w$.
This is because one can reexpress $f^{\bar{b} \bar{d}}_{\,\,\,\,\,\, e} \, \pa
V^e$ in terms of 
$ -[V^{\bar{b}}, V^{\bar{d}}]$ and the
factor $h^{\mu}_{\bar{a} \bar{b}} \, g^{\bar{c} \bar{a}} \,
d^{\nu}_{\bar{c} \bar{d}}$ under the condition $(\mu,\nu)=(0,0)$ or
$(i,i)$ is antisymmetric in the indices $\bar{b}$ and $\bar{d}$.
In other words, the expression $V^{\bar{b}} \, V^{\bar{d}}$
can be written in terms of $\frac{1}{2} [V^{\bar{b}}, V^{\bar{d}}]$
under the particular choice of $(\mu,\nu)$.
Furthermore,
one has the following property \cite{AK1411}
\bea
h^{\mu}_{\bar{a} \bar{b}} \, d^{\mu}_{\bar{c} \bar{d}} \, 
f^{\bar{d} \bar{b}}_{\,\,\,\,\,\, e} =
h^{\mu}_{\bar{c} \bar{b}} \, d^{\mu}_{\bar{a} \bar{d}} \, 
f^{\bar{d} \bar{b}}_{\,\,\,\,\,\, e}, \qquad
\mbox{no} \,\, \mbox{sum} \,\, \mbox{over} \,\, \mu=0,1,2,3,
\label{symmetric}
\eea
which leads to the fact that the $h d f$ factor is 
symmetric in the indices $\bar{a}$ and $\bar{c}$.
This implies that the third term of (\ref{poleone})
vanishes under the condition $\mu=\nu$. Therefore,
one can prove the Kronecker delta term in the
first order pole of (\ref{3half3half}). 

$\bullet$ $\mu \neq \nu$ case

One expects that
from (\ref{3half3half}), the $(\mu,\nu)=(0,1)$ case (which is given by
(\ref{poleone}) explicitly) provides
the higher spin current with $(2,3)$ component.
Similarly, one obtains the $(3,0)$ component higher spin current
for $(\mu,\nu)=(1,2)$. The remaining $(0,1)$, $(1,2)$, $(3,1)$ and $(0,2)$
components of higher spin currents can be obtained from
the cases $(\mu,\nu)=(2,3)$, $(3,0)$, $(0,2)$ and $(3,1)$
respectively \footnote{Note that $\varepsilon^{0123}=1$.}.

It turns out that the higher spin-$2$ currents in $SU(2) \times SU(2)$
basis can be written as 
\bea
\hat{\Phi}_{1}^{(1),\rho \si}(z)  & = &  -
\frac{1}{2(k+N+2)^2}  \, \varepsilon^{\rho \si \mu \nu} \,
h^{\mu}_{\bar{c} \bar{d}} \, d^{\nu}_{\bar{e} \bar{f}} \,  
\nonu \\
& \times &
\Bigg[ (k+N+2) \, g^{\bar{e} \bar{c}} \, V^{\bar{d}} \, V^{\bar{f}}
  +k \, g^{\bar{f} \bar{d}} \, Q^{\bar{e}} \, \pa Q^{\bar{c}} +
  f^{\bar{f} \bar{d}}_{\,\,\,\,\,\, g} \, Q^{\bar{e}} \, Q^{\bar{c}} \, V^{g}
  \Bigg](z).
\label{spin2higher}
\eea
One should make sure that
the expression in (\ref{poleone}) is antisymmetric
in the indices of $\mu$ and $\nu$.
For the $(\mu,\nu)=(0,i)$ case,
the first term of (\ref{poleone})
contains $ h^{0}_{\bar{a} \bar{b}} \, d^{i}_{\bar{c} \bar{d}}
 \, g^{\bar{c} \bar{a}}$ which becomes $ d^{i}_{\bar{b} \bar{d}}$.
On the other hand,
for the $(\mu,\nu)=(i,0)$ case,
one has $ h^{i}_{\bar{a} \bar{b}} \, d^{0}_{\bar{c} \bar{d}}
\, g^{\bar{c} \bar{a}}$ which is equal to
$ -h^{i}_{\bar{b} \bar{a}} \,  g^{\bar{a} \bar{c}}\,
d^{0}_{\bar{c} \bar{d}}$. Then this becomes $-d^{i}_{\bar{b} \bar{d}}$
from the defining relation (\ref{hgd-1}).
The second term of (\ref{poleone})
contains $ h^{0}_{\bar{a} \bar{b}} \, d^{i}_{\bar{c} \bar{d}}
\, g^{\bar{d} \bar{b}}$ which is given by $d^{i}_{\bar{c} \bar{a}}$
for $(\mu,\nu)=(0,i)$.
For the $(\mu,\nu)=(i,0)$ case,
one has $ h^{i}_{\bar{a} \bar{b}} \, d^{0}_{\bar{c} \bar{d}}
\, g^{\bar{d} \bar{b}}$ which can be written as
$ -h^{i}_{\bar{a} \bar{b}} \, g^{\bar{b} \bar{d}} \, d^{0}_{\bar{d} \bar{c}}$.
Then this is equal to $-d^{i}_{\bar{c} \bar{a}}$.
For $(\mu,\nu)=(i,j)$ case where $i \neq j$, the expressions
 $ h^{i}_{\bar{a} \bar{b}} \, d^{j}_{\bar{c} \bar{d}}
 \, g^{\bar{c} \bar{a}}$ and 
$ h^{i}_{\bar{a} \bar{b}} \, d^{j}_{\bar{c} \bar{d}}
 \, g^{\bar{d} \bar{b}}$ corresponding to the first and second terms
 of (\ref{poleone}) are antisymmetric in the indices $i$ and $j$
 from (\ref{higdi}).
 For the third term of (\ref{poleone}), one can show that
 the following identity holds
 \bea
 (h^{\mu}_{\bar{a} \bar{b}} \, d^{\nu}_{\bar{c} \bar{d}} +
h^{\nu}_{\bar{a} \bar{b}} \, d^{\mu}_{\bar{c} \bar{d}})
\,
  f^{\bar{d} \bar{b}}_{\,\,\,\,\,\, e} \, Q^{\bar{c}} \, Q^{\bar{a}} \, V^{e}
=0,
  \label{todayidentity}
 \eea
 which can be checked for several $N$ values. In (\ref{todayidentity}),
 the composite field is necessary.
Therefore, the higher spin-$2$ current in (\ref{spin2higher})
except epsilon tensor
is antisymmetric under the indices $\mu$ and $\nu$.
For fixed $\rho$ and $\si$ indices, the numerical factor
$\frac{1}{2}$ is cancelled.
 
 By identifying the correct indices in $SO(4)$ manifest way,
 one obtains the following six higher spin-$2$ currents
 which transform as $SO(4)$ adjoint representation, from (\ref{spin2higher}),
 {\small
   \bea
\Phi_{1}^{(1),12}(z)  & = &  
\frac{h^{1}_{\bar{c} \bar{d}} \, d^{2}_{\bar{e} \bar{f}}}{(k+N+2)^2}  \, 
\Bigg[ (k+N+2) \, g^{\bar{e} \bar{c}} \, V^{\bar{d}} \, V^{\bar{f}}
  +k \, g^{\bar{f} \bar{d}} \, Q^{\bar{e}} \, \pa Q^{\bar{c}} +
  f^{\bar{f} \bar{d}}_{\,\,\,\,\,\, g} \, Q^{\bar{e}} \, Q^{\bar{c}} \, V^{g}
  \Bigg](z),
\nonu \\
\Phi_{1}^{(1),13}(z)  & = &  
\frac{h^{2}_{\bar{c} \bar{d}} \, d^{0}_{\bar{e} \bar{f}}}{(k+N+2)^2}  \,   
\Bigg[ (k+N+2) \, g^{\bar{e} \bar{c}} \, V^{\bar{d}} \, V^{\bar{f}}
  +k \, g^{\bar{f} \bar{d}} \, Q^{\bar{e}} \, \pa Q^{\bar{c}} +
  f^{\bar{f} \bar{d}}_{\,\,\,\,\,\, g} \, Q^{\bar{e}} \, Q^{\bar{c}} \, V^{g}
  \Bigg](z),
\nonu \\
\Phi_{1}^{(1),14}(z)  & = &  
\frac{h^{1}_{\bar{c} \bar{d}} \, d^{0}_{\bar{e} \bar{f}}}{(k+N+2)^2}  \,  
\Bigg[ (k+N+2) \, g^{\bar{e} \bar{c}} \, V^{\bar{d}} \, V^{\bar{f}}
  +k \, g^{\bar{f} \bar{d}} \, Q^{\bar{e}} \, \pa Q^{\bar{c}} +
  f^{\bar{f} \bar{d}}_{\,\,\,\,\,\, g} \, Q^{\bar{e}} \, Q^{\bar{c}} \, V^{g}
  \Bigg](z),
\nonu \\
\Phi_{1}^{(1),23}(z)  & = &  
-\frac{h^{2}_{\bar{c} \bar{d}} \, d^{3}_{\bar{e} \bar{f}}}{(k+N+2)^2}  \,  
\Bigg[ (k+N+2) \, g^{\bar{e} \bar{c}} \, V^{\bar{d}} \, V^{\bar{f}}
  +k \, g^{\bar{f} \bar{d}} \, Q^{\bar{e}} \, \pa Q^{\bar{c}} +
  f^{\bar{f} \bar{d}}_{\,\,\,\,\,\, g} \, Q^{\bar{e}} \, Q^{\bar{c}} \, V^{g}
  \Bigg](z),
\nonu \\
\Phi_{1}^{(1),24}(z)  & = &  
-\frac{h^{1}_{\bar{c} \bar{d}} \, d^{3}_{\bar{e} \bar{f}}}{(k+N+2)^2}  \,   
\Bigg[ (k+N+2) \, g^{\bar{e} \bar{c}} \, V^{\bar{d}} \, V^{\bar{f}}
  +k \, g^{\bar{f} \bar{d}} \, Q^{\bar{e}} \, \pa Q^{\bar{c}} +
  f^{\bar{f} \bar{d}}_{\,\,\,\,\,\, g} \, Q^{\bar{e}} \, Q^{\bar{c}} \, V^{g}
  \Bigg](z),
\nonu \\
\Phi_{1}^{(1),34}(z)  & = &  
\frac{-h^{3}_{\bar{c} \bar{d}} \, d^{0}_{\bar{e} \bar{f}}}{(k+N+2)^2}  \, 
\Bigg[ (k+N+2) \, g^{\bar{e} \bar{c}} \, V^{\bar{d}} \, V^{\bar{f}}
  +k \, g^{\bar{f} \bar{d}} \, Q^{\bar{e}} \, \pa Q^{\bar{c}} +
  f^{\bar{f} \bar{d}}_{\,\,\,\,\,\, g} \, Q^{\bar{e}} \, Q^{\bar{c}} \, V^{g}
  \Bigg](z).
\label{spin2final}
\eea}
 Compared to the ones in \cite{AK1411},
 these are very simple forms in the sense
 that the $SO(4)$ adjoint indices are coming from the indices of almost
 complex structures and $d$ tensors.
They have the same WZW current dependent terms inside the bracket.
One can further simplify
the first two terms in (\ref{spin2final}) using (\ref{higdi}).

\subsection{Higher spin-$\frac{5}{2}$ currents}

Let us consider the OPE
between the spin-$\frac{3}{2}$ currents $\hat{G}^{\mu}(z)$
(\ref{11currents})
and the higher spin-$2$ currents $\hat{\Phi}_1^{(1),\nu \rho}(w)$
(\ref{spin2higher}).
The defining OPE is given by
{\footnotesize
  \bea
\hat{G}^{\mu}(z) \, \hat{\Phi}_1^{(1),\nu\rho}(w)  & = &
\frac{1}{(z-w)^2} \,  \Bigg[
  \frac{(N-k)}{(k+N+2)} \,(
  \delta^{\mu \nu} \, \hat{\Phi}_{\frac{1}{2}}^{(1),\rho} -
\delta^{\mu \rho} \, \hat{\Phi}_{\frac{1}{2}}^{(1),\nu})
  + \frac{(2+3k+3N)}{(k+N+2)} \, \varepsilon^{\mu \nu \rho \si} \,
  \hat{\Phi}_{\frac{1}{2}}^{(1),\si}
\Bigg](w)
  \nonu \\
  & + & \frac{1}{(z-w)} \, \Bigg[
  \frac{i}{(k+N+2)} \, (\varepsilon^{\si \alpha \mu \rho} \,
  T^{\si \alpha} \, \hat{\Phi}_{\frac{1}{2}}^{(1),\nu} -
\varepsilon^{\si \alpha \mu \nu} \,
  T^{\si \alpha} \, \hat{\Phi}_{\frac{1}{2}}^{(1),\rho}
  ) + \varepsilon^{\mu \nu \rho \si } \, \pa \,
  \hat{\Phi}_{\frac{1}{2}}^{(1),\si} 
  \nonu \\
  & - & \delta^{\mu \nu} \, ( \hat{\Phi}_{\frac{3}{2}}^{(1),\rho}
  + \frac{(k-N)}{3(k+N+2)} \, \pa \, \hat{\Phi}_{\frac{1}{2}}^{(1),\rho}
  + \frac{i}{(k+N+2)} \,  \varepsilon^{\alpha \beta \rho \si} \,
  T^{\alpha \beta} \, \hat{\Phi}_{\frac{1}{2}}^{(1),\si})
 \label{3halfandtwo}
   \\
  & + &
\delta^{\mu \rho} \, ( \hat{\Phi}_{\frac{3}{2}}^{(1),\nu}
  + \frac{(k-N)}{3(k+N+2)} \, \pa \, \hat{\Phi}_{\frac{1}{2}}^{(1),\nu}
  + \frac{i}{(k+N+2)} \,  \varepsilon^{\alpha \beta \nu \si} \,
  T^{\alpha \beta} \, \hat{\Phi}_{\frac{1}{2}}^{(1),\si}) 
  \Bigg](w)  +  \cdots.
 \nonu
\eea}
One can see the antisymmetric property between the indices $\nu$
and $\rho$ from the right hand side of this OPE.
It is obvious to see those property in the left hand side.

Let us check whether the realization of the higher spin currents
in terms of Wolf space coset currents satisfies each singular term in
(\ref{3halfandtwo}).

\subsubsection{The third order pole}

The third order pole of this OPE
contains $ f^{\bar{d} \bar{b}}_{\,\,\,\,\,\, e} \, g^{\bar{f} e}$ or
$f^{\bar{f} \bar{d}}_{\,\,\,\,\,\, g} \, g^{\bar{b} g}$ which vanishes
according to (\ref{fgrelation}).

\subsubsection{The second order pole}

Let us calculate the second order pole of the above OPE.
The OPE between $Q^{\bar{a}} \, V^{\bar{b}}(z)$
and $V^{\bar{d}} \, V^{\bar{f}}(w)$ provides the following
second order pole
\bea
(k \, g^{\bar{d} \bar{b}} \, Q^{\bar{a}} \, V^{\bar{f}} -
f^{\bar{d} \bar{b}}_{\,\,\,\,\,\, e} \, f^{\bar{f} e}_{\,\,\,\,\,\, g} \,
Q^{\bar{a}} \, V^{g} +
k \, g^{\bar{f} \bar{b}} \, Q^{\bar{a}} \, V^{\bar{d}})(w).
\label{firstfirst}
\eea
Similarly,
the OPE between $Q^{\bar{a}} \, V^{\bar{b}}(z)$
and $ Q^{\bar{e}} \, \pa Q^{\bar{c}}(w)$ leads to the following
second order pole
\bea
 (k+N+2) \,  g^{\bar{c} \bar{a}} \, Q^{\bar{e}} \, V^{\bar{b}}(w).
\label{secondsecond}
\eea
Finally,
the OPE between $Q^{\bar{a}} \, V^{\bar{b}}(z)$
and $ Q^{\bar{e}} \, Q^{\bar{c}} \, V^{g}(w)$ implies the following
second order pole
\bea
(k+N+2) ( g^{\bar{e} \bar{a}} \,  f^{\bar{b} g}_{\,\,\,\,\,\, h} \, Q^{\bar{c}} \,
V^{h} - g^{\bar{c} \bar{a}} \,  f^{\bar{b} g}_{\,\,\,\,\,\, h} \, Q^{\bar{e}} \,
V^{h})(w).
\label{thirdthird}
\eea
Although there exists the cubic term of spin-$\frac{1}{2}$ current,
the factor $f^{\bar{f} \bar{d}}_{\,\,\,\,\,\, g}$ with $g^{g\bar{b}}$ does not
contribute to the final second order pole.
See also Appendix (\ref{firstfirst-1}).

Then one obtains
the final second order pole after multiplying all the factors correctly
\bea
\frac{-i}{2(k+N+2)^3} \, \varepsilon^{\nu \rho \si \alpha} \,
h^{\mu}_{\bar{a} \bar{b}} \, h^{\si}_{\bar{c} \bar{d}} \,
d^{\alpha}_{\bar{e} \bar{f}}
\Bigg[
(k+N+2) \, g^{\bar{e} \bar{c}} \,  (\ref{firstfirst}) 
+k \, g^{\bar{f} \bar{d}} \,  (\ref{secondsecond})
+f^{\bar{f} \bar{d}}_{\,\,\,\,\,\, g} \,  (\ref{thirdthird})
\Bigg](w).
\label{Pole2}
\eea
As before, one can analyze this expression for five different cases
in order to see (\ref{3halfandtwo}).

$\bullet$ $(\mu,\nu)=(0,0)$

Let us consider the case where the indices $\mu$ and $\nu$
are equal to each other.
The first term of (\ref{Pole2}) contains
$ \varepsilon^{\nu \rho \si \alpha} \,
h^{\nu}_{\bar{a} \bar{b}} \, h^{\si}_{\bar{c} \bar{d}} \,
d^{\alpha}_{\bar{e} \bar{f}} \, g^{\bar{e} \bar{c}} \, g^{\bar{d} \bar{b}}$. 
There is no sum over the index $\nu$.
One can express this as $ \pm \varepsilon^{\nu \rho \si \alpha} \,
h^{\nu}_{\bar{a} \bar{b}} \, g^{\bar{b} \bar{d}}
\,  h^{\si}_{\bar{d} \bar{c}} \,  g^{\bar{c} \bar{e}} \, d^{\alpha}_{\bar{e} \bar{f}}$.
For the index $\si$ is $0$ (when the index $\nu$ is $i$),
then one has $+$ sign while
for the index $\si$ is equal to $i$ (when the index $\nu$ is $0$),
then one has
$-1$ sign.
When $\mu=\nu=0$, the factor $
 h^{\si}_{\bar{d} \bar{c}} \,  g^{\bar{c} \bar{e}} \, d^{\alpha}_{\bar{e} \bar{f}}$
 can be reduced to $\varepsilon^{\si \alpha k} \, d^k_{\bar{d} \bar{f}}$
 according to (\ref{higdi}). Then one obtains
 $ - \varepsilon^{\nu \rho \si \alpha} \, \varepsilon^{\si \alpha k} \,
 d^k_{\bar{a} \bar{f}}$ which is equal to $-2 d^{\rho}_{\bar{a} \bar{f}}$.
The 
third term of (\ref{Pole2}) contains
$ \varepsilon^{\nu \rho \si \alpha} \,
h^{\nu}_{\bar{a} \bar{b}} \, h^{\si}_{\bar{c} \bar{d}} \,
d^{\alpha}_{\bar{e} \bar{f}} \, g^{\bar{e} \bar{c}} \, g^{\bar{f} \bar{b}}$
which can be written as $\varepsilon^{\nu \rho \si \alpha} \,
h^{\nu}_{\bar{a} \bar{b}} \,  g^{\bar{b} \bar{f}} \, d^{\alpha}_{\bar{f} \bar{e}}
\, g^{\bar{e} \bar{c}} \, h^{\si}_{\bar{c} \bar{d}}$ with $(\mu,\nu)=(0,0)$.
As before, the factor
$h^{\nu}_{\bar{a} \bar{b}} \,  g^{\bar{b} \bar{f}} \, d^{\alpha}_{\bar{f} \bar{e}}$
becomes 
$ d^{\alpha}_{\bar{a} \bar{e}}$.
Furthermore, the expression $ d^{\alpha}_{\bar{a} \bar{e}} \,
g^{\bar{e} \bar{c}} \, h^{\si}_{\bar{c} \bar{d}}$ gives
$-\varepsilon^{\si  \alpha k} \, d^k_{\bar{a} \bar{d}}$.
Then, one obtains the final expression
$-\varepsilon^{\nu \rho \si \alpha}  \,
\varepsilon^{\si  \alpha k} \, d^k_{\bar{a} \bar{d}}$.
Again, this becomes $-2  d^{\rho}_{\bar{a} \bar{d}}$.
The 
fourth term of (\ref{Pole2}) contains
$ \varepsilon^{\nu \rho \si \alpha} \,
h^{\nu}_{\bar{a} \bar{b}} \, h^{\si}_{\bar{c} \bar{d}} \,
d^{\alpha}_{\bar{e} \bar{f}} \, g^{\bar{f} \bar{d}} \, g^{\bar{c} \bar{a}}$
which can be written as $\varepsilon^{\nu \rho \si \alpha} \,
 h^{\si}_{\bar{b} \bar{d}} \,  g^{\bar{d} \bar{f}} \,d^{\alpha}_{\bar{f} \bar{e}}$.
 This is equivalent to $\varepsilon^{\nu \rho \si \alpha} \,
 \varepsilon^{\si  \alpha k} \, d^k_{\bar{b} \bar{e}}$
 which leads to  $2  d^{\rho}_{\bar{b} \bar{e}}$.

When the second term of
 (\ref{Pole2}) is considered, one has
 $\varepsilon^{\nu \rho \si \alpha} \,
h^{\nu}_{\bar{a} \bar{b}} \, h^{\si}_{\bar{c} \bar{d}} \,
d^{\alpha}_{\bar{e} \bar{f}} \, g^{\bar{e} \bar{c}} \,
f^{\bar{d} \bar{b}}_{\,\,\,\,\,\, e} \, f^{\bar{f} e}_{\,\,\,\,\,\, g}$.
The factor $ h^{\si}_{\bar{c} \bar{d}} \,
d^{\alpha}_{\bar{e} \bar{f}} \, g^{\bar{e} \bar{c}}$,
which is $ -h^{\si}_{\bar{d} \bar{c}} \,  g^{\bar{c} \bar{e}} \,
d^{\alpha}_{\bar{e} \bar{f}}$,
can be simplified as $-\varepsilon^{\si  \alpha k} \, d^k_{\bar{d} \bar{f}}$.
Then one obtains $-\varepsilon^{\nu \rho \si \alpha} \,
\varepsilon^{\si  \alpha k} \,
h^{\nu}_{\bar{a} \bar{b}} \,
d^k_{\bar{d} \bar{f}} \, f^{\bar{d} \bar{b}}_{\,\,\,\,\,\, e} \,
f^{\bar{f} e}_{\,\,\,\,\,\, g}$.
Here,
one has the following identity
\bea
d^k_{\bar{f} \bar{d}} \, f^{\bar{d} \bar{b}}_{\,\,\,\,\,\, e} \,
f^{\bar{f} e}_{\,\,\,\,\,\, g} = 
N \,  d^k_{\bar{c} \bar{g}} \, g^{\bar{c} \bar{b}}. 
\label{dffrelation}
\eea
Then the final contribution is given by 
$-N \, \varepsilon^{\nu \rho \si \alpha} \,
\varepsilon^{\si  \alpha k} \,  d^k_{\bar{a} \bar{g}} = 2 N \,
d^{\rho}_{\bar{a} \bar{g}}$ \footnote{
Let us turn to the fifth term in (\ref{Pole2}).
One has $\varepsilon^{\nu \rho \si \alpha} \,
h^{\nu}_{\bar{a} \bar{b}} \, h^{\si}_{\bar{c} \bar{d}} \,
d^{\alpha}_{\bar{e} \bar{f}} \, g^{\bar{e} \bar{a}} \,
f^{\bar{f} \bar{d}}_{\,\,\,\,\,\, g} \,
f^{\bar{b} g}_{\,\,\,\,\,\, h}$. Then one sees that the factor
$h^{\nu}_{\bar{a} \bar{b}}  \,
d^{\alpha}_{\bar{e} \bar{f}} \, g^{\bar{e} \bar{a}}$ can be written as
$d^{\alpha}_{\bar{b} \bar{f}}$.
By using the identity (\ref{dffrelation}), one has
$N \, \varepsilon^{\nu \rho \si \alpha} \,  h^{\si}_{\bar{c} \bar{d}} \,
d^{\alpha}_{\bar{e} \bar{h}} \,  g^{\bar{d} \bar{e}}=
N \, \varepsilon^{\nu \rho \si \alpha} \, \varepsilon^{\si \alpha k }  \,
d^{k}_{\bar{c} \bar{h}} = 2N \, d^{\rho}_{\bar{c} \bar{h}}$.
One can consider the sixth term of (\ref{Pole2}).
In this case, one has
$\varepsilon^{\nu \rho \si \alpha} \,
h^{\nu}_{\bar{a} \bar{b}} \, h^{\si}_{\bar{c} \bar{d}} \,
d^{\alpha}_{\bar{e} \bar{f}} \, g^{\bar{c} \bar{a}} \,
f^{\bar{f} \bar{d}}_{\,\,\,\,\,\, g} \,
f^{\bar{b} g}_{\,\,\,\,\,\, h}$.
This can be simplified as
$\varepsilon^{\nu \rho \si \alpha} \,
 h^{\si}_{\bar{b} \bar{d}} \,
d^{\alpha}_{\bar{e} \bar{f}}  \,
f^{\bar{f} \bar{d}}_{\,\,\,\,\,\, g} \,
f^{\bar{b} g}_{\,\,\,\,\,\, h}$.
By using the following identity
$
 h^{\si}_{\bar{b} \bar{d}} \,
d^{\alpha}_{\bar{e} \bar{f}}  \,
f^{\bar{f} \bar{d}}_{\,\,\,\,\,\, g} \,
f^{\bar{b} g}_{\,\,\,\,\,\, h} = - N \, \varepsilon^{\si \alpha k} \, 
d^{k}_{\bar{e} \bar{h}}$,
one has
$-N\, \varepsilon^{\nu \rho \si \alpha} \, \varepsilon^{\si \alpha k} \, 
d^{k}_{\bar{e} \bar{h}} = 2N\, d^{\rho}_{\bar{e} \bar{h}}$.}.
By collecting the correct signs, one obtains
that the final contribution is
given by
\bea
-\frac{(k-N)}{(k+N+2)} \, \frac{-i}{(k+N+2)} \, d^{\rho}_{\bar{a} \bar{b}}
\, Q^{\bar{a}} \, V^{\bar{b}}(w) = -\frac{(k-N)}{(k+N+2)} \,
\delta^{\mu \nu} \,
\hat{\Phi}_{\frac{1}{2}}^{(1),\rho}(w),
\label{stage}
\eea
where at the final stage of (\ref{stage}), the previous expression
for the higher spin-$\frac{3}{2}$ current (\ref{3halfhat}) is inserted.
Therefore, one observes that this is coincident with the
second order pole of (\ref{3halfandtwo}).

$\bullet$ $(\mu,\nu)=(i,i)$

So far, the $(\mu,\nu)=(0,0)$ case is considered.
Let us move on the $(\mu,\nu)=(i,i)$ case.
One can see, for example,
Appendices (\ref{112}), (\ref{113}), (\ref{114}) or
(\ref{334}). One should have the corresponding expressions  
although the detailed calculations are not presented in this paper.

$\bullet$ $(\mu,\nu)=(0,i)$

Let us describe the case $\mu \neq \nu$.
In particular, for $(\mu,\nu)=(0,i)$ case,
The first term of (\ref{Pole2}) contains
$ \varepsilon^{\nu \rho \si \alpha} \,
h^{\mu}_{\bar{a} \bar{b}} \, h^{\si}_{\bar{c} \bar{d}} \,
d^{\alpha}_{\bar{e} \bar{f}} \, g^{\bar{e} \bar{c}} \, g^{\bar{d} \bar{b}}$. 
One can further simplify this as $\varepsilon^{\nu \rho \si \alpha} \,
 h^{\si}_{\bar{c} \bar{a}} \,
d^{\alpha}_{\bar{e} \bar{f}} \, g^{\bar{e} \bar{c}}$.
When the index $\si=0$, then this can be written as
$\varepsilon^{i \rho 0 \alpha} \,
d^{\alpha}_{\bar{a} \bar{f}}$. When  the index $\si=j$,
one has $-\varepsilon^{i \rho j 0} \,
d^{j}_{\bar{a} \bar{f}}$.
Then one obtains $2 \varepsilon^{0 i \rho  \alpha} \,
d^{\alpha}_{\bar{a} \bar{f}}(=2 \varepsilon^{i \rho \alpha} \,
d^{\alpha}_{\bar{a} \bar{f}})$.
Similarly, 
the 
third term of (\ref{Pole2}) is given by
$ \varepsilon^{\nu \rho \si \alpha} \,
h^{\mu}_{\bar{a} \bar{b}} \, h^{\si}_{\bar{c} \bar{d}} \,
d^{\alpha}_{\bar{e} \bar{f}} \, g^{\bar{e} \bar{c}} \, g^{\bar{f} \bar{b}}$.
This becomes
$ \varepsilon^{\nu \rho \si \alpha} \,
 h^{\si}_{\bar{c} \bar{d}} \,
d^{\alpha}_{\bar{e} \bar{a}} \, g^{\bar{e} \bar{c}}$.
One sees that this can be written in terms of
$2 \varepsilon^{0 i \rho \alpha} \,
d^{\alpha}_{\bar{d} \bar{a}}$ by realizing the role of indicies $\bar{d}$
and $\bar{a}$ (corresponding to $\bar{a}$ and $\bar{f}$ in the previous
case).
The 
fourth term of (\ref{Pole2}) is given by
$ \varepsilon^{\nu \rho \si \alpha} \,
h^{\mu}_{\bar{a} \bar{b}} \, h^{\si}_{\bar{c} \bar{d}} \,
d^{\alpha}_{\bar{e} \bar{f}} \, g^{\bar{f} \bar{d}} \, g^{\bar{c} \bar{a}}$.
This is equivalent to $\varepsilon^{\nu \rho \si \alpha} \,
 h^{\si}_{\bar{b} \bar{d}} \,
d^{\alpha}_{\bar{e} \bar{f}} \, g^{\bar{f} \bar{d}}$.
By interchanging the indices between $\bar{b}$ and ${\bar{d}}$
or between $\bar{e}$ and ${\bar{f}}$, there is no sign change
because the nonzero contributions arise when the index $\si=0$ and
the index $\alpha=j$
or the index $\si=j$ and the index $\alpha=0$.
Then one obtains that the contribution from the fourth term 
is $2 \varepsilon^{0 i \rho \alpha} \,
d^{\alpha}_{\bar{b} \bar{f}}$.

From the second term of (\ref{Pole2})
 $\varepsilon^{\nu \rho \si \alpha} \,
h^{\mu}_{\bar{a} \bar{b}} \, h^{\si}_{\bar{c} \bar{d}} \,
d^{\alpha}_{\bar{e} \bar{f}} \, g^{\bar{e} \bar{c}} \,
f^{\bar{d} \bar{b}}_{\,\,\,\,\,\, e} \, f^{\bar{f} e}_{\,\,\,\,\,\, g}$,
one can simplify this as follows.
There are $\varepsilon^{i j k 0} \,
h^{0}_{\bar{a} \bar{b}} \, h^{k}_{\bar{c} \bar{d}} \,
d^{0}_{\bar{e} \bar{f}} \, g^{\bar{e} \bar{c}} \,
f^{\bar{d} \bar{b}}_{\,\,\,\,\,\, e} \, f^{\bar{f} e}_{\,\,\,\,\,\, g}$
and $\varepsilon^{i j 0 k} \,
h^{0}_{\bar{a} \bar{b}} \, h^{0}_{\bar{c} \bar{d}} \,
d^{k}_{\bar{e} \bar{f}} \, g^{\bar{e} \bar{c}} \,
f^{\bar{d} \bar{b}}_{\,\,\,\,\,\, e} \, f^{\bar{f} e}_{\,\,\,\,\,\, g}$
for $\rho=j$ case.
For the former, one can use the definition of (\ref{dmu}).
One obtains $-\varepsilon^{ 0 i j k } \,
h^{0}_{\bar{a} \bar{b}} \, d^{k}_{\bar{d} \bar{f}} \, 
f^{\bar{d} \bar{b}}_{\,\,\,\,\,\, e} \, f^{\bar{f} e}_{\,\,\,\,\,\, g}$.
Now one can use the identity (\ref{dffrelation}) for the $d f f$ factor 
and obtains $-N \, \varepsilon^{ 0 i j k } \, d^{k}_{\bar{a} \bar{g}}$.
For the latter, one can use the Kronecker delta between the metric
tensor and one has the same contribution.
Therefore, the final contribution is given by 
$-2 N \, \varepsilon^{ 0 i j k } \, d^{k}_{\bar{a} \bar{g}}$.
One has $\varepsilon^{\nu \rho \si \alpha} \,
h^{\mu}_{\bar{a} \bar{b}} \, h^{\si}_{\bar{c} \bar{d}} \,
d^{\alpha}_{\bar{e} \bar{f}} \, g^{\bar{e} \bar{a}} \,
f^{\bar{f} \bar{d}}_{\,\,\,\,\,\, g} \,
f^{\bar{b} g}_{\,\,\,\,\,\, h}$ for the fifth term of (\ref{Pole2}).
Then one has
$\varepsilon^{i j k 0} \,
h^{0}_{\bar{a} \bar{b}} \, h^{k}_{\bar{c} \bar{d}} \,
d^{0}_{\bar{e} \bar{f}} \, g^{\bar{e} \bar{a}} \,
f^{\bar{f} \bar{d}}_{\,\,\,\,\,\, g} \,
f^{\bar{b} g}_{\,\,\,\,\,\, h}$ plus
$\varepsilon^{i j 0 k} \,
h^{0}_{\bar{a} \bar{b}} \, h^{0}_{\bar{c} \bar{d}} \,
d^{k}_{\bar{e} \bar{f}} \, g^{\bar{e} \bar{a}} \,
f^{\bar{f} \bar{d}}_{\,\,\,\,\,\, g} \,
f^{\bar{b} g}_{\,\,\,\,\,\, h}$.
The first one can be written as
$-\varepsilon^{0 i j k} \,
 h^{k}_{\bar{c} \bar{d}} \,
d^{0}_{\bar{b} \bar{f}} \,
f^{\bar{f} \bar{d}}_{\,\,\,\,\,\, g} \,
f^{\bar{b} g}_{\,\,\,\,\,\, h}$.
The $d f f$ term can be simplified 
as $-(N+2)\,  d^0_{\bar{i} \bar{h}} \, g^{\bar{i} \bar{d}}$
by using  the following identity
\bea
d^0_{\bar{f} \bar{d}} \, f^{\bar{d} \bar{b}}_{\,\,\,\,\,\, e} \,
f^{\bar{f} e}_{\,\,\,\,\,\, g} = 
-(N+2) \,  d^0_{\bar{c} \bar{g}} \, g^{\bar{c} \bar{b}}. 
\label{dffrelation1}
\eea 
Then this (\ref{dffrelation1})
leads to $(N+2) \, \varepsilon^{0 i j k} \,
d^k_{\bar{c} \bar{h}}$ with the help of (\ref{dmu}).
Combining with the second contribution
$N \, \varepsilon^{0 i j k} \,
d^k_{\bar{c} \bar{h}}$, one has
the final contribution $(2N+2) \, \varepsilon^{0 i j k} \,
d^k_{\bar{c} \bar{h}}$ \footnote{
For the sixth term of (\ref{Pole2}), one has
$\varepsilon^{\nu \rho \si \alpha} \,
h^{\nu}_{\bar{a} \bar{b}} \, h^{\si}_{\bar{c} \bar{d}} \,
d^{\alpha}_{\bar{e} \bar{f}} \, g^{\bar{c} \bar{a}} \,
f^{\bar{f} \bar{d}}_{\,\,\,\,\,\, g} \,
f^{\bar{b} g}_{\,\,\,\,\,\, h}$.
This is equivalent to the sum of
$\varepsilon^{i j k 0} \,
 h^{k}_{\bar{b} \bar{d}} \,
d^{0}_{\bar{e} \bar{f}}  \,
f^{\bar{f} \bar{d}}_{\,\,\,\,\,\, g} \,
f^{\bar{b} g}_{\,\,\,\,\,\, h}$ and $\varepsilon^{i j 0 k} \,
 h^{0}_{\bar{b} \bar{d}} \,
d^{k}_{\bar{e} \bar{f}} \,
f^{\bar{f} \bar{d}}_{\,\,\,\,\,\, g} \,
f^{\bar{b} g}_{\,\,\,\,\,\, h}$.
By using the identities
$
 h^{k}_{\bar{b} \bar{d}} \,
d^{0}_{\bar{e} \bar{f}}  \,
f^{\bar{f} \bar{d}}_{\,\,\,\,\,\, g} \,
f^{\bar{b} g}_{\,\,\,\,\,\, h} = N \, d^{k}_{\bar{e} \bar{h}}$
and $
 h^{0}_{\bar{b} \bar{d}} \,
d^{k}_{\bar{e} \bar{f}} \,
f^{\bar{f} \bar{d}}_{\,\,\,\,\,\, g} \,
f^{\bar{b} g}_{\,\,\,\,\,\, h} = -(N+2)\, d^{k}_{\bar{e} \bar{h}}$, 
the total contribution is given by $-(2N+2) \,
\varepsilon^{0 i j k} \, d^{k}_{\bar{e} \bar{h}}$.}.
Therefore, one has the final expression as follows
\bea
\frac{(2+3k+3N)}{(k+N+2)} \, \frac{-i}{(k+N+2)} \, \varepsilon^{\mu \nu \rho
  \si} \,
d^{\si}_{\bar{a} \bar{b}}
\, Q^{\bar{a}} \, V^{\bar{b}}(w) = \frac{(2+3k+3N)}{(k+N+2)} \,
\varepsilon^{\mu \nu \rho
  \si} \,
\hat{\Phi}_{\frac{1}{2}}^{(1),\si}(w),
\label{Exppp}
\eea
with (\ref{3halfhat}).
Therefore, this (\ref{Exppp}) coincides with the
second order pole of (\ref{3halfandtwo}).

One can analyze the following other cases also but
the detailed descriptions are not given in this paper. 

$\bullet$ $(\mu,\nu)=(i,0)$

For example, one sees Appendix (\ref{424}) and the
presence of $\Phi_{\frac{1}{2}}^{(1),2}(w)$ in $SO(4)$ basis is expected.

$\bullet$ $(\mu,\nu)=(i,j)$ with $i \neq j$

Appendices (\ref{313}), (\ref{414}) or (\ref{434})
correspond to this case.
The higher spin-$\frac{3}{2}$ current
$\Phi_{\frac{1}{2}}^{(1),j}(w)$ should appear.

\subsubsection{The first order pole}

In order to determine the higher spin-$\frac{5}{2}$ currents,
one should obtain the first oder pole of the OPE
between $\hat{G}^{\mu}(z)$ and $\hat{\Phi}^{(1),\nu \rho}_1(w)$.
By collecting Appendix $(F.1)$ of \cite{AK1411} or Appendix $B$
with the appropriate
factors, the first order pole is described as
{\small
\bea
&& \frac{i}{2(k+N+2)^3}\, \varepsilon^{\nu \rho \si \alpha} \,
h^{\mu}_{\bar{a} \bar{b}} \, h^{\si}_{\bar{c} \bar{d}} \,
d^{\alpha}_{\bar{e} \bar{f}}
\Bigg[  -(k+N+2) \, g^{\bar{e} \bar{c}} 
  \nonu \\
  && \times \,
  \Bigg( k \, (g^{\bar{d} \bar{b}} \, \pa \,
  Q^{\bar{a}} \, V^{\bar{f}} + g^{\bar{f} \bar{b}} \, \pa \, Q^{\bar{a}} \,
  V^{\bar{d}} ) 
 + f^{\bar{d} \bar{b}}_{\,\,\,\,\,\, g} \, Q^{\bar{a}} \, V^{\bar{f}} \, V^g  +
f^{\bar{f} \bar{b}}_{\,\,\,\,\,\, g} \, Q^{\bar{a}} \, V^{\bar{d}} \, V^g
+  f^{\bar{d} \bar{b}}_{\,\,\,\,\,\, g} \,
 f^{\bar{f} g}_{\,\,\,\,\,\, \bar{h}} \, \pa \, (Q^{\bar{a}} \, V^{\bar{h}})
 \Bigg) \nonu \\
&&  - k \, g^{\bar{f} \bar{d}} \, (k+N+2) \, (g^{\bar{c} \bar{a}}  \,
  Q^{\bar{e}} \, \pa \, V^{\bar{b}} - g^{\bar{e} \bar{a}} \, \pa \, Q^{\bar{c}} \,
  V^{\bar{b}} )
  \label{Pole1} \\
  && - f^{\bar{f} \bar{d}}_{\,\,\,\,\,\, g} \,
\Bigg( (k+N+2)( g^{\bar{c} \bar{a}}  \,
Q^{\bar{e}}  \, V^{\bar{b}} \, V^g
-  g^{\bar{e} \bar{a}}  \,
Q^{\bar{c}}  \, V^{\bar{b}} \, V^g) + k \, g^{g \bar{b}} \, Q^{\bar{e}} \,
Q^{\bar{c}} \, \pa \, Q^{\bar{a}} + f^{g \bar{b}}_{\,\,\,\,\,\, h} \,
Q^{\bar{e}} \, Q^{\bar{c}} \, Q^{\bar{a}} \, V^h \Bigg)
  \Bigg].
\nonu
\eea}
We would like to simplify this in order to extract the higher
spin-$\frac{5}{2}$ currents. 
In principle, the five cases can be considered.

$\bullet$ $(\mu,\nu)=(0,0)$

One can collect the quadratic terms with a derivative.
Let us analyze the first term of (\ref{Pole1}).
As described in previous subsection, the result is given by
$-2 d^{\rho}_{\bar{a} \bar{f}}$.
Similarly, the second term of (\ref{Pole1})
contributes to the similar quantity $-2 d^{\rho}_{\bar{a} \bar{d}}$
with different indices.
The sixth term of (\ref{Pole1}) leads to the similar
expression $2 d^{\rho}_{\bar{b} \bar{e}}$.
The seventh term has
$\varepsilon^{0 \rho \si \alpha} \,
h^{0}_{\bar{a} \bar{b}} \, h^{\si}_{\bar{c} \bar{d}} \,
d^{\alpha}_{\bar{e} \bar{f}} \, g^{\bar{f} \bar{d}} \, g^{\bar{e} \bar{a}}$
which can be simplified as
$\varepsilon^{0 \rho \si \alpha} \,
 h^{\si}_{\bar{c} \bar{d}} \,
d^{\alpha}_{\bar{b} \bar{f}} \, g^{\bar{f} \bar{d}} $.
Furthermore, one obtains
$\varepsilon^{0 \rho \si \alpha} \,
 h^{\si}_{\bar{c} \bar{d}} \,  g^{\bar{d} \bar{f}} \,
 d^{\alpha}_{\bar{f} \bar{b}}$. Using the identity (\ref{higdi}),
 one has 
$\varepsilon^{0 \rho \si \alpha} \,
 \varepsilon^{\si \alpha k}
 d^{k}_{\bar{c} \bar{b}} = 2  d^{\rho}_{\bar{c} \bar{b}}$ from previous analysis.

Let us look at the third term of (\ref{Pole1}).
 The expression $\varepsilon^{0 \rho \si \alpha} \,
h^{0}_{\bar{a} \bar{b}} \, h^{\si}_{\bar{c} \bar{d}} \,
d^{\alpha}_{\bar{e} \bar{f}} \, g^{\bar{e} \bar{c}} \,
f^{\bar{d} \bar{b}}_{\,\,\,\,\,\,g}$ can be 
expressed as
$-\varepsilon^{0 \rho \si \alpha} \,
h^{0}_{\bar{a} \bar{b}} \, h^{\si}_{\bar{d} \bar{c}} \, g^{\bar{c} \bar{e}} \,
d^{\alpha}_{\bar{e} \bar{f}} \, 
f^{\bar{d} \bar{b}}_{\,\,\,\,\,\,g}$.
According to (\ref{higdi}), one has
$-\varepsilon^{0 \rho \si \alpha} \,
h^{0}_{\bar{a} \bar{b}} \, \varepsilon^{\si \alpha k}  \,
d^k_{\bar{d} \bar{f}} \, f^{\bar{d} \bar{b}}_{\,\,\,\,\,\,g}$.
Therefore, one obtains
$ 2 h^{0}_{\bar{a} \bar{b}}  \,
d^{\rho}_{\bar{d} \bar{f}} \, f^{\bar{d} \bar{b}}_{\,\,\,\,\,\,g}$.
The fourth term is given by 
$\varepsilon^{0 \rho \si \alpha} \,
h^{0}_{\bar{a} \bar{b}} \, h^{\si}_{\bar{c} \bar{d}} \,
d^{\alpha}_{\bar{e} \bar{f}} \, g^{\bar{e} \bar{c}} \,
f^{\bar{f} \bar{b}}_{\,\,\,\,\,\,g}$ and can be rewritten
as
$-\varepsilon^{0 \rho \si \alpha} \,
h^{0}_{\bar{a} \bar{b}} \, h^{\si}_{\bar{d} \bar{c}} \, g^{\bar{c} \bar{e}} \,
d^{\alpha}_{\bar{e} \bar{f}} \, 
f^{\bar{f} \bar{b}}_{\,\,\,\,\,\,g}$.
Then one has
$ 2 h^{0}_{\bar{a} \bar{b}}  \,
d^{\rho}_{\bar{d} \bar{f}} \, f^{\bar{f} \bar{b}}_{\,\,\,\,\,\,g}$.
With the field $Q^{\bar{a}} \, V^{\bar{d}} \, V^g$, 
the fourth term is equal to the third term.
From the eighth term,
one has
$\varepsilon^{0 \rho \si \alpha} \,
h^{0}_{\bar{a} \bar{b}} \, h^{\si}_{\bar{c} \bar{d}} \,
d^{\alpha}_{\bar{e} \bar{f}} \, g^{\bar{c} \bar{a}} \,
f^{\bar{f} \bar{d}}_{\,\,\,\,\,\,g}$.
So this can be written as
$\varepsilon^{0 \rho \si \alpha} \,
 h^{\si}_{\bar{b} \bar{d}} \,
d^{\alpha}_{\bar{e} \bar{f}} \, 
f^{\bar{f} \bar{d}}_{\,\,\,\,\,\,g}$.
Similarly, the ninth term can be written as
$\varepsilon^{0 \rho \si \alpha} \,
 h^{\si}_{\bar{c} \bar{d}} \,
d^{\alpha}_{\bar{b} \bar{f}} \, 
f^{\bar{f} \bar{d}}_{\,\,\,\,\,\,g}$.
Note that one can check the following identity 
\bea
 (h^{i}_{\bar{b} \bar{d}} \, d^{j}_{\bar{e} \bar{f}} -
h^{j}_{\bar{e} \bar{d}} \, d^{i}_{\bar{b} \bar{f}})
\,
  f^{\bar{f} \bar{d}}_{\,\,\,\,\,\, g} \, Q^{\bar{e}} \, V^{\bar{b}} \, V^{g}
=0.
  \label{prop}
 \eea
 Using this property (\ref{prop}), one can simplify
 the eighth and ninth terms (the summation over the indices
 $\si$ and $\alpha$ is taken) further.
 The tenth term vanishes identically due to (\ref{fgrelation}) \footnote{
Let us consider the fifth term of (\ref{Pole1}). One has
$\varepsilon^{0 \rho \si \alpha} \,
h^{0}_{\bar{a} \bar{b}} \, h^{\si}_{\bar{c} \bar{d}} \,
d^{\alpha}_{\bar{e} \bar{f}} \, g^{\bar{e} \bar{c}} \,
f^{\bar{d} \bar{b}}_{\,\,\,\,\,\,g} \,
f^{\bar{f} g}_{\,\,\,\,\,\,h}$ which can be rewritten as
$-\varepsilon^{0 \rho \si \alpha} \,
h^{0}_{\bar{a} \bar{b}} \, h^{\si}_{\bar{d} \bar{c}} \,
 g^{\bar{c} \bar{e}} \, d^{\alpha}_{\bar{e} \bar{f}} \,
f^{\bar{d} \bar{b}}_{\,\,\,\,\,\,g} \,
f^{\bar{f} g}_{\,\,\,\,\,\,h}$.
From the identity (\ref{higdi}), one can have
$-\varepsilon^{0 \rho \si \alpha} \,
h^{0}_{\bar{a} \bar{b}} \,  \varepsilon^{\si \alpha k} \,
d^{k}_{\bar{d} \bar{f}} \,
f^{\bar{d} \bar{b}}_{\,\,\,\,\,\,g} \,
f^{\bar{f} g}_{\,\,\,\,\,\,h}$.
From (\ref{dffrelation}), the $ d f f$ factor
can be reduced to $N \, d^{k}_{\bar{c} \bar{h}} \, g^{\bar{c} \bar{b}}$.
This leads to $- N \, \varepsilon^{0 \rho \si \alpha} \,
\,  \varepsilon^{\si \alpha k} \,  d^{k}_{\bar{a} \bar{h}}=
-2N \,  d^{\rho}_{\bar{a} \bar{h}}$.
The last term of (\ref{Pole1}) is
$\varepsilon^{0 \rho \si \alpha} \,
h^{0}_{\bar{a} \bar{b}} \, h^{\si}_{\bar{c} \bar{d}} \,
d^{\alpha}_{\bar{e} \bar{f}} \,  f^{\bar{f} \bar{d}}_{\,\,\,\,\,\, g} \,
 f^{g \bar{b}}_{\,\,\,\,\,\, h}$.
One has the following identity
$
h_{\bar{a} \bar{b}}^0 \, (h^{i}_{\bar{c} \bar{d}} \, d^{j}_{\bar{e} \bar{f}} +
h^{j}_{\bar{c} \bar{d}} \, d^{i}_{\bar{e} \bar{f}})
\,
f^{\bar{f} \bar{d}}_{\,\,\,\,\,\, g} \,
f^{g \bar{b}}_{\,\,\,\,\,\, h} \,
Q^{\bar{e}} \, Q^{\bar{c}} \,
Q^{\bar{a}} \, V^{h}
=0$.
 This implies that the  summation over the indices
 $\si$ and $\alpha$ can be simplified further.}.

It turns out that the first order pole of (\ref{Pole1})
 with free index $\rho$ and fixed indices
 $(\mu,\nu)=(0,0)$ in $SU(2) \times SU(2)$ basis
 can be described as (the last term of (\ref{Pole1})
 remains)
\bea
&& \frac{i}{(k+N+2)^2}\, 
\Bigg[
  (3k+N) \,  d^{\rho}_{\bar{a} \bar{b}} \, \pa \,
  Q^{\bar{a}} \, V^{\bar{b}}  
+ (N-k) \,  d^{\rho}_{\bar{a} \bar{b}} \, 
  Q^{\bar{a}} \, \pa \, V^{\bar{b}}  
+ 2 h^{0}_{\bar{a} \bar{b}} \, d^{\rho}_{\bar{d} \bar{f}} \, 
f^{\bar{d} \bar{b}}_{\,\,\,\,\,\, g} \,  Q^{\bar{a}} \, V^{\bar{f}} \, V^g  
\nonu \\
&& -\frac{1}{2}\, \varepsilon^{0 \rho \si \alpha} \,
h^{\si}_{\bar{b} \bar{d}} \,  d^{\alpha}_{\bar{e} \bar{f}} \,
f^{\bar{f} \bar{d}}_{\,\,\,\,\,\, g} \,
 Q^{\bar{e}} \, V^{\bar{b}} \, V^g 
+ \frac{1}{2}\, \varepsilon^{0 \rho \si \alpha} \,
h^{\si}_{\bar{c} \bar{d}} \,  d^{\alpha}_{\bar{b} \bar{f}} \,
f^{\bar{f} \bar{d}}_{\,\,\,\,\,\, g} \,
 Q^{\bar{c}} \, V^{\bar{b}} \, V^g
 \nonu \\
 && -\frac{1}{(k+N+2)} \,
 \varepsilon^{0 \rho \si \alpha} \,
 h^{0}_{\bar{a} \bar{b}} \, h^{\si}_{\bar{c} \bar{d}} \,
 d^{\alpha}_{\bar{e} \bar{f}} \,
f^{\bar{f} \bar{d}}_{\,\,\,\,\,\, g} \,
f^{g \bar{b}}_{\,\,\,\,\,\, h} \, Q^{\bar{e}} \, Q^{\bar{c}} \, Q^{\bar{a}}
\, V^h
\Bigg].
\label{5halfcombination}
\eea  
Therefore, this quantity will be crucial to obtain the higher
spin-$3$ current eventually even in the fixed indices.
In particular, the last quartic term in (\ref{5halfcombination})
has rather complicated tensor indices compared to other terms.
We will fix the index $\rho$ later.

There are other cases one should consider as follows.
The detailed calculations on these cases are ignored in this paper.

$\bullet$ $(\mu,\nu)=(i,i)$

For the same $(\mu,\nu)=(1,1)$ but with $\rho=2$, one sees Appendix
(\ref{334}) in $SO(4)$ basis.
Appendix (\ref{112}) corresponds
to $(\mu,\nu)=(3,3)$ with $\rho=0$.
For the same $(\mu,\nu)=(3,3)$ but with $\rho=1$, one sees Appendix
(\ref{113}). Moreover,
for the same $(\mu,\nu)=(3,3)$ but with $\rho=2$, one sees Appendix
(\ref{114}).
The higher spin-$\frac{5}{2}$ current 
$\Phi_{\frac{3}{2}}^{(1),\rho}(z)$ can be determined explicitly.

$\bullet$ $(\mu,\nu)=(0,i)$

One observes this case in (\ref{212}) in $SO(4)$ basis when $i=3$.
The index $\rho$ is the same as the index $\mu$. 

$\bullet$ $(\mu,\nu)=(i,0)$

One sees the particular case with $i=2$ in Appendix (\ref{424})
in $SO(4)$ basis
where the index $\rho$ is the same as index $\mu$.
The higher spin-$\frac{5}{2}$ current 
$\Phi_{\frac{3}{2}}^{(1),\nu}(z)$ can be determined explicitly.

$\bullet$ $(\mu,\nu)=(i,j)$ with $i \neq j$

One can see Appendix (\ref{313}) in $SO(4)$ basis
corresponding to $(i,j)=(1,3)$.
Appendix (\ref{414}) corresponds to $(i,j)=(2,3)$.
Similarly, the $(\mu,\nu)=(2,1)$ case can be seen from Appendix
(\ref{434}).
All of these examples have the same index for $\mu$ and $\rho$.
The higher spin-$\frac{5}{2}$ current 
$\Phi_{\frac{3}{2}}^{(1),j}(z)$ can be seen explicitly.

Let us find the four higher spin-$\frac{5}{2}$ currents
explicitly. There are three different ways for writing down each higher
spin-$\frac{5}{2}$ current from (\ref{3halfandtwo}): the possibility of
$\rho$.
We will present one way for each higher spin-$\frac{5}{2}$ current
and the remaining two ways will appear in Appendix $C$.

\subsubsection{The first higher spin-$\frac{5}{2}$ current}

In order to proceed further, it is better to consider
the fixed $\rho$ case. Let us look at the final term of
(\ref{5halfcombination}) when $\rho=1$ (Other cases are described in Appendix
$C$). One has
\bea
&& \varepsilon^{0 1 2 3} \,
 h^{0}_{\bar{c} \bar{b}'} \, h^{2}_{\bar{b} \bar{d}'} \,
 d^{3}_{\bar{a} \bar{f}'} \,
f^{\bar{f}' \bar{d}'}_{\,\,\,\,\,\, g'} \,
f^{g' \bar{b}'}_{\,\,\,\,\,\, d} 
 = 
\nonu \\
&& -\frac{1}{2} \, d^{2}_{\bar{a} \bar{b}} \, h^3_{\bar{c} \bar{d}}
-\frac{1}{4} \,  d^{2}_{\bar{a} \bar{d}} \, h^3_{\bar{b} \bar{c}}
-\frac{1}{4} \, d^{2}_{\bar{a} \bar{c}} \, h^3_{\bar{d} \bar{b}}
+\frac{1}{4} \,  d^{2}_{\bar{d} \bar{b}} \, h^3_{\bar{a} \bar{c}}
+\frac{1}{4} \, d^{2}_{\bar{c} \bar{b}} \, h^3_{\bar{d} \bar{a}}
-\frac{1}{2} \,  d^{3}_{\bar{a} \bar{b}} \, h^2_{\bar{c} \bar{d}}
+\frac{1}{4} \, d^{3}_{\bar{a} \bar{d}} \, h^2_{\bar{b} \bar{c}}
\nonu \\
&& + \frac{1}{4} \,  d^{3}_{\bar{a} \bar{c}} \, h^2_{\bar{d} \bar{b}}
-\frac{1}{4} \,  d^{3}_{\bar{d} \bar{b}} \, h^2_{\bar{a} \bar{c}}
-\frac{1}{4} \, d^{3}_{\bar{c} \bar{b}} \, h^2_{\bar{d} \bar{a}}
-\frac{1}{2} \, d^{0}_{\bar{a} \bar{b}} \, h^1_{\bar{c} \bar{d}}
-\frac{1}{4} \,  d^{0}_{\bar{a} \bar{d}} \, h^1_{\bar{b} \bar{c}}
-\frac{1}{4} \, d^{0}_{\bar{a} \bar{c}} \, h^1_{\bar{d} \bar{b}}
-\frac{1}{2} \,  d^{0}_{\bar{d} \bar{c}} \, h^1_{\bar{a} \bar{b}}
\nonu \\
& & - \frac{1}{4} \,  d^{0}_{\bar{d} \bar{b}} \, h^1_{\bar{a} \bar{c}}
-\frac{1}{4} \,  d^{0}_{\bar{c} \bar{b}} \, h^1_{\bar{d} \bar{a}}
+\frac{1}{4} \, d^{1}_{\bar{a} \bar{d}} \, h^0_{\bar{b} \bar{c}}
-\frac{1}{4} \, d^{1}_{\bar{a} \bar{c}} \, h^0_{\bar{d} \bar{b}}
-\frac{1}{4} \,  d^{1}_{\bar{d} \bar{b}} \, h^0_{\bar{a} \bar{c}}
+\frac{1}{4} \, d^{1}_{\bar{b} \bar{c}} \, h^0_{\bar{a} \bar{d}}.
\label{dhtensor}
\eea
One can simplify the quartic term in (\ref{5halfcombination})
with (\ref{dhtensor}).
Among the first five terms of (\ref{dhtensor}),
the second and the fourth terms remain the nonzero contributions
because the $d^2$ tensor is symmetric while the quartic term
is antisymmetric in the indices $\bar{a}$, $\bar{b}$ and ${\bar{c}}$.
The second term can be written as $ d^{2}_{\bar{d} \bar{a}} \,
h^3_{\bar{b} \bar{c}}$. With quartic term, one can interchange the
indices ${\bar{a}}$ and $\bar{b}$ and obtains $ -d^{2}_{\bar{d} \bar{b}} \,
h^3_{\bar{a} \bar{c}}$. Therefore, one has the same contribution as the one in
fourth term.
For the next five terms in (\ref{dhtensor}),
one is left with the seventh and ninth terms according to the similar
analysis done before. The ninth term can be rewritten as
$ d^{3}_{\bar{b} \bar{d}} \,
h^2_{\bar{a} \bar{c}}$ with other factor.
By interchanging the indices ${\bar{a}}$ and $\bar{b}$, one has
$ -d^{3}_{\bar{a} \bar{d}} \,
h^2_{\bar{b} \bar{c}}$.  Then one has the same contribution as the one in
seventh term.
For the next six terms, 
the $11$th, $13$th and $16$th can be combined together and the remaining
terms vanish.
Finally, there are no contributions from the  last four terms. 
Then one obtains for the quartic term in (\ref{5halfcombination}) as follows
\bea
(\frac{1}{2} \, d^{2}_{\bar{d} \bar{b}} \, h^3_{\bar{a} \bar{c}} +
\frac{1}{2} \,  d^{3}_{\bar{a} \bar{d}} \, h^2_{\bar{b} \bar{c}}-
d^{0}_{\bar{a} \bar{b}} \, h^1_{\bar{c} \bar{d}}) \, Q^{\bar{a}} \, Q^{\bar{b}} \,
Q^{\bar{c}} \, V^{\bar{d}}(w).
\label{quarticterm}
\eea

Now one would like to determine the higher spin-$\frac{5}{2}$ current
which is a primary.
One can focus on the particular OPE of (\ref{3halfandtwo})
with the indices $\mu=0, \nu=0$ and $\rho=1$.
By noting that
$\hat{G}^0(z)=G^2(z)$ from (\ref{spin3halfrelation})
and $ \hat{\Phi}_{1}^{(1),01}(w) =\Phi_{1}^{(1),23}(w)$ from
(\ref{spin2higher}) and (\ref{spin2final}), one 
can have the following OPE in the $SO(4)$ manifest basis
{\small
  \bea
&& G^2(z) \, \Phi_{1}^{(1),23}(w) =
+ \cdots + \frac{1}{(z-w)} \Bigg[
-\Phi_{\frac{3}{2}}^{(1),3} 
-\frac{2 i}{(k+N+2)} \, T^{12} \, \Phi_{\frac{1}{2}}^{(1),4}
\nonu \\
&& -  \frac{2 i}{(k+N+2)} \, T^{24} \, \Phi_{\frac{1}{2}}^{(1),1}
+\frac{4 i}{(k+N+2)} \, T^{14} \, \Phi_{\frac{1}{2}}^{(1),2}
+  \frac{(N-k)}{3(k+N+2)} \, \pa \, \Phi_{\frac{1}{2}}^{(1),3}
\Bigg](w)+  \cdots,
\label{223}
\eea}
where the second order pole is ignored. All the hat notations
appearing in the right hand side of the OPE (\ref{3halfandtwo})
are gone.
See also the corresponding OPE in Appendix (\ref{1116}).

Then it is obvious that the higher spin-$\frac{5}{2}$ current
can be read off and is given by
\bea
\Phi_{\frac{3}{2}}^{(1),3}(z) & = &
-\frac{2 i}{(k+N+2)} \, T^{12} \, \Phi_{\frac{1}{2}}^{(1),4}
-\frac{2 i}{(k+N+2)} \, T^{24} \, \Phi_{\frac{1}{2}}^{(1),1}
+\frac{4 i}{(k+N+2)} \, T^{14} \, \Phi_{\frac{1}{2}}^{(1),2}
\nonu \\
& + & \frac{(N-k)}{3(k+N+2)} \, \pa \, \Phi_{\frac{1}{2}}^{(1),3}
-(\ref{5halfcombination}) 
\,\, \mbox{with} \,\, \rho=1,
\label{spin5halfrel}
\eea
where the relation (\ref{quarticterm}) is used.
Furthermore, one should rewrite the composite expressions
appearing in (\ref{spin5halfrel})
in terms of WZW adjoint currents. 
The spin-$1$ currents are presented in (\ref{TA}) and (\ref{11currents})
while the higher spin-$\frac{3}{2}$ currents are presented in
(\ref{spin3half}).
Because
the spin-$1$ currents $A^{-i}(z)$ are quadratic terms in
(\ref{11currents}),
one should consider the following normal ordered product
{\small
\bea
(Q^{\bar{a}} Q^{\bar{b}})( Q^{\bar{d}} \, V^{\bar{e}})  & = &
Q^{\bar{a}} Q^{\bar{b}} Q^{\bar{d}} \, V^{\bar{e}} -[  Q^{\bar{d}} \, V^{\bar{e}},
  Q^{\bar{a}} Q^{\bar{b}}] + \{ Q^{\bar{d}} \, V^{\bar{e}}, Q^{\bar{a}}\} \, Q^{\bar{b}}
-Q^{\bar{a}} \,  \{ Q^{\bar{d}} \, V^{\bar{e}}, Q^{\bar{b}}\}
\nonu \\
& = & Q^{\bar{a}} Q^{\bar{b}} Q^{\bar{d}} \, V^{\bar{e}} +(k+N+2) \,
g^{\bar{d} \bar{a}} \, \pa \, Q^{\bar{b}} \, V^{\bar{e}} -
(k+N+2) \,
g^{\bar{d} \bar{b}} \, \pa \, Q^{\bar{a}} \, V^{\bar{e}}.
\label{multiproduct}
\eea}
Note that there is a minus sign in the last term of the first line of
(\ref{multiproduct}) due to the exchange of the two fermionic
quantities. Each (anti)commutator can be obtained
from each defining OPE.

It turns out that one obtains the following relations 
\bea
T^{12} \, \Phi_{\frac{1}{2}}^{(1),4}  & = & -\frac{1}{4N(k+N+2)} \, 
f^{\bar{a} \bar{b}}_{\,\,\,\,\,\, c} \, h^3_{\bar{a} \bar{b}} \, d^2_{\bar{d} \bar{e}} \,
Q^{\bar{d}} \, V^{c} \, V^{\bar{e}} +
\frac{1}{4(k+N+2)^2} \,  h^3_{\bar{a} \bar{b}} \, d^2_{\bar{c} \bar{d}} \, 
Q^{\bar{a}} \, Q^{\bar{b}} \, Q^{\bar{c}} \, V^{\bar{d}}
\nonu \\
&+& \frac{1}{2(k+N+2)} \,  d^1_{\bar{a} \bar{b}} \, \pa \, Q^{\bar{a}} \,
V^{\bar{b}},
\nonu \\
T^{24} \, \Phi_{\frac{1}{2}}^{(1),1}  & = &
\frac{1}{4N(k+N+2)} \, 
f^{\bar{a} \bar{b}}_{\,\,\,\,\,\, c} \, h^2_{\bar{a} \bar{b}} \, d^3_{\bar{d} \bar{e}} \,
Q^{\bar{d}} \, V^{c} \, V^{\bar{e}} -
\frac{1}{4(k+N+2)^2} \,  h^2_{\bar{a} \bar{b}} \, d^3_{\bar{c} \bar{d}} \, 
Q^{\bar{a}} \, Q^{\bar{b}} \, Q^{\bar{c}} \, V^{\bar{d}}
\nonu \\
&+& \frac{1}{2(k+N+2)} \,  d^1_{\bar{a} \bar{b}} \, \pa \, Q^{\bar{a}} \,
V^{\bar{b}},
\nonu \\
T^{14} \, \Phi_{\frac{1}{2}}^{(1),2}  & = &
\frac{1}{4N(k+N+2)} \, 
f^{\bar{a} \bar{b}}_{\,\,\,\,\,\, c} \, h^1_{\bar{a} \bar{b}} \, d^0_{\bar{d} \bar{e}} \,
Q^{\bar{d}} \, V^{c} \, V^{\bar{e}} +
\frac{1}{4(k+N+2)^2} \,  h^1_{\bar{a} \bar{b}} \, d^0_{\bar{c} \bar{d}} \, 
Q^{\bar{a}} \, Q^{\bar{b}} \, Q^{\bar{c}} \, V^{\bar{d}}
\nonu \\
&-& \frac{1}{2(k+N+2)} \,  d^1_{\bar{a} \bar{b}} \, \pa \, Q^{\bar{a}} \,
V^{\bar{b}}.
\label{TPhirel}
\eea
The relations (\ref{higdi}) and (\ref{hgd-1}) are used in the first
two and the last one in (\ref{TPhirel}) respectively.
One can check that the quartic term of the first quantity in
(\ref{TPhirel}) corresponds to the first term of (\ref{quarticterm})
and those in the second quantity in (\ref{TPhirel})
does to the second term of (\ref{quarticterm}). Furthermore,
as one moves $V^g$ in (\ref{5halfcombination}) to the left at one step
(in order to combine with the cubic terms in (\ref{TPhirel})),
then the extra derivative terms can be reduced to the second term of
(\ref{5halfcombination}) with different coefficients by using the
identities (\ref{dffrelation}) and (\ref{higdi}).

Finally, one obtains the higher spin-$\frac{5}{2}$ current which will used
for the higher spin-$3$ current
\bea
\Phi_{\frac{3}{2}}^{(1),3}(z) & = &
\frac{i}{(k+N+2)^2} \, \Bigg[ \frac{1}{2N} \, h^3_{\bar{a} \bar{b}}\,
  d^2_{\bar{d} \bar{e}} -
\frac{1}{2N} \, h^2_{\bar{a} \bar{b}}\,
  d^3_{\bar{d} \bar{e}}
+ \frac{1}{N} \, h^1_{\bar{a} \bar{b}}\,
  d^0_{\bar{d} \bar{e}}
  \nonu \\
  & - & 2  \, h^0_{\bar{b} \bar{d}}\,
  d^1_{\bar{a} \bar{e}} -  h^2_{\bar{b} \bar{e}}\,
  d^3_{\bar{a} \bar{d}} +  h^2_{\bar{b} \bar{d}}\,
  d^3_{\bar{a} \bar{e}} \Bigg] f^{\bar{a} \bar{b}}_{\,\,\,\,\,\,c}
\, Q^{\bar{d}} \, V^c \, V^{\bar{e}}
-\frac{4i(3+2k+N)}{3(k+N+2)^2} \, d^{1}_{\bar{a} \bar{b}} \, \pa
Q^{\bar{a}} \, V^{\bar{b}}(z)
\nonu \\
&+& \frac{i}{(k+N+2)^3} \, \Bigg[ d^2_{\bar{b} \bar{d}} \, h^3_{\bar{a}
    \bar{c}} + d^3_{\bar{a} \bar{d}} \, h^2_{\bar{b} \bar{c}} -
  d^0_{\bar{a} \bar{b}} \, h^1_{\bar{c} \bar{d}} + d^0_{\bar{c} \bar{d}}
  \, h^1_{\bar{a} \bar{b}}\Bigg] Q^{\bar{a}} \, Q^{\bar{b}} \, Q^{\bar{c}} \,
V^{\bar{d}}(z) 
\nonu \\
& +& \frac{4i(k+2N)}{3(k+N+2)^2} \, d^{1}_{\bar{a} \bar{b}} \, 
Q^{\bar{a}} \, \pa \, V^{\bar{b}}(z).
\label{spin5halfone}
\eea
Note that the last term of the quartic term in (\ref{spin5halfone})
originates from the one in (\ref{TPhirel}).
Compared to (\ref{5halfcombination}), the coefficients of
all the derivative terms are changed as explained before.
One presents this higher spin-$\frac{5}{2}$ current in two different
ways in Appendix $C$ (The index $\mu$ in $SO(4)$ basis is $1$ or $4$).
They will differ from (\ref{spin5halfone}) only by the cubic term.

\subsubsection{The second higher spin-$\frac{5}{2}$ current}

Let us determine the second higher spin-$\frac{5}{2}$ current.
Let us consider the first order pole in (\ref{5halfcombination})
for the indices $\mu=0, \nu=0$ and $\rho=2$.
One has the following identity
\bea
&& \varepsilon^{0 2 3 1} \,
 h^{0}_{\bar{c} \bar{b}'} \, h^{3}_{\bar{b} \bar{d}'} \,
 d^{1}_{\bar{a} \bar{f}'} \,
f^{\bar{f}' \bar{d}'}_{\,\,\,\,\,\, g'} \,
f^{g' \bar{b}'}_{\,\,\,\,\,\, d} 
 = 
\nonu \\
&& -\frac{1}{4} \, d^{1}_{\bar{a} \bar{b}} \, h^3_{\bar{c} \bar{d}}
-\frac{1}{4} \,  d^{1}_{\bar{a} \bar{c}} \, h^3_{\bar{d} \bar{b}}
+\frac{1}{4} \, d^{1}_{\bar{d} \bar{c}} \, h^3_{\bar{a} \bar{b}}
-\frac{1}{4} \,  d^{1}_{\bar{d} \bar{b}} \, h^3_{\bar{a} \bar{c}}
+\frac{1}{2} \, d^{1}_{\bar{c} \bar{b}} \, h^3_{\bar{d} \bar{a}}
-\frac{1}{4} \,  d^{0}_{\bar{a} \bar{b}} \, h^2_{\bar{c} \bar{d}}
+\frac{1}{2} \, d^{0}_{\bar{a} \bar{d}} \, h^2_{\bar{b} \bar{c}}
\nonu \\
&& - \frac{1}{4} \,  d^{0}_{\bar{a} \bar{c}} \, h^2_{\bar{d} \bar{b}}
+\frac{1}{4} \,  d^{0}_{\bar{d} \bar{c}} \, h^2_{\bar{a} \bar{b}}
-\frac{1}{4} \, d^{0}_{\bar{d} \bar{b}} \, h^2_{\bar{a} \bar{c}}
-\frac{1}{2} \, d^{0}_{\bar{c} \bar{b}} \, h^2_{\bar{d} \bar{a}}
+\frac{1}{4} \,  d^{3}_{\bar{a} \bar{b}} \, h^1_{\bar{c} \bar{d}}
+\frac{1}{4} \, d^{3}_{\bar{a} \bar{c}} \, h^1_{\bar{d} \bar{b}}
-\frac{1}{4} \,  d^{3}_{\bar{d} \bar{c}} \, h^1_{\bar{a} \bar{b}}
\nonu \\
& & + \frac{1}{4} \,  d^{3}_{\bar{d} \bar{b}} \, h^1_{\bar{a} \bar{c}}
+\frac{1}{2} \,  d^{3}_{\bar{c} \bar{b}} \, h^1_{\bar{d} \bar{a}}
-\frac{1}{4} \, d^{2}_{\bar{a} \bar{b}} \, h^0_{\bar{c} \bar{d}}
+\frac{1}{4} \, d^{2}_{\bar{a} \bar{c}} \, h^0_{\bar{d} \bar{b}}
-\frac{1}{4} \,  d^{2}_{\bar{d} \bar{c}} \, h^0_{\bar{a} \bar{b}}
+\frac{1}{4} \, d^{2}_{\bar{d} \bar{b}} \, h^0_{\bar{a} \bar{c}}.
\label{hhdffrel}
\eea
There are no contributions for the terms having the $d^i$ tensor with
two indices among $\bar{a}, \bar{b}$ or $\bar{c}$ because they are symmetric
under the two indices while the composite field $-Q^{\bar{a}} \,
Q^{\bar{b}} \, Q^{\bar{c}} \, V^{\bar{d}}$ is antisymmetric \footnote{
That is, the first, the second, the fifth, the twelfth, the thirteenth,
the sixteenth, the seventeenth, the eighteenth
of (\ref{hhdffrel}), become zero.
One can combine the third and the fourth terms.
Similarly the fourteenth and the fifteenth terms can be combined each other. 
The last two terms vanish because the $h^0$ tensor is symmetric
while  the composite field is antisymmetric.
The sixth and the eighth terms can be combined.
The ninth and the tenth terms can contribute similarly.}.
Finally, one obtains the resulting expression
for the quartic term in (\ref{5halfcombination})
\bea
(-\frac{1}{2} \, d^{1}_{\bar{c} \bar{d}} \, h^3_{\bar{a} \bar{b}} +
\frac{1}{2} \,  d^{3}_{\bar{c} \bar{d}} \, h^1_{\bar{a} \bar{b}}+
d^{0}_{\bar{a} \bar{b}} \, h^2_{\bar{c} \bar{d}}) \, Q^{\bar{a}} \, Q^{\bar{b}} \,
Q^{\bar{c}} \, V^{\bar{d}}(w).
\label{quarticterm1}
\eea

As before, now one would like to determine the
second higher spin-$\frac{5}{2}$ current
which is a primary.
One can focus on the particular OPE of (\ref{3halfandtwo})
with the indices $\mu=0, \nu=0$ and $\rho=2$.
By noting that
$\hat{G}^0(z)=G^2(z)$ from (\ref{spin3halfrelation})
and $ \hat{\Phi}_{1}^{(1),02}(w) =-\Phi_{1}^{(1),24}(w)$ from
(\ref{spin2higher}) and (\ref{spin2final}), one 
can have the following OPE in the $SO(4)$ manifest basis
{\small
\bea
&& G^2(z) \, \Phi_{1}^{(1),24}(w) =
+ \cdots + \frac{1}{(z-w)} \Bigg[
-\Phi_{\frac{3}{2}}^{(1),4} 
+\frac{2 i}{(k+N+2)} \, T^{12} \, \Phi_{\frac{1}{2}}^{(1),3}
\nonu \\
&& +  \frac{2 i}{(k+N+2)} \, T^{23} \, \Phi_{\frac{1}{2}}^{(1),1}
-\frac{4 i}{(k+N+2)} \, T^{13} \, \Phi_{\frac{1}{2}}^{(1),2}
+  \frac{(N-k)}{3(k+N+2)} \, \pa \, \Phi_{\frac{1}{2}}^{(1),4}
\Bigg](w) +   \cdots,
\label{224}
\eea}
where the second order pole is ignored. All the hat notations
appearing in the right hand side of the OPE (\ref{3halfandtwo})
are gone.
See also the corresponding OPE in Appendix (\ref{1116}).

Then it is obvious that the second higher spin-$\frac{5}{2}$ current
can be read off and is given by
\bea
\Phi_{\frac{3}{2}}^{(1),4}(z) & = &
\frac{2 i}{(k+N+2)} \, T^{12} \, \Phi_{\frac{1}{2}}^{(1),3}
+\frac{2 i}{(k+N+2)} \, T^{23} \, \Phi_{\frac{1}{2}}^{(1),1}
-\frac{4 i}{(k+N+2)} \, T^{13} \, \Phi_{\frac{1}{2}}^{(1),2}
\nonu \\
& + & \frac{(N-k)}{3(k+N+2)} \, \pa \, \Phi_{\frac{1}{2}}^{(1),4}
+(\ref{5halfcombination}) 
\,\, \mbox{with} \,\, \rho=2,
\label{spin5halfrel1}
\eea
where the relation (\ref{quarticterm1}) is used.
Furthermore, one should rewrite the composite expressions
appearing in (\ref{spin5halfrel1})
in terms of WZW adjoint currents. 
One obtains
\bea
T^{12} \, \Phi_{\frac{1}{2}}^{(1),3}  & = & \frac{1}{4N(k+N+2)} \, 
f^{\bar{a} \bar{b}}_{\,\,\,\,\,\, c} \, h^3_{\bar{a} \bar{b}} \, d^1_{\bar{d} \bar{e}} \,
Q^{\bar{d}} \, V^{c} \, V^{\bar{e}} -
\frac{1}{4(k+N+2)^2} \,  h^3_{\bar{a} \bar{b}} \, d^1_{\bar{c} \bar{d}} \, 
Q^{\bar{a}} \, Q^{\bar{b}} \, Q^{\bar{c}} \, V^{\bar{d}}
\nonu \\
&+& \frac{1}{2(k+N+2)} \,  d^2_{\bar{a} \bar{b}} \, \pa \, Q^{\bar{a}} \,
V^{\bar{b}},
\nonu \\
T^{23} \, \Phi_{\frac{1}{2}}^{(1),1}  & = &
-\frac{1}{4N(k+N+2)} \, 
f^{\bar{a} \bar{b}}_{\,\,\,\,\,\, c} \, h^1_{\bar{a} \bar{b}} \, d^3_{\bar{d} \bar{e}} \,
Q^{\bar{d}} \, V^{c} \, V^{\bar{e}} +
\frac{1}{4(k+N+2)^2} \,  h^1_{\bar{a} \bar{b}} \, d^3_{\bar{c} \bar{d}} \, 
Q^{\bar{a}} \, Q^{\bar{b}} \, Q^{\bar{c}} \, V^{\bar{d}}
\nonu \\
&+& \frac{1}{2(k+N+2)} \,  d^2_{\bar{a} \bar{b}} \, \pa \, Q^{\bar{a}} \,
V^{\bar{b}},
\nonu \\
T^{13} \, \Phi_{\frac{1}{2}}^{(1),2}  & = &
\frac{1}{4N(k+N+2)} \, 
f^{\bar{a} \bar{b}}_{\,\,\,\,\,\, c} \, h^2_{\bar{a} \bar{b}} \, d^0_{\bar{d} \bar{e}} \,
Q^{\bar{d}} \, V^{c} \, V^{\bar{e}} +
\frac{1}{4(k+N+2)^2} \,  h^2_{\bar{a} \bar{b}} \, d^0_{\bar{c} \bar{d}} \, 
Q^{\bar{a}} \, Q^{\bar{b}} \, Q^{\bar{c}} \, V^{\bar{d}}
\nonu \\
&-& \frac{1}{2(k+N+2)} \,  d^2_{\bar{a} \bar{b}} \, \pa \, Q^{\bar{a}} \,
V^{\bar{b}}.
\label{TPhirel1}
\eea
The relations (\ref{higdi}) and (\ref{hgd-1}) are used in the first
two and the last one in (\ref{TPhirel1}) respectively.
One can check that the quartic term of the first quantity in
(\ref{TPhirel1}) corresponds to the first term of (\ref{quarticterm1})
and those in the second quantity in (\ref{TPhirel1})
does to the second term of (\ref{quarticterm1}). Furthermore,
as one moves $V^g$ in (\ref{5halfcombination}) to the left at one step
(in order to combine with the cubic terms in (\ref{TPhirel1})),
then the extra derivative terms can be reduced to the second term of
(\ref{5halfcombination}) with different coefficients by using the
identities (\ref{dffrelation}) and (\ref{higdi}).

Therefore, the final second higher spin-$\frac{5}{2}$ current
in $SO(4)$ basis
can be described as
\bea
\Phi_{\frac{3}{2}}^{(1),4}(z) & = &
\frac{i}{(k+N+2)^2} \, \Bigg[ \frac{1}{2N} \, h^3_{\bar{a} \bar{b}}\,
  d^1_{\bar{d} \bar{e}} -
\frac{1}{2N} \, h^1_{\bar{a} \bar{b}}\,
  d^3_{\bar{d} \bar{e}}
- \frac{1}{N} \, h^2_{\bar{a} \bar{b}}\,
  d^0_{\bar{d} \bar{e}}
  \nonu \\
  & + & 2  \, h^0_{\bar{b} \bar{d}}\,
  d^2_{\bar{a} \bar{e}} -  h^1_{\bar{b} \bar{e}}\,
  d^3_{\bar{a} \bar{d}} +  h^1_{\bar{b} \bar{d}}\,
  d^3_{\bar{a} \bar{e}} \Bigg] f^{\bar{a} \bar{b}}_{\,\,\,\,\,\,c}
\, Q^{\bar{d}} \, V^c \, V^{\bar{e}}
+\frac{4i(3+2k+N)}{3(k+N+2)^2} \, d^{2}_{\bar{a} \bar{b}} \, \pa
Q^{\bar{a}} \, V^{\bar{b}}(z)
\nonu \\
&+& \frac{i}{(k+N+2)^3} \, \Bigg[ -d^1_{\bar{c} \bar{d}} \, h^3_{\bar{a}
    \bar{b}} + d^0_{\bar{a} \bar{b}} \, h^2_{\bar{c} \bar{d}} -
  d^0_{\bar{a} \bar{d}} \, h^2_{\bar{b} \bar{c}} + d^3_{\bar{c} \bar{d}}
  \, h^1_{\bar{a} \bar{b}}\Bigg] Q^{\bar{a}} \, Q^{\bar{b}} \, Q^{\bar{c}} \,
V^{\bar{d}}(z) 
\nonu \\
& -& \frac{4i(k+2N)}{3(k+N+2)^2} \, d^{2}_{\bar{a} \bar{b}} \, 
Q^{\bar{a}} \, \pa \, V^{\bar{b}}(z).
\label{spin5halftwo}
\eea
Note that the third term of the quartic term in (\ref{spin5halftwo})
originates from the one in (\ref{TPhirel1}).
Compared to (\ref{5halfcombination}), the coefficients of
all the derivative terms are changed as explained before.
One presents this second higher spin-$\frac{5}{2}$ current
in two different
ways in Appendix $C$ (The index $\mu$ in $SO(4)$ basis is $1$ or $3$).
They will differ from (\ref{spin5halftwo}) only by the cubic term.

\subsubsection{The third higher spin-$\frac{5}{2}$ current}

Let us determine the third higher spin-$\frac{5}{2}$ current.
Let us consider the first order pole in (\ref{5halfcombination})
for the indices $\mu=0, \nu=0$ and $\rho=3$.
One has the following identity
\bea
&& \varepsilon^{0 3 1 2} \,
 h^{0}_{\bar{c} \bar{b}'} \, h^{1}_{\bar{b} \bar{d}'} \,
 d^{2}_{\bar{a} \bar{f}'} \,
f^{\bar{f}' \bar{d}'}_{\,\,\,\,\,\, g'} \,
f^{g' \bar{b}'}_{\,\,\,\,\,\, d} 
 = 
\nonu \\
&& -\frac{1}{4} \, d^{0}_{\bar{a} \bar{b}} \, h^3_{\bar{c} \bar{d}}
+\frac{1}{2} \,  d^{0}_{\bar{a} \bar{d}} \, h^3_{\bar{b} \bar{c}}
-\frac{1}{4} \, d^{0}_{\bar{a} \bar{c}} \, h^3_{\bar{d} \bar{b}}
+\frac{1}{4} \,  d^{0}_{\bar{d} \bar{c}} \, h^3_{\bar{a} \bar{b}}
+\frac{1}{4} \, d^{1}_{\bar{a} \bar{b}} \, h^2_{\bar{c} \bar{d}}
+\frac{1}{4} \,  d^{1}_{\bar{a} \bar{c}} \, h^2_{\bar{d} \bar{b}}
-\frac{1}{4} \, d^{1}_{\bar{d} \bar{c}} \, h^2_{\bar{a} \bar{b}}
\nonu \\
&& + \frac{1}{4} \,  d^{1}_{\bar{d} \bar{b}} \, h^2_{\bar{a} \bar{c}}
+\frac{1}{2} \,  d^{1}_{\bar{c} \bar{b}} \, h^2_{\bar{d} \bar{a}}
-\frac{1}{4} \, d^{2}_{\bar{a} \bar{b}} \, h^1_{\bar{c} \bar{d}}
-\frac{1}{4} \, d^{2}_{\bar{a} \bar{c}} \, h^1_{\bar{d} \bar{b}}
+\frac{1}{4} \,  d^{2}_{\bar{d} \bar{c}} \, h^1_{\bar{a} \bar{b}}
-\frac{1}{4} \, d^{2}_{\bar{d} \bar{b}} \, h^1_{\bar{a} \bar{c}}
+\frac{1}{2} \,  d^{2}_{\bar{c} \bar{b}} \, h^1_{\bar{d} \bar{a}}
\nonu \\
& & - \frac{1}{4} \,  d^{3}_{\bar{a} \bar{b}} \, h^0_{\bar{c} \bar{d}}
+\frac{1}{4} \,  d^{3}_{\bar{a} \bar{c}} \, h^0_{\bar{d} \bar{b}}
-\frac{1}{4} \, d^{3}_{\bar{d} \bar{c}} \, h^0_{\bar{a} \bar{b}}
+\frac{1}{4} \, d^{3}_{\bar{d} \bar{b}} \, h^0_{\bar{a} \bar{c}}
-\frac{1}{4} \,  d^{0}_{\bar{d} \bar{b}} \, h^3_{\bar{a} \bar{c}}
-\frac{1}{2} \, d^{0}_{\bar{c} \bar{b}} \, h^3_{\bar{d} \bar{a}}.
\label{hhdffrelation}
\eea
The fifth, sixth, ninth, tenth, eleventh, fourteenth, fifteenth,
sixteenth, seventeenth and eighteenth terms in (\ref{hhdffrelation})
vanish \footnote{
The sum of the second and fourth terms become zeroes.
The seventh and the eighth terms can combine each other.
Similarly, one can combine the twelfth and the thirteenth terms.
Finally, the first, the second and the last terms can be combined.}.
Then, one obtains the simple form as follows
\bea
(\frac{1}{2} \, d^{1}_{\bar{c} \bar{d}} \, h^2_{\bar{a} \bar{b}} -
\frac{1}{2} \,  d^{2}_{\bar{c} \bar{d}} \, h^1_{\bar{a} \bar{b}}+
d^{0}_{\bar{a} \bar{b}} \, h^3_{\bar{c} \bar{d}}) \, Q^{\bar{a}} \, Q^{\bar{b}} \,
Q^{\bar{c}} \, V^{\bar{d}}(w).
\label{quarticterm2}
\eea

As before, now one would like to determine the
third higher spin-$\frac{5}{2}$ current
which is a primary.
One can focus on the particular OPE of (\ref{3halfandtwo})
with the indices $\mu=0, \nu=0$ and $\rho=3$.
By noting that
$\hat{G}^0(z)=G^2(z)$ from (\ref{spin3halfrelation})
and $ \hat{\Phi}_{1}^{(1),03}(w) =-\Phi_{1}^{(1),12}(w)$ from
(\ref{spin2higher}) and (\ref{spin2final}), one 
can have the following OPE in the $SO(4)$ manifest basis
{\small
  \bea
&& G^2(z) \, \Phi_{1}^{(1),12}(w) =
+ \cdots + \frac{1}{(z-w)} \Bigg[
\Phi_{\frac{3}{2}}^{(1),1} 
+\frac{2 i}{(k+N+2)} \, T^{23} \, \Phi_{\frac{1}{2}}^{(1),4}
\nonu \\
&& -  \frac{2 i}{(k+N+2)} \, T^{24} \, \Phi_{\frac{1}{2}}^{(1),3}
+\frac{4 i}{(k+N+2)} \, T^{34} \, \Phi_{\frac{1}{2}}^{(1),2}
-  \frac{(N-k)}{3(k+N+2)} \, \pa \, \Phi_{\frac{1}{2}}^{(1),1}
\Bigg](w) +  \cdots,
\label{212}
\eea}
where the second order pole is ignored. All the hat notations
appearing in the right hand side of the OPE (\ref{3halfandtwo})
are gone.
See also the corresponding OPE in Appendix (\ref{1116}).

Then it is obvious that the third higher spin-$\frac{5}{2}$ current
can be read off and is given by
\bea
\Phi_{\frac{3}{2}}^{(1),1}(z) & = &
-\frac{2 i}{(k+N+2)} \, T^{23} \, \Phi_{\frac{1}{2}}^{(1),4}
+\frac{2 i}{(k+N+2)} \, T^{24} \, \Phi_{\frac{1}{2}}^{(1),3}
-\frac{4 i}{(k+N+2)} \, T^{34} \, \Phi_{\frac{1}{2}}^{(1),2}
\nonu \\
& + & \frac{(N-k)}{3(k+N+2)} \, \pa \, \Phi_{\frac{1}{2}}^{(1),1}
-(\ref{5halfcombination}) 
\,\, \mbox{with} \,\, \rho=3,
\label{spin5halfrel2}
\eea
where the relation (\ref{quarticterm2}) is used.
Furthermore, one should rewrite the composite expressions
appearing in (\ref{spin5halfrel2})
in terms of WZW adjoint currents. 
One obtains
\bea
T^{23} \, \Phi_{\frac{1}{2}}^{(1),4}  & = & \frac{1}{4N(k+N+2)} \, 
f^{\bar{a} \bar{b}}_{\,\,\,\,\,\, c} \, h^1_{\bar{a} \bar{b}} \, d^2_{\bar{d} \bar{e}} \,
Q^{\bar{d}} \, V^{c} \, V^{\bar{e}} -
\frac{1}{4(k+N+2)^2} \,  h^1_{\bar{a} \bar{b}} \, d^2_{\bar{c} \bar{d}} \, 
Q^{\bar{a}} \, Q^{\bar{b}} \, Q^{\bar{c}} \, V^{\bar{d}}
\nonu \\
&+& \frac{1}{2(k+N+2)} \,  d^3_{\bar{a} \bar{b}} \, \pa \, Q^{\bar{a}} \,
V^{\bar{b}},
\nonu \\
T^{24} \, \Phi_{\frac{1}{2}}^{(1),3}  & = &
\frac{1}{4N(k+N+2)} \, 
f^{\bar{a} \bar{b}}_{\,\,\,\,\,\, c} \, h^2_{\bar{a} \bar{b}} \, d^1_{\bar{d} \bar{e}} \,
Q^{\bar{d}} \, V^{c} \, V^{\bar{e}} -
\frac{1}{4(k+N+2)^2} \,  h^2_{\bar{a} \bar{b}} \, d^1_{\bar{c} \bar{d}} \, 
Q^{\bar{a}} \, Q^{\bar{b}} \, Q^{\bar{c}} \, V^{\bar{d}}
\nonu \\
&-& \frac{1}{2(k+N+2)} \,  d^3_{\bar{a} \bar{b}} \, \pa \, Q^{\bar{a}} \,
V^{\bar{b}},
\nonu \\
T^{34} \, \Phi_{\frac{1}{2}}^{(1),2}  & = &
-\frac{1}{4N(k+N+2)} \, 
f^{\bar{a} \bar{b}}_{\,\,\,\,\,\, c} \, h^3_{\bar{a} \bar{b}} \, d^0_{\bar{d} \bar{e}} \,
Q^{\bar{d}} \, V^{c} \, V^{\bar{e}} -
\frac{1}{4(k+N+2)^2} \,  h^3_{\bar{a} \bar{b}} \, d^0_{\bar{c} \bar{d}} \, 
Q^{\bar{a}} \, Q^{\bar{b}} \, Q^{\bar{c}} \, V^{\bar{d}}
\nonu \\
&+& \frac{1}{2(k+N+2)} \,  d^3_{\bar{a} \bar{b}} \, \pa \, Q^{\bar{a}} \,
V^{\bar{b}}.
\label{TPhirel2}
\eea
The relations (\ref{higdi}) and (\ref{hgd-1}) are used in the first
two and the last one in (\ref{TPhirel2}) respectively \footnote{
One can check that the quartic term of the first quantity in
(\ref{TPhirel1}) corresponds to the second term of (\ref{quarticterm2})
and those in the second quantity in (\ref{TPhirel2})
does to the first term of (\ref{quarticterm2}). Furthermore,
as one moves $V^g$ in (\ref{5halfcombination}) to the left at one step
(in order to combine with the cubic terms in (\ref{TPhirel2})),
then the extra derivative terms can be reduced to the second term of
(\ref{5halfcombination}) with different coefficients by using the
identities (\ref{dffrelation}) and (\ref{higdi}).}.

Therefore, the third higher spin-$\frac{5}{2}$ current
in $SO(4)$ basis is given by
\bea
\Phi_{\frac{3}{2}}^{(1),1}(z) & = &
\frac{i}{(k+N+2)^2} \, \Bigg[ -\frac{1}{2N} \, h^1_{\bar{a} \bar{b}}\,
  d^2_{\bar{d} \bar{e}} +
\frac{1}{2N} \, h^2_{\bar{a} \bar{b}}\,
  d^1_{\bar{d} \bar{e}}
+ \frac{1}{N} \, h^3_{\bar{a} \bar{b}}\,
  d^0_{\bar{d} \bar{e}}
  \nonu \\
  & - & 2  \, h^0_{\bar{b} \bar{d}}\,
  d^3_{\bar{a} \bar{e}} +  h^2_{\bar{b} \bar{e}}\,
  d^1_{\bar{a} \bar{d}} -  h^2_{\bar{b} \bar{d}}\,
  d^1_{\bar{a} \bar{e}} \Bigg] f^{\bar{a} \bar{b}}_{\,\,\,\,\,\,c}
\, Q^{\bar{d}} \, V^c \, V^{\bar{e}}
-\frac{4i(3+2k+N)}{3(k+N+2)^2} \, d^{3}_{\bar{a} \bar{b}} \, \pa
Q^{\bar{a}} \, V^{\bar{b}}(z)
\nonu \\
&+& \frac{i}{(k+N+2)^3} \, \Bigg[ -d^0_{\bar{a} \bar{b}} \, h^3_{\bar{c}
    \bar{d}} - d^1_{\bar{c} \bar{d}} \, h^2_{\bar{a} \bar{b}} +
  d^2_{\bar{c} \bar{d}} \, h^1_{\bar{a} \bar{b}} + d^0_{\bar{c} \bar{d}}
  \, h^3_{\bar{a} \bar{b}}\Bigg] Q^{\bar{a}} \, Q^{\bar{b}} \, Q^{\bar{c}} \,
V^{\bar{d}}(z) 
\nonu \\
& +& \frac{4i(k+2N)}{3(k+N+2)^2} \, d^{3}_{\bar{a} \bar{b}} \, 
Q^{\bar{a}} \, \pa \, V^{\bar{b}}(z).
\label{spin5halfthree}
\eea
Note that the last term of the quartic term in (\ref{spin5halfthree})
originates from the one in (\ref{TPhirel2}).
Compared to (\ref{5halfcombination}), the coefficients of
all the derivative terms are changed as explained before.
One presents this third higher spin-$\frac{5}{2}$ current
in two different
ways in Appendix $C$ (The index $\mu$ in $SO(4)$ basis is $3$ or $4$).
They will differ from (\ref{spin5halfthree}) only by the cubic term.

Let us describe the case of $(\mu,\nu)=(1,1)$ with $\rho=0$ in the
following subsection.

\subsubsection{The fourth higher spin-$\frac{5}{2}$ current}

Let us determine the fourth higher spin-$\frac{5}{2}$ current.
Let us consider the first order pole in (\ref{5halfcombination})
for the indices $\mu=1, \nu=1$ and $\rho=0$.
One has the following identity
\bea
&& \varepsilon^{0 1 2 3 } \,
 h^{1}_{\bar{c} \bar{b}'} \, h^{2}_{\bar{b} \bar{d}'} \,
 d^{3}_{\bar{a} \bar{f}'} \,
f^{\bar{f}' \bar{d}'}_{\,\,\,\,\,\, g'} \,
f^{g' \bar{b}'}_{\,\,\,\,\,\, d} 
 = 
\nonu \\
&& -\frac{1}{4} \, d^{3}_{\bar{a} \bar{b}} \, h^3_{\bar{c} \bar{d}}
-\frac{1}{4} \,  d^{3}_{\bar{a} \bar{c}} \, h^3_{\bar{d} \bar{b}}
-\frac{1}{4} \, d^{3}_{\bar{d} \bar{c}} \, h^3_{\bar{a} \bar{b}}
+\frac{1}{4} \,  d^{3}_{\bar{d} \bar{b}} \, h^3_{\bar{a} \bar{c}}
+\frac{1}{2} \, d^{3}_{\bar{c} \bar{b}} \, h^3_{\bar{d} \bar{a}}
-\frac{1}{4} \,  d^{2}_{\bar{a} \bar{b}} \, h^2_{\bar{c} \bar{d}}
-\frac{1}{4} \, d^{2}_{\bar{a} \bar{c}} \, h^2_{\bar{d} \bar{b}}
\nonu \\
&& - \frac{1}{4} \,  d^{2}_{\bar{d} \bar{c}} \, h^2_{\bar{a} \bar{b}}
+\frac{1}{4} \,  d^{2}_{\bar{d} \bar{b}} \, h^2_{\bar{a} \bar{c}}
-\frac{1}{2} \, d^{2}_{\bar{c} \bar{b}} \, h^2_{\bar{d} \bar{a}}
+\frac{1}{4} \, d^{1}_{\bar{a} \bar{b}} \, h^1_{\bar{c} \bar{d}}
-\frac{1}{2} \,  d^{1}_{\bar{a} \bar{d}} \, h^1_{\bar{b} \bar{c}}
+\frac{1}{4} \, d^{1}_{\bar{a} \bar{c}} \, h^1_{\bar{d} \bar{b}}
+\frac{1}{4} \,  d^{1}_{\bar{d} \bar{c}} \, h^1_{\bar{a} \bar{b}}
\nonu \\
& & - \frac{1}{4} \,  d^{1}_{\bar{d} \bar{b}} \, h^1_{\bar{a} \bar{c}}
-\frac{1}{4} \,  d^{0}_{\bar{a} \bar{b}} \, h^0_{\bar{c} \bar{d}}
+\frac{1}{4} \, d^{0}_{\bar{a} \bar{c}} \, h^0_{\bar{d} \bar{b}}
+\frac{1}{4} \, d^{0}_{\bar{d} \bar{c}} \, h^0_{\bar{a} \bar{b}}
-\frac{1}{4} \,  d^{0}_{\bar{d} \bar{b}} \, h^0_{\bar{a} \bar{c}}
+\frac{1}{2} \, d^{0}_{\bar{c} \bar{b}} \, h^0_{\bar{d} \bar{a}}.
\label{hhdffrelat}
\eea
Among the first five terms of (\ref{hhdffrelat}),
the nonzero contribution arises from the third and fourth terms \footnote{
Similarly among the next five terms of (\ref{hhdffrelat}),
the nonzero contribution arises from the third and fourth terms.
One sees that 
among the next five terms of (\ref{hhdffrelat}),
the nonzero contribution arises from the second, fourth and fifth terms.
However, this leads to zero.
For the last five terms, one sees that the third and fourth terms
do not contribute.
The remaining three terms can contribute.}.
Therefore, the simple form for the quartic term one obtains is given by
\bea
(\frac{1}{2} \, d^{3}_{\bar{c} \bar{d}} \, h^3_{\bar{a} \bar{b}} +
\frac{1}{2} \,  d^{2}_{\bar{c} \bar{d}} \, h^2_{\bar{a} \bar{b}}+
d^{0}_{\bar{a} \bar{b}} \, h^0_{\bar{c} \bar{d}}) \, Q^{\bar{a}} \, Q^{\bar{b}} \,
Q^{\bar{c}} \, V^{\bar{d}}(w).
\label{quarticterm3}
\eea

As before, now one would like to determine the
fourth higher spin-$\frac{5}{2}$ current
which is a primary.
One can focus on the particular OPE of (\ref{3halfandtwo})
with the indices $\mu=1, \nu=1$ and $\rho=0$.
By noting that
$\hat{G}^1(z)=G^3(z)$ from (\ref{spin3halfrelation})
and $ \hat{\Phi}_{1}^{(1),10}(w) =-\Phi_{1}^{(1),23}(w)$ from
(\ref{spin2higher}) and (\ref{spin2final}), one 
can have the following OPE in the $SO(4)$ manifest basis
{\small
  \bea
&& G^3(z) \, \Phi_{1}^{(1),23}(w) =
+ \cdots + \frac{1}{(z-w)} \Bigg[
\Phi_{\frac{3}{2}}^{(1),2} 
-\frac{2 i}{(k+N+2)} \, T^{13} \, \Phi_{\frac{1}{2}}^{(1),4}
\nonu \\
&& -  \frac{2 i}{(k+N+2)} \, T^{34} \, \Phi_{\frac{1}{2}}^{(1),1}
+\frac{4 i}{(k+N+2)} \, T^{14} \, \Phi_{\frac{1}{2}}^{(1),3}
-  \frac{(N-k)}{3(k+N+2)} \, \pa \, \Phi_{\frac{1}{2}}^{(1),2}
\Bigg](w)
+  \cdots,
\label{323}
\eea}
where the second order pole is ignored. All the hat notations
appearing in the right hand side of the OPE (\ref{3halfandtwo})
are gone.
See also the corresponding OPE in Appendix (\ref{1116}).

Then it is obvious that the fourth higher spin-$\frac{5}{2}$ current
can be read off and is given by
\bea
\Phi_{\frac{3}{2}}^{(1),2}(z) & = &
\frac{2 i}{(k+N+2)} \, T^{13} \, \Phi_{\frac{1}{2}}^{(1),4}
+\frac{2 i}{(k+N+2)} \, T^{34} \, \Phi_{\frac{1}{2}}^{(1),1}
-\frac{4 i}{(k+N+2)} \, T^{14} \, \Phi_{\frac{1}{2}}^{(1),3}
\nonu \\
& + & \frac{(N-k)}{3(k+N+2)} \, \pa \, \Phi_{\frac{1}{2}}^{(1),2}
-(\ref{5halfcombination}) 
\,\, \mbox{with} \,\, \mu=\nu=1, \rho=0,
\label{spin5halfrel3}
\eea
where the relation (\ref{quarticterm3}) is used.
Furthermore, one should rewrite the composite expressions
appearing in (\ref{spin5halfrel3})
in terms of WZW adjoint currents. 
One obtains
\bea
T^{13} \, \Phi_{\frac{1}{2}}^{(1),4}  & = & -\frac{1}{4N(k+N+2)} \, 
f^{\bar{a} \bar{b}}_{\,\,\,\,\,\, c} \, h^2_{\bar{a} \bar{b}} \, d^2_{\bar{d} \bar{e}} \,
Q^{\bar{d}} \, V^{c} \, V^{\bar{e}} -
\frac{1}{4(k+N+2)^2} \,  h^2_{\bar{a} \bar{b}} \, d^2_{\bar{c} \bar{d}} \, 
Q^{\bar{a}} \, Q^{\bar{b}} \, Q^{\bar{c}} \, V^{\bar{d}}
\nonu \\
&-& \frac{1}{2(k+N+2)} \,  d^0_{\bar{a} \bar{b}} \, \pa \, Q^{\bar{a}} \,
V^{\bar{b}},
\nonu \\
T^{34} \, \Phi_{\frac{1}{2}}^{(1),1}  & = &
-\frac{1}{4N(k+N+2)} \, 
f^{\bar{a} \bar{b}}_{\,\,\,\,\,\, c} \, h^3_{\bar{a} \bar{b}} \, d^3_{\bar{d} \bar{e}} \,
Q^{\bar{d}} \, V^{c} \, V^{\bar{e}} -
\frac{1}{4(k+N+2)^2} \,  h^3_{\bar{a} \bar{b}} \, d^3_{\bar{c} \bar{d}} \, 
Q^{\bar{a}} \, Q^{\bar{b}} \, Q^{\bar{c}} \, V^{\bar{d}}
\nonu \\
&-& \frac{1}{2(k+N+2)} \,  d^0_{\bar{a} \bar{b}} \, \pa \, Q^{\bar{a}} \,
V^{\bar{b}},
\nonu \\
T^{14} \, \Phi_{\frac{1}{2}}^{(1),3}  & = &
\frac{1}{4N(k+N+2)} \, 
f^{\bar{a} \bar{b}}_{\,\,\,\,\,\, c} \, h^1_{\bar{a} \bar{b}} \, d^1_{\bar{d} \bar{e}} \,
Q^{\bar{d}} \, V^{c} \, V^{\bar{e}} +
\frac{1}{4(k+N+2)^2} \,  h^1_{\bar{a} \bar{b}} \, d^1_{\bar{c} \bar{d}} \, 
Q^{\bar{a}} \, Q^{\bar{b}} \, Q^{\bar{c}} \, V^{\bar{d}}
\nonu \\
&+& \frac{1}{2(k+N+2)} \,  d^0_{\bar{a} \bar{b}} \, \pa \, Q^{\bar{a}} \,
V^{\bar{b}}.
\label{TPhirel3}
\eea
The relations (\ref{higdi}) and (\ref{hgd-1}) are used in the first
two and the last one in (\ref{TPhirel3}) respectively \footnote{
One can check that the quartic term of the first quantity in
(\ref{TPhirel3}) corresponds to the second term of (\ref{quarticterm3})
and those in the second quantity in (\ref{TPhirel3})
does to the first term of (\ref{quarticterm3}). Furthermore,
as one moves $V^g$ in (\ref{5halfcombination}) to the left at one step
(in order to combine with the cubic terms in (\ref{TPhirel3})),
then the extra derivative terms can be reduced to the second term of
(\ref{5halfcombination}) with different coefficients by using the
identities (\ref{dffrelation}) and (\ref{higdi}).}.

Therefore, the fourth higher spin-$\frac{5}{2}$ current
in $SO(4)$ basis
is given by
\bea
\Phi_{\frac{3}{2}}^{(1),2}(z) & = &
\frac{i}{(k+N+2)^2} \, \Bigg[ -\frac{1}{2N} \, h^2_{\bar{a} \bar{b}}\,
  d^2_{\bar{d} \bar{e}} -
\frac{1}{2N} \, h^3_{\bar{a} \bar{b}}\,
  d^3_{\bar{d} \bar{e}}
- \frac{1}{N} \, h^1_{\bar{a} \bar{b}}\,
  d^1_{\bar{d} \bar{e}}
  \nonu \\
  & - & 2  \, h^1_{\bar{b} \bar{d}}\,
  d^1_{\bar{a} \bar{e}} -  h^3_{\bar{b} \bar{d}}\,
  d^3_{\bar{a} \bar{e}} -  h^2_{\bar{b} \bar{d}}\,
  d^2_{\bar{a} \bar{e}} \Bigg] f^{\bar{a} \bar{b}}_{\,\,\,\,\,\,c}
\, Q^{\bar{d}} \, V^c \, V^{\bar{e}}
-\frac{4i(3+2k+N)}{3(k+N+2)^2} \, d^{0}_{\bar{a} \bar{b}} \, \pa
Q^{\bar{a}} \, V^{\bar{b}}(z)
\nonu \\
&+& \frac{i}{(k+N+2)^3} \, \Bigg[ -d^3_{\bar{c} \bar{d}} \, h^3_{\bar{a}
    \bar{b}} - d^2_{\bar{c} \bar{d}} \, h^2_{\bar{a} \bar{b}} -
  d^0_{\bar{a} \bar{b}} \, h^0_{\bar{c} \bar{d}} - 
  \, d^1_{\bar{c} \bar{d}} h^1_{\bar{a} \bar{b}}
  \Bigg] Q^{\bar{a}} \, Q^{\bar{b}} \, Q^{\bar{c}} \,
V^{\bar{d}}(z) 
\nonu \\
& +& \frac{4i(k+2N)}{3(k+N+2)^2} \, d^{0}_{\bar{a} \bar{b}} \, 
Q^{\bar{a}} \, \pa \, V^{\bar{b}}(z).
\label{spin5halffour}
\eea
Note that the last term of the quartic term in (\ref{spin5halffour})
originates from the one in (\ref{TPhirel3}).
Compared to (\ref{5halfcombination}), the coefficients of
all the derivative terms are changed as explained before.
One presents this fourth higher spin-$\frac{5}{2}$ current
in two different
ways in Appendix $C$ (The index $\mu$ in $SO(4)$ basis is $1$ or $4$).
They will differ from (\ref{spin5halffour}) only by the cubic term.

Then, one has the four higher spin-$\frac{5}{2}$ currents
in (\ref{spin5halfone}), (\ref{spin5halftwo}), (\ref{spin5halfthree})
and (\ref{spin5halffour}) in $SO(4)$ basis.
There are no quintic, cubic or linear terms in spin-$\frac{1}{2}$ current.
The next step is to obtain the higher spin-$3$ current using one of
these higher spin-$\frac{5}{2}$ currents.

\subsection{Higher spin-$3$ current}

Let us describe how one can obtain the final higher spin-$3$ current.
The following OPE (the first order pole) determines
the explicit higher spin-$3$ current in $SO(4)$ basis 
{\small
  \bea
&& G^{\mu}(z)\,\Phi_{\frac{3}{2}}^{(1),\nu}(w)  =  
\frac{1}{(z-w)^{3}}\Bigg[\frac{16(k-N)}{3(2+k+N)}
  \Bigg] \delta^{\mu \nu} \, \Phi_{0}^{(1)}(w)
\nonu \\
&& +  \frac{1}{(z-w)^{2}}\,\Bigg[-\,4\,\Phi_{1}^{(1),\mu \nu}
  -\frac{2(k-N)}{3(2+k+N)}\, \varepsilon^{\mu \nu \rho \si}
  \, \Phi_{1}^{(1),\rho \si}\,\Bigg](w)\nonu
\nonu \\ 
&& +  \frac{1}{(z-w)}\,\Bigg[\,\delta^{\mu \nu}\,(-\Phi_{2}^{(1)}
+\frac{16 (k-N)}{ (6 k N+5 k+5 N+4)}\, \Phi_{0}^{(1)} \, T
\,) - \frac{2 }{(2+k+N)}\,\varepsilon^{\mu \nu \rho \si}\, G^{\rho}
\, \Phi_{\frac{1}{2}}^{(1),\si}\,
\nonu \\ 
&& - 
\partial \, \Phi_{1}^{(1),\mu \nu}
-\frac{(k-N)}{6(2+k+N)}\, \varepsilon^{\mu \nu \rho \si} \,
\partial \, \Phi_{1}^{(1),\rho \si}
+\frac{2\, i}{(2+k+N)}\,\varepsilon^{\mu \nu \rho \si} \, \partial {T}^{\rho \si}
\, \Phi_{0}^{(1)}
\nonu \\ 
&& -  \frac{2\, i}{(2+k+N)}\,\varepsilon^{\mu \nu \rho \si} \, {T}^{\rho \si} \,
\partial\Phi_{0}^{(1)}
+ \frac{2\, i}{(2+k+N)}\,
(\, T^{\mu \rho} \, \Phi_{1}^{(1),\nu \rho}-
T^{\nu \rho} \, \Phi_{1}^{(1),\mu \rho} \,) \Bigg](w)
+ \cdots.
\label{process}
\eea}
That is, when one looks at the first order pole of this OPE
with the condition $\mu=\nu$,
one obtains the higher spin-$3$ current $\Phi_{2}^{(1)}(w)$
from the explicit calculation of the left hand side of this OPE.

Instead of calculating the left hand side of
this OPE directly, one considers some parts of
the higher spin-$\frac{5}{2}$ current. For example,
the higher spin-$\frac{5}{2}$ current is given by (\ref{spin5halfrel}).
One can focus on the last term of this expression.
Let us calculate the OPE between the spin-$\frac{3}{2}$
current and the expression from the first order pole in
(\ref{5halfcombination}) using the WZW currents.

\subsubsection{The fourth order pole}

According to Appendix
(\ref{finalfirst}), there is no fourth order pole.

\subsubsection{The third order pole}

Let us calculate the third order pole of the OPE
between $\hat{G}^{\mu=i}(z)$ and the expression
appearing the first order pole of (\ref{5halfcombination})
with $\rho=i$ (no sum over the index $i$). 
It turns out that
\bea
&& \frac{i}{(k+N+2)} \, \frac{i}{(k+N+2)^2}
\Bigg[ (3k+N) \, (k+N+2) \, h^i_{\bar{a} \bar{b}} \, d^{i}_{\bar{c} \bar{d}}\,
  g^{\bar{c} \bar{a}} \,  f^{\bar{b} \bar{d}}_{\,\,\,\,\,\,e} \, V^e
  \nonu \\
 && + (N-k) \,  h^i_{\bar{a} \bar{b}} \, d^{i}_{\bar{c} \bar{d}}\,
  ((k+N+2) \, g^{\bar{c} \bar{a}} \,  f^{\bar{b} \bar{d}}_{\,\,\,\,\,\,e} \,
  V^e - 2k \, g^{\bar{d} \bar{b}} \, Q^{\bar{c}} \, Q^{\bar{a}})
  \nonu \\
  && + (2  h^i_{\bar{a} \bar{b}} \,  h^0_{\bar{c} \bar{f}} \,
  d^{i}_{\bar{d} \bar{g}}\,   f^{\bar{g} \bar{f}}_{\,\,\,\,\,\,e} -\frac{1}{2} \,
  h^i_{\bar{a} \bar{b}} \, \varepsilon^{0 i \si \alpha} \,
h^{\si}_{\bar{d} \bar{f}} \,  d^{\alpha}_{\bar{c} \bar{g}} \,
f^{\bar{g} \bar{f}}_{\,\,\,\,\,\, e} \, + \frac{1}{2} \,  h^i_{\bar{a} \bar{b}} \,
 \varepsilon^{0 i \si \alpha} \,
h^{\si}_{\bar{c} \bar{f}} \,  d^{\alpha}_{\bar{g} \bar{d}} \,
f^{\bar{g} \bar{f}}_{\,\,\,\,\,\, e}
)
  \nonu \\
  && \times (-(k+N+2) g^{\bar{c} \bar{a}} \, ( k
  \, g^{\bar{b} \bar{d}} \, V^e +
  f^{\bar{b} \bar{d}}_{\,\,\,\,\,\,h} \,  f^{h e }_{\,\,\,\,\,\,i} \,
  V^i + k \, g^{\bar{b} e} \, V^{\bar{d}}
  ) - k \, f^{\bar{d} \bar{b}}_{\,\,\,\,\,\,h} \,
  g^{e h} \, Q^{\bar{c}} \, Q^{\bar{a}})
  \label{pole3} \\
  && -    h^i_{\bar{a} \bar{b}} \,
\varepsilon^{0 i \si \alpha} \,
 h^{0}_{\bar{e} \bar{b}'} \, h^{\si}_{\bar{d} \bar{d}'} \,
 d^{\alpha}_{\bar{c} \bar{f}'} \,
f^{\bar{f}' \bar{d}'}_{\,\,\,\,\,\, g} \,
f^{g \bar{b}'}_{\,\,\,\,\,\, f} \,
(-k\, g^{\bar{c} \bar{a}} \, g^{\bar{b} f} \, Q^{\bar{d}} \, Q^{\bar{e}}
+ k\, g^{\bar{d} \bar{a}} \, g^{\bar{b} f} \, Q^{\bar{c}} \, Q^{\bar{e}}-
k\, g^{\bar{e} \bar{a}} \, g^{\bar{b} f} \, Q^{\bar{c}} \, Q^{\bar{d}}) \Bigg](w).
\nonu
\eea
The first term of (\ref{pole3}) has
$h^i_{\bar{a} \bar{b}} \, g^{\bar{a} \bar{c}} \, d^i_{\bar{c} \bar{d}}$
which is written as
$d^0_{\bar{b} \bar{d}}$ using (\ref{higdi}).
The second term can be reduced to $d^0_{\bar{b} \bar{d}}$ similarly.
Let us consider $V^e(w)$ term appearing in the first term in the fourth line
of (\ref{pole3}).
The first contribution is related to
$ -d^{i}_{\bar{g} \bar{d}} \,
g^{\bar{d} \bar{b}} \, h^i_{\bar{b} \bar{a}}\, g^{\bar{a} \bar{c}} \,
h^0_{\bar{c} \bar{f}}$ which is equal to
$ -d^{i}_{\bar{g} \bar{d}} \,
g^{\bar{d} \bar{b}} \, h^i_{\bar{b} \bar{f}}$.
Again this can be written as $ h^i_{\bar{f} \bar{b}} \, g^{\bar{b} \bar{d}} \,
d^{i}_{\bar{d} \bar{g}}$. With (\ref{higdi}), this becomes
$ -d^0_{\bar{f} \bar{g}}$.
The second contribution contains
$ \varepsilon^{0 i \si \alpha} \, h^i_{\bar{a} \bar{b}} \, 
g^{\bar{b} \bar{d}} \, h^{\si}_{\bar{d} \bar{f}} \,
g^{\bar{a} \bar{c}} \, d^{\alpha}_{\bar{c} \bar{g}}$.
Then the $h g h $ factor is given by $\varepsilon^{i \si k} \,
h^k_{\bar{a} \bar{f}}$.
Then one has
$-\varepsilon^{0 i \si \alpha} \,
\varepsilon^{i \si k} \,
h^k_{\bar{f} \bar{a}} \, g^{\bar{a} \bar{c}} \, d^{\alpha}_{\bar{c} \bar{g}}$
which leads to $2 d^{0}_{\bar{f} \bar{g}}$ from (\ref{higdi}).
The third contribution has
$ -\varepsilon^{0 i \si \alpha} \, h^i_{\bar{b} \bar{a}} \, 
g^{\bar{a} \bar{c}} \, h^{\si}_{\bar{c} \bar{f}} \,
g^{\bar{b} \bar{d}} \, d^{\alpha}_{\bar{d} \bar{g}}$.
From the $h g h$ factor, one can simplify this as
$-\varepsilon^{0 i \si \alpha} \, \varepsilon^{i \si k} \,
h^k_{\bar{f} \bar{b}} \, g^{\bar{b} \bar{d}} \, d^{\alpha}_{\bar{d} \bar{g}}$.
This leads to the final result $2 d^{0}_{\bar{f} \bar{g}}$ as before.
Let us look at $V^i(w)$ term which is the second term of the
fourth line of (\ref{pole3}). The first contribution has
$-h^{i}_{\bar{b} \bar{a}} \, g^{\bar{a} \bar{c}} \, h^0_{\bar{c} \bar{f}}$
which is given by $-h^{i}_{\bar{b} \bar{f}}$. Together with the factor
$d^i_{\bar{d} \bar{g}} \, f^{\bar{b} \bar{d}}_{\,\,\,\,\,\, h}$,
one obtains $-h^{i}_{\bar{b} \bar{f}} \,
d^i_{\bar{d} \bar{g}} \, f^{\bar{b} \bar{d}}_{\,\,\,\,\,\, h}$
which is symmetric under the indicies $\bar{f}$ and $\bar{g}$ according to
the identity in (\ref{symmetric}).
On the other hand, there exists a factor $ f^{\bar{g} \bar{f}}_{\,\,\,\,\,\, e}$
which is antisymmetric in the indices $\bar{g}$ and $\bar{f}$.
Therefore, there is no contribution at all \footnote{
The second contribution has the factor
$-h^{i}_{\bar{b} \bar{a}} \, g^{\bar{a} \bar{c}} \, d^{\alpha}_{\bar{c} \bar{g}}$
which is equal to $\varepsilon^{i \alpha k} \, d^{k}_{\bar{b} \bar{g}}$.
Furthermore, one has $\varepsilon^{0 i \si \alpha } \,
\varepsilon^{i \alpha k} \,  f^{\bar{b} \bar{d}}_{\,\,\,\,\,\, h} \,
h^{\si}_{\bar{d} \bar{f}} \,
d^{k}_{\bar{b} \bar{g}}$ by adding the extra factor.
The nonzero contribution arises when the two indices are equal to
each other $\si =k$.
One sees that this is also symmetric in the indices
$\bar{f}$ and $\bar{g}$ in the presence of  $ f^{\bar{g} \bar{f}}_{\,\,\,\,\,\, e}$.
In this case also there is no contribution.
The third contribution has the factor
$-h^{i}_{\bar{b} \bar{a}} \, g^{\bar{a} \bar{c}} \, h^{\si}_{\bar{c} \bar{f}}$
which becomes $-\varepsilon^{i \si k}\, h^k_{\bar{b} \bar{f}}$.
Furthermore, with other factors
one has $-\varepsilon^{0 i \si \alpha } \,
\varepsilon^{i \si k} \,  f^{\bar{b} \bar{d}}_{\,\,\,\,\,\, h} \,
h^{k}_{\bar{b} \bar{f}} \,
d^{\alpha}_{\bar{g} \bar{d}}$.
Again,  the two indices are equal to
each other $\si =k$ for nonzero contribution.
Due to the symmetric property between the indices $\bar{f}$ and $\bar{g}$,
there is no contribution.
Let us focus on the $V^{\bar{d}}(w)$ term which contains
$f^{\bar{g} \bar{f}}_{\,\,\,\,\,\, e} \, g^{\bar{b} e}$.
According to the previous identity in (\ref{fgrelation}),
there is no contribution.}.
Let us collect the spin-$1$ dependent terms as follows:
{\small
  \bea
-\frac{1}{(k+N+2)^2}  \Bigg[ (N+3k) +(N-k)-2k-k-k \Bigg]
 d^0_{\bar{a} \bar{b}} 
f^{\bar{a} \bar{b}}_{\,\,\,\,\,\, c}  V^c
=
 \frac{-2(N-k)}{(k+N+2)^2}   d^0_{\bar{a} \bar{b}} 
f^{\bar{a} \bar{b}}_{\,\,\,\,\,\, c}  V^c. 
\label{spin1dep}
\eea}

Now let us consider the spin-$\frac{1}{2}$ dependent terms.
The second term of the second line of (\ref{pole3})
has $ h^i_{\bar{a} \bar{b}} \,  g^{\bar{b} \bar{d}} \,  d^i_{\bar{d} \bar{c}}$
which is equivalent to $- d^0_{\bar{a} \bar{c}}$.


The last term of the fourth line of (\ref{pole3})
contains
$ d^i_{\bar{d} \bar{g}} \, f^{\bar{g} \bar{f}}_{\,\,\,\,\,\, e} \,
f^{\bar{d} \bar{b}}_{\,\,\,\,\,\, h} \, g^{e h}$.
One has the following identity for the first contribution
\bea
 d^i_{\bar{d} \bar{g}} \, f^{\bar{g} \bar{f}}_{\,\,\,\,\,\, e} \,
 f^{\bar{d} \bar{b}}_{\,\,\,\,\,\, h} \, g^{e h} = - N \,
 d^i_{\bar{g} \bar{h}} \, g^{\bar{f} \bar{g}} \,  g^{\bar{b} \bar{h}}.
\label{dffgRel}
\eea
By combining the remaining factor, one obtains
$h^i_{\bar{a} \bar{b}} \, h^0_{\bar{c} \bar{f}} \,
d^i_{\bar{g} \bar{h}} \, g^{\bar{f} \bar{g}} \,  g^{\bar{b} \bar{h}}$
which becomes $h^i_{\bar{a} \bar{b}} \,
 g^{\bar{b} \bar{h}} \,
 d^i_{\bar{h} \bar{c}}$.
 By using the identity (\ref{pole3}), this can be simplified as
 $- d^0_{\bar{a} \bar{c}}$. 
 For the second contribution, one has the factor
 $ h^{\si}_{\bar{d} \bar{f}}  \,
 f^{\bar{g} \bar{f}}_{\,\,\,\,\,\, e} \,
  f^{\bar{d} \bar{b}}_{\,\,\,\,\,\,h} \,
  g^{e h}$. By using the identity
\bea
  h^{\si}_{\bar{d} \bar{f}} \, f^{\bar{g} \bar{f}}_{\,\,\,\,\,\, e} \,
 f^{\bar{d} \bar{b}}_{\,\,\,\,\,\, h} \, g^{e h} = - N \,
 h^{\si}_{\bar{i} \bar{h}} \, g^{\bar{g} \bar{i}} \,  g^{\bar{b} \bar{h}},
\label{idenideniden}
\eea
one can simplify further by combining the other factors.
Effectively, two $ff $ is gone in (\ref{idenideniden}). 
From the relation (\ref{hhrel}), one sees the factor
$  - h^{\si}_{\bar{i} \bar{h}} \,
g^{\bar{h} \bar{b}} \,  h^{i}_{\bar{b} \bar{a}} 
$ which becomes $-\varepsilon^{\si i k} \, h^{k}_{\bar{i} \bar{a}}$.
Finally one can combine this with the factor $ d^{\alpha}_{\bar{c}
  \bar{g}}$. Using the relation (\ref{higdi}),
one has 
$-\varepsilon^{0 i \si \alpha} \, \varepsilon^{\si i k} \,
h^k_{\bar{a} \bar{i}} \, g^{\bar{i} \bar{g}} \, d^{\alpha}_{\bar{g} \bar{c}}$ which
leads to the final result $2 d^{0}_{\bar{a} \bar{c}}$.
The third contribtuion can be calculated as done in the first contribution
and it turns out that  $- 2 d^0_{\bar{a} \bar{c}}$.

For the first term of the last line of (\ref{pole3}),
it is better to consider the particular indices.
One can use the relation (\ref{quarticterm}) for the factor
$\varepsilon h h d f f$ when the index \footnote{We will consider
this particular indices from now on. }
\bea
\rho=i=1, \qquad \mbox{in}  \,\, SU(2) \times SU(2) \,\, \mbox{basis}
\qquad (\mbox{or} \qquad \rho=i=3, \qquad \mbox{in}  \,\, SO(4) \,\, \mbox{basis}).
\label{rhoicon}
\eea
That is, the coefficient tensor of (\ref{quarticterm}) is given by 
\bea
(\frac{1}{2} \, d^{2}_{\bar{f} \bar{d}} \, h^3_{\bar{c} \bar{e}} +
\frac{1}{2} \,  d^{3}_{\bar{c} \bar{f}} \, h^2_{\bar{d} \bar{e}}-
d^{0}_{\bar{c} \bar{d}} \, h^1_{\bar{e} \bar{f}}).
\label{simpleExp}
\eea
Then it is obvious to see that with the factor
$h^{i=1}_{\bar{a} \bar{b}}$ and the factor $ g^{\bar{c} \bar{a}} \, g^{\bar{b} f}$,
one can simplify the corresponding terms.
It turns out that the first term and the third term of (\ref{simpleExp})
will
contribute to $-\frac{1}{2}\, d^{0}_{\bar{d} \bar{e}}$ and
$ d^{0}_{\bar{d} \bar{e}}$ respectively while the second term doesn't
contribute at all.
For the second term of the last line of (\ref{pole3}),
one can analyze similarly.
It turns out that the second term and the third term of (\ref{simpleExp})
will
contribute to $\frac{1}{2}\, d^{0}_{\bar{c} \bar{e}}$ and
$ d^{0}_{\bar{c} \bar{e}}$ respectively while the first term doesn't
contribute at all.
For the last term of the last line of (\ref{pole3}),
it turns out that
 the first term and the second term of (\ref{simpleExp})
will
contribute to $-\frac{1}{2} \, d^{0}_{\bar{c} \bar{d}}$ and
$ -\frac{1}{2} \, d^{0}_{\bar{c} \bar{d}}$ respectively.
The third term will contribute to $4N \,  d^{0}_{\bar{c} \bar{d}}$
because of $\delta^{\bar{a}}_{\bar{a}}=4N$.

By collecting the all contributions (from
spin-$\frac{1}{2}$ dependent parts) with the correct coefficients,
one obtains
{\small \bea
 -
\Bigg[
-2k(N-k) +2 k N+k N +k N+\frac{k}{2} +\frac{k}{2}-k (4N+1)
\Bigg]  d^{0}_{\bar{a} \bar{b}} \, Q^{\bar{a}} \, Q^{\bar{b}} =
 2k(N-k) \,
d^{0}_{\bar{a} \bar{b}} \, Q^{\bar{a}} \, Q^{\bar{b}},
\label{spinhalfdep}
\eea}
with the overall factor $\frac{1}{(k+N+2)^3}$.
Then,
one obtains the expression from the third order pole
as the sum of (\ref{spin1dep}) and (\ref{spinhalfdep}).
This leads to, by using the expression of (\ref{spinone}),
\bea
\frac{4(N-k)}{(k+N+2)} \, \Phi_0^{(1)}(w).
\label{firstfactor}
\eea
Furthermore, the derivative
term in (\ref{spin5halfrel}) has the following OPE
in $SO(4)$ basis with $G^3(z)$
{\small
  \bea
{G}^3(z) 
\pa \Phi_{\frac{1}{2}}^{(1),3}(w)  = 
-\frac{1}{(z-w)^3}  4  \Phi_0^{(1)}(w) 
- \frac{1}{(z-w)^2}  3  \pa  \Phi_0^{(1)}(w)
- \frac{1}{(z-w)}  \pa^2  \Phi_{0}^{(1)}(w) 
+  \cdots.
\label{extraope}
\eea}
Therefore, one can check the coefficient of the
third order pole of (\ref{process})
by realizing that, together with (\ref{firstfactor}) (there is a minus sign
in (\ref{spin5halfrel})),
\bea
\Bigg[
  -\frac{4(N-k)}{(k+N+2)} + \frac{(N-k)}{3(k+N+2)} \times (-4)
\Bigg] \,  \Phi_0^{(1)}(w)
  =
-\frac{16(N-k)}{3(k+N+2)} \,  \Phi_0^{(1)}(w).
\label{total}
\eea
Then the total contribution in (\ref{total}), from the
third order pole (\ref{pole3}) and the third order pole
(\ref{extraope}), coincides with the one in
(\ref{process}). One can easily check that there is no contribution
from the OPE between
the first three terms from (\ref{spin5halfrel}) and $G^3(z)$ current.

So far, the $\mu=\nu=1$ case in $SU(2) \times SU(2)$ basis
(or  the $\mu=\nu=3$ case in $SO(4)$ basis)
is considered.
What happens for $\mu \neq \nu$ case?
For example, when $\mu=4$ and $\nu=3$ in $SO(4)$ basis of (\ref{process})
(or $\mu=2$ and $\nu=1$ in $SU(2) \times SU(2)$ basis),
one expects that there is no contribution in the third order pole.
For the quadratic terms in the spin-$\frac{1}{2}$ current,
one obtains a symmetric $d^3$ tensor in the second and fourth lines
of (\ref{pole3}) where the $h$ tensor has the index $2$ and the
$d$ tensor or $\varepsilon$ tensor has the index $1$.
On the other hands, the composite field is antisymmetric.
Therefore, there is no contribution.
In the last line of (\ref{pole3}) with above
index assignment, each term has three contributions.
For the first term, one can use the traceless condition for the
$d^1$ tensor ($d^{1}_{\bar{a} \bar{b}} \, g^{\bar{a} \bar{b}}=0$)
in the second contribution
and  there is a symmetric $d^3$ tensor
in the first and last ones. For the second term,
 one can use the traceless condition for the
$d^1$ tensor in the first contribution
and  there is a symmetric $d^3$ tensor
in the  remaining ones.
For the last term,
 one can use the traceless condition for the
$d^1$ tensor in the third contribution
and  there is a symmetric $d^3$ tensor
in the  remaining ones \footnote{
One can analyze the linear terms in the spin-$1$ current. 
It turns out that  there is a symmetric $d^3$ tensor which shares
the two indices of the structure constant. This leads to zero.
Or there exists the factor $h d f f f$ which becomes identically zero.
Furthermore, one can check that there will be no contributions
from the first four terms of (\ref{spin5halfrel}).}.

\subsubsection{The second order pole}

The second order pole (of the OPE
between $\hat{G}^{\mu=i}(z)$ and the expression
appearing the first order pole of the OPE (\ref{5halfcombination})
with $\rho=i$ (no sum over the index $i$)) can be
described as
\bea
&& \frac{i}{(k+N+2)} \, \frac{i}{(k+N+2)^3}
\Bigg[ \nonu \\
  && (3k+N) \, (k+N+2) \, h^i_{\bar{a} \bar{b}} \, d^{i}_{\bar{c} \bar{d}}\,
  ( -(k+N+2)
  \, g^{\bar{c} \bar{a}} \,  V^{\bar{b}} \, V^{\bar{d}} - k \,
  g^{\bar{d} \bar{b}} \, \pa \, Q^{\bar{c}} \, Q^{\bar{a}})
  \nonu \\
 && + (N-k) \, (k+N+2)\,  h^i_{\bar{a} \bar{b}} \, d^{i}_{\bar{c} \bar{d}}\,
  ((k+N+2) \, g^{\bar{c} \bar{a}} \,  f^{\bar{b} \bar{d}}_{\,\,\,\,\,\,e} \,
  \pa \, V^e - 2k \, g^{\bar{d} \bar{b}} \, Q^{\bar{c}} \, \pa \, Q^{\bar{a}}-
   f^{\bar{d} \bar{b}}_{\,\,\,\,\,\,e} \, Q^{\bar{c}} \, Q^{\bar{a}} \, V^e)
  \nonu \\
  && + (2  h^i_{\bar{a} \bar{b}} \,  h^0_{\bar{c} \bar{f}} \,
  d^{i}_{\bar{d} \bar{g}}\,   f^{\bar{g} \bar{f}}_{\,\,\,\,\,\,e} -\frac{1}{2} \,
  h^i_{\bar{a} \bar{b}} \, \varepsilon^{0 i \si \alpha} \,
h^{\si}_{\bar{d} \bar{f}} \,  d^{\alpha}_{\bar{c} \bar{g}} \,
f^{\bar{g} \bar{f}}_{\,\,\,\,\,\, e} \, + \frac{1}{2} \,  h^i_{\bar{a} \bar{b}} \,
 \varepsilon^{0 i \si \alpha} \,
h^{\si}_{\bar{c} \bar{f}} \,  d^{\alpha}_{\bar{g} \bar{d}} \,
f^{\bar{g} \bar{f}}_{\,\,\,\,\,\, e}
)
  \nonu \\
  && \times ((k+N+2) g^{\bar{c} \bar{a}} \,
  f^{\bar{b} \bar{d}}_{\,\,\,\,\,\,f}  \, V^f \, V^e +
  (k+N+2) \, g^{\bar{c} \bar{a}} \, f^{\bar{b} e}_{\,\,\,\,\,\,f}
  \,  V^{\bar{d}} \, V^f - k \, g^{\bar{d} \bar{b}} \, Q^{\bar{c}} \,
  Q^{\bar{a}} \, V^{e}
  \nonu \\
  && - f^{\bar{d} \bar{b}}_{\,\,\,\,\,\,g} \, (k \, g^{eg} \, Q^{\bar{c}} \, \pa
  \, Q^{\bar{a}} +f^{e f}_{\,\,\,\,\,\,g} \, 
  Q^{\bar{c}} \, Q^{\bar{a}} \, V^g) - k \, g^{e\bar{b}} \, Q^{\bar{c}} \,
  V^{\bar{d}} \, Q^{\bar{a}})
  \label{poletwo} \\
  && -    h^i_{\bar{a} \bar{b}} \,
\varepsilon^{0 i \si \alpha} \,
 h^{0}_{\bar{e} \bar{b}'} \, h^{\si}_{\bar{d} \bar{d}'} \,
 d^{\alpha}_{\bar{c} \bar{f}'} \,
f^{\bar{f}' \bar{d}'}_{\,\,\,\,\,\, g} \,
f^{g \bar{b}'}_{\,\,\,\,\,\, f} \,
((k+N+2) \, g^{\bar{c} \bar{a}} \, f^{\bar{b} \bar{f}}_{\,\,\,\,\,\, g} \,
Q^{\bar{d}} \,
Q^{\bar{e}} \, V^g \nonu \\
&&
- (k+N+2)\, g^{\bar{d} \bar{a}} \,  f^{\bar{b} \bar{f}}_{\,\,\,\,\,\, g}  \,
Q^{\bar{c}} \, Q^{\bar{e}} \, V^g+
(k+N+2) \, g^{\bar{e} \bar{a}} \, f^{\bar{b} \bar{f}}_{\,\,\,\,\,\, g}  \,
Q^{\bar{c}} \, Q^{\bar{d}} \, V^g -k \, g^{\bar{f} \bar{b}} \, Q^{\bar{c}} \,
Q^{\bar{d}} \, Q^{\bar{e}} \, Q^{\bar{a}}) \Bigg](w).
\nonu
\eea
Let us simplify this expression.
First of all, let us look at the cubic terms.
The last term of the second line in (\ref{poletwo})
has the $h d f$ factor which is symmetric in the indices of
$\bar{a}$ and $\bar{c}$ according to the identity (\ref{symmetric})
while the $Q^{\bar{a}} \, Q^{\bar{c}}(w)$ term is antisymmetric in these
indices. Then there is no contribution from this term. 
Let us consider the last term of fourth line. The factor
$h^i_{\bar{a} \bar{b}} \, g^{\bar{b} \bar{d}} \, d^i_{\bar{d} \bar{g}}$
can be simplified as $-d^0_{\bar{a} \bar{g}}$. With other factor,
this will lead to $-h^0_{\bar{c} \bar{f}} \, d^0_{\bar{a} \bar{g}} \,
f^{\bar{g} \bar{f}}_{\,\,\,\,\,\,e}$ which is symmetric in the indices of
$\bar{a}$ and $\bar{c}$ according to the identity (\ref{symmetric}).
One has zero contribution from this term also.
The second contribution of this term has the factor
$h^i_{\bar{a} \bar{b}} \, g^{\bar{b} \bar{d}} \, h^{\si}_{\bar{d} \bar{f}}$
which is equal to $\varepsilon^{i \si k} \, h^k_{\bar{a} \bar{f}}$.
With other factor, this implies that one has
$ \varepsilon^{0 i \si \alpha} \,
\varepsilon^{i \si k} \, h^k_{\bar{a} \bar{f}} \, d^{\alpha}_{\bar{c} \bar{g}} \,
f^{\bar{g} \bar{f}}_{\,\,\,\,\,\,e}$ which is also symmetric in the indices
$\bar{a}$ and $\bar{c}$ and then there is no contribution.
Similarly, the third contribution contains the factor
$h^i_{\bar{a} \bar{b}} \, g^{\bar{b} \bar{d}} \, d^{\alpha}_{\bar{d} \bar{g}}$
which is given by $\varepsilon^{i \alpha k} \, d^k_{\bar{a} \bar{g}}$.
This gives no contribution by considering other factors also.
The second term of fifth line has 
the following property
\bea
 h^i_{\bar{a} \bar{b}} \,  h^0_{\bar{c} \bar{f}} \,
 d^{i}_{\bar{d} \bar{g}}\,   f^{\bar{g} \bar{f}}_{\,\,\,\,\,\,e} \,
 f^{\bar{d} \bar{b}}_{\,\,\,\,\,\,h} \, f^{e h}_{\,\,\,\,\,\,j},  
 \qquad
 \mbox{symmetric} \,\, \mbox{in} \,\, \mbox{the} \,\,
 \mbox{indices} \,\, \bar{a}, \,\, \bar{c}.
 \label{SYMM}
 \eea
 Moreover, the composite field $Q^{\bar{c}} \, Q^{\bar{a}}(w)$
 is antisymmetric
 in their indices.
 The second contribution of this term contains
 $ h h d f f f$ factor which is also symmetric under the indices
 $\bar{a}$ and $\bar{c}$. There is no contribution from the
 third contribution with same reason:
  the $ h h d f f f$ factor is also symmetric under the indices
 $\bar{a}$ and $\bar{c}$.
  The last term of fifth line vanishes due to the previous relation
  (\ref{fgrelation}).

Let us look at the sixth line.
Here   
 the relation (\ref{quarticterm}) for the factor
 $\varepsilon h h d f f$, when the index $\rho=i=1$ (\ref{rhoicon}),
 can be used as before in (\ref{simpleExp}).
 It is easy to see that by using the previous property in
 (\ref{symmetric}), there is no contribution because
 the composite field 
 $Q^{\bar{d}} \, Q^{\bar{e}}(w)$ is antisymmetric
 in their indices.
 There are first two terms in the seventh line.
One can check that there is no contribution from the same analysis.

Let us consider the quadratic terms in the spin-$\frac{1}{2}$
current. The last term of the first line contains
$ d^{0}_{\bar{c} \bar{a}}$.
Similarly,
the second term of the second line contains
$ d^{0}_{\bar{c} \bar{a}}$.
The first term in the fifth line
has three contributions.
The identity in (\ref{dffgRel}) can be used here.
The coefficients of the
contributions are given by $2 k N$, $k N$ and $k N$
respectively.
Then one obtains
{\small
  \bea
 -
\Bigg[ -k (3k+N)
  -2k(N-k) +2 k N+k N +k N
\Bigg]  d^{0}_{\bar{a} \bar{b}} \, Q^{\bar{a}} \, \pa \, Q^{\bar{b}} =
 -\frac{k(N-k)}{2} \,
\pa \, ( d^{0}_{\bar{a} \bar{b}} \, Q^{\bar{a}} \, Q^{\bar{b}}),
\label{contri1}
\eea}
with $\frac{1}{(k+N+2)^3}$ factor.
One observes that there is no contribution from the last term
in (\ref{poletwo}).

Let us describe the quadratic terms in the
spin-$1$ current.
Note that one can rewrite them in terms of the linear term with
derivative by using the following property
\bea
d^{0}_{\bar{a} \bar{b}} \, V^{\bar{a}} \, V^{\bar{b}} =
\frac{1}{2} \, d^{0}_{\bar{a} \bar{b}} \, [ V^{\bar{a}}, V^{\bar{b}}]
= - \frac{1}{2} \, d^{0}_{\bar{a} \bar{b}} \, 
f^{\bar{a} \bar{b}}_{\,\,\,\,\,\,c} \, \pa V^{\bar{c}}.
\label{Property}
\eea
The first term of (\ref{poletwo}) contains
$d^{0}_{\bar{a} \bar{b}}$. Similarly, the first term of the
second line has
$d^{0}_{\bar{a} \bar{b}} \, f^{\bar{a} \bar{b}}_{\,\,\,\,\,\,e}$.
The first term of fourth line has the three contributions.
There are no contributions in the first one and the last one.
For the second contribution, the $ h d f f $ factor leads to
$d^0_{\bar{d} \bar{h}}$ with $N$.
The second term of fourth line has the three contributions.
There is no contribution in the second one.
For the first and last contributions,
the $ h d f f $ factor leads to
$d^0_{\bar{d} \bar{h}}$ with $N$.
By collecting all the contributions,
one obtains the final result, with (\ref{Property}),  as follows:
\bea
-
\Bigg[ \frac{1}{2}\, (3k+N)
  +(N-k) - N-\frac{N}{2} -\frac{N }{2}
\Bigg]  d^{0}_{\bar{a} \bar{b}} \, 
f^{\bar{a} \bar{b}}_{\,\,\,\,\,\,c} \, \pa V^{\bar{c}}=
\frac{(N-k)}{2} \, d^{0}_{\bar{a} \bar{b}} \, 
f^{\bar{a} \bar{b}}_{\,\,\,\,\,\,c} \, \pa V^{\bar{c}},
\label{contri2}
\eea
with the $\frac{1}{(k+N+2)^2}$ factor.

Therefore, the total contribution from the second order pole,
the sum of (\ref{contri1}) and (\ref{contri2}), is
given by the derivative of the higher spin-$1$ current
\bea
-\frac{(N-k)}{(k+N+2)} \, \pa \, \Phi_0^{(1)}(w).
\label{express}
\eea
Again the explicit form of (\ref{spinone}) is used here (\ref{express}).
One can check that the coefficient of the
second order pole of (\ref{process})
vanishes
by realizing that
\bea
\Bigg[
  \frac{(N-k)}{(k+N+2)} + \frac{(N-k)}{3(k+N+2)} \times (-3)
\Bigg]
  = 0,
\label{vanishing}
\eea
where the second term comes from (\ref{extraope}).
Therefore, as we expect, this (\ref{vanishing})
is consistent with (\ref{process}).
As before,
there is no contribution
from the second order pole of the OPE between
the first three terms from (\ref{spin5halfrel}) and $G^3(z)$ current.

So far, the $\mu=\nu=3$ case in $SO(4)$ basis is considered.
What happens for $\mu \neq \nu$ case?
For example, when $\mu=4$ and $\nu=3$ in (\ref{process}),
one expects that there is a nontrivial contribution in the second
order pole.

First of all, one can calculate the following OPE
\bea
-{G}^4(z) \,
\tilde{\Phi}_{\frac{3}{2}}^{(1),3}(w) & = &
\frac{1}{(z-w)^2} \, \Bigg[ -\frac{(k-N)}{3(k+N+2)}
  \, \Phi_{1}^{(1),12}-\frac{6}{(k+N+2)} \,  \Phi_{1}^{(1),34} 
  \nonu \\
  & - &  \frac{4 i}{(k+N+2)} \,  T^{12} \, \Phi_0^{(1)} \Bigg](w)
+ {\cal O} (\frac{1}{(z-w)}). 
\label{temp}
\eea
Here $\tilde{\Phi}_{\frac{3}{2}}^{(1),3}(w)$ is introduced by
the first four terms in (\ref{spin5halfrel}).
The quartic term appearing in the last term of (\ref{poletwo})
corresponds to the one in the last term in the second order pole
of (\ref{temp}).  
With the help of the tensor structure appearing
in the coefficient of
(\ref{quarticterm}), one can check that the second term does not contribute
and the first and the last terms can contribute. 
It turns out that
one obtains, by multiplying the quartic terms of spin-$\frac{1}{2}$
current,
\bea
-\frac{k}{2(k+N+2)^4} \,  d^{0}_{\bar{a} \bar{b}} \,
 h^{3}_{\bar{c} \bar{d}} \, Q^{\bar{a}} \, Q^{\bar{b}} \, Q^{\bar{c}} \, Q^{\bar{d}},
\label{fourq-1}
\eea
which is exactly same as the one in the last term appearing in the
second order pole of (\ref{temp}) after some calculation.
There is also the derivative term
in $ T^{12} \, \Phi_0^{(1)}(w)$ which contains (\ref{fourq-1}).
For the quadratic terms in the spin-$\frac{1}{2}$ current,
one can collect all the contributions and obtain
\bea
&& - \Bigg[ -(3k+N) k + 2k (N-k) -2k N -k N -k N
  \Bigg]  \, d^{3}_{\bar{a} \bar{b}} \,
Q^{\bar{a}} \, \pa \, Q^{\bar{b}} =
 k(5k+3N)
\, d^{3}_{\bar{a} \bar{b}} \,
Q^{\bar{a}} \, \pa \, Q^{\bar{b}},
\nonu
\eea
together with $\frac{1}{(k+N+2)^3}$
which is exactly same as the corresponding terms  in
\bea
\frac{(k-N)}{(k+N+2)}\, \Phi_1^{(1),12} + \frac{2(1+2k+2N)}{(k+N+2)} \,
\Phi_1^{(1),34} -\frac{4 i}{(k+N+2)} \,  T^{12} \, \Phi_0^{(1)}.
\label{EExp}
\eea
Note that the factor $h d g$ in the higher spin-$2$ currents
in (\ref{EExp})
reduces to the $d^3$ tensor
\footnote{One can easily see that the second order pole of the OPE
  between $G^4(z)$ and $\Phi_{\frac{3}{2}}^{(1),3}(w)$
  is given by $ \Bigg[ \frac{4(k-N)}{3(k+N+2)} \, \Phi_1^{(1),12} + 4
    \Phi_1^{(1),34} \Bigg](w)$.
  By combining this with (\ref{temp}), one obtains
the relation (\ref{EExp}).}. 

For the quadratic terms in spin-$1$ current,
one has the final expression
\bea
\frac{1}{(k+N+2)^2} \Bigg[ 3k+N +2 N +N +N+2\Bigg]
 d^{3}_{\bar{a} \bar{b}} \,
 V^{\bar{a}}  \, V^{\bar{b}} =
 \frac{(3k+5N+2)}{(k+N+2)^2} \,  d^{3}_{\bar{a} \bar{b}} \,
 V^{\bar{a}}  \, V^{\bar{b}}, 
\label{vv}
\eea
which corresponds to the ones for the first two terms in (\ref{EExp}).
There are also some contributions from the first term in the fourth line
of (\ref{poletwo}) in addition to (\ref{vv}). The first contribution
can be simplified as $h d f f$ factor which does not vanish but
when one multiples the composite field $V^h \, V^e$, then this becomes zero.
Similarly,
the second contribution
can be simplified as $h d f f$ factor which does not vanish but
when the composite field $V^h \, V^e$ is added, then
this becomes zero.
Finally, the second contribution
has nontrivial nonzero expression. In this case,
one arrives at the factor $h d f f$.
\bea
d^{3}_{\bar{g} \bar{d}} \,  h^{0}_{\bar{b} \bar{f}} \,
f^{\bar{g} \bar{f}}_{\,\,\,\,\,\,e} \, f^{\bar{b} \bar{d}}_{\,\,\,\,\,\,h} 
= \frac{1}{2N} \,
h^{3}_{\bar{b} \bar{d}} \,  d^{0}_{\bar{g} \bar{f}} \,
f^{\bar{g} \bar{f}}_{\,\,\,\,\,\,e} \, f^{\bar{b} \bar{d}}_{\,\,\,\,\,\,h} -
d^{1}_{\bar{d} \bar{g}} \,  h^{2}_{\bar{f} \bar{b}} \,
f^{\bar{g} \bar{f}}_{\,\,\,\,\,\,e} \, f^{\bar{b} \bar{d}}_{\,\,\,\,\,\,h}.
\label{nontrivial}
\eea
By acting on the composite field 
$V^h \, V^e$ into (\ref{nontrivial}), then
the second term vanishes and the remaining one is exactly
same as the corresponding one in the last term in (\ref{EExp}) \footnote{
One can also analyze the mixed cubic terms. There is nontrivial
factor $ h h d f f f $ in front of the composite field
$Q^{\bar{c}} Q^{\bar{a}} V^g(w)$ in the fifth line of (\ref{poletwo}).
One expects that this will be written in terms of the factor $h d f$.
Although the explicit details for this calculation are ignored in this
paper, the similar features will arise at the end of this section. }.

Let us consider the final first order pole terms
which will generate the higher spin-$3$ current in terms of
WZW currents.

\subsubsection{The first order pole}

The first order pole (of the OPE
between $\hat{G}^{\mu=i}(z)$ and the expression
appearing the first order pole of the OPE (\ref{5halfcombination})
with $\rho=i$) can be written as 
{\small
  \bea
&& \frac{i}{(k+N+2)} \, \frac{i}{(k+N+2)^3}
\Bigg[ \nonu \\
  && (3k+N) \, (k+N+2) \, h^i_{\bar{a} \bar{b}} \, d^{i}_{\bar{c} \bar{d}}\,
  ( (k+N+2)
  \, (-g^{\bar{c} \bar{a}} \,  V^{\bar{d}} \, \pa \, V^{\bar{b}} +\frac{1}{2} \,
  g^{\bar{c} \bar{a}} \,  f^{\bar{b} \bar{d}}_{\,\,\,\,\,\,e} \,
  \pa^2 \, V^{e}) \nonu \\
  && - k \, g^{\bar{d} \bar{b}} \, \pa \, Q^{\bar{c}} \, \pa
  Q^{\bar{a}}-  f^{\bar{b} \bar{d}}_{\,\,\,\,\,\,e} \, Q^{\bar{a}} \, \pa \,
  Q^{\bar{c}} \, V^e)
  \nonu \\
 && + (N-k) \, (k+N+2)\,  h^i_{\bar{a} \bar{b}} \, d^{i}_{\bar{c} \bar{d}}\,
  (-(k+N+2) \, g^{\bar{c} \bar{a}}  \,
  V^{\bar{b}} \, \pa \, V^d - k \, g^{\bar{d} \bar{b}} \, Q^{\bar{c}} \, \pa^2
  \, Q^{\bar{a}}-
   f^{\bar{d} \bar{b}}_{\,\,\,\,\,\,e} \, Q^{\bar{c}} \, \pa \, (Q^{\bar{a}} \, V^e))
  \nonu \\
  && + (2  h^i_{\bar{a} \bar{b}} \,  h^0_{\bar{i} \bar{f}} \,
  d^{i}_{\bar{j} \bar{g}}\,   f^{\bar{g} \bar{f}}_{\,\,\,\,\,\,k} -\frac{1}{2} \,
  h^i_{\bar{a} \bar{b}} \, \varepsilon^{0 i \si \alpha} \,
h^{\si}_{\bar{j} \bar{f}} \,  d^{\alpha}_{\bar{i} \bar{g}} \,
f^{\bar{g} \bar{f}}_{\,\,\,\,\,\, k} \, + \frac{1}{2} \,  h^i_{\bar{a} \bar{b}} \,
 \varepsilon^{0 i \si \alpha} \,
h^{\si}_{\bar{i} \bar{f}} \,  d^{\alpha}_{\bar{g} \bar{j}} \,
f^{\bar{g} \bar{f}}_{\,\,\,\,\,\, k}
)
  \nonu \\
  && \times (-(k+N+2) g^{\bar{i} \bar{a}} \,
  V^{\bar{b}} \, V^{\bar{j}} \, V^k -
  k \, g^{\bar{j} \bar{b}} \, Q^{\bar{i}}
  \,  \pa \, Q^{\bar{a}} \, V^k - f^{\bar{j} \bar{b}}_{\,\,\,\,\,\,l}
  \, Q^{\bar{i}} \,
  Q^{\bar{a}} \, V^{k} \, V^l
  \nonu \\
  && - f^{\bar{j} \bar{b}}_{\,\,\,\,\,\,l} \,
f^{k l}_{\,\,\,\,\,\,m} \,
 Q^{\bar{i}} \, \pa
 \, ( Q^{\bar{a}} \, V^{m})
 -f^{k \bar{b}}_{\,\,\,\,\,\,l} \, 
 Q^{\bar{i}} \, Q^{\bar{a}} \, V^{\bar{j}} \, V^l -
 \frac{k}{2} \, f^{\bar{j} \bar{b}}_{\,\,\,\,\,\,l} \,
 g^{k l} \, Q^{\bar{i}} \,
 \pa^2 \, Q^{\bar{a}} - k\, g^{k \bar{b}} \, Q^{\bar{i}} \, \pa Q^{\bar{a}} \,
 V^{\bar{j}})
  \label{poleoneone} \\
  && -    h^i_{\bar{a} \bar{b}} \,
\varepsilon^{0 i \si \alpha} \,
 h^{0}_{\bar{k} \bar{b}'} \, h^{\si}_{\bar{j} \bar{d}'} \,
 d^{\alpha}_{\bar{i} \bar{f}'} \,
f^{\bar{f}' \bar{d}'}_{\,\,\,\,\,\, g} \,
f^{g \bar{b}'}_{\,\,\,\,\,\, l} \,
((k+N+2) \, g^{\bar{k} \bar{a}}  \,
Q^{\bar{j}} \,
Q^{\bar{i}} \, V^{\bar{b}} \, V^{\bar{l}} \nonu \\
&&
- (k+N+2)\, g^{\bar{j} \bar{a}}   \,
Q^{\bar{k}} \, Q^{\bar{i}} \, V^{\bar{b}}\, V^{\bar{l}}+
(k+N+2) \, g^{\bar{i} \bar{a}} \, 
Q^{\bar{k}} \, Q^{\bar{j}} \, V^{\bar{b}} \, V^{\bar{l}} +
f^{\bar{l} \bar{b}}_{\,\,\,\,\,\, m} \, Q^{\bar{k}} \, Q^{\bar{j}} \, Q^{\bar{i}}
\, Q^{\bar{a}} \, V^m
\nonu \\
&& +k \, g^{\bar{l} \bar{b}} \, Q^{\bar{k}} \,
Q^{\bar{j}} \, Q^{\bar{i}} \, \pa \, Q^{\bar{a}}) \Bigg](w), \qquad
\mbox{no sum over the index} \,\, i.
\nonu
\eea}
Let us simplify this first order pole of (\ref{poleoneone}). 
The first term can be written as
$-h^i_{\bar{b} \bar{a}} \, g^{\bar{a} \bar{c}} \, d^i_{\bar{c} \bar{d}}$ with other
factors. This is equal to $d^0_{\bar{b} \bar{d}}$ from (\ref{higdi}). 
Similarly, the second term has  $d^0_{\bar{b} \bar{d}}$.
The third term also has $-d^0_{\bar{a} \bar{c}}$ from the  $ h g d$ factor.
The fourth term cannot be simplified further.
The fifth term contains $d^0_{\bar{b} \bar{d}}$.
The sixth term contains $-d^0_{\bar{a} \bar{c}}$.
The seventh term cannot be simplified further.
Note that the term with derivative acting on $V^e$
becomes zero.

Let us look at the cubic in spin-$1$ currents.
The first contribution is given by $h^i_{\bar{f} \bar{b}}$.
The second contribution has the factor
$-h^i_{\bar{b} \bar{a}} \, g^{\bar{a} \bar{i}} \, d^{\alpha}_{\bar{i}
  \bar{g}}$ which can be written as $-\varepsilon^{i \alpha k}\,
 d^{k}_{\bar{b}
   \bar{g}}$. Combining with other piece, one obtains
   $-\varepsilon^{0 i  \alpha \si }\, \varepsilon^{i \alpha k} \,
 h^{\si}_{\bar{j} \bar{f}} \, d^{k}_{\bar{b}
   \bar{g}}\,  f^{\bar{f} \bar{g}}_{\,\,\,\,\,\,k}$.
 The third contribution has the factor
 $ -h^{i}_{\bar{b} \bar{a}} \,  g^{\bar{a} \bar{i}} \,
 h^{\si}_{\bar{i} \bar{f}}$ which leads to $- \varepsilon^{i  \si k  }\,
 h^{k}_{\bar{b} \bar{f}}$. Then with other factor, one has
 $\varepsilon^{0 i  \si \alpha  }\, \varepsilon^{i \si k} \,
 h^{k}_{\bar{b} \bar{f}} \, d^{\alpha}_{\bar{i}
   \bar{g}}\,  f^{\bar{f} \bar{g}}_{\,\,\,\,\,\,k}$.
One can easily see that by considering the additional signs,
the final contributions from the second and third terms are equal to 
each other.
Let us move on the next term $Q^{\bar{i}} \, \pa \, Q^{\bar{a}} \, V^{k}(w)$.
The first contribution has the factor $h^{i}_{\bar{a} \bar{b}} \,
g^{\bar{b} \bar{j}} \, d^i_{\bar{j} \bar{g}}$ which is equal to
$- d^0_{\bar{a} \bar{g}}$. Then one obtains
the final form $h^0_{\bar{i} \bar{f}} \, d^0_{\bar{a} \bar{g}} \,
f^{\bar{f} \bar{g}}_{\,\,\,\,\,\,k}$.
One can consider the second contribution. The $h h g $
factor leads to $h^{i}_{\bar{a} \bar{b}} \, g^{\bar{b} \bar{j}} \,
h^{\si}_{\bar{j} \bar{f}}$ which is given by $\varepsilon^{i \si k}\,
h^{k}_{\bar{a} \bar{f}}$. With other factor, one has
$-\varepsilon^{0 i \si \alpha}\, \varepsilon^{i \si k}\,
h^{k}_{\bar{a} \bar{f}} \,
d^{\alpha}_{\bar{i} \bar{g}} \, f^{\bar{f} \bar{g}}_{\,\,\,\,\,\,k}$.
Similarly, the third contribution can be
obtained and one has the factor
$h^{i}_{\bar{a} \bar{b}} \, g^{\bar{b} \bar{j}} \,
d^{\alpha}_{\bar{j} \bar{g}}$. This becomes
$\varepsilon^{i \alpha k}\,
d^{k}_{\bar{a} \bar{g}}$. Then one obtains 
$\varepsilon^{0 i  \alpha \si}\, \varepsilon^{i \alpha k}\,
h^{\si}_{\bar{i} \bar{f}} \,
d^{k}_{\bar{a} \bar{g}} \, f^{\bar{f} \bar{g}}_{\,\,\,\,\,\,k}$.
One realizes that
the final contributions from the second and third terms are equal to 
each other.
Let us consider the term $Q^{\bar{i}} \, \pa^2 \, Q^{\bar{a}}(w)$.
One can see the factor $d^i_{\bar{j} \bar{g}} \,
f^{\bar{g} \bar{f}}_{\,\,\,\,\,\,k} \, f^{\bar{j} \bar{b}}_{\,\,\,\,\,\,l}
\, g^{kl}$.
From the identity
\bea
d^i_{\bar{j} \bar{g}} \,
f^{\bar{g} \bar{f}}_{\,\,\,\,\,\,k} \, f^{\bar{j} \bar{b}}_{\,\,\,\,\,\,l}
\, g^{k l} = -N \, d^i_{\bar{c} \bar{d}} \,
 g^{\bar{c} \bar{f}} \,  g^{\bar{d} \bar{b}},
\label{Idenone}
\eea
one can simplify further.
With other factor, one obtains
$ -N \, h^i_{\bar{a} \bar{b}}  \,  h^0_{\bar{i} \bar{f}} \,
d^i_{\bar{c} \bar{d}} \,
 g^{\bar{c} \bar{f}} \,  g^{\bar{d} \bar{b}}
$. Then one has $ -N \, h^i_{\bar{a} \bar{b}}  \,  g^{\bar{b} \bar{d}} \,
d^i_{\bar{d} \bar{i}}$ which leads to $N \, d^0_{\bar{a} \bar{i}}$.
The second contribution can be obtained similarly.
By realizing that the factor $h^{\si}_{\bar{j} \bar{f}} \,
f^{\bar{g} \bar{f}}_{\,\,\,\,\,\,k} \, f^{\bar{j} \bar{b}}_{\,\,\,\,\,\,l} \,
g^{k l}$ can be written as
further simple form, by using the identity
\bea
h^{\si}_{\bar{j} \bar{f}} \,
f^{\bar{g} \bar{f}}_{\,\,\,\,\,\,k} \, f^{\bar{j} \bar{b}}_{\,\,\,\,\,\,l} \,
g^{k l} = -N \, h^{\si}_{\bar{c} \bar{d}} \, g^{\bar{c} \bar{g}} \, g^{\bar{d}
  \bar{b}},
\label{Identwo}
\eea
with other factor, one has
$N \, \varepsilon^{0 i \si \alpha} \, 
h^{i}_{\bar{a} \bar{b}} \, d^{\alpha}_{\bar{i} \bar{g}}
 h^{\si}_{\bar{d} \bar{c}} \, g^{\bar{c} \bar{g}} \, g^{\bar{b} \bar{d}}
 $. Then from the relation (\ref{hhrel}), one can replace
 the $h  g  h$ factor with $\varepsilon^{i \si k} \, h^k_{\bar{a} \bar{c}}$.
 Then the above expression leads to $N \, \varepsilon^{0 i \si \alpha} \,
 \varepsilon^{i \si k} \, h^k_{\bar{a} \bar{c}} \,   g^{\bar{c} \bar{g}} \,
 d^{\alpha}_{\bar{g} \bar{i}}
$. This is equal to $-2N \, d^{0}_{\bar{a} \bar{i}}$ from (\ref{higdi}).
 One can analyze the third contribution similarly. One can consider the
 factor $d^{\alpha}_{\bar{g} \bar{j}} \,
f^{\bar{g} \bar{f}}_{\,\,\,\,\,\,k} \, f^{\bar{j} \bar{b}}_{\,\,\,\,\,\,l} \,
g^{k l}$. With the help of the following identity
\bea
d^{\alpha}_{\bar{g} \bar{j}} \,
f^{\bar{g} \bar{f}}_{\,\,\,\,\,\,k} \, f^{\bar{j} \bar{b}}_{\,\,\,\,\,\,l} \,
g^{k l} = -N \, d^{\alpha}_{\bar{c} \bar{d}} \, g^{\bar{c} \bar{f}} \, g^{\bar{d}
  \bar{b}}, 
\label{Identhree}
\eea
one obtains $-N \, \varepsilon^{0 i \si \alpha} \,
h^i_{\bar{a} \bar{b}} \, h^{\si}_{\bar{i} \bar{f}} \, d^{\alpha}_{\bar{d} \bar{c}}
\, g^{\bar{c} \bar{f}} \, g^{\bar{b} \bar{d}}$. Then the $h g d $
factor becomes $\varepsilon^{i \alpha k} \, d^k_{\bar{a} \bar{c}}$ and
one obtains $N \, \varepsilon^{0 i  \alpha \si} \, \varepsilon^{i \alpha k} \,
h^{\si}_{\bar{i} \bar{f}} \,  g^{\bar{f} \bar{c}} \,  d^k_{\bar{c} \bar{a}}$
which leads to $2 N\,  d^0_{\bar{i} \bar{a}}$ according to (\ref{higdi}).
The contribution from the term $Q^{\bar{i}} \, \pa \, Q^{\bar{a}} \,
V^{\bar{j}}(w)$ vanishes because of (\ref{fgrelation}).
It would be interesting to observe the above identities (\ref{Idenone}),
(\ref{Identwo}), and
(\ref{Identhree}) in the general context.

Let us simplify the first term of the seventh line of (\ref{poleoneone}).
Again by using the coefficient tensor appearing in (\ref{quarticterm}),
one obtains the following factor
\bea
(k+N+2) \Bigg[-\frac{1}{2}\, h^2_{\bar{b} \bar{c}} \, d^2_{\bar{d} \bar{a}}
  +\frac{1}{2}\, h^3_{\bar{a} \bar{c}} \, d^3_{\bar{b} \bar{d}}
+  h^0_{\bar{c} \bar{d}} \, d^0_{\bar{b} \bar{a}}
  \Bigg] \, Q^{\bar{a}} \, Q^{\bar{b}} \, V^{\bar{c}} \, V^{\bar{d}}.
\label{qqvvexp}
\eea
It turns out that the
the first term of the eighth line of (\ref{poleoneone}) contains the
factor
\footnote{
Similarly, it is easy to see that
the second term of the eighth line
has the factor
$
(k+N+2) \left[-\frac{1}{2}\, h^2_{\bar{c} \bar{a}} \, d^2_{\bar{d} \bar{b}}
  -\frac{1}{2}\, h^2_{\bar{b} \bar{a}} \, d^2_{\bar{c} \bar{d}}
-  h^1_{\bar{a} \bar{d}} \, d^1_{\bar{c} \bar{b}}
  \right] \, Q^{\bar{a}} \, Q^{\bar{b}} \, V^{\bar{c}} \, V^{\bar{d}}$.
One cannot simplify the third term of the eight line further.
The last term of (\ref{poleone})
contains the factor
$
\left[ -\frac{1}{2} \, h^3_{\bar{a} \bar{b}} \, d^3_{\bar{c} \bar{d}}
  + \frac{1}{2} \, h^2_{\bar{b} \bar{a}} \, d^2_{\bar{d} \bar{c}}
  + h^0_{\bar{d} \bar{a}} \, d^0_{\bar{c} \bar{b}}
  \right] \,  Q^{\bar{a}} \, Q^{\bar{b}} \,  Q^{\bar{c}} \, \pa \,
Q^{\bar{d}}$. 
}
\bea
(k+N+2) \Bigg[-\frac{1}{2}\, h^3_{\bar{b} \bar{a}} \, d^3_{\bar{d} \bar{c}}
  -\frac{1}{2}\, h^3_{\bar{c} \bar{a}} \, d^3_{\bar{b} \bar{d}}
+  h^1_{\bar{a} \bar{d}} \, d^1_{\bar{b} \bar{c}}
  \Bigg] \, Q^{\bar{a}} \, Q^{\bar{b}} \, V^{\bar{c}} \, V^{\bar{d}}.
\label{qqvvexp1}
\eea
This (\ref{qqvvexp1}) looks similar to (\ref{qqvvexp}).

Let us consider the third term of the fifth line.
As before, it has three contributions.
One cannot simplify them further.
Let us look at the first term of the sixth line.
According to the property of (\ref{SYMM}), the  first contribution
occurs when the derivative acts on $Q^{\bar{a}}$.
The corresponding tensor structure has the following form
\bea
\Bigg[ N \,  h^1_{\bar{c} \bar{a}}
    \, d^1_{\bar{d} \bar{b}} 
+(\frac{1}{4}+\frac{1}{2N}) \,  h^1_{\bar{a} \bar{b}}
    \, d^1_{\bar{c} \bar{d}} 
-\frac{1}{4} \,  h^0_{\bar{c} \bar{d}}
    \, d^0_{\bar{a} \bar{b}} 
  -\frac{1}{4} \,  h^2_{\bar{a} \bar{b}}
    \, d^2_{\bar{c} \bar{d}} 
- \frac{1}{4} \,  h^3_{\bar{a} \bar{b}}
    \, d^3_{\bar{c} \bar{d}} \Bigg] \, f^{\bar{a} \bar{b}}_{\,\,\,\,\,\, e},
\label{qdqvexp}
\eea
which acts on the composite field $Q^{\bar{d}} \, \pa \, Q^{\bar{c}}
\, V^e(w)$.
Of course, the $N$-dependence in here
is nontrivial and one can
observe by trying to determine from the
several $N$ values cases.
We will see that the above terms (\ref{qdqvexp}) contribute to the
final higher spin-$3$ current which will appear in next subsection.
Similarly, the second contribution
contains
\bea
\Bigg[ -N \,  h^1_{\bar{c} \bar{a}}
    \, d^1_{\bar{d} \bar{b}} 
-\frac{1}{4} \,  h^1_{\bar{a} \bar{b}}
    \, d^1_{\bar{c} \bar{d}} 
+\frac{1}{4} \,  h^0_{\bar{c} \bar{d}}
    \, d^0_{\bar{a} \bar{b}} 
  + (\frac{1}{4}+\frac{1}{2N} ) \,  h^2_{\bar{a} \bar{b}}
    \, d^2_{\bar{c} \bar{d}} 
+ \frac{1}{4} \,  h^3_{\bar{a} \bar{b}}
    \, d^3_{\bar{c} \bar{d}} \Bigg] \, f^{\bar{a} \bar{b}}_{\,\,\,\,\,\, e}.
\label{qdqvexp1}
\eea
This looks like as (\ref{qdqvexp}) but
the tensorial structure is different from each other.
One can see the $N$-dependence from the explicit expressions
for the several $N$ cases.
The corresponding quantity for the third contribution
is given by
\bea
\Bigg[ N \,  h^1_{\bar{c} \bar{a}}
    \, d^1_{\bar{d} \bar{b}} 
+\frac{1}{4} \,  h^1_{\bar{a} \bar{b}}
    \, d^1_{\bar{c} \bar{d}} 
-\frac{1}{4} \,  h^0_{\bar{c} \bar{d}}
    \, d^0_{\bar{a} \bar{b}} 
  - \frac{1}{4} \,  h^2_{\bar{a} \bar{b}}
    \, d^2_{\bar{c} \bar{d}} 
-( \frac{1}{4} +\frac{1}{2N} ) \,  h^3_{\bar{a} \bar{b}}
    \, d^3_{\bar{c} \bar{d}} \Bigg] \, f^{\bar{a} \bar{b}}_{\,\,\,\,\,\, e}.
\label{qdqvexp2}
\eea
As in (\ref{qdqvexp}) and (\ref{qdqvexp1}), the tensorial
structure of (\ref{qdqvexp2}) looks similar but
is different from those expressions.

Let us move the next term.
The first contribution can have the following tensor structure
{\footnotesize
  \bea
h^{1}_{\bar{a} \bar{b'}} \,
h^{0}_{\bar{b} \bar{c'}} \,
d^{3}_{\bar{c} \bar{d'}} \,
f^{\bar{d'} \bar{c'}}_{\,\,\,\,\,\, k} \,
 f^{\bar{k} \bar{b'}}_{\,\,\,\,\,\, d}
& = &
-\frac{1}{4} \, d^{3}_{\bar{a} \bar{b}} \, h^3_{\bar{c} \bar{d}}
-\frac{1}{4} \,  d^{3}_{\bar{a} \bar{c}} \, h^3_{\bar{d} \bar{b}}
-\frac{1}{4} \, d^{3}_{\bar{d} \bar{c}} \, h^3_{\bar{a} \bar{b}}
+\frac{1}{4} \,  d^{3}_{\bar{d} \bar{b}} \, h^3_{\bar{a} \bar{c}}
+\frac{1}{2} \, d^{3}_{\bar{c} \bar{b}} \, h^3_{\bar{d} \bar{a}}
-\frac{1}{4} \,  d^{2}_{\bar{a} \bar{b}} \, h^2_{\bar{c} \bar{d}}
-\frac{1}{4} \, d^{2}_{\bar{a} \bar{c}} \, h^2_{\bar{d} \bar{b}}
\nonu \\
& - & \frac{1}{4} \,  d^{2}_{\bar{d} \bar{c}} \, h^2_{\bar{a} \bar{b}}
+\frac{1}{4} \,  d^{2}_{\bar{d} \bar{b}} \, h^2_{\bar{a} \bar{c}}
+\frac{1}{2} \, d^{2}_{\bar{c} \bar{b}} \, h^2_{\bar{d} \bar{a}}
+\frac{1}{4} \, d^{1}_{\bar{a} \bar{b}} \, h^1_{\bar{c} \bar{d}}
-\frac{1}{2} \,  d^{1}_{\bar{a} \bar{d}} \, h^1_{\bar{b} \bar{c}}
+\frac{1}{4} \, d^{1}_{\bar{a} \bar{c}} \, h^1_{\bar{d} \bar{b}}
+\frac{1}{4} \,  d^{1}_{\bar{d} \bar{c}} \, h^1_{\bar{a} \bar{b}}
\nonu \\
&  - & \frac{1}{4} \,  d^{1}_{\bar{d} \bar{b}} \, h^1_{\bar{a} \bar{c}}
-\frac{1}{4} \,  d^{0}_{\bar{a} \bar{b}} \, h^0_{\bar{c} \bar{d}}
+\frac{1}{4} \, d^{0}_{\bar{a} \bar{c}} \, h^0_{\bar{d} \bar{b}}
+\frac{1}{4} \, d^{0}_{\bar{d} \bar{c}} \, h^0_{\bar{a} \bar{b}}
-\frac{1}{4} \,  d^{0}_{\bar{d} \bar{b}} \, h^0_{\bar{a} \bar{c}}
-\frac{1}{2} \, d^{0}_{\bar{c} \bar{b}} \, h^0_{\bar{d} \bar{a}},
\label{Someidentity}
\eea}
which acts on the composite field $Q^{\bar{b}} \, Q^{\bar{a}} \, V^{\bar{c}}
\, V^{d}(w)$. One can further simplify this (\ref{Someidentity})
using the property of the indices
$\bar{i}$ and $\bar{a}$ \footnote{
Similarly, one obtains the following structure
for the second contribution
{\footnotesize
\bea
h^{1}_{\bar{b} \bar{b'}} \,
h^{2}_{\bar{c} \bar{c'}} \,
d^{3}_{\bar{a} \bar{d'}} \,
f^{\bar{d'} \bar{c'}}_{\,\,\,\,\,\, k} \,
 f^{\bar{k} \bar{b'}}_{\,\,\,\,\,\, d}
& = &
 \frac{1}{4} \, d^{3}_{\bar{a} \bar{b}} \, h^3_{\bar{c} \bar{d}}
-\frac{1}{4} \,  d^{3}_{\bar{a} \bar{d}} \, h^3_{\bar{b} \bar{c}}
+\frac{1}{2} \, d^{3}_{\bar{a} \bar{c}} \, h^3_{\bar{d} \bar{b}}
-\frac{1}{4} \,  d^{3}_{\bar{d} \bar{c}} \, h^3_{\bar{a} \bar{b}}
+\frac{1}{4} \, d^{3}_{\bar{c} \bar{b}} \, h^3_{\bar{d} \bar{a}}
+\frac{1}{4} \,  d^{2}_{\bar{a} \bar{b}} \, h^2_{\bar{c} \bar{d}}
-\frac{1}{4} \, d^{2}_{\bar{a} \bar{d}} \, h^2_{\bar{b} \bar{c}}
\nonu \\
& - & \frac{1}{2} \,  d^{2}_{\bar{a} \bar{c}} \, h^2_{\bar{d} \bar{b}}
-\frac{1}{4} \,  d^{2}_{\bar{d} \bar{c}} \, h^2_{\bar{a} \bar{b}}
+\frac{1}{4} \, d^{2}_{\bar{c} \bar{b}} \, h^2_{\bar{d} \bar{a}}
-\frac{1}{4} \, d^{1}_{\bar{a} \bar{b}} \, h^1_{\bar{c} \bar{d}}
+\frac{1}{4} \,  d^{1}_{\bar{a} \bar{d}} \, h^1_{\bar{b} \bar{c}}
+\frac{1}{4} \, d^{1}_{\bar{d} \bar{c}} \, h^1_{\bar{a} \bar{b}}
+\frac{1}{2} \,  d^{1}_{\bar{d} \bar{b}} \, h^1_{\bar{a} \bar{c}}
\nonu \\
&  - & \frac{1}{4} \,  d^{1}_{\bar{c} \bar{b}} \, h^1_{\bar{d} \bar{a}}
-\frac{1}{4} \,  d^{0}_{\bar{a} \bar{b}} \, h^0_{\bar{c} \bar{d}}
-\frac{1}{4} \, d^{0}_{\bar{a} \bar{d}} \, h^0_{\bar{b} \bar{c}}
+\frac{1}{2} \, d^{0}_{\bar{a} \bar{c}} \, h^0_{\bar{d} \bar{b}}
-\frac{1}{4} \,  d^{0}_{\bar{d} \bar{c}} \, h^0_{\bar{a} \bar{b}}
+\frac{1}{4} \, d^{0}_{\bar{c} \bar{b}} \, h^0_{\bar{d} \bar{a}},
\nonu
\eea}
which acts on the composite field $Q^{\bar{a}} \, Q^{\bar{b}} \, V^{\bar{c}}
\, V^{d}(w)$.
Finally, the third contribution contains the following tensor structure
{\footnotesize
  \bea
  h^{1}_{\bar{b} \bar{b'}} \,
h^{2}_{\bar{a} \bar{c'}} \,
d^{3}_{\bar{d'} \bar{c}} \,
f^{\bar{d'} \bar{c'}}_{\,\,\,\,\,\, k} \,
 f^{\bar{k} \bar{b'}}_{\,\,\,\,\,\, d}
& = &
 -\frac{1}{4} \, d^{3}_{\bar{a} \bar{b}} \, h^3_{\bar{c} \bar{d}}
-\frac{1}{4} \,  d^{3}_{\bar{a} \bar{c}} \, h^3_{\bar{d} \bar{b}}
-\frac{1}{4} \, d^{3}_{\bar{d} \bar{c}} \, h^3_{\bar{a} \bar{b}}
+\frac{1}{4} \,  d^{3}_{\bar{d} \bar{b}} \, h^3_{\bar{a} \bar{c}}
+\frac{1}{2} \, d^{3}_{\bar{c} \bar{b}} \, h^3_{\bar{d} \bar{a}}
-\frac{1}{4} \,  d^{2}_{\bar{a} \bar{b}} \, h^2_{\bar{c} \bar{d}}
-\frac{1}{4} \, d^{2}_{\bar{a} \bar{c}} \, h^2_{\bar{d} \bar{b}}
\nonu \\
& - & \frac{1}{4} \,  d^{2}_{\bar{d} \bar{c}} \, h^2_{\bar{a} \bar{b}}
+\frac{1}{4} \,  d^{2}_{\bar{d} \bar{b}} \, h^2_{\bar{a} \bar{c}}
-\frac{1}{2} \, d^{2}_{\bar{c} \bar{b}} \, h^2_{\bar{d} \bar{a}}
+\frac{1}{4} \, d^{1}_{\bar{a} \bar{b}} \, h^1_{\bar{c} \bar{d}}
-\frac{1}{2} \,  d^{1}_{\bar{a} \bar{d}} \, h^1_{\bar{b} \bar{c}}
+\frac{1}{4} \, d^{1}_{\bar{d} \bar{c}} \, h^1_{\bar{a} \bar{b}}
+\frac{1}{4} \,  d^{1}_{\bar{d} \bar{c}} \, h^1_{\bar{a} \bar{b}}
\nonu \\
&  - & \frac{1}{4} \,  d^{1}_{\bar{d} \bar{b}} \, h^1_{\bar{a} \bar{c}}
-\frac{1}{4} \,  d^{0}_{\bar{a} \bar{b}} \, h^0_{\bar{c} \bar{d}}
+\frac{1}{4} \, d^{0}_{\bar{a} \bar{c}} \, h^0_{\bar{d} \bar{b}}
+\frac{1}{4} \, d^{0}_{\bar{d} \bar{c}} \, h^0_{\bar{a} \bar{b}}
-\frac{1}{4} \,  d^{0}_{\bar{d} \bar{b}} \, h^0_{\bar{a} \bar{c}}
+\frac{1}{2} \, d^{0}_{\bar{c} \bar{b}} \, h^0_{\bar{d} \bar{a}},
\nonu
\eea}
which acts on the composite field $Q^{\bar{b}} \, Q^{\bar{a}} \, V^{\bar{c}}
\, V^{d}(w)$. The $hff$ factor is gone.}.

Therefore, one can summarize the field contents appearing in the
first order pole in (\ref{poleoneone}).
The spin-$\frac{1}{2}$ dependent terms of 
(\ref{poleoneone}) is given by
\bea
&& \frac{k}{(k+N+2)^4} \,
(d^0_{\bar{a} \bar{b}} \, h^0_{\bar{c} \bar{d}}
+
\frac{1}{2} \, h^2_{\bar{a} \bar{b}} \, d^2_{\bar{c} \bar{d}}
+\frac{1}{2} \, h^3_{\bar{a} \bar{b}} \, d^3_{\bar{c} \bar{d}}) \,
Q^{\bar{a}} \, Q^{\bar{b}} \, Q^{\bar{c}} \, \pa \, Q^{\bar{d}}(z)
\nonu \\
& &- \frac{k(k+N)}{
  (k+N+2)^3} \, d^0_{\bar{a} \bar{b}} \, Q^{\bar{a}} \,
\pa^2 \, Q^{\bar{b}}(z) 
+  \frac{k(3k+N)}{
  (k+N+2)^3} \, d^0_{\bar{a} \bar{b}} \, \pa \, Q^{\bar{a}} \,
\pa \, Q^{\bar{b}}(z). 
\label{q}
\eea
The expression which depends on the spin-$1$ current
of (\ref{poleoneone}) only 
  is given by
\bea
& &
 \frac{1}{(k+N+2)^2}
(2 h^1_{\bar{a} \bar{b}} \, d^1_{\bar{c} \bar{d}}
+
h^2_{\bar{a} \bar{b}} \, d^2_{\bar{c} \bar{d}}
+h^3_{\bar{a} \bar{b}} \, d^3_{\bar{c} \bar{d}})
f^{\bar{b}
  \bar{d}}_{\,\,\,\,\,\, k} \, V^{\bar{a}} \,
  V^{\bar{c}} \, V^{k}(z)
  \nonu \\
  && -  \frac{4k}{(k+N+2)^2} \, d^0_{\bar{a} \bar{b}} \, V^{\bar{a}} \,
  \pa \, V^{\bar{b}}(z)
  -\frac{(3k+N)}{2(k+N+2)^2} \, d^0_{\bar{a} \bar{b}}
  \, f^{\bar{a}
  \bar{b}}_{\,\,\,\,\,\, c} \, \pa^2 \, V^{c}(z). 
  \label{v}
  \eea
The first two terms appearing only in this particular first order pole
are new in the sense that they cannot be obtained from the known
(higher spin) currents. It is obvious that the
last term can be seen from the
second derivative of the higher spin-$1$ current.
  The remaining mixed terms of (\ref{poleoneone}) are given by 
  \bea
  & &  
- \frac{1}{(k+N+2)^4} \,
(-\frac{1}{2} \,
d^2_{\bar{b} \bar{g}} \, h^3_{\bar{c} \bar{d}} 
+\frac{1}{2} \,
d^3_{\bar{b} \bar{g}} \, h^2_{\bar{c} \bar{d}} 
-h^1_{\bar{b} \bar{g}} \, d^0_{\bar{c} \bar{d}}) \, h^1_{\bar{a} \bar{f}}
\, f^{\bar{f}
  \bar{g}}_{\,\,\,\,\,\, e} \,
Q^{\bar{a}} \, Q^{\bar{b}} \, Q^{\bar{c}}  \, Q^{\bar{d}}
\, V^e(z)
\nonu \\
    &&- \frac{1}{(k+N+2)^3} \,
    ( -2 d^3_{\bar{a} \bar{c}} \, h^3_{\bar{d} \bar{b}}+
    2 d^3_{\bar{d} \bar{b}} \, h^3_{\bar{a} \bar{c}}-
    2 d^2_{\bar{a} \bar{c}} \, h^2_{\bar{d} \bar{b}}+
    2 d^2_{\bar{d} \bar{b}} \, h^2_{\bar{a} \bar{c}}
+ 
    d^1_{\bar{d} \bar{c}} \, h^1_{\bar{a} \bar{b}}
    \nonu \\
    && - 
    d^1_{\bar{a} \bar{d}} \, h^1_{\bar{b} \bar{c}}-
    d^1_{\bar{a} \bar{c}} \, h^1_{\bar{d} \bar{b}}-
    2 d^0_{\bar{a} \bar{b}} \, h^0_{\bar{c} \bar{d}}+
    d^0_{\bar{a} \bar{c}} \, h^0_{\bar{d} \bar{b}}-
    d^0_{\bar{d} \bar{b}} \, h^0_{\bar{a} \bar{c}} -\frac{1}{2} \,
    d^3_{\bar{d} \bar{c}} \, h^3_{\bar{a} \bar{b}} -
\frac{1}{2} \,
    d^2_{\bar{d} \bar{c}} \, h^2_{\bar{a} \bar{b}}
    )\,
    Q^{\bar{a}} \, Q^{\bar{b}} \, V^{\bar{c}}
    \, V^{\bar{d}}(z)
    \nonu \\
    & &+
    \frac{2}{(k+N+2)^3} \,
    h^1_{\bar{a} \bar{b}} \, h^0_{\bar{i} \bar{c}}
    \, d^1_{\bar{j} \bar{d}}  
 f^{\bar{d} \bar{c}}_{\,\,\,\,\,\, k} \,
 f^{\bar{j} \bar{b}}_{\,\,\,\,\,\, l} \,
  Q^{\bar{i}} \, Q^{\bar{a}} \, V^{k}
    \, V^{l}(z)
    \nonu \\
  & & -
    \frac{1}{(k+N+2)^3} \,
    h^1_{\bar{a} \bar{b}} \, h^2_{\bar{j} \bar{c}}
    \, d^3_{\bar{i} \bar{d}}  
 f^{\bar{d} \bar{c}}_{\,\,\,\,\,\, k} \,
 f^{\bar{j} \bar{b}}_{\,\,\,\,\,\, l} \,
  Q^{\bar{i}} \, Q^{\bar{a}} \, V^{k}
    \, V^{l}(z)
    \nonu  \\
     & & +
    \frac{1}{(k+N+2)^3} \,
    h^1_{\bar{a} \bar{b}} \, h^2_{\bar{i} \bar{c}}
    \, d^3_{\bar{d} \bar{j}}  
 f^{\bar{d} \bar{c}}_{\,\,\,\,\,\, k} \,
 f^{\bar{j} \bar{b}}_{\,\,\,\,\,\, l} \,
  Q^{\bar{i}} \, Q^{\bar{a}} \, V^{k}
    \, V^{l}(z)
    \nonu \\
  & &+ \frac{k}{(k+N+2)^3} \, ( 4 h^1_{\bar{a} \bar{b}}
    \, d^1_{\bar{c} \bar{d}}+ 2  h^0_{\bar{a} \bar{b}}
    \, d^0_{\bar{c} \bar{d}}+  h^3_{\bar{a} \bar{b}}
    \, d^3_{\bar{c} \bar{d}} +  h^2_{\bar{a} \bar{b}}
    \, d^2_{\bar{c} \bar{d}}) \,  f^{\bar{b} \bar{d}}_{\,\,\,\,\,\, e} \,
 Q^{\bar{a}} \, \pa \, Q^{\bar{c}} \, V^{e}(z)
 \nonu \\
  & &- \frac{1}{(k+N+2)^3} \, \Bigg[
-4N \,  h^1_{\bar{c} \bar{a}}
    \, d^1_{\bar{d} \bar{b}} \,  f^{\bar{a} \bar{b}}_{\,\,\,\,\,\, e}
+(-1-\frac{1}{N}) \,  h^1_{\bar{a} \bar{b}}
    \, d^1_{\bar{c} \bar{d}} \,  f^{\bar{a} \bar{b}}_{\,\,\,\,\,\, e}
+  h^0_{\bar{c} \bar{d}}
    \, d^0_{\bar{a} \bar{b}} \,  f^{\bar{a} \bar{b}}_{\,\,\,\,\,\, e}
    \nonu \\
    & &+ (1+\frac{1}{2N}) \,  h^2_{\bar{a} \bar{b}}
    \, d^2_{\bar{c} \bar{d}} \,  f^{\bar{a} \bar{b}}_{\,\,\,\,\,\, e}
+ (1+\frac{1}{2N}) \,  h^3_{\bar{a} \bar{b}}
    \, d^3_{\bar{c} \bar{d}} \,  f^{\bar{a} \bar{b}}_{\,\,\,\,\,\, e}
    \Bigg] \, 
 Q^{\bar{d}} \, \pa \, Q^{\bar{c}} \, V^{e}(z).
 \label{qv}
 \eea
 Therefore, one obtains the first order pole
 (\ref{poleoneone}) in terms of three simplified terms
 as follows:
 \bea
(\ref{poleoneone}) = (\ref{q}) + (\ref{v}) + (\ref{qv}).
 \label{threecombi}
 \eea
 Recall that the expression appearing in (\ref{poleoneone})
 was obtained from the OPE between the spin-$\frac{3}{2}$ current 
 and the expression in (\ref{5halfcombination}) which is {\it not}
 a higher spin-$\frac{5}{2}$ current. In next subsection, the extra
 terms will be considered in order to determine
 the final higher spin-$3$ current.
 
It would be interesting to observe any nontrivial relations
between the various tensors when $\mu \neq \nu$ by looking at
(\ref{process}), although the higher
spin-$3$ current is not involved.
 
\subsubsection{The higher spin-$3$ current}

From the first order pole of the OPE in (\ref{223}), 
one introduces the known quantity as 
\bea
\tilde{\Phi}_{\frac{3}{2}}^{(1),3}
& \equiv &
-\frac{2 i}{(k+N+2)} \, T^{12} \, \Phi_{\frac{1}{2}}^{(1),4}
-\frac{2 i}{(k+N+2)} \, T^{24} \, \Phi_{\frac{1}{2}}^{(1),1}
+\frac{4 i}{(k+N+2)} \, T^{14} \, \Phi_{\frac{1}{2}}^{(1),2}
\nonu \\
& + & \frac{(N-k)}{3(k+N+2)} \, \pa \, \Phi_{\frac{1}{2}}^{(1),3}.
\label{ExtraExtra}
\eea
In order to determine the higher spin-$3$ current, one should
also calculate the OPE between the spin-$\frac{3}{2}$ current
and this quantity (\ref{ExtraExtra}). 
By realizing the defining OPEs in Appendix
(\ref{1116}), one can calculate the
following OPE
\bea
{G}^3(z) \,
\tilde{\Phi}_{\frac{3}{2}}^{(1),3}(w) & = &
\frac{1}{(z-w)^3} \, \Bigg[ \frac{4(k-N)}{3(k+N+2)}
  \Bigg] \, \Phi_0^{(1)}(w) 
+ \frac{1}{(z-w)^2} \, \Bigg[ \frac{(k-N)}{(k+N+2)}
  \Bigg] \, \pa \, \Phi_0^{(1)}(w)
\nonu \\
&+& \frac{1}{(z-w)} \Bigg[
-\frac{2 i}{(k+N+2)} \, T^{12} \, \Phi_{1}^{(1),12}
-\frac{2 i}{(k+N+2)} \, T^{24} \, \Phi_{1}^{(1),24}
\nonu \\
& - & \frac{4 i}{(k+N+2)} \, T^{14} \, \Phi_{1}^{(1),14}
+  \frac{(k-N)}{3(k+N+2)} \, \pa^2 \, \Phi_{0}^{(1)}
  \Bigg](w) 
+\cdots.
\label{OPEabove}
\eea
Let us focus on the first order pole of (\ref{OPEabove}).
On the other hands, according to the first order pole of (\ref{process})
when $\mu=\nu=3$ in $SO(4)$ basis, one obtains
the very simple terms
\bea
-\Phi_{2}^{(1)} +
\frac{16(k-N)}{(4+5k+5N+6k N)} \, \Phi_0^{(1)} \, T.
\label{spin3location}
\eea
Compared to the one \cite{AK1411} in $SU(2) \times SU(2)$
basis, there are not too many composite fields except the
higher spin-$3$ current  itself in this pole. 
This implies that it turns out the higher spin-$3$ current can be
described as  
{\small  \bea
\Phi_{2}^{(1)}(z) & = & (\ref{threecombi})(z) +
\frac{16(k-N)}{(4+5k+5N+6k N)} \, \Phi_0^{(1)} \, T(z)+
\frac{2 i}{(k+N+2)} \, T^{12} \, \Phi_{1}^{(1),12}(z)
\label{simplespin3}
\\
& + & \frac{2 i}{(k+N+2)} \, T^{24} \, \Phi_{1}^{(1),24}(z)
+  \frac{4 i}{(k+N+2)} \, T^{14} \, \Phi_{1}^{(1),14}(z)
+  \frac{(N-k)}{3(k+N+2)} \, \pa^2 \, \Phi_{0}^{(1)}(z),
\nonu
\eea}
where the last four terms come from the first order pole
of (\ref{OPEabove}).
Therefore, the remaining things are to write down and simplify
the last five terms appearing in (\ref{simplespin3}) using the
WZW currents. One should use the explicit expressions in (\ref{11currents}),
(\ref{spinone}) and (\ref{spin2final}). 

Let us consider the third term of (\ref{simplespin3}).
One should write down the known (higher spin) currents
  in terms of WZW currents.
  The nontrivial parts arise from the quadratic term in $T^{12}(z)$
  and the spin-$\frac{1}{2}$ dependent parts in $\Phi^{(1),12}_{1}(z)$.
  One should use the following identity from the normal ordering product
  \cite{BBSS1,BS,Fuchs}
   \bea
(Q^{\bar{a}} Q^{\bar{b}})( Q^{\bar{f}} \, \pa \, Q^{\bar{d}})  & = &
    Q^{\bar{a}} Q^{\bar{b}} Q^{\bar{f}} \, \pa \, Q^{\bar{d}} -
    [  Q^{\bar{f}} \, \pa \, Q^{\bar{d}},
      Q^{\bar{a}} Q^{\bar{b}}] + [ Q^{\bar{f}} \, \pa \, Q^{\bar{d}}, Q^{\bar{a}}] \,
    Q^{\bar{b}}
+ Q^{\bar{a}} \,  [ Q^{\bar{f}} \, \pa \, Q^{\bar{d}}, Q^{\bar{b}} ]
\nonu \\
& = & Q^{\bar{a}} Q^{\bar{b}} Q^{\bar{f}} \, \pa \, Q^{\bar{d}} +(k+N+2) \,(
\frac{1}{2} \, g^{\bar{a} \bar{d}}  \, Q^{\bar{f}} \, \pa^2 \, Q^{\bar{b}} -
g^{\bar{a} \bar{f}} \, \pa \, Q^{\bar{d}} \, \pa \, Q^{\bar{b}}
\nonu \\
&+& \frac{1}{2} \, g^{\bar{b} \bar{d}}  \, \pa^2 \, Q^{\bar{a}} \, Q^{\bar{f}} -
g^{\bar{b} \bar{f}} \, \pa \, Q^{\bar{a}} \, \pa \, Q^{\bar{d}}).
\label{qqqdq}
\eea
In this case, there are no anticommutators
and the signs are coming from the usual bosonic case.
In the second line, one substitutes the various commutators
and obtains the final result \footnote{
Similarly, the following identity is also used
{\footnotesize
  \bea
(Q^{\bar{a}} Q^{\bar{b}})( Q^{\bar{f}} \, Q^{\bar{d}} \, V^h)  & = &
    Q^{\bar{a}} Q^{\bar{b}} Q^{\bar{f}} \, Q^{\bar{d}} \, V^h -
    [  Q^{\bar{f}}  \, Q^{\bar{d}} \, V^h,
      Q^{\bar{a}} Q^{\bar{b}}] +
    [ Q^{\bar{f}}  \, Q^{\bar{d}} \, V^h, Q^{\bar{a}}] \,
    Q^{\bar{b}}
+ Q^{\bar{a}} \,  [ Q^{\bar{f}}  \, Q^{\bar{d}} \, V^h, Q^{\bar{b}} ]
\nonu \\
& = & Q^{\bar{a}} Q^{\bar{b}} Q^{\bar{f}}  \, Q^{\bar{d}} \, V^h +(k+N+2) \,(
 \, g^{\bar{b} \bar{d}}  \, \pa \, Q^{\bar{a}} \, Q^{\bar{f}} \, V^h +
g^{\bar{a} \bar{f}} \, \pa \, Q^{\bar{b}} \, Q^{\bar{d}} \, V^h
\nonu \\
&-&  g^{\bar{a} \bar{d}}  \, \pa \, Q^{\bar{b}} \, Q^{\bar{f}} \, V^h -
g^{\bar{b} \bar{f}} \, \pa \, Q^{\bar{a}} \, Q^{\bar{d}}\, V^h),
\label{qqqqv-2}
\eea}
for the other nontrivial part of the composite field.
Again, there are no anticommutators in the first line of
this normal ordered product (\ref{qqqqv-2}).
One should calculate
the various rather complicated OPEs.
}.
It turns out that the following composite field  
can be expressed in terms of WZW currents
{\footnotesize
  \bea
&&  T^{12} \, \Phi^{(1),12}_{1}(z)  = 
  -\frac{i}{4N(k+N+2)} \, h^3_{\bar{a} \bar{b}} \, d^3_{\bar{c} \bar{d}}
f^{\bar{a}
  \bar{b}}_{\,\,\,\,\,\, e} \, V^{e} \,
  V^{\bar{c}} \, V^{\bar{d}}(z)
+\frac{i k}{4N(k+N+2)^2} \, h^3_{\bar{a} \bar{b}}
    \, d^3_{\bar{c} \bar{d}} \,  f^{\bar{a} \bar{b}}_{\,\,\,\,\,\, e} \,
 Q^{\bar{c}} \, \pa \, Q^{\bar{d}} \, V^{e}(z)
 \nonu \\
& &+ 
  \frac{i}{4N(k+N+2)^2} \, h^3_{\bar{a} \bar{b}} \, h^1_{\bar{d} \bar{e}}
    \, d^2_{\bar{f} \bar{g}} \,  
 f^{\bar{a} \bar{b}}_{\,\,\,\,\,\, c} \,
 f^{\bar{g} \bar{e}}_{\,\,\,\,\,\, h} \,
 Q^{\bar{f}} \, Q^{\bar{d}} \, V^{c}
    \, V^{h}(z)
+\frac{i}{4(k+N+2)^2} \,
    h^3_{\bar{a} \bar{b}} \, d^3_{\bar{c} \bar{d}}\,
    Q^{\bar{a}} \, Q^{\bar{b}} \, V^{\bar{c}}
    \, V^{\bar{d}}(z)
    \nonu \\
    &&-
    \frac{i k}{4(k+N+2)^3}
h^3_{\bar{a} \bar{b}} \, d^3_{\bar{c} \bar{d}} \,
Q^{\bar{a}} \, Q^{\bar{b}} \, Q^{\bar{c}} \, \pa \, Q^{\bar{d}}(z)
+ \frac{i k}{
  4(k+N+2)^2} \, d^0_{\bar{a} \bar{b}} \, Q^{\bar{a}} \,
\pa^2 \, Q^{\bar{b}}(z) 
\nonu \\
&&-  \frac{i k}{
  2(k+N+2)^2} \, d^0_{\bar{a} \bar{b}} \, \pa \, Q^{\bar{a}} \,
\pa \, Q^{\bar{b}}(z) 
-
\frac{i}{4(k+N+2)^3} \,
h^3_{\bar{a} \bar{b}} \, h^1_{\bar{d} \bar{e}} \, d^2_{\bar{f} \bar{g}}
\,
f^{\bar{g}
  \bar{e}}_{\,\,\,\,\,\, h} \,
Q^{\bar{a}} \, Q^{\bar{b}} \, Q^{\bar{f}}  \, Q^{\bar{d}}
\, V^h(z)
\nonu \\
&& - 
 \frac{i}{2(k+N+2)^2} \, h^1_{\bar{d} \bar{e}}
 \, d^1_{\bar{b} \bar{g}}
  \,  f^{\bar{g} \bar{e}}_{\,\,\,\,\,\, h} \,
 \pa \, Q^{\bar{b}}  \, Q^{\bar{d}} \, V^{h}(z)
+ \frac{i}{2(k+N+2)^2} \, h^2_{\bar{e} \bar{b}}
 \, d^2_{\bar{f} \bar{g}}
  \,  f^{\bar{g} \bar{e}}_{\,\,\,\,\,\, h} \,
 \pa \, Q^{\bar{b}}  \, Q^{\bar{f}} \, V^{h}(z).
 \label{1212}
    \eea}
    In doing this calculation, the identities (\ref{hgd-1})
    and (\ref{hhrel}) are used.
    There are also three kinds of composite fields:
    spin-$\frac{1}{2}$ dependent part, spin-$1$ dependent part and
    mixed ones.
    
    Similarly, one can obtain the following result
    for the fourth term of (\ref{simplespin3}) as follows:
{\footnotesize   \bea
&& T^{14} \, \Phi^{(1),14}_{1}(z)  =  
-\frac{i}{4N(k+N+2)} \, h^1_{\bar{a} \bar{b}} \, d^1_{\bar{c} \bar{d}}
f^{\bar{a}
  \bar{b}}_{\,\,\,\,\,\, e} \, V^{e} \,
  V^{\bar{c}} \, V^{\bar{d}}(z)
-\frac{i k}{4N(k+N+2)^2} \, h^1_{\bar{a} \bar{b}}
    \, d^1_{\bar{c} \bar{d}} \,  f^{\bar{a} \bar{b}}_{\,\,\,\,\,\, e} \,
 Q^{\bar{c}} \, \pa \, Q^{\bar{d}} \, V^{e}(z)
 \nonu \\
 && + 
  \frac{i}{4N(k+N+2)^2} \, h^1_{\bar{a} \bar{b}} \, h^1_{\bar{d} \bar{e}}
    \, d^0_{\bar{f} \bar{g}} \,  
 f^{\bar{a} \bar{b}}_{\,\,\,\,\,\, c} \,
 f^{\bar{g} \bar{e}}_{\,\,\,\,\,\, h} \,
 Q^{\bar{f}} \, Q^{\bar{d}} \, V^{c}
    \, V^{h}(z)
-\frac{i}{4(k+N+2)^2} \,
    h^1_{\bar{a} \bar{b}} \, d^1_{\bar{c} \bar{d}}\,
    Q^{\bar{a}} \, Q^{\bar{b}} \, V^{\bar{c}}
    \, V^{\bar{d}}(z)
    \nonu \\
    && -
    \frac{i k}{4(k+N+2)^3}
h^1_{\bar{a} \bar{b}} \, d^1_{\bar{c} \bar{d}} \,
Q^{\bar{a}} \, Q^{\bar{b}} \, Q^{\bar{c}} \, \pa \, Q^{\bar{d}}(z)
+ \frac{i k}{
  4(k+N+2)^2} \, d^0_{\bar{a} \bar{b}} \, Q^{\bar{a}} \,
\pa^2 \, Q^{\bar{b}}(z) 
\nonu \\
&&-  \frac{i k}{
  2(k+N+2)^2} \, d^0_{\bar{a} \bar{b}} \, \pa \, Q^{\bar{a}} \,
\pa \, Q^{\bar{b}}(z) 
+
\frac{i}{4(k+N+2)^3} \,
h^1_{\bar{a} \bar{b}} \, h^1_{\bar{d} \bar{e}} \, d^0_{\bar{f} \bar{g}}
\,
f^{\bar{g}
  \bar{e}}_{\,\,\,\,\,\, h} \,
Q^{\bar{a}} \, Q^{\bar{b}} \, Q^{\bar{f}}  \, Q^{\bar{d}}
\, V^h(z)
\nonu \\
&& - 
 \frac{i}{2(k+N+2)^2} \, h^1_{\bar{d} \bar{e}}
 \, d^1_{\bar{b} \bar{g}}
  \,  f^{\bar{g} \bar{e}}_{\,\,\,\,\,\, h} \,
 \pa \, Q^{\bar{b}}  \, Q^{\bar{d}} \, V^{h}(z)
- \frac{i}{2(k+N+2)^2} \, h^0_{\bar{e} \bar{b}}
 \, d^0_{\bar{f} \bar{g}}
  \,  f^{\bar{g} \bar{e}}_{\,\,\,\,\,\, h} \,
 \pa \, Q^{\bar{b}}  \, Q^{\bar{f}} \, V^{h}(z),
\label{1414}
\eea}
where
 the identities (\ref{hgd-1})
    and (\ref{hhrel}) can be used.
Finally, one obtains the following expression
for the fifth term of (\ref{simplespin3})
{\footnotesize \bea
&& T^{24} \, \Phi^{(1),24}_{1}(z)  =  
-\frac{i}{4N(k+N+2)} \, h^2_{\bar{a} \bar{b}} \, d^2_{\bar{c} \bar{d}}
f^{\bar{a}
  \bar{b}}_{\,\,\,\,\,\, e} \, V^{e} \,
  V^{\bar{c}} \, V^{\bar{d}}(z)
+\frac{i k}{4N(k+N+2)^2} \, h^2_{\bar{a} \bar{b}}
    \, d^2_{\bar{c} \bar{d}} \,  f^{\bar{a} \bar{b}}_{\,\,\,\,\,\, e} \,
 Q^{\bar{c}} \, \pa \, Q^{\bar{d}} \, V^{e}(z)
 \nonu \\
& &- 
  \frac{i}{4N(k+N+2)^2} \, h^2_{\bar{a} \bar{b}} \, h^1_{\bar{d} \bar{e}}
    \, d^3_{\bar{f} \bar{g}} \,  
 f^{\bar{a} \bar{b}}_{\,\,\,\,\,\, c} \,
 f^{\bar{g} \bar{e}}_{\,\,\,\,\,\, h} \,
 Q^{\bar{f}} \, Q^{\bar{d}} \, V^{c}
    \, V^{h}(z)
+\frac{i}{4(k+N+2)^2} \,
    h^2_{\bar{a} \bar{b}} \, d^2_{\bar{c} \bar{d}}\,
    Q^{\bar{a}} \, Q^{\bar{b}} \, V^{\bar{c}}
    \, V^{\bar{d}}(z)
    \nonu \\
   & &-
    \frac{i k}{4(k+N+2)^3}
h^2_{\bar{a} \bar{b}} \, d^2_{\bar{c} \bar{d}} \,
Q^{\bar{a}} \, Q^{\bar{b}} \, Q^{\bar{c}} \, \pa \, Q^{\bar{d}}(z)
+ \frac{i k}{
  4(k+N+2)^2} \, d^0_{\bar{a} \bar{b}} \, Q^{\bar{a}} \,
\pa^2 \, Q^{\bar{b}}(z) 
\nonu \\
&& -  \frac{i k}{
  2(k+N+2)^2} \, d^0_{\bar{a} \bar{b}} \, \pa \, Q^{\bar{a}} \,
\pa \, Q^{\bar{b}}(z) 
+
\frac{i}{4(k+N+2)^3} \,
h^2_{\bar{a} \bar{b}} \, h^1_{\bar{d} \bar{e}} \, d^3_{\bar{f} \bar{g}}
\,
f^{\bar{g}
  \bar{e}}_{\,\,\,\,\,\, h} \,
Q^{\bar{a}} \, Q^{\bar{b}} \, Q^{\bar{f}}  \, Q^{\bar{d}}
\, V^h(z)
\nonu \\
&& - 
 \frac{i}{2(k+N+2)^2} \, h^1_{\bar{d} \bar{e}}
 \, d^1_{\bar{b} \bar{g}}
  \,  f^{\bar{g} \bar{e}}_{\,\,\,\,\,\, h} \,
 \pa \, Q^{\bar{b}}  \, Q^{\bar{d}} \, V^{h}(z)
+ \frac{i}{2(k+N+2)^2} \, h^3_{\bar{e} \bar{a}}
 \, d^3_{\bar{f} \bar{g}}
  \,  f^{\bar{g} \bar{e}}_{\,\,\,\,\,\, h} \,
 \pa \, Q^{\bar{a}}  \, Q^{\bar{f}} \, V^{h}(z).
\label{2424}
  \eea}
Here  the identities (\ref{hgd-1})
    and (\ref{hhrel}) can be used to simplify the various expressions.

Let us move on the second term of (\ref{simplespin3}).    
In order to determine the composite field
$\Phi_0^{(1)} \, T$ (from (\ref{spinone}), (\ref{11currents}) and
(\ref{Spin2}))
in terms of WZW currents,
  one should use the following identity
{\small \bea
  (Q^{\bar{a}} Q^{\bar{b}})( Q^{\bar{c}}  \, Q^{\bar{d}} \, Q^{\bar{e}} \,
  Q^{\bar{f}} )  & = &
    Q^{\bar{a}} Q^{\bar{b}} Q^{\bar{c}}  \, Q^{\bar{d}} \, Q^{\bar{e}} \,
  Q^{\bar{f}} -
    [  Q^{\bar{c}}  \, Q^{\bar{d}} \, Q^{\bar{e}} \,
  Q^{\bar{f}},
  Q^{\bar{a}} Q^{\bar{b}}] \nonu \\
    & + & [ Q^{\bar{c}} \, Q^{\bar{d}}
      \, Q^{\bar{e}} \,
  Q^{\bar{f}}, Q^{\bar{a}}] \,
    Q^{\bar{b}}
    + Q^{\bar{a}} \,  [ Q^{\bar{c}} \,  Q^{\bar{d}}
\, Q^{\bar{e}} \,
  Q^{\bar{f}}, Q^{\bar{b}} ]
\nonu \\
& = & Q^{\bar{a}} Q^{\bar{b}} Q^{\bar{c}} \,  Q^{\bar{d}}
\, Q^{\bar{e}} \,
  Q^{\bar{f}}
+(k+N+2) \,(
\, g^{\bar{c} \bar{a}}  \, \pa \, Q^{\bar{b}}  \, Q^{\bar{d}}
\, Q^{\bar{e}} \,
  Q^{\bar{f}}
-
g^{\bar{c} \bar{b}} \, \pa \, Q^{\bar{a}} \,  Q^{\bar{d}}
\, Q^{\bar{e}} \,
  Q^{\bar{f}}
\nonu \\
&-&  g^{\bar{d} \bar{a}}  \, \pa \, Q^{\bar{b}} \, Q^{\bar{c}}
\, Q^{\bar{e}} \,
  Q^{\bar{f}}
+
g^{\bar{d} \bar{b}} \, \pa \, Q^{\bar{a}}  \, Q^{\bar{c}}
\, Q^{\bar{e}} \,
  Q^{\bar{f}}
+g^{\bar{e} \bar{a}} \, \pa \, Q^{\bar{b}}  \, Q^{\bar{c}}
\, Q^{\bar{d}} \,
Q^{\bar{f}} \nonu \\
& - &
g^{\bar{e} \bar{b}} \, \pa \, Q^{\bar{a}}  \, Q^{\bar{c}}
\, Q^{\bar{d}} \,
  Q^{\bar{f}}
-g^{\bar{f} \bar{a}} \, \pa \, Q^{\bar{b}}  \, Q^{\bar{c}}
\, Q^{\bar{d}} \,
  Q^{\bar{e}}
 +g^{\bar{f} \bar{b}} \, \pa \, Q^{\bar{a}}  \, Q^{\bar{c}}
\, Q^{\bar{d}} \,
  Q^{\bar{e}} ),
  \label{qidentity}
  \eea}
as well as (\ref{qqqdq}), and (\ref{qqqqv-2}).
In obtaining this expression (\ref{qidentity}),
one should calculate the various
complicated OPEs from which the commutators can be determined \cite{BS}.
It turns out that  the composite field is 
{\footnotesize
  \bea
&&  \Phi_0^{(1)} \, T(z)   =  
-\frac{k}{32(k+N+2)^5}
\, d^0_{\bar{a} \bar{b}} \, h^i_{\bar{c} \bar{d}} \, h^i_{\bar{e} \bar{f}} \,
Q^{\bar{a}} \, Q^{\bar{b}} \, Q^{\bar{c}} \, Q^{\bar{d}} \, Q^{\bar{e}} \,
Q^{\bar{f}}
+ \frac{k(3+2k)}{8(k+N+2)^4} \,
d^0_{\bar{a} \bar{b}} \, h^0_{\bar{c} \bar{d}} \,
Q^{\bar{a}} \, Q^{\bar{b}} \, Q^{\bar{c}} \, \pa \, Q^{\bar{d}}
\nonu \\
&&-\frac{k}{4(k+N+2)^4} \,
(h^1_{\bar{a} \bar{b}} \, d^1_{\bar{c} \bar{d}}
+
h^2_{\bar{a} \bar{b}} \, d^2_{\bar{c} \bar{d}}
+h^3_{\bar{a} \bar{b}} \, d^3_{\bar{c} \bar{d}}) \,
Q^{\bar{a}} \, Q^{\bar{b}} \, Q^{\bar{c}} \, \pa \, Q^{\bar{d}}
+ \frac{k(3+2k)}{
  8(k+N+2)^3} \, d^0_{\bar{a} \bar{b}} \, Q^{\bar{a}} \,
\pa^2 \, Q^{\bar{b}}
\nonu \\
&&-  \frac{k(3+2k)}{
  4(k+N+2)^3} \, d^0_{\bar{a} \bar{b}} \, \pa \, Q^{\bar{a}} \,
\pa \, Q^{\bar{b}}   
\nonu \\
& & -  \frac{1}{4(k+N+2)^2} \, d^0_{\bar{a} \bar{b}} \, f^{\bar{a}
    \bar{b}}_{\,\,\,\,\,\, c} \, h^0_{\bar{d} \bar{e}} \, V^{c} \,
  V^{\bar{d}} \, V^{\bar{e}}
  +  \frac{1}{32N^2(k+N+2)^2 }
\, d^0_{\bar{a} \bar{b}} \, f^{\bar{a}
  \bar{b}}_{\,\,\,\,\,\, c} \, h^{i}_{\bar{d} \bar{e}} \,
f^{\bar{d}
  \bar{e}}_{\,\,\,\,\,\, f} \, h^{i}_{\bar{g} \bar{h}} \,
f^{\bar{g}
  \bar{h}}_{\,\,\,\,\,\, j} \, V^{c} \,
V^{f} \, V^{j}
\nonu \\
&&+   \frac{1}{32(k+N+2)^4} \, d^0_{\bar{a} \bar{b}} \,
    h^i_{\bar{d} \bar{e}} \, h^i_{\bar{f} \bar{g}} \, 
f^{\bar{a}
  \bar{b}}_{\,\,\,\,\,\, c} \,
Q^{\bar{d}} \, Q^{\bar{e}} \, Q^{\bar{f}}  \, Q^{\bar{g}}
\, V^c
+  \frac{k}{4(k+N+2)^4} \,
 d^0_{\bar{a} \bar{b}} \,
    h^0_{\bar{c} \bar{f}} \, h^0_{\bar{d} \bar{g}} \, 
f^{\bar{f}
  \bar{g}}_{\,\,\,\,\,\, e} \,
Q^{\bar{a}} \, Q^{\bar{b}} \, Q^{\bar{c}}  \, Q^{\bar{d}}
\, V^e
\nonu \\
&& - 
\frac{k}{16N(k+N+2)^4} \,
 d^0_{\bar{a} \bar{b}} \,
    h^i_{\bar{c} \bar{d}} \, h^i_{\bar{f} \bar{g}} \, 
f^{\bar{f}
  \bar{g}}_{\,\,\,\,\,\, e} \,
Q^{\bar{a}} \, Q^{\bar{b}} \, Q^{\bar{c}}  \, Q^{\bar{d}}
\, V^e
    \nonu \\
   & &+ \frac{1}{16N(k+N+2)^3} \,
    d^0_{\bar{a} \bar{b}} \, h^i_{\bar{d} \bar{e}}
    \, h^i_{\bar{g} \bar{h}} \, f^{\bar{a}
  \bar{b}}_{\,\,\,\,\,\, c} \, f^{\bar{d}
      \bar{e}}_{\,\,\,\,\,\, f} \,
 Q^{\bar{g}} \, Q^{\bar{h}} \, V^{c}
    \, V^{f}
+ \frac{k}{4(k+N+2)^3} \,
    d^0_{\bar{a} \bar{b}} \, h^0_{\bar{c} \bar{d}}
    \,  Q^{\bar{a}} \, Q^{\bar{b}} \, V^{\bar{c}}
    \, V^{\bar{d}}
    \nonu \\
    &&- \frac{k}{32N^2(k+N+2)^3} \,
    d^0_{\bar{a} \bar{b}} \, h^i_{\bar{c} \bar{d}}
    \, h^i_{\bar{f} \bar{g}} \, f^{\bar{c}
  \bar{d}}_{\,\,\,\,\,\, e} \, f^{\bar{f}
      \bar{g}}_{\,\,\,\,\,\, h} \,
 Q^{\bar{a}} \, Q^{\bar{b}} \, V^{e}
 \, V^{h}
 \nonu \\
 && -  \frac{(3+2k)}{8(k+N+2)^3} \,
       d^0_{\bar{a} \bar{b}}
    \, h^0_{\bar{d} \bar{e}}  
 f^{\bar{a} \bar{b}}_{\,\,\,\,\,\, c} \,
 Q^{\bar{d}} \, \pa \, Q^{\bar{e}} \, V^{c}
-  \frac{k}{(k+N+2)^3} \,
       d^0_{\bar{a} \bar{b}}
    \, h^0_{\bar{c} \bar{d}}  
 f^{\bar{d} \bar{b}}_{\,\,\,\,\,\, e} \,
 Q^{\bar{c}} \, \pa \, Q^{\bar{a}} \, V^{e}
    \nonu \\
  &&  -  \frac{k}{4N(k+N+2)^3} \,
       d^i_{\bar{a} \bar{c}}
    \, h^i_{\bar{f} \bar{g}}  
 f^{\bar{f} \bar{g}}_{\,\,\,\,\,\, e} \,
 Q^{\bar{c}} \, \pa \, Q^{\bar{a}} \, V^{e}
-   \frac{1}{4(k+N+2)^3} \, d^0_{\bar{a} \bar{b}} \, h^0_{\bar{d} \bar{f}}
    \, h^0_{\bar{e} \bar{g}} \, f^{\bar{a}
  \bar{b}}_{\,\,\,\,\,\, c} \, f^{\bar{f}
      \bar{g}}_{\,\,\,\,\,\, h} \, Q^{\bar{d}} \, Q^{\bar{e}} \, V^{c}
    \, V^{h}.
 \label{spin1spin2}
  \eea}
Because the composite field $\Phi_0^{(1)} \, T(z)$ is written in terms of
adjoint spin-$1,\frac{1}{2}$ currents, one obtains
some eigenvalues for the higher spin-$1$ current in the higher
representations indirectly (the conformal dimension, which
is the zeromode eigenvalue of $T(z)$, in any representations
are known explicitly). In next section, we will use this property
very frequently.

From the
previous results, (\ref{simplespin3}), (\ref{1212}), (\ref{1414}),
(\ref{2424}), (\ref{spin1spin2}), (\ref{threecombi}) and
second derivative term of the higher spin-$1$ current (\ref{spinone}),
one obtains the final higher spin-$3$ current which consists of
the three independent pieces as follows:
   \bea
  \Phi_{2}^{(1)}(z) =
\Phi_{2, Q}^{(1)}(z) + \Phi_{2, V}^{(1)}(z) +\Phi_{2, Q, V}^{(1)}(z).
  \label{spin3expression}
  \eea
  Here the higher spin-$3$  current
  which depends on the spin-$\frac{1}{2}$ current
  is given by
{\small  \bea
\Phi_{2, Q}^{(1)}(z) & \equiv & -\frac{k(k-N)}{2(4+5k+5N+6k N)(k+N+2)^5}
\, d^0_{\bar{a} \bar{b}} \, h^i_{\bar{c} \bar{d}} \, h^i_{\bar{e} \bar{f}} \,
Q^{\bar{a}} \, Q^{\bar{b}} \, Q^{\bar{c}} \, Q^{\bar{d}} \, Q^{\bar{e}} \,
Q^{\bar{f}}(z)
\nonu \\
&+& \frac{k(4+11k+4k^2-N+2k N)}{(4+5k+5N+6k N)(k+N+2)^4} \,
d^0_{\bar{a} \bar{b}} \, h^0_{\bar{c} \bar{d}} \,
Q^{\bar{a}} \, Q^{\bar{b}} \, Q^{\bar{c}} \, \pa \, Q^{\bar{d}}(z)
\label{qspin3}
 \\
&+&\frac{k(4+k+9N+6 k N)}{(4+5k+5N+6k N)(k+N+2)^4} \,
(h^1_{\bar{a} \bar{b}} \, d^1_{\bar{c} \bar{d}}
+
h^2_{\bar{a} \bar{b}} \, d^2_{\bar{c} \bar{d}}
+h^3_{\bar{a} \bar{b}} \, d^3_{\bar{c} \bar{d}}) \,
Q^{\bar{a}} \, Q^{\bar{b}} \, Q^{\bar{c}} \, \pa \, Q^{\bar{d}}(z)
\nonu \\
&-& \frac{2k(12+14k+4k^2+28N+39 k N+12 k^2 N+5 N^2+6 k N^2)}{
  3(4+5k+5N+6k N)(k+N+2)^3} \, d^0_{\bar{a} \bar{b}} \, Q^{\bar{a}} \,
\pa^2 \, Q^{\bar{b}}(z) 
\nonu \\
&+&  \frac{4k(12+14k+4k^2+28N+39 k N+12 k^2 N+5 N^2+6 k N^2)}{
  3(4+5k+5N+6k N)(k+N+2)^3} \, d^0_{\bar{a} \bar{b}} \, \pa \, Q^{\bar{a}} \,
\pa \, Q^{\bar{b}}(z). 
\nonu
\eea}
The first term of this expression originates from (\ref{spin1spin2}).
The second and third terms come from the one in
(\ref{q}), the corresponding term in (\ref{spin1spin2}),
those in (\ref{1212}), (\ref{1414}) and (\ref{2424}).
The remaining derivative terms originate from
the ones in (\ref{q}), (\ref{spin1spin2}),
(\ref{1212}), (\ref{1414}), (\ref{2424}) and the second derivative
of the higher spin-$1$ current. All the coefficients
which depend on $N$ and $k$ appearing in (\ref{qspin3}) are
very important to obtain the eigenvalues in the Wolf space coset
representations in next section.
Furthermore, the fourth term of (\ref{qspin3}) will play an important
role for the eigenvalues under the large $(N, k)$ 't Hooft-like limit.

The higher spin-$3$ current which depends on the spin-$1$ current
  is given by
\bea
\Phi_{2, V}^{(1)}(z)
& \equiv & \frac{16(k-N)}{(4+5k+5N+6k N)} \Bigg[
  -\frac{1}{4(k+N+2)^2} \, d^0_{\bar{a} \bar{b}} \, f^{\bar{a}
    \bar{b}}_{\,\,\,\,\,\, c} \, h^0_{\bar{d} \bar{e}} \, V^{c} \,
  V^{\bar{d}} \, V^{\bar{e}}
  \nonu \\
  & + & \frac{1}{32N^2(k+N+2)^2 }
\, d^0_{\bar{a} \bar{b}} \, f^{\bar{a}
  \bar{b}}_{\,\,\,\,\,\, c} \, h^{i}_{\bar{d} \bar{e}} \,
f^{\bar{d}
  \bar{e}}_{\,\,\,\,\,\, f} \, h^{i}_{\bar{g} \bar{h}} \,
f^{\bar{g}
  \bar{h}}_{\,\,\,\,\,\, j} \, V^{c} \,
  V^{f} \, V^{j}
\Bigg](z)
\nonu \\
&+& \frac{1}{2N(k+N+2)^2}
(2 h^1_{\bar{a} \bar{b}} \, d^1_{\bar{c} \bar{d}}
+
h^2_{\bar{a} \bar{b}} \, d^2_{\bar{c} \bar{d}}
+h^3_{\bar{a} \bar{b}} \, d^3_{\bar{c} \bar{d}})
f^{\bar{a}
  \bar{b}}_{\,\,\,\,\,\, e} \, V^{e} \,
  V^{\bar{c}} \, V^{\bar{d}}(z)
  \nonu \\
  &+& \frac{1}{(k+N+2)^2}
(2 h^1_{\bar{a} \bar{b}} \, d^1_{\bar{c} \bar{d}}
+
h^2_{\bar{a} \bar{b}} \, d^2_{\bar{c} \bar{d}}
+h^3_{\bar{a} \bar{b}} \, d^3_{\bar{c} \bar{d}})
f^{\bar{b}
  \bar{d}}_{\,\,\,\,\,\, k} \, V^{\bar{a}} \,
  V^{\bar{c}} \, V^{k}(z)
  \nonu \\
  & - & \frac{4k}{(k+N+2)^2} \, d^0_{\bar{a} \bar{b}} \, V^{\bar{a}} \,
  \pa \, V^{\bar{b}}(z)
  -\frac{2(2k+N)}{3(k+N+2)^2} \, d^0_{\bar{a} \bar{b}}
  \, f^{\bar{a}
  \bar{b}}_{\,\,\,\,\,\, c} \, \pa^2 \, V^{\bar{c}}(z). 
  \label{vspin3}
  \eea
  The first two terms of this expression
  come from (\ref{spin1spin2}).
  The third term comes from the corresponding terms in
  (\ref{1212}), (\ref{1414}) and (\ref{2424}).
  The fourth term can be seen from (\ref{v}).
  The fifth term originates from the one in (\ref{v}).
  The last term was combined from the one in (\ref{v}) and
  the derivative term of the higher spin-$1$ current.
  Note that the composite fields in the fourth and the fifth terms
occur only in the first order pole (\ref{poleoneone}).  
The other ones can be seen from the known composite fields
made of the known (higher spin) currents. 
 All the coefficients
depending on $N$ and $k$ in (\ref{vspin3}) are
crucial to obtain the eigenvalues in the Wolf space coset
representations in next section.
Furthermore, the fourth, the fifth and the sixth terms of (\ref{vspin3})
will play an important role for the eigenvalues
under the large $(N,k)$ 't Hooft-like limit. 

The higher spin-$3$ current
which contain the remaining mixed terms is given by 
{\footnotesize  \bea
&&   \Phi_{2, Q, V}^{(1)}(z)
   \equiv  
  \frac{16(k-N)}{(4+5k+5N+6k N)} \Bigg[
    \frac{1}{32(k+N+2)^4} \, d^0_{\bar{a} \bar{b}} \,
    h^i_{\bar{d} \bar{e}} \, h^i_{\bar{f} \bar{g}} \, 
f^{\bar{a}
  \bar{b}}_{\,\,\,\,\,\, c} \,
Q^{\bar{d}} \, Q^{\bar{e}} \, Q^{\bar{f}}  \, Q^{\bar{g}}
\, V^c
\nonu \\
& &+  \frac{k}{4(k+N+2)^4} \,
 d^0_{\bar{a} \bar{b}} \,
    h^0_{\bar{c} \bar{f}} \, h^0_{\bar{d} \bar{g}} \, 
f^{\bar{f}
  \bar{g}}_{\,\,\,\,\,\, e} \,
Q^{\bar{a}} \, Q^{\bar{b}} \, Q^{\bar{c}}  \, Q^{\bar{d}}
\, V^e
\nonu \\
&& - 
\frac{k}{16N(k+N+2)^4} \,
 d^0_{\bar{a} \bar{b}} \,
    h^i_{\bar{c} \bar{d}} \, h^i_{\bar{f} \bar{g}} \, 
f^{\bar{f}
  \bar{g}}_{\,\,\,\,\,\, e} \,
Q^{\bar{a}} \, Q^{\bar{b}} \, Q^{\bar{c}}  \, Q^{\bar{d}}
\, V^e
\Bigg](z)
  \nonu \\
 & &+ \frac{1}{2(k+N+2)^4} \,
(h^3_{\bar{a} \bar{b}} \, h^1_{\bar{c} \bar{d}} \, d^2_{\bar{e} \bar{f}}
-2
h^1_{\bar{a} \bar{b}} \, h^1_{\bar{c} \bar{d}} \, d^0_{\bar{e} \bar{f}}
-h^2_{\bar{a} \bar{b}} \, h^1_{\bar{c} \bar{d}} \, d^3_{\bar{e} \bar{f}})
f^{\bar{f}
  \bar{d}}_{\,\,\,\,\,\, g} \,
Q^{\bar{a}} \, Q^{\bar{b}} \, Q^{\bar{e}}  \, Q^{\bar{c}}
\, V^g(z)
\nonu \\
 &&- \frac{1}{(k+N+2)^4} \,
(-\frac{1}{2} \,
d^2_{\bar{b} \bar{g}} \, h^3_{\bar{c} \bar{d}} 
+\frac{1}{2} \,
d^3_{\bar{b} \bar{g}} \, h^2_{\bar{c} \bar{d}} 
-h^1_{\bar{b} \bar{g}} \, d^0_{\bar{c} \bar{d}}) \, h^1_{\bar{a} \bar{f}}
\, f^{\bar{f}
  \bar{g}}_{\,\,\,\,\,\, e} \,
Q^{\bar{a}} \, Q^{\bar{b}} \, Q^{\bar{c}}  \, Q^{\bar{d}}
\, V^e(z)
\nonu \\
 & &+ 
  \frac{16(k-N)}{(4+5k+5N+6k N)} \Bigg[
    -\frac{1}{4(k+N+2)^3} \, d^0_{\bar{a} \bar{b}} \, h^0_{\bar{d} \bar{f}}
    \, h^0_{\bar{e} \bar{g}} \, f^{\bar{a}
  \bar{b}}_{\,\,\,\,\,\, c} \, f^{\bar{f}
      \bar{g}}_{\,\,\,\,\,\, h} \, Q^{\bar{d}} \, Q^{\bar{e}} \, V^{c}
    \, V^{h}
    \nonu \\
    &&+ \frac{1}{16N(k+N+2)^3} \,
    d^0_{\bar{a} \bar{b}} \, h^i_{\bar{d} \bar{e}}
    \, h^i_{\bar{g} \bar{h}} \, f^{\bar{a}
  \bar{b}}_{\,\,\,\,\,\, c} \, f^{\bar{d}
      \bar{e}}_{\,\,\,\,\,\, f} \,
 Q^{\bar{g}} \, Q^{\bar{h}} \, V^{c}
    \, V^{f}
    + \frac{k}{4(k+N+2)^3} \,
    d^0_{\bar{a} \bar{b}} \, h^0_{\bar{c} \bar{d}}
    \,  Q^{\bar{a}} \, Q^{\bar{b}} \, V^{\bar{c}}
    \, V^{\bar{d}}
    \nonu \\
    &&- \frac{k}{32N^2(k+N+2)^3} \,
    d^0_{\bar{a} \bar{b}} \, h^i_{\bar{c} \bar{d}}
    \, h^i_{\bar{f} \bar{g}} \, f^{\bar{c}
  \bar{d}}_{\,\,\,\,\,\, e} \, f^{\bar{f}
      \bar{g}}_{\,\,\,\,\,\, h} \,
 Q^{\bar{a}} \, Q^{\bar{b}} \, V^{e}
    \, V^{h} \Bigg](z)
    \nonu \\
    &&+
    \frac{1}{2N(k+N+2)^3} \, (
    -2 h^1_{\bar{a} \bar{b}} \, h^1_{\bar{c} \bar{d}}
    \, d^0_{\bar{e} \bar{f}} +
   h^2_{\bar{a} \bar{b}} \, h^1_{\bar{c} \bar{d}}
    \, d^3_{\bar{e} \bar{f}}  - h^3_{\bar{a} \bar{b}} \, h^1_{\bar{c} \bar{d}}
    \, d^2_{\bar{e} \bar{f}}  )
 f^{\bar{a} \bar{b}}_{\,\,\,\,\,\, g} \,
 f^{\bar{f} \bar{d}}_{\,\,\,\,\,\, h} \,
 Q^{\bar{e}} \, Q^{\bar{c}} \, V^{g}
    \, V^{h}(z)
    \nonu \\
   & &- \frac{1}{(k+N+2)^3} \,
    ( -2 d^3_{\bar{a} \bar{c}} \, h^3_{\bar{d} \bar{b}}+
    2 d^3_{\bar{d} \bar{b}} \, h^3_{\bar{a} \bar{c}}-
    2 d^2_{\bar{a} \bar{c}} \, h^2_{\bar{d} \bar{b}}+
    2 d^2_{\bar{d} \bar{b}} \, h^2_{\bar{a} \bar{c}}
    \nonu \\
    && - 
    d^1_{\bar{a} \bar{d}} \, h^1_{\bar{b} \bar{c}}-
    d^1_{\bar{a} \bar{c}} \, h^1_{\bar{d} \bar{b}}-
    2 d^0_{\bar{a} \bar{b}} \, h^0_{\bar{c} \bar{d}}+
    d^0_{\bar{a} \bar{c}} \, h^0_{\bar{d} \bar{b}}-
    d^0_{\bar{d} \bar{b}} \, h^0_{\bar{a} \bar{c}})\,
    Q^{\bar{a}} \, Q^{\bar{b}} \, V^{\bar{c}}
    \, V^{\bar{d}}(z)
    \nonu \\
   & &+
    \frac{2}{(k+N+2)^3} \,
    h^1_{\bar{a} \bar{b}} \, h^0_{\bar{i} \bar{c}}
    \, d^1_{\bar{j} \bar{d}}  
 f^{\bar{d} \bar{c}}_{\,\,\,\,\,\, k} \,
 f^{\bar{j} \bar{b}}_{\,\,\,\,\,\, l} \,
  Q^{\bar{i}} \, Q^{\bar{a}} \, V^{k}
    \, V^{l}(z)
    \nonu \\
 & &-
    \frac{1}{(k+N+2)^3} \,
    h^1_{\bar{a} \bar{b}} \, h^2_{\bar{j} \bar{c}}
    \, d^3_{\bar{i} \bar{d}}  
 f^{\bar{d} \bar{c}}_{\,\,\,\,\,\, k} \,
 f^{\bar{j} \bar{b}}_{\,\,\,\,\,\, l} \,
  Q^{\bar{i}} \, Q^{\bar{a}} \, V^{k}
    \, V^{l}(z)
    \nonu  \\
    & &+
    \frac{1}{(k+N+2)^3} \,
    h^1_{\bar{a} \bar{b}} \, h^2_{\bar{i} \bar{c}}
    \, d^3_{\bar{d} \bar{j}}  
 f^{\bar{d} \bar{c}}_{\,\,\,\,\,\, k} \,
 f^{\bar{j} \bar{b}}_{\,\,\,\,\,\, l} \,
  Q^{\bar{i}} \, Q^{\bar{a}} \, V^{k}
    \, V^{l}(z)
    \nonu \\
   & &+ \frac{16(k-N)}{(4+5k+5N+6k N)} \Bigg[
      -\frac{(3+2k)}{8(k+N+2)^3} \,
       d^0_{\bar{a} \bar{b}}
    \, h^0_{\bar{d} \bar{e}}  
 f^{\bar{a} \bar{b}}_{\,\,\,\,\,\, c} \,
 Q^{\bar{d}} \, \pa \, Q^{\bar{e}} \, V^{c}
 \nonu \\
& & -  \frac{k}{(k+N+2)^3} \,
       d^0_{\bar{a} \bar{b}}
    \, h^0_{\bar{c} \bar{d}}  
 f^{\bar{d} \bar{b}}_{\,\,\,\,\,\, e} \,
 Q^{\bar{c}} \, \pa \, Q^{\bar{a}} \, V^{e}
     +  \frac{k}{4N(k+N+2)^3} \,
       d^i_{\bar{a} \bar{c}}
    \, h^i_{\bar{f} \bar{g}}  
 f^{\bar{f} \bar{g}}_{\,\,\,\,\,\, e} \,
 Q^{\bar{c}} \, \pa \, Q^{\bar{a}} \, V^{e}
 \Bigg](z)
    \nonu  \\
   & &+ \frac{k}{2N(k+N+2)^3} \, ( 2 h^1_{\bar{a} \bar{b}}
    \, d^1_{\bar{c} \bar{d}}-  h^2_{\bar{a} \bar{b}}
    \, d^2_{\bar{c} \bar{d}}-  h^3_{\bar{a} \bar{b}}
    \, d^3_{\bar{c} \bar{d}}) \,  f^{\bar{a} \bar{b}}_{\,\,\,\,\,\, e} \,
 Q^{\bar{c}} \, \pa \, Q^{\bar{d}} \, V^{e}(z)
 \nonu \\
 & &- \frac{1}{(k+N+2)^3} \, ( - h^2_{\bar{a} \bar{b}}
    \, d^2_{\bar{c} \bar{d}}+ 2  h^0_{\bar{a} \bar{b}}
    \, d^0_{\bar{c} \bar{d}}-  h^3_{\bar{a} \bar{b}}
    \, d^3_{\bar{c} \bar{d}} - 4  h^1_{\bar{a} \bar{b}}
    \, d^1_{\bar{c} \bar{d}}) \,  f^{\bar{d} \bar{a}}_{\,\,\,\,\,\, e} \,
 Q^{\bar{c}} \, \pa \, Q^{\bar{b}} \, V^{e}(z)
 \nonu \\
  &&+ \frac{k}{(k+N+2)^3} \, ( 4 h^1_{\bar{a} \bar{b}}
    \, d^1_{\bar{c} \bar{d}}+ 2  h^0_{\bar{a} \bar{b}}
    \, d^0_{\bar{c} \bar{d}}+  h^3_{\bar{a} \bar{b}}
    \, d^3_{\bar{c} \bar{d}} +  h^2_{\bar{a} \bar{b}}
    \, d^2_{\bar{c} \bar{d}}) \,  f^{\bar{b} \bar{d}}_{\,\,\,\,\,\, e} \,
 Q^{\bar{a}} \, \pa \, Q^{\bar{c}} \, V^{e}(z)
 \nonu \\
& &- \frac{1}{(k+N+2)^3} \, \Bigg[
-4N \,  h^1_{\bar{c} \bar{a}}
    \, d^1_{\bar{d} \bar{b}} \,  f^{\bar{a} \bar{b}}_{\,\,\,\,\,\, e}
+(-1-\frac{1}{N}) \,  h^1_{\bar{a} \bar{b}}
    \, d^1_{\bar{c} \bar{d}} \,  f^{\bar{a} \bar{b}}_{\,\,\,\,\,\, e}
+  h^0_{\bar{c} \bar{d}}
    \, d^0_{\bar{a} \bar{b}} \,  f^{\bar{a} \bar{b}}_{\,\,\,\,\,\, e}
    \nonu \\
  &  &+ (1+\frac{1}{2N}) \,  h^2_{\bar{a} \bar{b}}
    \, d^2_{\bar{c} \bar{d}} \,  f^{\bar{a} \bar{b}}_{\,\,\,\,\,\, e}
+ (1+\frac{1}{2N}) \,  h^3_{\bar{a} \bar{b}}
    \, d^3_{\bar{c} \bar{d}} \,  f^{\bar{a} \bar{b}}_{\,\,\,\,\,\, e}
    \Bigg] \, 
 Q^{\bar{d}} \, \pa \, Q^{\bar{c}} \, V^{e}(z).
 \label{qvspin3}
  \eea}
First of all, the expressions having the factor $16(k-N)$
are coming from the ones in (\ref{spin1spin2}).
The fourth and the ninth lines of
(\ref{qvspin3}) can be seen from the ones in
(\ref{1212}), (\ref{1414}) and (\ref{2424}).
The fifth line comes from the first term of (\ref{qv}).
The tenth and eleventh lines are coming from 
 the ones in
 (\ref{1212}), (\ref{1414}) and (\ref{2424}) and
 the corresponding terms in (\ref{qv}).
 The twelfth to fourteenth lines are the same as the ones in
 (\ref{qv}).
 The fourth and fifth lines from the below of (\ref{qvspin3})
 are the same as the ones in (\ref{1212}), (\ref{1414}) and (\ref{2424}).
 Finally, the last three lines come from the ones in (\ref{qv}).
 There are terms which are obtained from the first order pole in
 (\ref{poleoneone}).
 For the eigenvalue calculation in next section,
 this mixed terms do not contribute to the eigenvalues for the minimal
 (and higher) representations.

 It would be interesting to determine the higher spin-$3$ current in
 the bais of \cite{AK1411} and see any differences with the above
 higher spin-$3$ current.
 
\subsubsection{Other way of  obtaining the higher spin-$3$ current
using the  higher spin-$\frac{5}{2}$ current (\ref{spin5halfone})}

  One can determine the above higher spin-$3$ current by considering
  the higher spin-$\frac{5}{2}$ current (\ref{spin5halfone})
  directly. In order to use previous relations, one can move
  $V^c$ appearing in the cubic terms of (\ref{spin5halfone})
  to the right.
  Then one sees that the extra
  terms are given by $-f^{c \bar{e}}_{\,\,\,\,\,\,g} \, Q^{\bar{d}} \, \pa \, V^g$.
  One can check the following identity, by multiplying the structure
  constant to the cubic term of (\ref{spin5halfone}),
  \bea
&& -\Bigg[ \frac{1}{2N} \, h^3_{\bar{a} \bar{b}}\,
  d^2_{\bar{d} \bar{e}} -
\frac{1}{2N} \, h^2_{\bar{a} \bar{b}}\,
  d^3_{\bar{d} \bar{e}}
+ \frac{1}{N} \, h^1_{\bar{a} \bar{b}}\,
  d^0_{\bar{d} \bar{e}}
 -  2  \, h^0_{\bar{b} \bar{d}}\,
  d^1_{\bar{a} \bar{e}} -  h^2_{\bar{b} \bar{e}}\,
  d^3_{\bar{a} \bar{d}} +  h^2_{\bar{b} \bar{d}}\,
  d^3_{\bar{a} \bar{e}} \Bigg] \, f^{\bar{a} \bar{b}}_{\,\,\,\,\,\,c} \,
f^{c \bar{e}}_{\,\,\,\,\,\,g}
\nonu \\
&& =( 1 + 1 + 2 -2N -N -N) \, d^1_{\bar{d} \bar{g}} =
4(1-N) \, d^1_{\bar{d} \bar{g}}.
\label{extraexp}
  \eea

  Then one can write down the (intermediate) first order pole in the OPE
  between $G^3(z)$ and $\Phi_{\frac{3}{2}}^{(1),3}(w)$
  which should be equal to (\ref{spin3location}).
  By adding the contribution from (\ref{extraexp}),
 the first order pole of the OPE
  between $G^3(z)$ and $\Phi_{\frac{3}{2}}^{(1),3}(w)$
  is given by
{\small  \bea
&& -\frac{1}{(k+N+2)^3} \, \Bigg[ \frac{1}{2N} \, h^3_{\bar{a'} \bar{b'}}\,
  d^2_{\bar{i} \bar{j}} -
\frac{1}{2N} \, h^2_{\bar{a'} \bar{b'}}\,
  d^3_{\bar{i} \bar{j}}
+ \frac{1}{N} \, h^1_{\bar{a'} \bar{b'}}\,
  d^0_{\bar{i} \bar{j}}
 -  2  \, h^0_{\bar{b'} \bar{i}}\,
  d^1_{\bar{a'} \bar{j}} -  h^2_{\bar{b'} \bar{j}}\,
  d^3_{\bar{a'} \bar{i}} +  h^2_{\bar{b'} \bar{i}}\,
  d^3_{\bar{a'} \bar{j}} \Bigg]
\nonu \\
&& \times f^{\bar{a'} \bar{b'}}_{\,\,\,\,\,\,k} \, h^1_{\bar{a} \bar{b}} \,
\Bigg[-(k+N+2) g^{\bar{i} \bar{a}} \,
  V^{\bar{b}} \, V^{\bar{j}} \, V^k -
  k \, g^{\bar{j} \bar{b}} \, Q^{\bar{i}}
  \,  \pa \, Q^{\bar{a}} \, V^k - f^{\bar{j} \bar{b}}_{\,\,\,\,\,\,l}
  \, Q^{\bar{i}} \,
  Q^{\bar{a}} \, V^{k} \, V^l
  \nonu \\
  && - f^{\bar{j} \bar{b}}_{\,\,\,\,\,\,l} \,
f^{k l}_{\,\,\,\,\,\,m} \,
 Q^{\bar{i}} \, \pa
 \, ( Q^{\bar{a}} \, V^{m})
 -f^{k \bar{b}}_{\,\,\,\,\,\,l} \, 
 Q^{\bar{i}} \, Q^{\bar{a}} \, V^{\bar{j}} \, V^l -
 \frac{k}{2} \, f^{\bar{j} \bar{b}}_{\,\,\,\,\,\,l} \,
 g^{k l} \, Q^{\bar{i}} \,
 \pa^2 \, Q^{\bar{a}} - k\, g^{k \bar{b}} \, Q^{\bar{i}} \, \pa Q^{\bar{a}} \,
 V^{\bar{j}} \Bigg]
\nonu \\
&& -\frac{1}{(k+N+2)^3} \, 4 (1-N) \,  h^1_{\bar{a} \bar{b}}\,
d^1_{\bar{c} \bar{d}} \nonu \\
&& \times \Bigg[
-(k+N+2) \, g^{\bar{c} \bar{a}}  \,
V^{\bar{b}} \, \pa \, V^{\bar{d}} -
k \, g^{\bar{d} \bar{b}} \, Q^{\bar{c}} \, \pa^2
  \, Q^{\bar{a}}-
   f^{\bar{d} \bar{b}}_{\,\,\,\,\,\,e} \, Q^{\bar{c}} \, \pa \, (Q^{\bar{a}} \, V^e)
  \Bigg]
\nonu \\
&& +\frac{4(3+2k+N)}{3(k+N+2)^2} \, h^1_{\bar{a} \bar{b}}\,
d^1_{\bar{c} \bar{d}}
\nonu \\
&& \times \Bigg[
(k+N+2)
  \, (-g^{\bar{c} \bar{a}} \,  V^{\bar{d}} \, \pa \, V^{\bar{b}} +\frac{1}{2} \,
  g^{\bar{c} \bar{a}} \,  f^{\bar{b} \bar{d}}_{\,\,\,\,\,\,e} \,
  \pa^2 \, V^{e})  - k \, g^{\bar{d} \bar{b}} \, \pa \, Q^{\bar{c}} \, \pa
  Q^{\bar{a}}-  f^{\bar{b} \bar{d}}_{\,\,\,\,\,\,e} \, Q^{\bar{a}} \, \pa \,
  Q^{\bar{c}} \, V^e
  \Bigg]
\nonu \\
&&- \frac{1}{(k+N+2)^4} \, \Bigg[ d^2_{\bar{l} \bar{j}} \, h^3_{\bar{i}
    \bar{k}} + d^3_{\bar{i} \bar{l}} \, h^2_{\bar{j} \bar{k}} -
  d^0_{\bar{i} \bar{j}} \, h^1_{\bar{k} \bar{l}} + h^1_{\bar{i} \bar{j}}
  \, d^0_{\bar{k} \bar{l}}\Bigg] \, h^1_{\bar{a} \bar{b}} 
\nonu \\
&& \times
\Bigg[
  (k+N+2) \, ( g^{\bar{k} \bar{a}}  \,
Q^{\bar{j}} \,
Q^{\bar{i}} \, V^{\bar{b}} \, V^{\bar{l}} 
-  g^{\bar{j} \bar{a}}   \,
Q^{\bar{k}} \, Q^{\bar{i}} \, V^{\bar{b}}\, V^{\bar{l}}+
 g^{\bar{i} \bar{a}} \, 
 Q^{\bar{k}} \, Q^{\bar{j}} \, V^{\bar{b}} \, V^{\bar{l}})
 \nonu \\
 && +
f^{\bar{l} \bar{b}}_{\,\,\,\,\,\, m} \, Q^{\bar{k}} \, Q^{\bar{j}} \, Q^{\bar{i}}
\, Q^{\bar{a}} \, V^m
+k \, g^{\bar{l} \bar{b}} \, Q^{\bar{k}} \,
Q^{\bar{j}} \, Q^{\bar{i}} \, \pa \, Q^{\bar{a}}
  \Bigg]
\label{newpoleone}
\\
&& - \frac{4(k+2N)}{3(k+N+2)^3} \, h^1_{\bar{a} \bar{b}}\,
d^1_{\bar{c} \bar{d}} \,\Bigg[
-(k+N+2) \, g^{\bar{c} \bar{a}}  \,
V^{\bar{b}} \, \pa \, V^{\bar{d}} -
k \, g^{\bar{d} \bar{b}} \, Q^{\bar{c}} \, \pa^2
  \, Q^{\bar{a}}-
   f^{\bar{d} \bar{b}}_{\,\,\,\,\,\,e} \, Q^{\bar{c}} \, \pa \, (Q^{\bar{a}} \, V^e)
   \Bigg].
\nonu
\eea}  
One would like to see that
(\ref{newpoleone}) should be equal to (\ref{spin3location}) together with
(\ref{spin3expression}), (\ref{qspin3}), (\ref{vspin3}), (\ref{qvspin3})
and (\ref{spin1spin2}).

For the quartic terms in the spin-$\frac{1}{2}$ currents,
the identities (\ref{hgd-1}) and (\ref{hhrel}) can be used
and it turns out that
\bea
-\frac{k}{(k+N+2)^4} \,
\Bigg[ \, h^3_{\bar{a} \bar{b}} \, d^3_{\bar{c} \bar{d}} +
 h^2_{\bar{a} \bar{b}} \, d^2_{\bar{c} \bar{d}}
 + h^0_{\bar{a} \bar{b}} \, d^0_{\bar{c} \bar{d}}
 + h^1_{\bar{a} \bar{b}} \, d^1_{\bar{c} \bar{d}}
  \, \Bigg]
Q^{\bar{a}} \, Q^{\bar{b}} \, Q^{\bar{c}} \, \pa \, Q^{\bar{d}}(z),
\label{fourq}
\eea
which is exactly the same as the corresponding terms in
(\ref{spin3location}) by using (\ref{qspin3}) and (\ref{spin1spin2}).
Note that according to (\ref{spin3location}), there is a minus sign
in front of higher spin-$3$ current and can be seen in (\ref{fourq}). 

For the quadratic term $Q^{\bar{a}} \, \pa^2 \, Q^{\bar{b}}$, 
the contribution from the third line of (\ref{newpoleone})
can be written in terms of
{\footnotesize
  \bea
-\frac{1}{(k+N+2)^3} \, \frac{k}{2} \, (1+1+2-2N-N-N) \, d^0_{\bar{a} \bar{b}} \,
Q^{\bar{a}} \, \pa^2
  \, Q^{\bar{b}} = -\frac{2k(1-N)}{(k+N+2)^3} \, d^0_{\bar{a} \bar{b}} \,
Q^{\bar{a}} \, \pa^2
  \, Q^{\bar{b}},
\label{qddqexp}
\eea}
by identifying the $ h d f f h g$ factors for generic $N$ correctly.
By collecting other contributions from (\ref{newpoleone}), it turns out that
one obtains, together with (\ref{qddqexp}),  the final result
\bea
\frac{2k(3+2k+N)}{3(k+N+2)^3}  \, d^0_{\bar{a} \bar{b}} \,
Q^{\bar{a}} \, \pa^2
  \, Q^{\bar{b}},
\label{qfirst}
\eea
which is exactly the same as the corresponding terms in
(\ref{spin3location}) together with (\ref{qspin3}) and (\ref{spin1spin2}).
The other quadratic terms
can be simplified as
\bea
-\frac{4k(3+2k+N)}{3(k+N+2)^3}   \, d^0_{\bar{a} \bar{b}} \,
\pa \, Q^{\bar{a}} \, \pa
  \, Q^{\bar{b}},
\label{qsecond}
\eea
where the identity (\ref{hgd-1}) is used.
One sees that this is exactly the same as the corresponding terms in
(\ref{spin3location}) together with
(\ref{qspin3}) and (\ref{spin1spin2}).
Then, one obtains the total expression in (\ref{qfirst}) and (\ref{qsecond}).

For the spin-$1$ dependent terms,
one obtains
\bea
 \frac{4(k+2)}{(k+N+2)^2} \, d^0_{\bar{a} \bar{b}} \, V^{\bar{a}} \,
  \pa \, V^{\bar{b}}(z)
  +\frac{2(3+2k+N)}{3(k+N+2)^2} \, d^0_{\bar{a} \bar{b}}
  \, f^{\bar{a}
  \bar{b}}_{\,\,\,\,\,\, c} \, \pa^2 \, V^{\bar{c}}(z),
\label{derivativespin1}
\eea
by using the identity (\ref{hgd-1}). 
Furthermore,
the cubic terms are described as
\bea
&& -\frac{1}{2N(k+N+2)^2}
(2 h^1_{\bar{a} \bar{b}} \, d^1_{\bar{c} \bar{d}}
+
h^2_{\bar{a} \bar{b}} \, d^2_{\bar{c} \bar{d}}
+h^3_{\bar{a} \bar{b}} \, d^3_{\bar{c} \bar{d}})
f^{\bar{a}
  \bar{b}}_{\,\,\,\,\,\, e} \,
  V^{\bar{c}} \, V^{\bar{d}} \,  V^{e}(z)
  \nonu \\
  &&- \frac{1}{(k+N+2)^2}
(2 h^1_{\bar{a} \bar{b}} \, d^1_{\bar{c} \bar{d}}
+
h^2_{\bar{a} \bar{b}} \, d^2_{\bar{c} \bar{d}}
+h^3_{\bar{a} \bar{b}} \, d^3_{\bar{c} \bar{d}})
f^{\bar{b}
  \bar{d}}_{\,\,\,\,\,\, e} \, V^{\bar{a}} \,
  V^{\bar{c}} \, V^{e}(z),
\label{cubicspin1}
\eea
by using the identities (\ref{hgd-1}) and (\ref{hhrel}).
In order to compare the previous expressions 
(\ref{spin3location}) with (\ref{derivativespin1}) and (\ref{cubicspin1}),
one should move $V^e$ in the first term of (\ref{cubicspin1}) to the left.
This will give the additional contributions to (\ref{derivativespin1})
and it turns out that
they are exactly the same as the corresponding terms in
(\ref{spin3location}) together with
(\ref{vspin3}) and (\ref{spin1spin2}).

For the quintic terms,
one obtains
\bea
-\frac{1}{(k+N+2)^4} \, \Bigg[ d^2_{\bar{l} \bar{c}} \, h^3_{\bar{b}
    \bar{d}} + d^3_{\bar{b} \bar{l}} \, h^2_{\bar{c} \bar{d}} -
  d^0_{\bar{b} \bar{c}} \, h^1_{\bar{d} \bar{l}} + h^1_{\bar{b} \bar{c}}
  \, d^0_{\bar{d} \bar{l}} \, \Bigg] \, h^1_{\bar{a} \bar{b'}} \,
f^{\bar{l} \bar{b'}}_{\,\,\,\,\,\, e} \, Q^{\bar{a}} \, Q^{\bar{b}} \, Q^{\bar{c}}
\, Q^{\bar{d}} \, V^e,
\label{qqqqv}
\eea
which is equal to the previous expressions (\ref{spin3location})
corresponding to the quintic terms together with 
(\ref{qvspin3}) and (\ref{spin1spin2}).
One can simplify this (\ref{qqqqv}) further by using the indices $\bar{a},
\cdots, \bar{d}$. 

For the quartic terms appearing in the third line from the below of
(\ref{newpoleone}), one can use the identities (\ref{hgd-1}) and
(\ref{hhrel}).
For the quartic terms appearing in the third line of
(\ref{newpoleone}), one can use some identity between the $ h h f f$
factors and the $h$ factors with some $N$ dependent coefficients for the
first three terms.
For the last three terms one has the following identity
\bea
&& \Bigg[
 -  2  \, h^0_{\bar{b'} \bar{i}}\,
  d^1_{\bar{a'} \bar{j}} -  h^2_{\bar{b'} \bar{j}}\,
  d^3_{\bar{a'} \bar{i}} +  h^2_{\bar{b'} \bar{i}}\,
  d^3_{\bar{a'} \bar{j}} \Bigg]  
f^{\bar{a'} \bar{b'}}_{\,\,\,\,\,\,k} \, h^1_{\bar{a} \bar{b}} \,
f^{k \bar{b}}_{\,\,\,\,\,\,l}  \,   Q^{\bar{i}} \, Q^{\bar{a}} \, V^{\bar{j}}
\, V^{l} =
\nonu \\
&& -\Bigg[ d^3_{\bar{a} \bar{b}}\,
  h^3_{\bar{c} \bar{d}} +  d^3_{\bar{a} \bar{c}}\,
  h^3_{\bar{d} \bar{b}}+ d^3_{\bar{d} \bar{c}}\,
  h^3_{\bar{a} \bar{b}}-
   d^3_{\bar{d} \bar{b}}\,
  h^3_{\bar{a} \bar{c}} -  d^3_{\bar{c} \bar{b}}\,
  h^3_{\bar{d} \bar{a}} + d^2_{\bar{a} \bar{b}}\,
  h^2_{\bar{c} \bar{d}}  + d^2_{\bar{a} \bar{c}}\,
  h^2_{\bar{d} \bar{b}} \nonu \\
  && +  d^2_{\bar{d} \bar{c}}\,
  h^2_{\bar{a} \bar{b}}- d^2_{\bar{d} \bar{b}}\,
  h^2_{\bar{a} \bar{c}} - d^2_{\bar{c} \bar{b}}\,
  h^2_{\bar{d} \bar{a}} - d^1_{\bar{a} \bar{b}}\,
  h^1_{\bar{c} \bar{d}}  + 2 d^1_{\bar{a} \bar{d}}\,
  h^1_{\bar{b} \bar{c}} -  d^1_{\bar{a} \bar{c}}\,
  h^1_{\bar{d} \bar{b}}- d^1_{\bar{d} \bar{c}}\,
  h^1_{\bar{a} \bar{b}} \nonu \\
  && + d^1_{\bar{d} \bar{b}}\,
  h^1_{\bar{a} \bar{c}}
 +  d^0_{\bar{a} \bar{b}}\,
  h^0_{\bar{c} \bar{d}}  -  d^0_{\bar{a} \bar{c}}\,
  h^0_{\bar{d} \bar{b}} -  d^0_{\bar{d} \bar{c}}\,
  h^0_{\bar{a} \bar{b}}+ d^0_{\bar{d} \bar{b}}\,
  h^0_{\bar{a} \bar{c}}
  \Bigg] Q^{\bar{a}} \, Q^{\bar{b}} \, V^{\bar{c}}
\, V^{\bar{d}}.
\label{propro}
\eea
It turns out, together with (\ref{propro}), that by collecting
with all the contributions from $Q^{\bar{a}} \, Q^{\bar{b}} \, V^{\bar{c}}
\, V^{\bar{d}}$, one can check
the coincidence of the corresponding terms in (\ref{spin3location})
together with (\ref{qvspin3}) and (\ref{spin1spin2}).

Let us consider the quartic terms appearing in the second line of
(\ref{newpoleone}). Then the first three contributions are equal to
the ninth line of (\ref{qvspin3}) with minus sign.
Moreover, the contribution from the last three terms 
is equal to the ones in the twelfth, thirteenth, and the fourteenth
line of (\ref{qvspin3}) with minus sign respectively.

Let us describe the final terms, the cubic terms with one derivative.
The previous identities (\ref{hgd-1}) and (\ref{hhrel}) can be used also.
There are nontrivial identities 
corresponding to the first terms in the third line of (\ref{newpoleone}).
That is,
\bea
  h^3_{\bar{a'} \bar{b'}}\,
  d^2_{\bar{i} \bar{j}} \, f^{\bar{a'} \bar{b'}}_{\,\,\,\,\,\,k} \,
  h^1_{\bar{a} \bar{b}} \,f^{\bar{j} \bar{b}}_{\,\,\,\,\,\,l} \,
  f^{k l}_{\,\,\,\,\,\,m} & = &
( -h^1_{\bar{a} \bar{b}}\,
  d^1_{\bar{c} \bar{d}} +
 h^2_{\bar{a} \bar{b}}\,
d^2_{\bar{c} \bar{d}} \, ) f^{\bar{a} \bar{b}}_{\,\,\,\,\,\,e},
  \nonu \\
 h^2_{\bar{a'} \bar{b'}}\,
 d^3_{\bar{i} \bar{j}} \,
f^{\bar{a'} \bar{b'}}_{\,\,\,\,\,\,k} \, h^1_{\bar{a} \bar{b}} \,
 f^{\bar{j} \bar{b}}_{\,\,\,\,\,\,l} \,
  f^{k l}_{\,\,\,\,\,\,m} & = & ( \, h^1_{\bar{a} \bar{b}}\,
  d^1_{\bar{c} \bar{d}} -
 h^3_{\bar{a} \bar{b}}\,
d^3_{\bar{c} \bar{d}} \, ) f^{\bar{a} \bar{b}}_{\,\,\,\,\,\,e},
  \nonu \\
h^1_{\bar{a'} \bar{b'}}\,
d^0_{\bar{i} \bar{j}} \,
f^{\bar{a'} \bar{b'}}_{\,\,\,\,\,\,k} \, h^1_{\bar{a} \bar{b}} \,
f^{\bar{j} \bar{b}}_{\,\,\,\,\,\,l} \,
f^{k l}_{\,\,\,\,\,\,m}  & = &
( \, h^2_{\bar{a} \bar{b}}\,
  d^2_{\bar{c} \bar{d}} +
 h^3_{\bar{a} \bar{b}}\,
d^3_{\bar{c} \bar{d}} \, ) f^{\bar{a} \bar{b}}_{\,\,\,\,\,\,e}.
\label{nontrivialrel}
\eea
Then one can check that these coincide with the ones in
(\ref{spin3location}) together with (\ref{qvspin3})
and (\ref{spin1spin2}).

The remaining two subsections maybe be skipped without any discontinuity
and the readers can go to the next section $4$ directly.  

\subsubsection{Other way of  obtaining the higher spin-$3$ current
  using  the higher spin-$\frac{5}{2}$ current in Appendix
  (\ref{spin5halfone-1})}

In Appendix $C$, there are other two different 
expressions Appendix (\ref{spin5halfone-1}) and Appendix
(\ref{spin5halfone-2}),
corresponding to the above
higher spin-$\frac{5}{2}$ current in (\ref{spin5halfone}).
Let us see how
the higher spin-$\frac{5}{2}$ current (\ref{spin5halfone})
and the same quantity given in Appendix
(\ref{spin5halfone-1}) differ from each
other.
The difference appears in the cubic terms (the first six terms).
Then
after one calculates the OPE between the spin-$\frac{3}{2}$
current $G^3(z)$ and the higher spin-$\frac{5}{2}$ current in
(\ref{spin5halfone-1}), one obtains
{\footnotesize
\bea
&& -\frac{1}{(k+N+2)^3} \, \Bigg[ -
  \frac{1}{2N} \, h^2_{\bar{a'} \bar{b'}}\,
  d^3_{\bar{i} \bar{j}} +
\frac{1}{2N} \, h^1_{\bar{a'} \bar{b'}}\,
  d^0_{\bar{i} \bar{j}}
+ \frac{1}{N} \, h^3_{\bar{a'} \bar{b'}}\,
  d^2_{\bar{i} \bar{j}}
 +  2  \, h^2_{\bar{b'} \bar{i}}\,
  d^3_{\bar{a'} \bar{j}} -  h^2_{\bar{b'} \bar{j}}\,
  d^3_{\bar{a'} \bar{i}} -  h^0_{\bar{b'} \bar{i}}\,
  d^1_{\bar{a'} \bar{j}} \Bigg]
\nonu \\
&& \times f^{\bar{a'} \bar{b'}}_{\,\,\,\,\,\,k} \, h^1_{\bar{a} \bar{b}} \,
\Bigg[-(k+N+2) g^{\bar{i} \bar{a}} \,
  V^{\bar{b}} \, V^{\bar{j}} \, V^k -
  k \, g^{\bar{j} \bar{b}} \, Q^{\bar{i}}
  \,  \pa \, Q^{\bar{a}} \, V^k - f^{\bar{j} \bar{b}}_{\,\,\,\,\,\,l}
  \, Q^{\bar{i}} \,
  Q^{\bar{a}} \, V^{k} \, V^l
  \nonu \\
  && - f^{\bar{j} \bar{b}}_{\,\,\,\,\,\,l} \,
f^{k l}_{\,\,\,\,\,\,m} \,
 Q^{\bar{i}} \, \pa
 \, ( Q^{\bar{a}} \, V^{m})
 -f^{k \bar{b}}_{\,\,\,\,\,\,l} \, 
 Q^{\bar{i}} \, Q^{\bar{a}} \, V^{\bar{j}} \, V^l -
 \frac{k}{2} \, f^{\bar{j} \bar{b}}_{\,\,\,\,\,\,l} \,
 g^{k l} \, Q^{\bar{i}} \,
 \pa^2 \, Q^{\bar{a}} - k\, g^{k \bar{b}} \, Q^{\bar{i}} \, \pa Q^{\bar{a}} \,
 V^{\bar{j}} \Bigg] \nonu \\
&& + \mbox{the remaining terms starting from the fourth line to the last }
(\ref{newpoleone}). 
\label{secondcase}
\eea}
One should check whether the three lines of (\ref{secondcase})
are equal to those (the first three lines) in
(\ref{newpoleone}) or not.

One can simplify the cubic terms in the spin-$1$ current as before.
It turns out that
\bea
&& -\frac{1}{2N(k+N+2)^2}
(h^1_{\bar{a} \bar{b}} \, d^1_{\bar{c} \bar{d}}
+
h^2_{\bar{a} \bar{b}} \, d^2_{\bar{c} \bar{d}}
+2 h^3_{\bar{a} \bar{b}} \, d^3_{\bar{c} \bar{d}})
f^{\bar{a}
  \bar{b}}_{\,\,\,\,\,\, e} \,
  V^{\bar{c}} \, V^{\bar{d}} \,  V^{e}(z)
  \nonu \\
  &&- \frac{1}{(k+N+2)^2}
( h^1_{\bar{c} \bar{a}} \, d^1_{\bar{d} \bar{b}}
+
h^2_{\bar{c} \bar{a}} \, d^2_{\bar{d} \bar{b}}
+2 h^3_{\bar{c} \bar{a}} \, d^3_{\bar{d} \bar{b}})
f^{\bar{b}
  \bar{d}}_{\,\,\,\,\,\, e} \, V^{\bar{a}} \,
  V^{\bar{c}} \, V^{e}(z).
\label{cubicspin1-new}
\eea
By identifying (\ref{cubicspin1}) with (\ref{cubicspin1-new}),
one obtains the nontrivial relation between the various tensors
as follows:
\bea
 \frac{1}{2N} ( \, h^1_{\bar{a} \bar{b}}\,
  d^1_{\bar{c} \bar{d}} -
 h^3_{\bar{a} \bar{b}}\,
d^3_{\bar{c} \bar{d}} \, ) f^{\bar{a} \bar{b}}_{\,\,\,\,\,\,e}
  & = &  -( \, h^1_{\bar{c} \bar{a}}\,
  d^1_{\bar{d} \bar{b}} -
 h^3_{\bar{c} \bar{a}}\,
 d^3_{\bar{d} \bar{b}} \, ) f^{\bar{a} \bar{b}}_{\,\,\,\,\,\,e}.
 \label{1133}
 \eea
 This identity can be also observed in the different context of Appendix
 $C$.
 
 From the contributions of the cubic terms in the second line of
 (\ref{secondcase})
 and the corresponding terms in (\ref{newpoleone}), one obtains the
 following nontrivial relation between the tensors
 \bea
\frac{1}{2N} ( \, -h^1_{\bar{a} \bar{b}}\,
  d^1_{\bar{c} \bar{d}} -
 h^3_{\bar{a} \bar{b}}\,
d^3_{\bar{c} \bar{d}} \, ) f^{\bar{a} \bar{b}}_{\,\,\,\,\,\,e}
  & = &  ( \, h^0_{\bar{b} \bar{c}}\,
  d^0_{\bar{a} \bar{d}} -
 h^2_{\bar{b} \bar{c}}\,
 d^2_{\bar{a} \bar{d}} \, ) f^{\bar{a} \bar{b}}_{\,\,\,\,\,\,e}.
 \label{1133-1}
 \eea
 Due to the relative sign change in the left hand sides of
 (\ref{1133}) and (\ref{1133-1}), one can obtain
that each $h d f$ factor can be written in terms of other quantities 
in the right hand side by adding or subtracting them.

One can check the coincidence of the quadratic terms.  
For the cubic terms with one derivative, one can use the
previous relations in (\ref{nontrivialrel}) and
the other remaining terms can be checked explicitly.

\subsubsection{Other way of  obtaining the higher spin-$3$ current
  using  the higher spin-$\frac{5}{2}$ current in Appendix
  (\ref{spin5halfone-2})}

After one calculates the OPE between the spin-$\frac{3}{2}$
current $G^3(z)$ and the higher spin-$\frac{5}{2}$ current in
Appendix (\ref{spin5halfone-2}), one obtains the first order pole of
this OPE as follows:
{\footnotesize
\bea
&& -\frac{1}{(k+N+2)^3} \, \Bigg[ 
  \frac{1}{2N} \, h^3_{\bar{a'} \bar{b'}}\,
  d^2_{\bar{i} \bar{j}} +
\frac{1}{2N} \, h^1_{\bar{a'} \bar{b'}}\,
  d^0_{\bar{i} \bar{j}}
- \frac{1}{N} \, h^2_{\bar{a'} \bar{b'}}\,
  d^3_{\bar{i} \bar{j}}
 -  2  \, h^3_{\bar{b'} \bar{i}}\,
  d^2_{\bar{a'} \bar{j}} +  h^3_{\bar{b'} \bar{j}}\,
  d^2_{\bar{a'} \bar{i}} -  h^0_{\bar{b'} \bar{i}}\,
  d^1_{\bar{a'} \bar{j}} \Bigg]
\nonu \\
&& \times f^{\bar{a'} \bar{b'}}_{\,\,\,\,\,\,k} \, h^1_{\bar{a} \bar{b}} \,
\Bigg[-(k+N+2) g^{\bar{i} \bar{a}} \,
  V^{\bar{b}} \, V^{\bar{j}} \, V^k -
  k \, g^{\bar{j} \bar{b}} \, Q^{\bar{i}}
  \,  \pa \, Q^{\bar{a}} \, V^k - f^{\bar{j} \bar{b}}_{\,\,\,\,\,\,l}
  \, Q^{\bar{i}} \,
  Q^{\bar{a}} \, V^{k} \, V^l
  \nonu \\
  && - f^{\bar{j} \bar{b}}_{\,\,\,\,\,\,l} \,
f^{k l}_{\,\,\,\,\,\,m} \,
 Q^{\bar{i}} \, \pa
 \, ( Q^{\bar{a}} \, V^{m})
 -f^{k \bar{b}}_{\,\,\,\,\,\,l} \, 
 Q^{\bar{i}} \, Q^{\bar{a}} \, V^{\bar{j}} \, V^l -
 \frac{k}{2} \, f^{\bar{j} \bar{b}}_{\,\,\,\,\,\,l} \,
 g^{k l} \, Q^{\bar{i}} \,
 \pa^2 \, Q^{\bar{a}} - k\, g^{k \bar{b}} \, Q^{\bar{i}} \, \pa Q^{\bar{a}} \,
 V^{\bar{j}} \Bigg] \nonu \\
&& + \mbox{the remaining terms of
 starting from the fourth line to the last
} (\ref{newpoleone}). 
\label{secondcase-1}
\eea}
One should check that the three lines of (\ref{secondcase-1})
are equal to those (the first three lines) in
(\ref{newpoleone}) as before.

One can simplify the cubic terms in the spin-$1$ current as before.
It turns out that
\bea
&& -\frac{1}{2N(k+N+2)^2}
( h^1_{\bar{a} \bar{b}} \, d^1_{\bar{c} \bar{d}}
+
2 h^2_{\bar{a} \bar{b}} \, d^2_{\bar{c} \bar{d}}
+ h^3_{\bar{a} \bar{b}} \, d^3_{\bar{c} \bar{d}})
f^{\bar{a}
  \bar{b}}_{\,\,\,\,\,\, e} \,
  V^{\bar{c}} \, V^{\bar{d}} \,  V^{e}(z)
  \nonu \\
  &&- \frac{1}{(k+N+2)^2}
( h^1_{\bar{b} \bar{c}} \, d^1_{\bar{a} \bar{d}}
+
2 h^2_{\bar{b} \bar{c}} \, d^2_{\bar{a} \bar{d}}
+ h^3_{\bar{b} \bar{c}} \, d^3_{\bar{a} \bar{d}})
f^{\bar{b}
  \bar{d}}_{\,\,\,\,\,\, e} \, V^{\bar{a}} \,
  V^{\bar{c}} \, V^{e}(z).
\label{cubicspin1-new-1}
\eea
By identifying (\ref{cubicspin1}) with (\ref{cubicspin1-new-1}),
one obtains the nontrivial relation between the tensors
 \bea
  \frac{1}{2N} ( \, -h^1_{\bar{a} \bar{b}}\,
  d^1_{\bar{c} \bar{d}} +
 h^2_{\bar{a} \bar{b}}\,
d^2_{\bar{c} \bar{d}} \, ) f^{\bar{a} \bar{b}}_{\,\,\,\,\,\,e}
  & = &  ( \, h^1_{\bar{c} \bar{a}}\,
  d^1_{\bar{d} \bar{b}} -
 h^2_{\bar{c} \bar{a}}\,
 d^2_{\bar{d} \bar{b}} \, ) f^{\bar{a} \bar{b}}_{\,\,\,\,\,\,e}.
 \label{1122}
 \eea
 This identity can be also observed in the different context of Appendix
 $C$.
 
From the contributions of the cubic terms in the second line of
 (\ref{secondcase-1})
 and the corresponding terms in (\ref{newpoleone}), one obtains the
 following nontrivial relation
 \bea
\frac{1}{2N} ( \, -h^2_{\bar{a} \bar{b}}\,
  d^2_{\bar{c} \bar{d}} -
 h^1_{\bar{a} \bar{b}}\,
d^1_{\bar{c} \bar{d}} \, ) f^{\bar{a} \bar{b}}_{\,\,\,\,\,\,e}
  & = &  ( \, h^0_{\bar{b} \bar{c}}\,
  d^0_{\bar{a} \bar{d}} -
 h^3_{\bar{b} \bar{c}}\,
 d^3_{\bar{a} \bar{d}} \, ) f^{\bar{a} \bar{b}}_{\,\,\,\,\,\,e}. 
\label{1122-1}
 \eea
 Due to the relative sign change in the left hand sides of
 (\ref{1122}) and (\ref{1122-1}),
each $h d f$ factor can be written in terms of other quantities 
in the right hand side by adding or subtracting them.
One can check that all the other remaining terms in (\ref{secondcase-1})
are equal to those in (\ref{newpoleone}) each other.

In Appendix $D$, there are also other possibilities to express
the higher spin-$3$ current using the different $(\mu,\nu)$ cases.
It would be interesting to check whether the other cases also lead to
the previous higher spin-$3$ current with nontrivial relations
between the various tensors.


\section{Eigenvalues and Three-point functions}

We reinterpret the previous results in \cite{Ahn1711}
by examining the structure of (\ref{qspin3}) and (\ref{vspin3})
carefully, calculate the eigenvalues of the higher spin-$3$ current
acting on the other higher representations and extract the
corresponding three-point functions. We also analyze its large
$(N,k)$ 't Hooft-like limit in terms of two-point functions
of two scalar operators with
't Hooft coupling constant-dependent coefficients (and the number of
boxes of Young tableaux). 

\subsection{ The $(0; \La_{-})$ representations up to three boxes}

The relevant part of the higher spin-$3$ current
acting on this representation is given by (\ref{qspin3})
where there are four kinds of independent terms. 
It turns out that
there is no third order pole between the six multiple product of
spin-$\frac{1}{2}$ current in the higher spin-$3$ current
(corresponding to the first term in (\ref{qspin3}))
and the spin-$\frac{1}{2}$ current.
Similarly, 
there is no third order pole between the six multiple product of
spin-$\frac{1}{2}$ current in the higher spin-$3$ current
and the quadratic term in the spin-$\frac{1}{2}$ current.

However, there exist
third order poles between the six multiple product of
spin-$\frac{1}{2}$ current in the higher spin-$3$ current
(\ref{qspin3})
and the cubic term in the spin-$\frac{1}{2}$ current as follows:
\bea
&&
 d^0_{\bar{a} \bar{b}} \, h^i_{\bar{c} \bar{d}} \, h^i_{\bar{e} \bar{f}} \,
Q^{\bar{a}} \, Q^{\bar{b}} \, Q^{\bar{c}} \, Q^{\bar{d}} \, Q^{\bar{e}} \, Q^{\bar{f}}
(z) \, Q^{\bar{g}} \, Q^{\bar{h}} \, Q^{\bar{i}}(w) =
\nonu \\
&& -\frac{1}{(z-w)^3} \, 8 \, (k+N+2)^3 \,
 d^0_{\bar{a} \bar{b}} \, h^i_{\bar{c} \bar{d}} \, h^i_{\bar{e} \bar{f}} \,
\Bigg[ g^{\bar{g} \bar{a}} \, 
  g^{\bar{h} \bar{b}}\,
  g^{\bar{i} \bar{c}} \, Q^{\bar{d}} \, Q^{\bar{e}} \, Q^{\bar{f}}
   -
 g^{\bar{g} \bar{a}} \, 
  g^{\bar{h} \bar{c}}\,
  g^{\bar{i} \bar{b}} \, Q^{\bar{d}} \, Q^{\bar{e}} \, Q^{\bar{f}}
  \nonu \\
  && 
  + g^{\bar{g} \bar{a}} \, 
  g^{\bar{h} \bar{c}}\,
  g^{\bar{i} \bar{d}} \, Q^{\bar{b}} \, Q^{\bar{e}} \, Q^{\bar{f}}
- g^{\bar{g} \bar{a}} \, 
  g^{\bar{h} \bar{c}}\,
  g^{\bar{i} \bar{e}} \, Q^{\bar{b}} \, Q^{\bar{d}} \, Q^{\bar{f}}
  + 
 g^{\bar{g} \bar{a}} \, 
  g^{\bar{h} \bar{e}}\,
  g^{\bar{i} \bar{c}} \, Q^{\bar{b}} \, Q^{\bar{d}} \, Q^{\bar{f}}  
  \nonu \\
  && +  g^{\bar{g} \bar{c}} \, 
  g^{\bar{h} \bar{a}}\,
  g^{\bar{i} \bar{b}} \, Q^{\bar{d}} \, Q^{\bar{e}} \, Q^{\bar{f}}
  - g^{\bar{g} \bar{c}} \, 
  g^{\bar{h} \bar{a}}\,
  g^{\bar{i} \bar{d}} \, Q^{\bar{b}} \, Q^{\bar{e}} \, Q^{\bar{f}}
  +   g^{\bar{g} \bar{c}} \, 
  g^{\bar{h} \bar{a}}\,
  g^{\bar{i} \bar{e}} \, Q^{\bar{b}} \, Q^{\bar{d}} \, Q^{\bar{f}}
  \nonu \\
  &&
  + g^{\bar{g} \bar{c}} \, 
  g^{\bar{h} \bar{d}}\,
  g^{\bar{i} \bar{a}} \, Q^{\bar{b}} \, Q^{\bar{e}} \, Q^{\bar{f}} +
   g^{\bar{g} \bar{c}} \, 
  g^{\bar{h} \bar{d}}\,
  g^{\bar{i} \bar{e}} \, Q^{\bar{a}} \, Q^{\bar{b}} \, Q^{\bar{f}}
  -
 g^{\bar{g} \bar{c}} \, 
  g^{\bar{h} \bar{e}}\,
  g^{\bar{i} \bar{a}} \, Q^{\bar{b}} \, Q^{\bar{d}} \, Q^{\bar{f}}
\nonu \\
&&  - g^{\bar{g} \bar{c}} \, 
  g^{\bar{h} \bar{e}}\,
  g^{\bar{i} \bar{d}} \, Q^{\bar{a}} \, Q^{\bar{b}} \, Q^{\bar{f}}
  +  g^{\bar{g} \bar{c}} \, 
  g^{\bar{h} \bar{e}}\,
  g^{\bar{i} \bar{f}} \, Q^{\bar{a}} \, Q^{\bar{b}} \, Q^{\bar{d}}
   -  g^{\bar{g} \bar{e}} \, 
  g^{\bar{h} \bar{a}}\,
  g^{\bar{i} \bar{c}} \, Q^{\bar{b}} \, Q^{\bar{d}} \, Q^{\bar{f}}
\nonu \\
&&  +  g^{\bar{g} \bar{e}} \, 
  g^{\bar{h} \bar{c}}\,
  g^{\bar{i} \bar{a}} \, Q^{\bar{b}} \, Q^{\bar{d}} \, Q^{\bar{f}}
  \Bigg](w) + {\cal O}(\frac{1}{(z-w)^2}).
\label{q6q3}
\eea
We focus on the third order pole only for the calculation of
the eigenvalues.
Only after one specifies the indices $\bar{g}, \bar{h}$ and $\bar{i}$,
one can determine the coefficient of the composite field
$ Q^{\bar{g}} \, Q^{\bar{h}} \, Q^{\bar{i}}(w)$ in the right hand side of
(\ref{q6q3}).

It turns out that
there is no third order pole between the quartic product of
spin-$\frac{1}{2}$ current (corresponding to
the second term of (\ref{qspin3})) in the higher spin-$3$ current
and the spin-$\frac{1}{2}$ current.
There are third order poles between the quartic product of
spin-$\frac{1}{2}$ current in the higher spin-$3$ current
(\ref{qspin3})
and the quadratic term in the spin-$\frac{1}{2}$ current,
by considering the second term of (\ref{qspin3}), as follows:
\bea
&&
 d^0_{\bar{a} \bar{b}} \, h^0_{\bar{c} \bar{d}}  \,
Q^{\bar{a}} \, Q^{\bar{b}} \, Q^{\bar{c}} \, \pa \, Q^{\bar{d}} 
(z) \, Q^{\bar{e}} \, Q^{\bar{f}}(w) =
\nonu \\
&& \frac{1}{(z-w)^3}  \, (k+N+2)^2 \,
 d^0_{\bar{a} \bar{b}} \, h^0_{\bar{c} \bar{d}} \, 
\Bigg[ - g^{\bar{e} \bar{d}} \, 
  g^{\bar{f} \bar{a}} \, Q^{\bar{b}} \, Q^{\bar{c}} 
+ g^{\bar{e} \bar{d}} \, 
g^{\bar{f} \bar{b}} \, Q^{\bar{a}} \, Q^{\bar{c}} -
g^{\bar{e} \bar{d}} \, 
  g^{\bar{f} \bar{c}} \, Q^{\bar{a}} \, Q^{\bar{b}} 
  \nonu \\
  && +
g^{\bar{e} \bar{a}} \, 
g^{\bar{f} \bar{d}} \, Q^{\bar{b}} \, Q^{\bar{c}}
- g^{\bar{e} \bar{b}} \, 
g^{\bar{f} \bar{d}} \, Q^{\bar{a}} \, Q^{\bar{c}}
+ g^{\bar{e} \bar{c}} \, 
  g^{\bar{f} \bar{d}} \, Q^{\bar{a}} \, Q^{\bar{b}} 
  \Bigg](w) + {\cal O}(\frac{1}{(z-w)^2}).  
  \label{q4qq}
  \eea
As before, the third order pole is the relevant terms \footnote{
Moreover,
one has, for the third term of (\ref{qspin3}) with the sum over the
index $i$,
 \bea
&&
 h^i_{\bar{a} \bar{b}} \, d^i_{\bar{c} \bar{d}}  \,
Q^{\bar{a}} \, Q^{\bar{b}} \, Q^{\bar{c}} \, \pa \, Q^{\bar{d}} 
(z) \, Q^{\bar{e}} \, Q^{\bar{f}}(w) =
\nonu \\
&& \frac{1}{(z-w)^3}  \, (k+N+2)^2 \,
 h^i_{\bar{a} \bar{b}} \, d^i_{\bar{c} \bar{d}} \, 
\Bigg[ - g^{\bar{e} \bar{d}} \, 
  g^{\bar{f} \bar{a}} \, Q^{\bar{b}} \, Q^{\bar{c}} 
+ g^{\bar{e} \bar{d}} \, 
g^{\bar{f} \bar{b}} \, Q^{\bar{a}} \, Q^{\bar{c}} -
g^{\bar{e} \bar{d}} \, 
  g^{\bar{f} \bar{c}} \, Q^{\bar{a}} \, Q^{\bar{b}} 
  \nonu \\
  && +
g^{\bar{e} \bar{a}} \, 
g^{\bar{f} \bar{d}} \, Q^{\bar{b}} \, Q^{\bar{c}}
- g^{\bar{e} \bar{b}} \, 
g^{\bar{f} \bar{d}} \, Q^{\bar{a}} \, Q^{\bar{c}}
+ g^{\bar{e} \bar{c}} \, 
  g^{\bar{f} \bar{d}} \, Q^{\bar{a}} \, Q^{\bar{b}} 
  \Bigg](w) + {\cal O}(\frac{1}{(z-w)^2}).  
  \label{q4qq-1}
  \eea}. 
  Similarly,
   there are third order poles between the quartic product of
spin-$\frac{1}{2}$ current in the higher spin-$3$ current
(\ref{qspin3})
and the cubic term in the spin-$\frac{1}{2}$ current as follows
\footnote{ Similarly,
 one obtains (with the sum over the index $i$)
  \bea
&&
 h^i_{\bar{a} \bar{b}} \, d^i_{\bar{c} \bar{d}}  \,
Q^{\bar{a}} \, Q^{\bar{b}} \, Q^{\bar{c}} \, \pa \, Q^{\bar{d}} 
(z) \, Q^{\bar{e}} \, Q^{\bar{f}} \, Q^{\bar{g}}(w) =
\nonu \\
&& \frac{1}{(z-w)^3} \,  (k+N+2)^2 \,
 h^i_{\bar{a} \bar{b}} \, d^i_{\bar{c} \bar{d}}
\Bigg[ -2 g^{\bar{e} \bar{d}} \, 
  g^{\bar{f} \bar{a}}\,
 (Q^{\bar{d}} \, Q^{\bar{c}}) \, Q^{\bar{g}}
+ 2 g^{\bar{e} \bar{d}} \, 
  g^{\bar{g} \bar{a}}\,
 Q^{\bar{f}} \, Q^{\bar{b}} \, Q^{\bar{c}}
\nonu \\
&& + 2 g^{\bar{e} \bar{a}} \, 
  g^{\bar{f} \bar{d}}\,
 (Q^{\bar{b}} \, Q^{\bar{c}}) \, Q^{\bar{g}}
+ 2 g^{\bar{e} \bar{a}} \, 
  g^{\bar{f} \bar{c}}\, g^{\bar{g} \bar{d}} \, \pa \, Q^{\bar{b}}
  + 2 g^{\bar{e} \bar{a}} \, 
  g^{\bar{f} \bar{c}}\, g^{\bar{g} \bar{b}} \, \pa \, Q^{\bar{d}}
-  2 g^{\bar{e} \bar{a}} \, 
  g^{\bar{f} \bar{b}}\, g^{\bar{g} \bar{d}} \, \pa \, Q^{\bar{c}}
  \nonu \\
  & &
  -2   g^{\bar{e} \bar{a}} \, 
  g^{\bar{f} \bar{b}}\, g^{\bar{g} \bar{c}} \, \pa \, Q^{\bar{d}}
  -2 g^{\bar{e} \bar{a}} \, 
  g^{\bar{g} \bar{d}}\,
  Q^{\bar{f}} \, Q^{\bar{b}} \, Q^{\bar{c}}
  +  2 g^{\bar{e} \bar{c}} \, 
  g^{\bar{f} \bar{b}}\, g^{\bar{g} \bar{d}} \, \pa \, Q^{\bar{a}}
  + 2   g^{\bar{e} \bar{c}} \, 
  g^{\bar{f} \bar{b}}\, g^{\bar{g} \bar{a}} \, \pa \, Q^{\bar{d}}
  \nonu \\
  &&  -2 g^{\bar{f} \bar{d}} \, 
  g^{\bar{g} \bar{a}}\,
  Q^{\bar{e}} \, Q^{\bar{b}} \, Q^{\bar{c}}
  +  2 g^{\bar{f} \bar{a}} \, 
  g^{\bar{g} \bar{d}}\,
  Q^{\bar{e}} \, Q^{\bar{b}} \, Q^{\bar{c}}
  \Bigg](w) + {\cal O}(\frac{1}{(z-w)^2}).  
 \label{q4q3-1}
 \eea}
:
  \bea
&&
 d^0_{\bar{a} \bar{b}} \, h^0_{\bar{c} \bar{d}}  \,
Q^{\bar{a}} \, Q^{\bar{b}} \, Q^{\bar{c}} \, \pa \, Q^{\bar{d}} 
(z) \, Q^{\bar{e}} \, Q^{\bar{f}} \, Q^{\bar{g}}(w) =
\nonu \\
&& \frac{1}{(z-w)^3} \,  (k+N+2)^2 \,
 d^0_{\bar{a} \bar{b}} \, h^0_{\bar{c} \bar{d}}
\Bigg[ -2 g^{\bar{e} \bar{d}} \, 
  g^{\bar{f} \bar{a}}\,
 (Q^{\bar{b}} \, Q^{\bar{c}}) \, Q^{\bar{g}}
+ 2 g^{\bar{e} \bar{d}} \, 
  g^{\bar{g} \bar{a}}\,
 Q^{\bar{f}} \, Q^{\bar{b}} \, Q^{\bar{c}}
\nonu \\
&& + 2 g^{\bar{e} \bar{a}} \, 
  g^{\bar{f} \bar{d}}\,
 (Q^{\bar{b}} \, Q^{\bar{c}}) \, Q^{\bar{g}}
+ 2 g^{\bar{e} \bar{a}} \, 
  g^{\bar{f} \bar{c}}\, g^{\bar{g} \bar{d}} \, \pa \, Q^{\bar{b}}
  + 2 g^{\bar{e} \bar{a}} \, 
  g^{\bar{f} \bar{c}}\, g^{\bar{g} \bar{b}} \, \pa \, Q^{\bar{d}}
-  2 g^{\bar{e} \bar{a}} \, 
  g^{\bar{f} \bar{b}}\, g^{\bar{g} \bar{d}} \, \pa \, Q^{\bar{c}}
  \nonu \\
  & &
  -2   g^{\bar{e} \bar{a}} \, 
  g^{\bar{f} \bar{b}}\, g^{\bar{g} \bar{c}} \, \pa \, Q^{\bar{d}}
  -2 g^{\bar{e} \bar{a}} \, 
  g^{\bar{g} \bar{d}}\,
  Q^{\bar{f}} \, Q^{\bar{b}} \, Q^{\bar{c}}
  +  2 g^{\bar{e} \bar{c}} \, 
  g^{\bar{f} \bar{b}}\, g^{\bar{g} \bar{d}} \, \pa \, Q^{\bar{a}}
  + 2   g^{\bar{e} \bar{c}} \, 
  g^{\bar{f} \bar{b}}\, g^{\bar{g} \bar{a}} \, \pa \, Q^{\bar{d}}
  \nonu \\
  &&  -2 g^{\bar{f} \bar{d}} \, 
  g^{\bar{g} \bar{a}}\,
  Q^{\bar{e}} \, Q^{\bar{b}} \, Q^{\bar{c}}
  +  2 g^{\bar{f} \bar{a}} \, 
  g^{\bar{g} \bar{d}}\,
  Q^{\bar{e}} \, Q^{\bar{b}} \, Q^{\bar{c}}
  \Bigg](w) + {\cal O}(\frac{1}{(z-w)^2}).  
 \label{q4q3}
 \eea

 It turns out that
 there are third order poles between the quadratic product of
 spin-$\frac{1}{2}$ current (corresponding to the fourth term of
 (\ref{qspin3})) in the higher spin-$3$ current
(\ref{qspin3})
and the spin-$\frac{1}{2}$ current as follows: 
 \bea
 d^0_{\bar{a} \bar{b}}   \,
Q^{\bar{a}} \, \pa^2\, Q^{\bar{b}} 
(z) \, Q^{\bar{c}}(w) & =
& -\frac{1}{(z-w)^3} \,  2\, (k+N+2) \,
d^0_{\bar{a} \bar{b}} \, g^{\bar{c} \bar{b}} \, Q^{\bar{a}}(w)
+ {\cal O}(\frac{1}{(z-w)^2}).  
\label{qd2qq}
\eea
Moreover, there are third order poles between the quadratic product of
spin-$\frac{1}{2}$ current in the higher spin-$3$ current
(\ref{qspin3})
and the quadratic terms in the spin-$\frac{1}{2}$ current as follows:
 \bea
 d^0_{\bar{a} \bar{b}}   \,
Q^{\bar{a}} \, \pa^2\, Q^{\bar{b}}  
(z) \, Q^{\bar{c}} \, Q^{\bar{d}}(w) & =
& -\frac{1}{(z-w)^3} \,  2\, (k+N+2) \,
d^0_{\bar{a} \bar{b}} \, \Bigg[ g^{\bar{c} \bar{b}} \, Q^{\bar{a}} \, Q^{\bar{d}}
+ g^{\bar{d} \bar{b}} \, Q^{\bar{c}} \, Q^{\bar{a}}
\Bigg](w)
\nonu \\
& + & {\cal O}(\frac{1}{(z-w)^2}).  
\label{qd2qq2}
\eea
There are third order poles between the quadratic product of
spin-$\frac{1}{2}$ current in the higher spin-$3$ current
(\ref{qspin3})
and the cubic terms in the spin-$\frac{1}{2}$ current as follows:
 \bea
&& d^0_{\bar{a} \bar{b}}   \,
Q^{\bar{a}} \, \pa^2\, Q^{\bar{b}}  
(z) \, Q^{\bar{c}} \, Q^{\bar{d}} \, Q^{\bar{e}}(w)  =
\nonu \\
& &  -\frac{1}{(z-w)^3} \,  2\, (k+N+2) \,
d^0_{\bar{a} \bar{b}} \, \Bigg[ g^{\bar{c} \bar{b}} \, Q^{\bar{a}} \, Q^{\bar{d}}
  \, Q^{\bar{e}}
  + g^{\bar{d} \bar{b}} \, Q^{\bar{c}} \, Q^{\bar{a}} \, Q^{\bar{e}}
  + g^{\bar{e} \bar{b}} \, Q^{\bar{c}} \, Q^{\bar{d}} \, Q^{\bar{a}}
\Bigg](w)
\nonu \\
& & +  {\cal O}(\frac{1}{(z-w)^2}).  
\label{q2q3}
 \eea

One can easily check that
 there are no contributions from the last term of higher spin-$3$
 current in (\ref{qspin3}) with cubic, quadratic or linear terms in the
 spin-$\frac{1}{2}$ currents.

From the third order poles of the above OPEs, one can  
determine the various eigenvalues in the representations as follows.

\subsubsection{ The eigenvalue in $(0;f)$ representation }

Because the nontrivial third order pole in the OPE
between the higher spin-$3$
current and the spin-$\frac{1}{2}$ current
arises in (\ref{qd2qq}), by combining the
coefficient of the fourth line of (\ref{qspin3}) and
the eigenvalue in (\ref{qd2qq}), one can determine
the final eigenvalue in this representation as follows:
\bea
&& \phi_2^{(1)}(0;\tiny\yng(1)) =
\label{Eigen0f} \\
&& -\frac{2 k ({\bf 12 k^2 N}+4 k^2{\bf +6 k N^2}+
  39 k N+14 k+5 N^2+28 N+12)}{
   3 (k+N+2)^3 (6 k N+5 k+5 N+4)} \times
 \Bigg[{\bf -2} (k+N+2) \Bigg]
 = \nonu \\
 && \frac{4 k ({\bf 12 k^2 N}+4 k^2{\bf +6 k N^2}+39 k N+14 k+5 N^2+28 N+12)}{
   3 (k+N+2)^2 (6 k N+5 k+5 N+4)} \longrightarrow
 \frac{4}{3} (2-\lambda) (1-\lambda).
 \nonu
 \eea
Here the large $(N,k)$ 't Hooft-like limit
is defined by \cite{GG1305}
\bea
N,k \rightarrow \infty, \qquad \la \equiv \frac{(N+1)}{(N+k+2)}
\qquad \mbox{fixed}.
\label{largenk}
\eea 
The leading behavior term is denoted by the boldface notation. 

Note that the nontrivial contribution from the composite
field $\Phi_0^{(1)} \, T$ occurs in the
fourth term of (\ref{spin1spin2}) according to (\ref{qd2qq}).
Then the eigenvalue, by combining the coefficient with
the eigenvalue, is given by
\bea
\frac{k (2 k+3)}{8 (k+N+2)^3} \times \Bigg[-2 (k+N+2) \Bigg]
=
-\frac{k (2 k+3)}{4 (k+N+2)^2}.
\label{eigenproduct}
 \eea
 Then the zeromode \cite{BBSS1} of the composite field
 $\Phi_0^{(1)} \, T$ is given by $ T_0 \, (\Phi_0^{(1)})_0$ plus other
 terms $\sum_{p \leq -1} \, (\Phi_0^{(1)})_p \, T_{-p} +
 \sum_{p>0} \, T_{-p} \, (\Phi_0^{(1)})_p$ which do not contibute
 to the eigenvalue because the positive mode $ T_{-p}$ or
 $(\Phi_0^{(1)})_p$ acting on the state corresponding to this
 representation vanishes.
 Once one knows one of the eigenvalue, then the other one can be
 determined from (\ref{eigenproduct}). 
 For example, the conformal dimension for this representation
 can be obtained from the formula \cite{GG1305}
\bea
h(\Lambda_{+}; \Lambda_{-})= \frac{C^{(N+2)}(\Lambda_{+})}{(k+N+2)} -
\frac{C^{(N)}(\Lambda_{-})}{(k+N+2)}- \frac{\hat{u}^2}{N(N+2)(k+N+2)} +n.
\label{conformaldimension}
\eea
It turns out that $h (0;\tiny\yng(1))= \frac{(2 k+3)}{4 (k+N+2)}$
\cite{GG1305}.
Then from (\ref{eigenproduct}),  
one can split as follows
 \bea
\phi_0^{(1)}(0;\tiny\yng(1)) \,
h (0;\tiny\yng(1))
 =
 \Bigg[ -\frac{k}{(k+N+2)} \Bigg]
 \times \Bigg[ \frac{(2 k+3)}{4 (k+N+2)} \Bigg].
 \label{120f}
 \eea
 In this way, one can conclude indirectly that
 the eigenvalue of the higher spin-$1$ current
 is given by $\phi_0^{(1)}(0;\tiny\yng(1))=-\frac{k}{(k+N+2)}$
 which was found in \cite{Ahn1711} by calculating the OPE between
 the second term of (\ref{spinone}) and the spin-$\frac{1}{2}$
 current $Q^{\bar{A}^{\ast}}$ where $\bar{A}^{\ast}=1^{\ast}, \cdots,
 (2N)^{\ast}$ and reading off the first order pole.

 It is straightforward to write down the three-point function
of the higher spin-$3$ current 
 with the scalar operator corresponding to $(0;f)$
 and the scalar operator corresponding to $(0;\overline{f})$
 from (\ref{Eigen0f}).
 
 \subsubsection{ The eigenvalue in $(0;\mbox{symm})$ representation
 with two boxes}

 From the nontrivial results for the OPEs between the
 quartic terms and the quadratic term in (\ref{q4qq}) and (\ref{q4qq-1}) 
 and the OPE
  between the
 quadratic terms and the quadratic term in (\ref{qd2qq2}), 
 the corresponding eigenvalue can be obtained
 by considering the coefficients in (\ref{qspin3}) (and fixing the
 indices $\bar{e}, \bar{f}$ and $\bar{c},\bar{d}$ for the symmetric
 representation)
 as follows:
 \bea
 && \phi_2^{(1)}(0;\tiny\yng(2)) =
 \frac{k (4 k^2+2 k N+11 k-N+4)}{(k+N+2)^4 (6 k N+5 k+5 N+4)}
 \times \Bigg[ -4 (k+N+2)^2 \Bigg]
 \nonu \\
 && + \frac{k (6 k N+k+9 N+4)}{(k+N+2)^4 (6 k N+5 k+5 N+4)}
 \times
 \Bigg[-12 (k+N+2)^2\Bigg]
 \nonu \\
 && -\frac{2 k ({\bf 12 k^2 N}+4 k^2 {\bf +6 k N^2}+
   39 k N+14 k+5 N^2+28 N+12)}{
   3 (k+N+2)^3 (6 k N+5 k+5 N+4)} \times \Bigg[{\bf - 4} (k+N+2) \Bigg] =
 \nonu \\
 && \frac{8 k ({\bf 12 k^2 N}-2 k^2 {\bf +6 k N^2}+
   9 k N-7 k+5 N^2-11 N-12)}{
   3 (k+N+2)^2 (6 k N+5 k+5 N+4)} \longrightarrow
 \frac{8}{3} (2-\lambda) (1-\lambda),
 \label{Eigen0symm}
 \eea
where the limit in (\ref{largenk}) is taken and
one sees that the leading contribution in this limit
occurs in the last term denoted by the boldface notation.
The corresponding eigenvalue is the twice of the one in
previous subsection and this leads to the fact that the large
$(N,k)$ 't Hooft like behavior in this representation is the
twice of the one in (\ref{Eigen0f}).

Again, by collecting the second, the third and the fourth terms
in (\ref{spin1spin2}) together with their eigenvalues,
(\ref{q4qq}), (\ref{q4qq-1}) and (\ref{qd2qq2}) where one of the 
quadratic terms is given by $Q^{1^{\ast}} \, Q^{(N+1)^{\ast}}$,
one obtains the following eigenvalue corresponding to the composite
field $\Phi_0^{(1)} \, T$
\bea
&& \frac{k (2 k+3)}{8 (k+N+2)^4} \times \Bigg[
   -4 (k+N+2)^2\Bigg]
  -\frac{k}{4 (k+N+2)^4} \times \Bigg[ -12 (k+N+2)^2 \Bigg]
  \nonu \\
  && + \frac{k (2 k+3)}{8 (k+N+2)^3} \times
  \Bigg[ -4 (k+N+2) \Bigg]
  = -\frac{2 k^2}{(k+N+2)^2}.
  \label{2times2}
 \eea
 By substituting the conformal dimension in this representation
 using the formula (\ref{conformaldimension}) or the equation $(2.17)$
 of \cite{Ahn1711}, the following decomposition is, from (\ref{2times2}),
 valid
 \bea
\phi_0^{(1)}(0;\tiny\yng(2)) \,
h (0;\tiny\yng(2)) = \Bigg[ -\frac{2 k}{(k+N+2)} \Bigg]
  \times \Bigg[ \frac{k}{(k+N+2)} \Bigg]. 
 \label{todayspin1}
 \eea
 One can see that the eigenvalue for the higher spin-$1$ current
 in (\ref{todayspin1}) is equal to the one in $(5.20)$ of \cite{Ahn1711}.

The three-point function
of the higher spin-$3$ current 
 with the scalar operator corresponding to $(0;\tiny\yng(2))$
 and the scalar operator corresponding to $(0;\overline{\tiny\yng(2)})$
 can be read off from (\ref{Eigen0symm}).
 
 \subsubsection{ The eigenvalue in $(0;\mbox{antisymm})$ representation
 with two boxes}

As done in previous subsection,  by using the relations
 (\ref{q4qq}), (\ref{q4qq-1}) and (\ref{qd2qq2}),
the following eigenvalue is obtained
 \bea
 &&
\phi_2^{(1)}(0;\tiny\yng(1,1)) =
 \frac{k (4 k^2+2 k N+11 k-N+4)}{(k+N+2)^4 (6 k N+5 k+5 N+4)}
\times \Bigg[- 4 (k+N+2)^2\Bigg]
\nonu \\
&&+ \frac{k (6 k N+k+9 N+4)}{(k+N+2)^4 (6 k N+5 k+5 N+4)}
\times \Bigg[4 (k+N+2)^2 \Bigg]
\nonu \\
&& -\frac{2 k ({\bf 12 k^2 N}+4 k^2 {\bf +6 k N^2}+
  39 k N+14 k+5 N^2+28 N+12)}{3 (k+N+2)^3 (6 k N+5 k+5 N+4)} \times
\Bigg[{\bf - 4} (k+N+2) \Bigg] =
\nonu \\
&& \frac{8 k ({\bf 12 k^2 N}-2 k^2{\bf +6 k N^2}+45 k N-k+5 N^2+43 N+12)}{
  3 (k+N+2)^2 (6 k N+5 k+5 N+4)} \longrightarrow
\frac{8}{3} (2-\lambda) (1-\lambda).
\label{Eigen0antisymm}
 \eea
where the limit in (\ref{largenk}) is taken and
the leading contribution in this limit
occurs in the last term denoted by the boldface notation.
The eigenvalues from (\ref{qd2qq2}) in this subsection
and previous subsection are the same as each other and the large
$(N,k)$ 't Hooft like behavior in (\ref{Eigen0antisymm})
and (\ref{Eigen0symm}) is the same. 

By combining the second, the third and the fourth terms
in (\ref{spin1spin2}) together with their eigenvalues,
(\ref{q4qq}), (\ref{q4qq-1}) and (\ref{qd2qq2}) correctly
where one of the 
quadratic terms is given by $Q^{1^{\ast}} \, Q^{2^{\ast}}$,
one obtains the following eigenvalue corresponding to the composite
field $\Phi_0^{(1)} \, T$
  \bea
&& \frac{k (2 k+3)}{8 (k+N+2)^4} \times \Bigg[
   -4 (k+N+2)^2\Bigg]
  -\frac{k}{4 (k+N+2)^4} \times \Bigg[ 4 (k+N+2)^2 \Bigg]
  \nonu \\
  && + \frac{k (2 k+3)}{8 (k+N+2)^3} \times
  \Bigg[ -4 (k+N+2) \Bigg]=
  -\frac{2 k (k+2)}{(k+N+2)^2}. 
  \label{2Times2}
  \eea
After substituting the conformal dimension in this representation
 using the formula (\ref{conformaldimension}) or the equation $(2.19)$
 of \cite{Ahn1711}, it is obvious see that
 the following decomposition from (\ref{2Times2}) is valid
  \bea
\phi_0^{(1)}(0;\tiny\yng(1,1)) \,
h (0;\tiny\yng(1,1))  
= \Bigg[ -\frac{2 k}{(k+N+2)} \Bigg]
  \times \Bigg[ \frac{(k+2)}{(k+N+2)} \Bigg].
  \label{120anti2}
  \eea
The eigenvalue for the higher spin-$1$ current
 is equal to the one in $(5.21)$ of \cite{Ahn1711}. 

For the three-point function
of the higher spin-$3$ current 
 with the scalar operator corresponding to $(0;\tiny\yng(1,1))$
 and the scalar operator corresponding to $(0;\overline{\tiny\yng(1,1)})$
one can use the relation (\ref{Eigen0antisymm}). 
 
 \subsubsection{ The eigenvalue in $(0;\mbox{antisymm})$ representation
 with three boxes}

Because the nontrivial third order pole in the OPE
between the higher spin-$3$
current and the cubic terms in the spin-$\frac{1}{2}$ current
arises in (\ref{q6q3}), (\ref{q4q3}), (\ref{q4q3-1}) and (\ref{q2q3}),
by combining the
coefficients of (\ref{qspin3}) and
the eigenvalues, one can determine
the final eigenvalue in this representation as follows: 
 \bea
 &&
\phi_2^{(1)}(0;\tiny\yng(1,1,1)) =
 -\frac{k (k-N)}{2 (6 k N+5 k+5 N+4) (k+N+2)^5} \times
 \Bigg[ 48 (k+N+2)^3 \Bigg]
 \nonu \\
 && + \frac{k (4 k^2+2 k N+11 k-N+4)}{(k+N+2)^4 (6 k N+5 k+5 N+4)}
 \times
 \Bigg[- 12 (k+N+2)^2 \Bigg]
 \nonu \\
 && + \frac{k (6 k N+k+9 N+4)}{(k+N+2)^4 (6 k N+5 k+5 N+4)}
 \times \Bigg[12 (k+N+2)^2\Bigg]
 \nonu \\
 && -\frac{2 k ({\bf 12 k^2 N}+4 k^2 {\bf +6 k N^2}+
   39 k N+14 k+5 N^2+28 N+12)}{
   3 (k+N+2)^3 (6 k N+5 k+5 N+4)}
 \times \Bigg[{\bf - 6} (k+N+2)\Bigg] =
 \nonu \\
 && \frac{4 k ({\bf 12 k^2 N}-8 k^2 {\bf +6 k N^2}+51 k N-22 k+5 N^2+64 N+12)}{
   (k+N+2)^2 (6 k N+5 k+5 N+4)} \longrightarrow
 4 (2-\lambda) (1-\lambda),
 \label{Eigen0antisymmthree}
 \eea
where the limit in (\ref{largenk}) is taken and
the leading contribution in this limit
occurs in the last term denoted by the boldface notation.
Note that
 the large
$(N,k)$ 't Hooft like behavior in this representation is the
three times of the one in (\ref{Eigen0f}).

By combining the first, second, the third and the fourth terms
in (\ref{spin1spin2}) together with their eigenvalues,
(\ref{q6q3}), (\ref{q4q3}), (\ref{q4q3-1}) and (\ref{q2q3}) correctly
where one of the 
quadratic terms is given by $Q^{1^{\ast}} \, Q^{2^{\ast}} \, Q^{3^{\ast}}$,
one obtains the following eigenvalue corresponding to the composite
field $\Phi_0^{(1)} \, T$
\bea
 && -\frac{k}{32 (k+N+2)^5} \times \Bigg[48 (k+N+2)^3 \Bigg]
 + \frac{k (2 k+3)}{8 (k+N+2)^4} \times \Bigg[
   -12 (k+N+2)^2\Bigg]
 \nonu \\
&& -\frac{k}{4 (k+N+2)^4} \times \Bigg[ 12 (k+N+2)^2 \Bigg]
   + \frac{k (2 k+3)}{8 (k+N+2)^3} \times
  \Bigg[ -6 (k+N+2) \Bigg]= 
  \nonu \\
  && -\frac{9 k (2 k+5)}{4 (k+N+2)^2}.
  \label{3times3}
 \eea
 After substituting the conformal dimension in this representation
 using the formula (\ref{conformaldimension}) or the equation $(2.20)$
 of \cite{Ahn1711}, it is obvious see that
 the following decomposition from (\ref{3times3}) satisfies
 \bea
\phi_0^{(1)}(0;\tiny\yng(1,1,1)) \,
h (0;\tiny\yng(1,1,1))=
\Bigg[  -\frac{3 k}{(k+N+2)} \Bigg] \times
\Bigg[ \frac{3 (2 k+5)}{4 (k+N+2)} \Bigg].
 \label{120anti3}
 \eea
The eigenvalue for the higher spin-$1$ current
 is equal to the one in $(9.1)$ of \cite{Ahn1711}. 
 
For the three-point function
of the higher spin-$3$ current 
 with the scalar operator corresponding to $(0;\tiny\yng(1,1,1))$
 and the scalar operator corresponding to
 $(0;\overline{\tiny\yng(1,1,1)})$
 one can use the relation (\ref{Eigen0antisymmthree})
 explicitly.
 
\subsubsection{ The eigenvalue in $(0;\mbox{mixed})$ representation }

By using the relations
(\ref{q6q3}), (\ref{q4q3}), (\ref{q4q3-1}) and (\ref{q2q3}),
one obtains the following eigenvalue
\bea
&&
\phi_2^{(1)}(0;\tiny\yng(2,1)) =
-\frac{k (k-N)}{2 (6 k N+5 k+5 N+4) (k+N+2)^5}
  \times \Bigg[
- 48 (k+N+2)^3 \Bigg]
    \nonu \\
    && +\frac{k (4 k^2+2 k N+11 k-N+4)}{
      (k+N+2)^4 (6 k N+5 k+5 N+4)}
\times  \Bigg[ - 12 (k+N+2)^2 \Bigg]
\nonu \\
&& + \frac{k (6 k N+k+9 N+4)}{(k+N+2)^4 (6 k N+5 k+5 N+4)}
\times \Bigg[ - 12 (k+N+2)^2 \Bigg]
\nonu \\
&& -\frac{2 k ({\bf 12 k^2 N}+4 k^2 {\bf +6 k N^2}+
  39 k N+14 k+5 N^2+28 N+12)}{
  3 (k+N+2)^3 (6 k N+5 k+5 N+4)} \times
\Bigg[ {\bf - 6} (k+N+2) \Bigg] =
\nonu \\
&& \frac{4 k ({\bf 12 k^2 N}-8 k^2 {\bf +6 k N^2}+15 k N-16 k+5 N^2-2 N-12)}{
  (k+N+2)^2 (6 k N+5 k+5 N+4)} \longrightarrow
4 (2-\lambda) (1-\lambda),
\label{Eigenmixed}
  \eea
where the limit in (\ref{largenk}) is taken and
the leading contribution in this limit
occurs in the last term denoted by the boldface notation.
Because the eigenvalue for the last term is the same as
the one in previous subsection, the large $(N,k)$ 't Hooft like
behavior is the same as the one in (\ref{Eigen0antisymmthree}).

By combining the first, second, the third and the fourth terms
in (\ref{spin1spin2}) together with their eigenvalues,
(\ref{q6q3}), (\ref{q4q3}), (\ref{q4q3-1}) and (\ref{q2q3}) correctly
where one of the 
quadratic terms is given by $Q^{2^{\ast}} \, Q^{(N+1)^{\ast}} \, Q^{1^{\ast}}$,
one obtains the following eigenvalue corresponding to the composite
field $\Phi_0^{(1)} \, T$
\bea
 && -\frac{k}{32 (k+N+2)^5} \times \Bigg[-48 (k+N+2)^3 \Bigg]
 + \frac{k (2 k+3)}{8 (k+N+2)^4} \times \Bigg[
   -12 (k+N+2)^2\Bigg]
 \nonu \\
&& -\frac{k}{4 (k+N+2)^4} \times \Bigg[ -12 (k+N+2)^2 \Bigg]
   + \frac{k (2 k+3)}{8 (k+N+2)^3} \times
   \Bigg[ -6 (k+N+2) \Bigg]=
   \nonu \\
   &&  \frac{9 k (2 k+1)}{4 (k+N+2)^2}.
   \label{mixedtimesmixed}
   \eea
 After substituting the conformal dimension in this representation
 using the formula (\ref{conformaldimension}) or the equation $(2.23)$
 of \cite{Ahn1711}, it is obvious see that
 the following decomposition from (\ref{mixedtimesmixed}) satisfies   
   \bea
\phi_2^{(1)}(0;\tiny\yng(2,1)) \,
h (0;\tiny\yng(2,1))    =
\Bigg[ -\frac{3 k}{(k+N+2)} \Bigg]
\times \Bigg[ \frac{3 (2 k+1)}{4 (k+N+2)} \Bigg].
   \label{ttodayspin1}
   \eea
The eigenvalue for the higher spin-$1$ current
 in (\ref{ttodayspin1}) is equal to the one in section $9$ of \cite{Ahn1711}. 
 
The three-point function
of the higher spin-$3$ current 
 with the scalar operator corresponding to $(0;\tiny\yng(2,1))$
 and the scalar operator corresponding to
 $(0;\overline{\tiny\yng(2,1)})$
can be obtained from the relation (\ref{Eigenmixed})
explicitly.
   
\subsection{ The $(0; \La_{-})$ representations with more than four boxes}

So far, the eigenvalues are obtained in the representations
up to three boxes.
One can consider the higher representations with more than
four boxes. Th simplest cases are
the antisymmetric representations with more than four boxes.

  \subsubsection{ The eigenvalue in $(0;\mbox{antisymm})$
    representation with more than four boxes }

For the four boxes in the antisymmetric representation, 
the minimum value for $N$ is given by $N=5$.
One can visualize the spin-$\frac{1}{2}$ currents in the following
$(N+2) \times (N+2)$ matrix
\bea
\left(
\begin{array}{ccccc|cc}
  0 & 0 & 0 & \cdots & 0 &
   Q^{1^{\ast}} &  Q^{(N+1)^{\ast}}  \\
  0 & 0 & 0 &\cdots & 0 &
   Q^{2^{\ast}} &  Q^{(N+2)^{\ast}} \\
  0 & 0 & 0 & \cdots & 0 &
  Q^{3^{\ast}} &  Q^{(N+3)^{\ast}} \\
  \vdots & \vdots & \vdots & \vdots & \vdots &
  \vdots & \vdots \\
 0 & 0 & 0 & \cdots & 0 &
  Q^{N^{\ast}}   &   Q^{(2N)^{\ast}} \\
  \hline  
 Q^1 & Q^2 & Q^3 & \cdots &  Q^{N}  & 0 & 0 \\
 Q^{N+1} & Q^{N+2} & Q^{N+3} & \cdots  &  Q^{2N}  & 0 & 0 \\
\end{array}
\right).
\label{matrix}
\eea
Then one of the four boxes in the antisymmetric representations
is given by $Q^{1^{\ast}} \, Q^{2^{\ast}} \, Q^{3^{\ast}} \, Q^{4^{\ast}}$ in the
view of (\ref{matrix}).

It turns out (the explicit form for the
OPE between the higher spin-$3$ current (\ref{qspin3})
and the quartic term in the
spin-$\frac{1}{2}$ current is not given)
that the eigenvalue is given by
\bea
&& \phi_2^{(1)}(0;\tiny\yng(1,1,1,1))  =
-\frac{k (k-N)}{2 (6 k N+5 k+5 N+4) (k+N+2)^5}
  \times \Bigg[
192 (k+N+2)^3 \Bigg]
    \nonu \\
    && +\frac{k (4 k^2+2 k N+11 k-N+4)}{
      (k+N+2)^4 (6 k N+5 k+5 N+4)}
\times  \Bigg[ - 24 (k+N+2)^2 \Bigg]
\nonu \\
&& + \frac{k (6 k N+k+9 N+4)}{(k+N+2)^4 (6 k N+5 k+5 N+4)}
\times \Bigg[ 24 (k+N+2)^2 \Bigg]
\label{Eigenantisymmfour}
\\
&& -\frac{2 k ({\bf 12 k^2 N}+4 k^2 {\bf +6 k N^2}+
  39 k N+14 k+5 N^2+28 N+12)}{
  3 (k+N+2)^3 (6 k N+5 k+5 N+4)} \times
\Bigg[ {\bf - 8} (k+N+2) \Bigg] =
\nonu \\
&& \frac{16 k ({\bf 12 k^2 N}-14 k^2 {\bf +6 k N^2}+
  57 k N-49 k+5 N^2+91 N+12)}{
  3 (k+N+2)^2 (6 k N+5 k+5 N+4)} \longrightarrow
\frac{16}{3} (2-\lambda) (1-\lambda),
\nonu
\eea
where the limit in (\ref{largenk}) is taken and
the leading contribution in this limit
occurs in the last term denoted by the boldface notation.
Because the eigenvalue for the last term is four times of 
the one in previous subsection $4.1.1$,
the large $(N,k)$ 't Hooft like
behavior is four times of the one in (\ref{Eigen0f}).

By combining the first, second, the third and the fourth terms
in (\ref{spin1spin2}) together with their eigenvalues correctly
where one of the 
quadratic terms is given by $Q^{1^{\ast}} \, Q^{2^{\ast}} \, Q^{3^{\ast}} \,
Q^{4^{\ast}}$,
one obtains the following eigenvalue corresponding to the composite
field $\Phi_0^{(1)} \, T$
  \bea
 && -\frac{k}{32 (k+N+2)^5} \times \Bigg[192 (k+N+2)^3 \Bigg]
 + \frac{k (2 k+3)}{8 (k+N+2)^4} \times \Bigg[
   -24 (k+N+2)^2\Bigg]
 \nonu \\
&& -\frac{k}{4 (k+N+2)^4} \times \Bigg[ 24 (k+N+2)^2 \Bigg]
   + \frac{k (2 k+3)}{8 (k+N+2)^3} \times
   \Bigg[ -8 (k+N+2) \Bigg]=
   \nonu \\
   &&  -\frac{8 k (k+3)}{ (k+N+2)^2}.
   \label{composite}
   \eea

   One can calculate the conformal dimension in this representation
   by using the formula (\ref{conformaldimension}).
   \bea
   h(0;\tiny\yng(1,1,1,1)) = -\frac{4(N-4)(1+\frac{1}{N})}{2(k+N+2)}
   -\frac{4(N+2)^2}{N(N+2)(k+N+2)} +2 =\frac{2 (k+3)}{ (k+N+2)}. 
   \label{dimensionfour}
   \eea
This calculation was not done in \cite{Ahn1711}.
   Then one has the following decomposition from (\ref{composite})
   and (\ref{dimensionfour})   
   \bea   
\phi_0^{(1)}(0;\tiny\yng(1,1,1,1)) \,
h(0;\tiny\yng(1,1,1,1)) =
\Bigg[  -\frac{4 k}{(k+N+2)} \Bigg] \times
\Bigg[ \frac{2 (k+3)}{ (k+N+2)} \Bigg].
   \label{120anti4}
   \eea
   Therefore, one obtains the eigenvalue for the higher spin-$1$
   current in this representation indirectly.
See also $(5.50)$ of \cite{Ahn1711}.
On the other hand,  one
can check by calculating the OPE between the higher spin-$1$
   current and $Q^{1^{\ast}} \, Q^{2^{\ast}} \, Q^{3^{\ast}} \,
Q^{4^{\ast}}$ and reading off the first order pole.
   
   The three-point function
of the higher spin-$3$ current 
 with the scalar operator corresponding to $(0;\tiny\yng(1,1,1,1))$
 and the scalar operator corresponding to
 $(0;\overline{\tiny\yng(1,1,1,1)})$
can be obtained from the relation (\ref{Eigenantisymmfour})
explicitly as follows:
\bea
&& <\overline{\cal O}(0;\overline{\tiny\yng(1,1,1,1)})
   \,  {\cal O} (0;\tiny\yng(1,1,1,1)) \, \Phi_2^{(1)} > \nonu \\
    && =  \Bigg[
\frac{16 k ({\bf 12 k^2 N}-14 k^2 {\bf +6 k N^2}+
  57 k N-49 k+5 N^2+91 N+12)}{
  3 (k+N+2)^2 (6 k N+5 k+5 N+4)} 
      \Bigg]
    <\overline{\cal O} (0;\overline{\tiny\yng(1,1,1,1)})
    \, {\cal O} (0;\tiny\yng(1,1,1,1))>
    \nonu \\
    && \longrightarrow
    \frac{16}{3} (2-\lambda) (1-\lambda)
    <\overline{\cal O} (0;\overline{\tiny\yng(1,1,1,1)})
    \, {\cal O} (0;\tiny\yng(1,1,1,1))>=
-\frac{4}{3} (2-\la)\,
    <\overline{\cal O}(0;\overline{\tiny\yng(1,1,1,1)})
   \,  {\cal O} (0;\tiny\yng(1,1,1,1)) \, \Phi_0^{(1)} >,
\label{3point}
\eea
where the limit in (\ref{largenk}) is taken at the final stage.
Furthermore, one can use the relation of the three-point function
for the higher spin-$1$ current (containing the factor $-4(1-\la)$)
and obtain the explicit relation
between two three-point functions, which is simple linear expression
of the 't Hooft coupling constant $\la$.

  \subsubsection{ The eigenvalue in $(0;\mbox{antisymm})$
    representation with $p \equiv |\La_-|$ boxes }

  As in previous subsection,  one can consider
the $p$ multiple product    
$Q^{1^{\ast}} \, Q^{2^{\ast}} \, Q^{3^{\ast}} \, \cdots \, Q^{p^{\ast}}$
(see the matrix (\ref{matrix}))
as one of the antisymmetric representation with $p$ boxes.
One should calculate the corresponding eigenvalues as done in
(\ref{Eigenantisymmfour}). 
In particular, the eigenvalue for the
operator $ d^0_{\bar{a} \bar{b}}   \,
Q^{\bar{a}} \, \pa^2\, Q^{\bar{b}}$ is the $p$ times of
the one for the fundamental representation from the observations
in (\ref{qd2qq}), (\ref{qd2qq2}) and (\ref{q2q3}).
One expects that the generalization of three-point function 
of the higher spin-$3$ current in (\ref{3point})
 with the scalar operator corresponding to $(0;\mbox{antisymm})$
 and the scalar operator corresponding to
 $(0;\overline{\mbox{antisymm}})$
 can be written in terms of $
 \frac{4 p}{3} (2-\lambda) (1-\lambda)$ multiplied by the
 two-point function between the two scalar operators under the
 large $(N, k)$ 't Hooft like limit (\ref{largenk}). 
That is, one has
 \bea
<\overline{\cal O}(0;\overline{\La}_{-})
\,  {\cal O} (0;\La_{-}) \, \Phi_2^{(1)} > & \longrightarrow &
\frac{4 |\La_{-}|}{3} (2-\lambda) (1-\lambda) \,
 <\overline{\cal O}(0;\overline{\La}_{-})
\,  {\cal O} (0;\La_{-})>
\nonu \\
& = & -\frac{4}{3} (2-\la)\,
    <\overline{\cal O}(0;\overline{\La}_{-})
   \,  {\cal O} (0;\La_{-}) \, \Phi_0^{(1)} >.
\label{threepointforq}
\eea
The $|\La_-|$ is the number of boxes of Young tableaux.
Here the relation for the three-point function of the higher spin-$1$
current with two scalar operators under the large $(N,k)$ 't Hooft like
limit  (from its finite result) is used.
In other words, 
the eigenvalue is given by
$-\frac{ |\La_{-}| k}{(k+N+2)} \longrightarrow -  |\La_{-}|(1-\la)$
which generalizes the cases in (\ref{120f}),
(\ref{120anti2}), (\ref{120anti3}) or
(\ref{120anti4}).

Note that although the above analysis is for the antisymmetric
representations, this result holds (under the large
$(N, k)$ 't Hooft-like limit) for any representations which
have the number of total boxes $|\La_-|$. One can check that
the OPE between the quadratic term in the
fourth line of (\ref{qspin3}) corresponding to the leading behavior
and 
the $|\La_-|$ multiple product of spin-$\frac{1}{2}$ current
implies the coefficient, $-2 (k+N+2) \times |\La_-|$, in the
third order pole. 

   \subsection{ The $(\La_{+}; \La_{-})$ representations
   up to two boxes for $\La_+$}

   Let us consider the case where
   the representation $\La_-$ appears in the
branching of $\La_+$ under the $SU(N)_k \times SU(2)_k \times U(1)$.
There is a trivial $l^-=0$ quantum number.
The $l^+$ quantum number and $\hat{u}$ charge
can be read off from the multiple product of
$  ({\tiny\yng(1)},{\bf 1})_1
    + ({\bf 1},{\bf 2})_{-\frac{N}{2}}$ in \cite{GG1305}.
It is known that
the branching rules for the symmetric and antisymmetric
representations 
satisfy \cite{GG1305,Slansky,FK}
\bea
    {\tiny\yng(2)} & \rightarrow &
     ({\tiny\yng(2)},{\bf 1})_2 +
     ({\tiny\yng(1)},{\bf 2})_{1-\frac{N}{2}}
      +  ({\bf 1},{\bf 3})_{-N},
      \nonu \\
\tiny\yng(1,1) & \rightarrow &
({\tiny\yng(1,1)},{\bf 1})_2 +
      ({\tiny\yng(1)},{\bf 2})_{1-\frac{N}{2}} +
      ({\bf 1},{\bf 1})_{-N}.
  \label{transtrans}
\eea

In this subsection, the previous known results are
reinterpreted and some new features are presented based on the
results of section $3$.

\subsubsection{ The eigenvalue in
(f;0) representation  }

The eigenvalue of the zeromode for the higher spin current
appearing in the first term of (\ref{vspin3}) in
this representation is given by
$(18,50,98)$ for $N=3,5,7$ respectively. 
Then the general $N$ dependence is
given by $2 N^2$.
Similarly,
From the eigenvalues $(-648,-3000,-8232,-17496)$
of the zeromode for the higher spin current
appearing in the second term of (\ref{vspin3})
for $N=3,5,7,9$,  the general $N$ behavior
is given by $-24 N^3$.
The eigenvalue of the zeromode for the higher spin current
appearing in the third term of (\ref{vspin3}) in
this representation is given by
$(-72,-200,-392)$ for $N=3,5,7$ respectively. 
Then the general $N$ dependence can be read off and it is
given by $-8 N^2$.
From the eigenvalues $(0,0,0)$
of the zeromode for the higher spin current
appearing in the fourth term of (\ref{vspin3})
for $N=3,5,7$,  the general $N$ behavior
is given by $0$.
The eigenvalue of the zeromode for the higher spin current
appearing in the fifth term of (\ref{vspin3}) in
this representation is given by
$(-3,-5,-7)$
for $N=3,5,7$ respectively. 
Then the general $N$ dependence can be read off and it is
given by $- N$.
Finally, from the eigenvalues $(12,20,28)$
of the zeromode for the higher spin current
appearing in the last term of (\ref{vspin3})
for $N=3,5,7$,  the general $N$ behavior
is given by $4 N$.
Then one obtains the following eigenvalue by collecting all the
contributions with correct coefficients
{\small
  \bea
&& \phi_2^{(1)}(\tiny\yng(1);0)  = 
 \frac{16 (k-N)}{(6 k N+5 k+5 N+4)} \Bigg[
  -\frac{1}{4 (k+N+2)^2} \times 2 N^2
  + \frac{1}{32 N^2 (k+N+2)^2} \times 
(- 24 N^3)
  \Bigg]
\nonu \\
&& +\frac{1}{2 N (k+N+2)^2} \times
(- 8 N^2) +\frac{1}{(k+N+2)^2} \times 0
-\frac{4 k}{(k+N+2)^2} \times ({\bf -  N})
\nonu \\
&& -\frac{2 (2 k+N)}{3 (k+N+2)^2} \times {\bf 4 N}
=
-\frac{4 N ({\bf 6 k^2 N}+5 k^2{\bf +12 k N^2}+
  39 k N+28 k+4 N^2+14 N+12)}{3 (k+N+2)^2 (6 k N+5 k+5 N+4)}
\nonu \\
&& \longrightarrow - \frac{4}{3} \, \lambda  \, (\lambda +1),
\label{eigenfzero}
\eea}
where the limit in (\ref{largenk}) is taken at the final stage.
It is obvious to see that
the leading contributions in this limit come from the
fifth and the sixth terms of (\ref{eigenfzero}).

By noting that the eigenvalue for the zeromode of
$\Phi_0^{(1)} \, T$ can be read off from the first two terms in
(\ref{eigenfzero}),
it is given by
\bea
\Bigg[
  -\frac{1}{4 (k+N+2)^2} \times 2 N^2
  + \frac{1}{32 N^2 (k+N+2)^2} \times 
(- 24 N^3)
  \Bigg] =-\frac{N (2 N+3)}{4 (k+N+2)^2}.
\label{this}
\eea
Then one can decompose this (\ref{this}) as the product of
two eigenvalues 
\bea
\phi_0^{(1)}(\tiny\yng(1);0) \, h(\tiny\yng(1);0)=
\Bigg[ -\frac{N }{ (k+N+2)} \Bigg]
\times \Bigg[ \frac{ (2 N+3)}{4 (k+N+2)} \Bigg],
\label{12f0}
\eea
where the conformal dimension is substituted.
For example, the equation $(2.9)$ of \cite{Ahn1711} can be used or
one can use the formula in (\ref{conformaldimension}).
Now one can see the equation $(4.4)$ of \cite{Ahn1711} for the eigenvalue
of the higher spin-$1$ current in this representation.

The three-point function
of the higher spin-$3$ current 
 with the scalar operator corresponding to $(\tiny\yng(1);0)$
 and the scalar operator corresponding to
 $(\overline{\tiny\yng(1)};0)$
can be obtained from the relation (\ref{eigenfzero})
explicitly.

\subsubsection{ The eigenvalue in
(f;f) representation  }

The eigenvalue of the zeromode for the higher spin current
appearing in the first term of (\ref{vspin3}) in
this representation is given by
$(-8,-8,-8)$ for $N=3,5,7$ respectively. 
Similarly,
The eigenvalues are given by $(0,0,0)$
of the zeromode for the higher spin current
appearing in the second or third term 
for $N=3,5,7$.
From the eigenvalues $(-24,-40,-56)$
of the zeromode for the higher spin current
appearing in the fourth term
for $N=3,5,7$,  the general $N$ behavior
is given by $-8N$.
The eigenvalues are $(2,2,2)$
of the zeromode for the higher spin current
appearing in the fifth term
for $N=3,5,7$.
The eigenvalues
of the zeromode for the higher spin current
appearing in the last term  are $(-8,-8,-8)$
for $N=3,5,7$.
By collecting all the contributions with correct
coefficients, one determines the final eigenvalue as follows:
{\small
  \bea
&& \phi_2^{(1)}(\tiny\yng(1);\tiny\yng(1))  = 
 \frac{16 (k-N)}{(6 k N+5 k+5 N+4)} \Bigg[
  -\frac{1}{4 (k+N+2)^2} \times (-8)
  + \frac{1}{32 N^2 (k+N+2)^2} \times 
0
  \Bigg]
\nonu \\
&& +\frac{1}{2 N (k+N+2)^2} \times
0 +\frac{1}{(k+N+2)^2} \times ({\bf -8 N})
-\frac{4 k}{(k+N+2)^2} \times {\bf 2}
\nonu \\
&& -\frac{2 (2 k+N)}{3 (k+N+2)^2} \times ({\bf -8})
=
\frac{8 (k-N) ({\bf 6 k N}+5 k+5 N+16)}{3 (k+N+2)^2 (6 k N+5 k+5 N+4)}
 \longrightarrow - \frac{8}{3N} \, \lambda  \, (2\lambda -1),
\label{eigenff}
\eea}
where the limit in (\ref{largenk}) is taken at the final stage.
It is obvious to see that
the leading contributions ($\frac{1}{N}$ term)
in this limit come from the fourth, the
fifth and the sixth terms of (\ref{eigenff}).

One can decompose the zeromode of
$\Phi_0^{(1)} \, T$ appearing in the first and second terms
of (\ref{eigenff})
as
\bea
\phi_0^{(1)}(\tiny\yng(1);\tiny\yng(1)) \,
h(\tiny\yng(1);\tiny\yng(1))=
\Bigg[ \frac{2}{(k+N+2)} \Bigg] \times \Bigg[ \frac{1}{(k+N+2)} \Bigg],
\label{12ff}
\eea
where the equation $(2.10)$ of \cite{Ahn1711}
for the conformal dimension is substituted.
Then one sees the coincidence with the equation $(5.1)$ of \cite{Ahn1711}
for the eigenvalue of
the higher
spin-$1$ current.

 It is straightforward to write down the three-point function
of the higher spin-$3$ current 
 with the scalar operator corresponding to $(f;0)$
 and the scalar operator corresponding to $(\overline{f};0)$
 from (\ref{eigenff}) as before.

\subsubsection{ The eigenvalue in
(\mbox{symm};0) representation  
with two boxes}

The eigenvalue of the zeromode for the higher spin current
appearing in the first term of (\ref{vspin3}) in
this representation is given by
$(72,200,392)$ for $N=3,5,7$ respectively. 
Then the general $N$ dependence is
given by $8 N^2$ \footnote{
From the eigenvalues $(-3456,-16000,-43904,-93312)$
of the zeromode for the higher spin current
appearing in the second term
for $N=3,5,7,9$,  the general $N$ behavior
is given by $-128 N^3$.
From the eigenvalues $(-144,-400,-784)$
of the zeromode for the higher spin current
appearing in the third term
for $N=3,5,7$,  the general $N$ behavior
is given by $-16N^2$.
From the eigenvalues $(0,0,0)$
of the zeromode for the higher spin current
appearing in the fourth term
for $N=3,5,7$,  the general $N$ behavior
is given by $0$.
From the eigenvalues $(-6,-10,-14)$
of the zeromode for the higher spin current
appearing in the fifth term 
for $N=3,5,7$,  the general $N$ behavior
is given by $-2 N$.
From the eigenvalues $(24,40,56)$
of the zeromode for the higher spin current
appearing in the last term
for $N=3,5,7$,  the general $N$ behavior
is given by $8N $.}.
By collecting all the contributions with correct
coefficients, one determines the final eigenvalue as follows:
{\footnotesize
  \bea
&& \phi_2^{(1)}(\tiny\yng(2);0)  = 
 \frac{16 (k-N)}{(6 k N+5 k+5 N+4)} \Bigg[
  -\frac{1}{4 (k+N+2)^2} \times 8 N^2
  + \frac{1}{32 N^2 (k+N+2)^2} \times 
(- 128 N^3)
  \Bigg]
\nonu \\
&& +\frac{1}{2 N (k+N+2)^2} \times
(- 16 N^2) +\frac{1}{(k+N+2)^2} \times 0
-\frac{4 k}{(k+N+2)^2} \times ({\bf -  2N})
\nonu \\
&& -\frac{2 (2 k+N)}{3 (k+N+2)^2} \times {\bf 8 N}
=
-\frac{8 N ({\bf 6 k^2 N}+5 k^2 {\bf +12 k N^2}+45 k N+
  43 k-2 N^2-N+12)}{3 (k+N+2)^2 (6 k N+5 k+5 N+4)}
\nonu \\
&& \longrightarrow - \frac{8}{3} \, \lambda  \, (\lambda +1),
\label{eigensymm2zero}
\eea}
where the limit in (\ref{largenk}) is taken at the final stage.
It is obvious that
the leading contributions in this limit come from the
fifth and the sixth terms of (\ref{eigensymm2zero}) as before.
The eigenvalue in this limit is the twice of the one in (\ref{eigenfzero}).

In particular, the first and second terms of (\ref{eigensymm2zero})
contain
\bea
\Bigg[
  -\frac{1}{4 (k+N+2)^2} \times 8 N^2
  + \frac{1}{32 N^2 (k+N+2)^2} \times 
(- 128 N^3)
  \Bigg] =-\frac{2 N (N+2)}{(k+N+2)^2}.
\label{this1}
\eea
One can also decompose this (\ref{this1})
as follows:
\bea
\phi_0^{(1)}(\tiny\yng(2);0) \,
h(\tiny\yng(2);0) =
\Bigg[ -\frac{2 N}{(k+N+2)} \Bigg]
\times \Bigg[ \frac{ (N+2)}{(k+N+2)} \Bigg],
\label{12symm0}
\eea
where one uses the equation $(2.13)$ of \cite{Ahn1711}.
This implies that one sees the eigenvalue for the higher spin-$1$
current given in the equation $(5.18)$ of \cite{Ahn1711}.

The third and fourth terms of (\ref{eigensymm2zero}) are given by
\bea
\Bigg[ \frac{1}{2 N (k+N+2)^2} \times
  (- 16 N^2) +\frac{1}{(k+N+2)^2} \times 0 \Bigg]
= -\frac{8 N}{(k+N+2)^2}.
\label{trieigen1}
\eea
Moreover, there are also other eigenvalues corresponding to
these two terms on this representation
\bea
\Bigg[ \frac{1}{2 N (k+N+2)^2} \times
(- 24 N^2) +\frac{1}{(k+N+2)^2} \times 4N \Bigg] =
-\frac{8 N}{(k+N+2)^2},
\label{trieigen2}
\eea
which is equal to (\ref{trieigen1}).
According to the first equation of
(\ref{transtrans}), there are triplet states
under the $SU(2)_k$.
This leads to the same eigenvalue, (\ref{trieigen1}) or (\ref{trieigen2}),
for the higher spin-$3$ current
in this representation although the eigenvalues corresponding to
the third and fourth terms are different from each other.

The three-point function
of the higher spin-$3$ current 
 with the scalar operator corresponding to $(\tiny\yng(2);0)$
 and the scalar operator corresponding to $(\overline{\tiny\yng(2)};0)$
 can be read off from (\ref{eigensymm2zero}).
 
\subsubsection{ The eigenvalue in
(\mbox{symm};\mbox{symm}) representation  
with two, and two boxes}

The eigenvalue of the zeromode for the higher spin current
appearing in the first term of (\ref{vspin3}) in
this representation is given by
$(-32,-32, -32)$ for $N=3,5,7$ respectively \footnote{
Similarly,
from the eigenvalues $(0,0,0)$
appearing in the second or third term
for $N=3,5,7$,  the general $N$ behavior
can be read off.
From the eigenvalues $(-64,-96,-128)$
appearing in the fourth term
for $N=3,5,7$,  the general $N$ behavior
is given by $(-16 N-16)$.
The eigenvalue
appearing in the fifth term  in
this representation is given by
$(4,4,4)$
for $N=3,5,7$ respectively. 
Finally, from the eigenvalues $(-16,-16,-16)$
appearing in the last term
for $N=3,5,7$,  the general $N$ behavior
can be obtained.}.
Then one obtains the following eigenvalue
{\small
  \bea
  && \phi_2^{(1)}(\tiny\yng(2);
  \tiny\yng(2))  = 
 \frac{16 (k-N)}{(6 k N+5 k+5 N+4)} \Bigg[
  -\frac{1}{4 (k+N+2)^2} \times (-32)
  + \frac{1}{32 N^2 (k+N+2)^2} \times 
0
  \Bigg]
\nonu \\
&& +\frac{1}{2 N (k+N+2)^2} \times
0 +\frac{1}{(k+N+2)^2} \times ({\bf -16N}-16)
-\frac{4 k}{(k+N+2)^2} \times {\bf 4}
\nonu \\
&& -\frac{2 (2 k+N)}{3 (k+N+2)^2} \times ({\bf -16})
=
\frac{16 ({\bf 6 k^2 N}+5 k^2{\bf -6 k N^2}-18 k N+13 k-5 N^2-43 N-12)}{
  3 (k+N+2)^2 (6 k N+5 k+5 N+4)}
\nonu \\
&& \longrightarrow - \frac{16}{3N} \, \lambda  \, (2\lambda -1),
\label{eigensymmsymm22}
\eea}
where the limit in (\ref{largenk}) is taken at the final stage.
It is obvious that
the leading contributions in this limit come from the fourth,
fifth and the sixth terms of (\ref{eigensymmsymm22}) as before.
The eigenvalue in this limit is the twice of the one in
(\ref{eigenff}).
Note that the number of boxes is increased by two respectively.

From the first and the second terms of (\ref{eigensymmsymm22}),
one can decompose the corresponding eigenvalue
as
\bea
 \phi_0^{(1)}(\tiny\yng(2);
 \tiny\yng(2)) \,
  h(\tiny\yng(2);
  \tiny\yng(2)) = \Bigg[ \frac{4}{(k+N+2)} \Bigg]
  \times \Bigg[ \frac{2}{(k+N+2)} \Bigg],
\label{12symmsymm}
\eea
where the equation $(3.5)$ of \cite{Ahn1711} can be used.
Then one sees the coincidence with the equation $(5.22)$ of \cite{Ahn1711}
for the eigenvalue of the higher spin-$1$ current.

The three-point function
of the higher spin-$3$ current 
 with the scalar operator corresponding to $(\tiny\yng(2);\tiny\yng(2))$
 and the scalar operator corresponding
 to $(\overline{\tiny\yng(2)};\overline{\tiny\yng(2)})$
 can be read off from (\ref{eigensymmsymm22}) as before.

\subsubsection{ The eigenvalue in
(\mbox{symm};f) representation  
with two boxes}

The eigenvalue of the zeromode for the higher spin current
appearing in the first term of (\ref{vspin3}) in
this representation is given by
$(14,54, 110)$ for $N=3,5,7$ respectively. 
Then the general $N$ dependence is
given by $(2N^2+4N-16)$ \footnote{
Similarly,
from the eigenvalues $(-216,-1800,-5880,-13608)$
appearing in the second
 term 
for $N=3,5,7,9$,  the general $N$ behavior
is given by $(-24N^3+48 N^2)$.
From the eigenvalues $(-120,-280,-504)$
appearing in the third
 term 
for $N=3,5,7$,  the general $N$ behavior
is given by $(-8 N^2-16 N)$.
From the eigenvalues $(-48,-80,-112)$
appearing in the fourth term 
for $N=3,5,7$,  the general $N$ behavior
is given by $(-16 N)$.
The eigenvalue
appearing in the fifth term in
this representation is given by
$(-1,-3,-5)$
for $N=3,5,7$ respectively. 
Then the general $N$ dependence  is
given by $(-N+2)$.
Finally, from the eigenvalues $(4,12,20)$
appearing in the last term 
for $N=3,5,7$,  the general $N$ behavior
is given by $(4N-8)$.}.
Then one obtains the following eigenvalue
{\footnotesize
  \bea
  && \phi_2^{(1)}(\tiny\yng(2);
  \tiny\yng(1))  = 
\nonu \\
&& \frac{16 (k-N)}{(6 k N+5 k+5 N+4)} \Bigg[
  -\frac{1}{4 (k+N+2)^2} \times (2N^2+4N-16)
  + \frac{1}{32 N^2 (k+N+2)^2} \times 
(-24N^3+48N^2)
  \Bigg]
\nonu \\
&& +\frac{1}{2 N (k+N+2)^2} \times
(-8N^2-16N) +\frac{1}{(k+N+2)^2} \times (-16N)
-\frac{4 k}{(k+N+2)^2} \times ({\bf -N}+2)
\nonu \\
&& -\frac{2 (2 k+N)}{3 (k+N+2)^2} \times ({\bf 4N}-8)
=
\nonu \\
&& -\frac{4 ({\bf 6 k^2 N^2}-7 k^2 N-10 k^2{\bf +12 k N^3}+
  87 k N^2+106 k N-44 k+4 N^3+42 N^2+140 N+24)}{3 (k+N+2)^2 (6 k N+5 k+5 N+4)}
\nonu \\
&& \longrightarrow - \frac{4}{3} \, \lambda  \, (\lambda +1),
\label{eigensymmf}
\eea}
where the limit in (\ref{largenk}) is taken at the final stage.
It is obvious that
the leading contributions in this limit come from the 
fifth and the sixth terms of (\ref{eigensymmf}) as before.
The eigenvalue in this limit is the same as the one in
(\ref{eigenfzero}). In other words, the boldface parts of two
expressions are the same as each other.

The first and second terms of (\ref{eigensymmf}) contain
\bea
&& \Bigg[
  -\frac{1}{4 (k+N+2)^2} \times (2N^2+4N-16)
  + \frac{1}{32 N^2 (k+N+2)^2} \times 
(-24N^3+48N^2)
  \Bigg]= \nonu \\
&& -\frac{(N-2) (2 N+11)}{4 (k+N+2)^2}.
\label{this2}
\eea
One can decompose this (\ref{this2})
as
\bea
\phi_0^{(1)}(\tiny\yng(2);
\tiny\yng(1)) \,
h(\tiny\yng(2);
\tiny\yng(1)) = \Bigg[ -\frac{(N-2)}{ (k+N+2)} \Bigg]
\times \Bigg[ \frac{(2 N+11)}{4(k+N+2)} \Bigg],
\label{12symmf}
\eea
where the equation $(3.6)$ of \cite{Ahn1711} is used.
Then one can see the equation $(5.23)$ of \cite{Ahn1711}
in this expression. 

The three-point function
of the higher spin-$3$ current 
 with the scalar operator corresponding to $(\tiny\yng(2);\tiny\yng(1))$
 and the scalar operator corresponding
 to $(\overline{\tiny\yng(2)};\overline{\tiny\yng(1)})$
 can be read off from (\ref{eigensymmf}).

\subsubsection{ The eigenvalue in
(\mbox{antisymm};0) representation  
with two boxes}

The eigenvalue of the zeromode for the higher spin current
appearing in the first term of (\ref{vspin3}) in
this representation is given by
$(72,200,392)$ for $N=3,5,7$ respectively. 
Then the general $N$ dependence is
given by $8 N^2$ \footnote{
Similarly,
there are the eigenvalues $(0,0,0)$
appearing in the second or third term 
for $N=3,5,7$.
From the eigenvalues $(24,40,56)$
appearing in the fourth term
for $N=3,5,7$,  the general $N$ behavior
is given by $8N$.
From the eigenvalues $(-6,-10,-14)$
appearing in the fifth term 
for $N=3,5,7$,  the general $N$ behavior
is given by $-2 N$.
From the eigenvalues $(24,40,56)$
appearing in the last term 
for $N=3,5,7$,  the general $N$ behavior
is given by $8N $.}.
By collecting all the contributions with correct
coefficients, one determines the final eigenvalue as follows:
{\small
  \bea
  && \phi_2^{(1)}(\tiny\yng(1,1);0)  = 
 \frac{16 (k-N)}{(6 k N+5 k+5 N+4)} \Bigg[
  -\frac{1}{4 (k+N+2)^2} \times (8N^2)
  + \frac{1}{32 N^2 (k+N+2)^2} \times 
0
  \Bigg]
\nonu \\
&& +\frac{1}{2 N (k+N+2)^2} \times
0 +\frac{1}{(k+N+2)^2} \times 8N
-\frac{4 k}{(k+N+2)^2} \times ({\bf -2N})
\nonu \\
&& -\frac{2 (2 k+N)}{3 (k+N+2)^2} \times {\bf 8N}
=
\nonu \\
&& -\frac{8 N ({\bf 6 k^2 N}+5 k^2 {
    \bf +12 k N^2}+9 k N-11 k-2 N^2-7 N-12)}{
  3 (k+N+2)^2 (6 k N+5 k+5 N+4)}
 \longrightarrow - \frac{8}{3} \, \lambda  \, (\lambda +1),
\label{eigenantisymmzero}
\eea}
where the limit in (\ref{largenk}) is taken at the final stage.
It is obvious that
the leading contributions in this limit come from the 
fifth and the sixth terms of (\ref{eigenantisymmzero}) as before.
The eigenvalue in this limit is the twice of the one in
(\ref{eigenfzero}). The number of boxes is increased by two.

The following decomposition corresponding to the first and second terms
of (\ref{eigenantisymmzero}) can be seen from the previous results
in the equations $(2.14)$ and $(5.19)$ of \cite{Ahn1711}
\bea
\phi_0^{(1)}(\tiny\yng(1,1);0)
\, h(\tiny\yng(1,1);0) = \Bigg[ -\frac{2N}{(k+N+2)} \Bigg]
\times
\Bigg[ \frac{N}{(k+N+2)} \Bigg].
\label{12anti0}
\eea

The three-point function
of the higher spin-$3$ current 
 with the scalar operator corresponding to $(\tiny\yng(1,1);0)$
 and the scalar operator corresponding
 to $(\overline{\tiny\yng(1,1)};0)$
 can be read off from (\ref{eigenantisymmzero}).

\subsubsection{ The eigenvalue in
(\mbox{antisymm};\mbox{antisymm}) representation  
with two, and two boxes}

The eigenvalue of the zeromode for the higher spin current
appearing in the first term of (\ref{vspin3}) in
this representation is given by
$(-32,-32, -32)$ for $N=3,5,7$ respectively \footnote{
Similarly,
there exist the eigenvalues $(0,0,0)$
appearing in the second or third
 term 
for $N=3,5,7$.
From the eigenvalues $(-32,-64,-96)$
appearing in the fourth term 
for $N=3,5,7$,  the general $N$ behavior
is given by $(-16 N+16)$.
The eigenvalue
appearing in the fifth term  in
this representation is given by
$(4,4,4)$
for $N=3,5,7$ respectively. 
Finally, the eigenvalues are given by $(-16,-16,-16)$
appearing in the last term 
for $N=3,5,7$.}.
Then one obtains the following eigenvalue
{\small
  \bea
  && \phi_2^{(1)}(\tiny\yng(1,1);
  \tiny\yng(1,1))  = 
 \frac{16 (k-N)}{(6 k N+5 k+5 N+4)} \Bigg[
  -\frac{1}{4 (k+N+2)^2} \times (-32)
  + \frac{1}{32 N^2 (k+N+2)^2} \times 
0
  \Bigg]
\nonu \\
&& +\frac{1}{2 N (k+N+2)^2} \times
0 +\frac{1}{(k+N+2)^2} \times ({\bf -16N}+16)
-\frac{4 k}{(k+N+2)^2} \times {\bf 4}
\nonu \\
&& -\frac{2 (2 k+N)}{3 (k+N+2)^2} \times ({\bf -16})
=
\nonu \\
&& \frac{16 \left({\bf 6 k^2 N}+5 k^2 {\bf -6 k N^2}+
  18 k N+43 k-5 N^2-13 N+12\right)}{3 (k+N+2)^2 (6 k N+5 k+5 N+4)}
 \longrightarrow - \frac{16}{3N} \, \lambda  \, (2\lambda -1),
\label{eigenantisymmantisymm22}
\eea}
where the limit in (\ref{largenk}) is taken at the final stage.
It is obvious that
the leading contributions in this limit come from the 
fourth, the fifth and
the sixth terms of (\ref{eigenantisymmantisymm22}) as before.
The eigenvalue in this limit is the twice of the one in
(\ref{eigenff}). In other words, the boldface parts are
increased by two.

One decomposes the expression appearing in the first two terms of
(\ref{eigenantisymmantisymm22})
as
\bea
\phi_0^{(1)}(\tiny\yng(1,1);
\tiny\yng(1,1)) \,
h(\tiny\yng(1,1);
\tiny\yng(1,1)) = \Bigg[ \frac{4}{(k+N+2)} \Bigg]
\times \Bigg[ \frac{2}{(k+N+2)} \Bigg],
\label{12antianti}
\eea
by using the previous results in the equations $(3.16)$
and $(5.36)$ of \cite{Ahn1711}.

The three-point function
of the higher spin-$3$ current 
 with the scalar operator corresponding to $(\tiny\yng(1,1);\tiny\yng(1,1))$
 and the scalar operator corresponding to
 $(\overline{\tiny\yng(1,1)};\overline{\tiny\yng(1,1)})$
can be obtained from the relation (\ref{eigenantisymmantisymm22})
explicitly.

\subsubsection{ The eigenvalue in
(\mbox{antisymm};f) representation  
with two boxes}

The eigenvalue of the zeromode for the higher spin current
appearing in the first term of (\ref{vspin3}) in
this representation is given by
$(6,30,70)$ for $N=3,5,7$ respectively \footnote{ 
Then the general $N$ dependence is
given by $(2N^2-4N)$.
Similarly,
from the eigenvalues $(-216,-1800,-5880,-13608)$
appearing in the second  term 
for $N=3,5,7,9$,  the general $N$ behavior
is given by $(-24N^3+48N^2)$.
From the eigenvalues $(-24,-120,-280)$
appearing in the third  term 
for $N=3,5,7$,  the general $N$ behavior
is given by $(-8N^2+16N)$.
From the eigenvalues $(0,0,0)$
appearing in the fourth term
for $N=3,5,7$,  the general $N$ behavior
is given by $0$.
From the eigenvalues $(-1,-3,-5)$
appearing in the fifth term 
for $N=3,5,7$,  the general $N$ behavior
is given by $(-N+2)$.
From the eigenvalues $(4,12,20)$
appearing in the last term 
for $N=3,5,7$,  the general $N$ behavior
is given by $(4N-8)$.}.
By collecting all the contributions with correct
coefficients, one determines the final eigenvalue as follows:
{\footnotesize
  \bea
  && \phi_2^{(1)}(\tiny\yng(1,1);
  \tiny\yng(1))  = 
\nonu \\
&& \frac{16 (k-N)}{(6 k N+5 k+5 N+4)} \Bigg[
  -\frac{1}{4 (k+N+2)^2} \times (2N^2-4N)
  + \frac{1}{32 N^2 (k+N+2)^2} \times 
(-24N^3+48N^2)
  \Bigg]
\nonu \\
&& +\frac{1}{2 N (k+N+2)^2} \times
(-8N^2+16N) +\frac{1}{(k+N+2)^2} \times 0
-\frac{4 k}{(k+N+2)^2} \times ({\bf -N}+2)
\nonu \\
&& -\frac{2 (2 k+N)}{3 (k+N+2)^2} \times ({\bf 4N}-8)
=
\nonu \\
&& -\frac{4 (N-2) ({\bf 6 k^2 N}+5 k^2{\bf +12 k N^2}+
  39 k N+28 k+4 N^2+14 N+12)}{
  3 (k+N+2)^2 (6 k N+5 k+5 N+4)}
 \longrightarrow - \frac{4}{3} \, \lambda  \, (\lambda +1),
\label{eigenantisymmf}
\eea}
where the limit in (\ref{largenk}) is taken at the final stage.
It is obvious that
the leading contributions in this limit come from the 
fifth and the sixth terms of (\ref{eigenantisymmf}) as before.
The eigenvalue in this limit is the same as the one in
(\ref{eigenfzero}).
The boldface parts are the same as each other.

From the expression coming from the first two terms of
(\ref{eigenantisymmf})
\bea
&& \Bigg[
  -\frac{1}{4 (k+N+2)^2} \times (2N^2-4N)
  + \frac{1}{32 N^2 (k+N+2)^2} \times 
(-24N^3+48N^2)
  \Bigg] =
\nonu \\
&& -\frac{(N-2) (2 N+3)}{4 (k+N+2)^2},
\label{this3}
\eea
one decomposes this (\ref{this3}) as
\bea
\phi_0^{(1)}(\tiny\yng(1,1);
\tiny\yng(1)) \,
h(\tiny\yng(1,1);
\tiny\yng(1)) = \Bigg[ -\frac{(N-2)}{(k+N+2)} \Bigg]
\times \Bigg[ \frac{(2N+3)}{4(k+N+2)} \Bigg],
\label{12antif}
\eea
by using the previous results in the equations $(3.17)$
and $(5.37)$ of \cite{Ahn1711}.

The three-point function
of the higher spin-$3$ current 
 with the scalar operator corresponding to $(\tiny\yng(1,1);\tiny\yng(1))$
 and the scalar operator corresponding to
 $(\overline{\tiny\yng(1,1)};\overline{\tiny\yng(1)})$
can be obtained from the relation (\ref{eigenantisymmf})
explicitly.

\subsection{ The $(\La_{+}; \La_{-})$ representations
with three boxes for $\La_+$}

It is known that the following branching rules
for symmetric, mixed and antisymmetric representations
hold \cite{Ahn1711}
 \bea
        {\tiny\yng(3)} & \rightarrow &
          ({\tiny\yng(3)},{\bf 1})_3+
      ({\tiny\yng(2)},{\bf 2})_{2-\frac{N}{2}} +
      ({\tiny\yng(1)},{\bf 3})_{1-N}
+  ({\bf 1},{\bf 4})_{-\frac{3N}{2}},
        \nonu \\
              {\tiny\yng(2,1)} & \rightarrow &
({\tiny\yng(2,1)},{\bf 1})_3+
      ({\tiny\yng(1,1)},{\bf 2})_{2-\frac{N}{2}} +
       ({\tiny\yng(2)},{\bf 2})_{2-\frac{N}{2}} +
      ({\tiny\yng(1)},{\bf 3})_{1-N} +({\tiny\yng(1)},{\bf 1})_{1-N}
+  ({\bf 1},{\bf 2})_{-\frac{3N}{2}},              
\nonu \\
              {\tiny\yng(1,1,1)}
              & \rightarrow &
({\tiny\yng(1,1,1)},{\bf 1})_3+
      ({\tiny\yng(1,1)},{\bf 2})_{2-\frac{N}{2}} +
      ({\tiny\yng(1)},{\bf 1})_{1-N}.              
              \label{transtranstrans}
    \eea

    In this subsection, some new features in the representations with more than
    three boxes are presented.    
    
\subsubsection{ The eigenvalue in
(\mbox{symm};0) representation  
with three boxes}

The eigenvalue of the zeromode for the higher spin current
appearing in the first term of (\ref{vspin3}) in
this representation is given by
$(162,450,882)$ for $N=3,5,7$ respectively. 
Then the general $N$ dependence can be read off and it is
given by $18 N^2$ \footnote{
Similarly,
from the eigenvalues $(-9720,-45000,-123480,-262440)$
appearing in the second term 
for $N=3,5,7,9$,  the general $N$ behavior
is given by $-360N^3$.
The eigenvalue 
appearing in the third term  in
this representation is given by
$(-432,-1200,-2352)$ for $N=3,5,7$ respectively. 
Then the general $N$ dependence can be read off and it is
given by $-48 N^2$.
From the eigenvalues $(36,60,84)$
appearing in the fourth term 
for $N=3,5,7$,  the general $N$ behavior
is given by $12N$.
The eigenvalue 
appearing in the fifth term  in
this representation is given by
$(-9,-15,-21)$
for $N=3,5,7$ respectively. 
Then the general $N$ dependence can be read off and it is
given by $-3 N$.
Finally, from the eigenvalues $(36,60,84)$
appearing in the last term 
for $N=3,5,7$,  the general $N$ behavior
is given by $12N$.}.
One obtains the final eigenvalue as follows:
{\small
  \bea
&& \phi_2^{(1)}(\tiny\yng(3);0)  = 
\nonu \\
&& \frac{16 (k-N)}{(6 k N+5 k+5 N+4)} \Bigg[
  -\frac{1}{4 (k+N+2)^2} \times 18 N^2
  + \frac{1}{32 N^2 (k+N+2)^2} \times 
(- 360 N^3)
  \Bigg]
\nonu \\
&& +\frac{1}{2 N (k+N+2)^2} \times
(- 48 N^2) +\frac{1}{(k+N+2)^2} \times 12 N
-\frac{4 k}{(k+N+2)^2} \times ({\bf - 3 N})
\nonu \\
&& -\frac{2 (2 k+N)}{3 (k+N+2)^2} \times {\bf 12 N}
=
-\frac{4 N ({\bf 6 k^2 N}+5 k^2 {\bf +12 k N^2}+
  51 k N+64 k-8 N^2-22 N+12)}{
  (k+N+2)^2 (6 k N+5 k+5 N+4)}
\nonu \\
&& \longrightarrow -4 \, \lambda  \, (\lambda +1),
\label{eigensymmzero}
\eea}
where the limit in (\ref{largenk}) is taken at the final stage.
It is obvious that
the leading contributions in this limit come from the 
fifth and the sixth terms of (\ref{eigensymmzero}).
The eigenvalue in this limit is the third times of the one in
(\ref{eigenfzero}).
The boldface parts are increased by three.

On the other hand,
the other eigenvalue of the zeromode for the higher spin current
appearing in the third term of (\ref{vspin3}) in
this representation is given by
$(-288,-800,-1568)$ for $N=3,5,7$ respectively. 
Then the general $N$ dependence can be read off and it is
given by $-32 N^2$.
From the other eigenvalue $(12,20,28)$
of the zeromode for the higher spin current
appearing in the fourth term of (\ref{vspin3})
for $N=3,5,7$,  the general $N$ behavior
is given by $4N$.
From the previous results of third and fourth terms of (\ref{eigensymmzero}),
one has the following value 
\bea
\Bigg[ \frac{1}{2 N (k+N+2)^2} \times
  (- 48 N^2) +\frac{1}{(k+N+2)^2} \times 12 N \Bigg]=
-\frac{12 N}{(k+N+2)^2}.
\label{Eigen1}
\eea
Moreover, one has the same total eigenvalue for
each different eigenvalue (according to the first equation of
(\ref{transtranstrans}) there are quartic states under $SU(2)_k$
in this representation)
as follows:
\bea
\Bigg[ \frac{1}{2 N (k+N+2)^2} \times
  (- 32 N^2) +\frac{1}{(k+N+2)^2} \times 4 N \Bigg]=
-\frac{12 N}{(k+N+2)^2}.
\label{Eigen2}
\eea
Therefore, one obtains the same eigenvalue in (\ref{eigensymmzero})
because of the coincidence between (\ref{Eigen1}) and (\ref{Eigen2}).

By realizing that the eigenvalue of the zeromode
of $\Phi_0^{(1)} \, T$ is given by 
\bea
\Bigg[ -\frac{1}{4 (k+N+2)^2} \times 18 N^2
  + \frac{1}{32 N^2 (k+N+2)^2} \times 
(- 360 N^3) \Bigg] =-\frac{9 N (2 N+5)}{4 (k+N+2)^2},
  \label{this4}
  \eea
from the first two terms in (\ref{eigensymmzero}),
one can decompose this eigenvalue (\ref{this4}) as the following form  
  \bea
\phi_0^{(1)} (\tiny\yng(3);0)\,
h (\tiny\yng(3);0) = \Bigg[ -\frac{3N}{(k+N+2)} \Bigg]
\times \Bigg[ \frac{3(2N+5)}{4(k+N+2)} \Bigg],
\label{12Symm0}
\eea
where the conformal dimension for this representation
appearing in the equation $(3.29)$ of \cite{Ahn1711} is substituted.
Then one observes that
the eigenvalue for the zeromode of the higher spin-$1$ current
in this representation can be read off and it is given by
\bea
\phi_0^{(1)} (\tiny\yng(3);0) = -\frac{3N}{(k+N+2)}
= 3 \phi_0^{(1)} (\tiny\yng(1);0),
\label{spinonethreebox}
\eea
where the relation in (\ref{12f0}) is used.
On the other hand,
the other eigenvalue of the zeromode for the higher spin-$1$ current
appearing in the first term of (\ref{spinone}) in
this representation is given by
$(-\frac{9}{(k+5)},-\frac{15}{(k+7)},-\frac{21}{(k+9)})$
for $N=3,5,7$ respectively.
The result of (\ref{spinonethreebox}) implies that
the multiplicity $3$ on the eigenvalue
$\phi_0^{(1)} (\tiny\yng(1);0)$ is valid at finite $(N,k)$
(as well as under the large $(N,k)$ 't Hooft like limit).
The three-point function of the higher spin-$1$ current
with two scalars is 
{\small
\bea
&& <\overline{\cal O}
(\overline{\tiny\yng(3)};0)
  \,  {\cal O}
    (\tiny\yng(3);0) \, \Phi_0^{(1)} >  =  \Bigg[
-\frac{3N}{(k+N+2)}
      \Bigg]
     <\overline{\cal O} (\overline{\tiny\yng(3)};0)
      \, {\cal O} (\tiny\yng(3);0) >
      \nonu \\
      && \longrightarrow
      - 3 \, \lambda \,
<\overline{\cal O}
(\overline{\tiny\yng(3)};0)
   \, {\cal O} (\tiny\yng(3);0) >,
\label{threespin1}
\eea}
where the limit in (\ref{largenk}) is taken at the final stage.

Note that one observes
the relation between two eigenvalues
\bea
\phi_2^{(1)} (\tiny\yng(3);0) = 3 \phi_2^{(1)} (\tiny\yng(1);0)
= 3 \times \Bigg[ -\frac{4}{3} \, \lambda  \, (\lambda +1)\Bigg],
\label{tttday}
\eea
under the large $(N,k)$ 't Hooft like limit, as anticipated in
\cite{Ahn1711}. In (\ref{tttday}), the previous result (\ref{eigenfzero})
is used.

The three-point function
of the higher spin-$3$ current 
 with the scalar operator corresponding to $(\tiny\yng(3);0)$
 and the scalar operator corresponding to
 $(\overline{\tiny\yng(3)};0)$
can be obtained from the relation (\ref{eigensymmzero})
explicitly
{\footnotesize
  \bea
&& <\overline{\cal O} (\overline{\tiny\yng(3)};0) \,
    {\cal O} (\tiny\yng(3);0) \, \Phi_2^{(1)} > \nonu \\
    && =  \Bigg[
-\frac{4 N ({\bf 6 k^2 N}+5 k^2 {\bf +12 k N^2}+
  51 k N+64 k-8 N^2-22 N+12)}{
  (k+N+2)^2 (6 k N+5 k+5 N+4)}
      \Bigg]
    <\overline{\cal O} (\overline{\tiny\yng(3)};0)\,
    {\cal O} (\tiny\yng(3);0) >
    \nonu \\
    && \longrightarrow -4 \la (\la+1)
    <\overline{\cal O} (\overline{\tiny\yng(3)};0) \,
    {\cal O} (\tiny\yng(3);0)>  =
    \frac{4}{3} \, (\la+1) \,
 <\overline{\cal O}
(\overline{\tiny\yng(3)};0)
  \,  {\cal O}
  (\tiny\yng(3);0) \, \Phi_0^{(1)} >,
    \label{threetodaytoday}
\eea}
where the limit in (\ref{largenk}) is taken at the final stage
and the relation (\ref{threespin1}) is used.
One can see that the three-point function (\ref{threetodaytoday})
is a multiple of the one in the subsection $4.3.1$ because of
(\ref{tttday}).

\subsubsection{ The eigenvalue in (\mbox{symm}; \mbox{symm})
representation with  three, and three boxes}

The eigenvalue of the zeromode for the higher spin current
appearing in the first term of (\ref{vspin3}) in
this representation is given by
$(-72,-72,-72)$ for $N=3,5,7$ respectively \footnote{
Similarly,
the eigenvalues appearing in the second or third term
are given by $(0,0,0)$
for $N=3,5,7,9$.
The eigenvalue 
appearing in the fourth term  in
this representation is given by
$(-120,-168,-216)$ for $N=3,5,7$ respectively. 
Then the general $N$ dependence can be read off and it is
given by $(-24N-48)$.
The eigenvalues are $(6,6,6)$
appearing in the fifth term 
for $N=3,5,7$.
The eigenvalues appearing in the last term
are given by $(-24,-24,-24)$
for $N=3,5,7$.}.
By collecting all the contributions with correct
coefficients, one determines the final eigenvalue as follows:
{\small
  \bea
&& \phi_2^{(1)}(\tiny\yng(3);\tiny\yng(3)) =
\frac{16 (k-N)}{(6 k N+5 k+5 N+4)} \Bigg[
  -\frac{1}{4 (k+N+2)^2} \times (-72)
  + \frac{1}{32 N^2 (k+N+2)^2} \times 0
  \Bigg]
\nonu \\
&& +  \frac{1}{2 N (k+N+2)^2} \times
0 +\frac{1}{(k+N+2)^2} \times ({\bf -24 N}-48)
-\frac{4 k}{(k+N+2)^2} \times {\bf 6}
\nonu \\
& &-  \frac{2 (2 k+N)}{3 (k+N+2)^2} \times ({\bf -24})
 =  
\frac{8 ( {\bf 6 k^2 N}+5 k^2 {\bf -6 k N^2}-36 k N+10 k-5 N^2-70 N-24)}{
  (k+N+2)^2 (6 k N+5 k+5 N+4)}
\nonu \\
& & \longrightarrow  -\frac{8}{N} \, \lambda  \, (2 \lambda -1),
\label{eigensymmsymm33}
\eea}
where the limit in (\ref{largenk}) is taken at the final stage.
It is obvious that
the leading contributions in this limit come from the 
fourth, the fifth and the sixth terms of (\ref{eigensymmsymm33}).
The eigenvalue in this limit is the third times of the one in
(\ref{eigenff}).
The number of boxes is increased by three.

By realizing that the eigenvalue of the zeromode
of $\Phi_0^{(1)} \, T$ is given by 
\bea
\Bigg[ \frac{1}{2 N (k+N+2)^2} \times
  (-72) +\frac{1}{(k+N+2)^2} \times 0 \Bigg]=
\frac{18}{(k+N+2)^2},
\label{this5}
\eea
from the first two terms of (\ref{eigensymmsymm33}),
one can decompose this eigenvalue (\ref{this5}) as the following form  
\bea
\phi_0^{(1)} (\tiny\yng(3);\tiny\yng(3))\,
h (\tiny\yng(3);\tiny\yng(3)) = \Bigg[ \frac{6}{(k+N+2)} \Bigg]
\times \Bigg[ \frac{3}{(k+N+2)} \Bigg],
\label{12SymmSymm}
\eea
where the conformal dimension for this representation  appearing in
the subsection $3.3.1$ of \cite{Ahn1711} is substituted.
Then one sees that
the eigenvalue for the zeromode of the higher spin-$1$ current
in this representation can be read off and it is given by, together with
(\ref{12ff}),
\bea
\phi_0^{(1)} (\tiny\yng(3);\tiny\yng(3))= 
\frac{6}{(k+N+2)} = 3 \phi_0^{(1)} (\tiny\yng(1);\tiny\yng(1)),
\label{maytoday}
\eea
which
implies that
the multiplicity $3$ on the eigenvalue
$\phi_0^{(1)} (\tiny\yng(1);\tiny\yng(1))$ in (\ref{maytoday})
is valid at finite $(N,k)$
(as well as under the large $(N,k)$ 't Hooft like limit).
On the other hand,
the other eigenvalue of the zeromode for the higher spin-$1$ current
appearing in the first term of (\ref{spinone}) in
this representation is given by
$(\frac{6}{(k+5)},\frac{6}{(k+7)},\frac{6}{(k+9)})$
for $N=3,5,7$ respectively. 
The three-point function of the higher spin-$1$ current
with scalars is given by
{\small
\bea
&& <\overline{\cal O}
(\overline{\tiny\yng(3)};\overline{\tiny\yng(3)})
  \,  {\cal O}
    (\tiny\yng(3);\tiny\yng(3)) \, \Phi_0^{(1)} >  =  \Bigg[
\frac{6}{(k+N+2)}
      \Bigg]
     <\overline{\cal O} (\overline{\tiny\yng(3)};
      \overline{\tiny\yng(3)})
      \, {\cal O} (\tiny\yng(3);\tiny\yng(3)) > \nonu \\
      && \longrightarrow
      \frac{6 \lambda}{N} \,
<\overline{\cal O}
(\overline{\tiny\yng(3)};\overline{\tiny\yng(3)})
   \, {\cal O} (\tiny\yng(3);\tiny\yng(3)) >,
\label{threepointspin1}
\eea}
where the limit in (\ref{largenk}) is taken at the final stage.

Note that one observes
the relation between two eigenvalues
\bea
\phi_2^{(1)} (\tiny\yng(3);\tiny\yng(3)) = 3 \phi_2^{(1)}
(\tiny\yng(1);\tiny\yng(1))
= 3 \times \Bigg[ -\frac{8}{3 N} \, \lambda  \, (2 \lambda -1) \Bigg],
\label{today1}
\eea
under the large $(N,k)$ 't Hooft like limit, as anticipated in
\cite{Ahn1711}. In (\ref{today1}), the previous relation
(\ref{eigenff}) is used. 

The three-point function
of the higher spin-$3$ current 
 with the scalar operator corresponding to $(\tiny\yng(3);\tiny\yng(3))$
 and the scalar operator corresponding to
 $(\overline{\tiny\yng(3)};\overline{\tiny\yng(3)})$
can be obtained from the relation (\ref{eigensymmsymm33})
explicitly
{\footnotesize
\bea
&& <\overline{\cal O} (\overline{\tiny\yng(3)};\overline{\tiny\yng(3)})
  \,  {\cal O} (\tiny\yng(3);\tiny\yng(3)) \, \Phi_2^{(1)} > \nonu \\
    && =  \Bigg[
\frac{8 ( {\bf 6 k^2 N}+5 k^2 {\bf -6 k N^2}-36 k N+10 k-5 N^2-70 N-24)}{
  (k+N+2)^2 (6 k N+5 k+5 N+4)}
      \Bigg]
    <\overline{\cal O} (\overline{\tiny\yng(3)};\overline{\tiny\yng(3)})
   \, {\cal O} (\tiny\yng(3);\tiny\yng(3))>
    \nonu \\
     && \longrightarrow
 -\frac{8}{N} \, \lambda  \, (2 \lambda -1) 
    <\overline{\cal O} (\overline{\tiny\yng(3)};\overline{\tiny\yng(3)})
    \, {\cal O} (\tiny\yng(3);\tiny\yng(3))>    =
    \nonu \\
    && -\frac{4}{3} \, (2\la-1) \,
 <\overline{\cal O}
(\overline{\tiny\yng(3)};\overline{\tiny\yng(3)})
  \,  {\cal O}
    (\tiny\yng(3);\tiny\yng(3)) \, \Phi_0^{(1)} >, 
\label{today2}
\eea}
where the limit in (\ref{largenk}) is taken at the final stage.
Moreover, the relation in (\ref{threepointspin1}) is used.
The three-point function (\ref{today2}) is a multiple of the one in
the subsection $4.3.2$ according to (\ref{today1}).

\subsubsection{ The eigenvalue in (\mbox{symm}; \mbox{symm})
representation with  three, and two boxes}

The eigenvalue of the zeromode for the higher spin current
appearing in the first term of (\ref{vspin3}) in
this representation is given by
$(-22,26,90)$ for $N=3,5,7$ respectively. 
Then the general $N$ dependence can be read off and it is
given by $(2 N^2+8N-64)$ \footnote{
Similarly,
from the eigenvalues $(216,-600,-3528,-9720)$
appearing in the second term 
for $N=3,5,7,9$,  the general $N$ behavior
is given by $(-24N^3+96N^2)$.
The eigenvalue 
appearing in the third term  in
this representation is given by
$(-168,-360,-616)$ for $N=3,5,7$ respectively. 
Then the general $N$ dependence can be read off and it is
given by $(-8 N^2-32N)$.
From the eigenvalues $(1,-1,-3)$
appearing in the fourth term 
for $N=3,5,7$,  the general $N$ behavior
is given by $(-N+4)$.
The eigenvalue 
appearing in the fifth term  in
this representation is given by
$(-9,-15,-21)$
for $N=3,5,7$ respectively. 
Then the general $N$ dependence can be read off and it is
given by $-3 N$.
Finally, from the eigenvalues $(-4,4,12)$
appearing in the last term 
for $N=3,5,7$,  the general $N$ behavior
is given by $(4N-16)$.}.
One obtains the final eigenvalue as follows:
{\footnotesize
\bea
&& \phi_2^{(1)}(\tiny\yng(3);\tiny\yng(2)) = \nonu \\
&& \frac{16 (k-N)}{(6 k N+5 k+5 N+4)} \Bigg[
  -\frac{1}{4 (k+N+2)^2} \times (2 N^2+8 N-64)
  + \frac{1}{32 N^2 (k+N+2)^2} \times (96 N^2-24 N^3)
  \Bigg]
\nonu \\
&& +\frac{1}{2 N (k+N+2)^2} \times
(-8 N^2-32 N) +\frac{1}{(k+N+2)^2} \times (-32 N-32)
-\frac{4 k}{(k+N+2)^2} \times ({\bf -N} +4)
\nonu \\
&& -\frac{2 (2 k+N)}{3 (k+N+2)^2} \times ({\bf 4 N}-16)
\nonu \\
&& =
-\frac{4 ({\bf 6  k^2 N^2}-19 k^2 N-20 k^2 {\bf +12 k N^3}+135 k N^2+
  328 k N-64 k+4 N^3+70 N^2+484 N+144)}{3 (k+N+2)^2 (6 k N+5 k+5 N+4)}
\nonu \\
&& \longrightarrow -\frac{4}{3} \, \lambda \, (\lambda +1),
\label{eigensymmsymm32}
\eea}
where the limit in (\ref{largenk}) is taken at the final stage.
It is obvious that
the leading contributions in this limit come from the 
fifth and the sixth terms of (\ref{eigensymmsymm32}).
The eigenvalue in this limit is the same as  the one in
(\ref{eigenfzero}).
The boldface parts are the same as each other.

By realizing that the eigenvalue of the zeromode
of $\Phi_0^{(1)} \, T$ is given by 
{\footnotesize
  \bea
 \Bigg[ \frac{1}{2 N (k+N+2)^2} \times
  (2 N^2+8 N-64) +\frac{1}{(k+N+2)^2} \times (96 N^2-24 N^3) \Bigg]
=
-\frac{(N-4) (2 N+19)}{4 (k+N+2)^2},
\label{this6}
\eea}
from the first two terms of (\ref{eigensymmsymm32}),
one can decompose this eigenvalue (\ref{this6})
as the following form  
\bea
\phi_0^{(1)} (\tiny\yng(3);\tiny\yng(2))\,
h (\tiny\yng(3);\tiny\yng(2)) = \Bigg[ \frac{(4-N)}{(k+N+2)} \Bigg]
\times
\Bigg[ \frac{(2N+19)}{
  4(k+N+2)} \Bigg],
\label{12product}
\eea
where the conformal dimension for this representation
appearing in the subsection $3.3.2$  of \cite{Ahn1711} is substituted
into (\ref{12product}).
Then one sees that
the eigenvalue for the zeromode of the higher spin-$1$ current
in this representation can be read off and it is given by
\bea
\phi_0^{(1)} (\tiny\yng(3);\tiny\yng(2)) & = & 
\frac{(4-N)}{(k+N+2)}
\longrightarrow  -\la = \phi_0^{(1)} (\tiny\yng(1);0),
\label{1Symmsymm}
\eea
under the large $(N,k)$ 't Hooft like limit.
On the other hand,
the other eigenvalue of the zeromode for the higher spin current
appearing in the first term of (\ref{spinone}) in
this representation is given by
$(\frac{1}{(k+5)},-\frac{1}{(k+7)},-\frac{3}{(k+9)})$
for $N=3,5,7$ respectively. 
The three-point function of the higher spin-$1$ current
with scalars is given by
{\small
\bea
&& <\overline{\cal O}
(\overline{\tiny\yng(3)};\overline{\tiny\yng(2)})
  \,  {\cal O}
    (\tiny\yng(3);\tiny\yng(2)) \, \Phi_0^{(1)} >  =  \Bigg[
\frac{(4-N)}{(k+N+2)}
      \Bigg]
     <\overline{\cal O} (\overline{\tiny\yng(3)};
      \overline{\tiny\yng(2)})
      \, {\cal O} (\tiny\yng(3);\tiny\yng(2)) >
      \nonu \\
      && \longrightarrow
      - \lambda \,
<\overline{\cal O}
(\overline{\tiny\yng(3)};\overline{\tiny\yng(2)})
   \, {\cal O} (\tiny\yng(3);\tiny\yng(2)) >,
\label{threepointspinone2}
\eea}
where the limit in (\ref{largenk}) is taken at the final stage.

Note that there is a relation 
between the two eigenvalues
\bea
\phi_2^{(1)} (\tiny\yng(3);\tiny\yng(2)) =  \phi_2^{(1)}
(\tiny\yng(1);0),
\label{Today}
\eea
under the large $(N,k)$ 't Hooft like limit, as anticipated in
\cite{Ahn1711}. The previous relation (\ref{eigenfzero}) is used in
(\ref{Today}).

 It is straightforward to write down the three-point function
of the higher spin-$3$ current 
with the scalar operator corresponding to
 $(\tiny\yng(3);\tiny\yng(2))$
 and the scalar operator corresponding to
$ (\overline{\tiny\yng(3)};\overline{\tiny\yng(2)})$
 from (\ref{eigensymmsymm32})
{\footnotesize
\bea
&& <\overline{\cal O} (\overline{\tiny\yng(3)};\overline{\tiny\yng(2)})
  \,  {\cal O} (\tiny\yng(3);\tiny\yng(2)) \, \Phi_2^{(1)} > \nonu \\
    && =  \Bigg[
-\frac{4 ({\bf 6  k^2 N^2}-19 k^2 N-20 k^2 {\bf +12 k N^3}+135 k N^2+
  328 k N-64 k+4 N^3+70 N^2+484 N+144)}{3 (k+N+2)^2 (6 k N+5 k+5 N+4)}
      \Bigg]
    \nonu \\
    && \times <\overline{\cal O} (\overline{\tiny\yng(3)};
    \overline{\tiny\yng(2)})
   \, {\cal O} (\tiny\yng(3);\tiny\yng(2))>
     \rightarrow -\frac{4}{3} \, \lambda \, (\lambda +1)
    <\overline{\cal O} (\overline{\tiny\yng(3)};\overline{\tiny\yng(2)})
    \, {\cal O} (\tiny\yng(3);\tiny\yng(2))>
    \nonu \\
   && =
    \frac{4}{3} \, (\la+1) \,
 <\overline{\cal O}
(\overline{\tiny\yng(3)};\overline{\tiny\yng(2)})
  \,  {\cal O}
    (\tiny\yng(3);\tiny\yng(2)) \, \Phi_0^{(1)} >,
\label{TTday}
\eea}
where the limit in (\ref{largenk}) is taken at the final stage
and the relation (\ref{threepointspinone2}) is used.
Again, the three-point function in (\ref{TTday}) is equal to
the one in the subsection $4.3.1$.

\subsubsection{ The eigenvalue in (\mbox{symm}; f)
representation with  three boxes}

The eigenvalue of the zeromode for the higher spin current
appearing in the first term of (\ref{vspin3}) in
this representation is given by
$(96,256,480)$ for $N=3,5,7$ respectively \footnote{
Then the general $N$ dependence  is
given by $(8 N^2+16N-24)$.
Similarly,
from the eigenvalues $(-2304,-12800,-37632,-82944)$
appearing in the second term 
for $N=3,5,7,9$,  the general $N$ behavior
is given by $(-128N^3+128N^2)$.
The eigenvalue 
appearing in the third term  in
this representation is given by
$(-260,-840,-1512)$ for $N=3,5,7$ respectively. 
Then the general $N$ dependence can be read off and it is
given by $(-24 N^2-48N)$.
From the eigenvalues $(-72,-120,-168)$
appearing in the fourth term
for $N=3,5,7$,  the general $N$ behavior
is given by $-24N$.
The eigenvalue 
appearing in the fifth term  in
this representation is given by
$(-4,-8,-12)$
for $N=3,5,7$ respectively. 
Then the general $N$ dependence can be read off and it is
given by $(-2 N+2)$.
Finally, from the eigenvalues $(16,32,48)$
appearing in the last term 
for $N=3,5,7$,  the general $N$ behavior
is given by $(8N-8)$.}.
One obtains the final eigenvalue
{\footnotesize
\bea
&& \phi_2^{(1)}(\tiny\yng(3);\tiny\yng(1)) = 
 \frac{16 (k-N)}{(6 k N+5 k+5 N+4)} \Bigg[
  -\frac{1}{4 (k+N+2)^2} \times (8 N^2+16 N-24)
  \nonu \\
  && + \frac{1}{32 N^2 (k+N+2)^2} \times (128 N^2-128 N^3)
  \Bigg]
\nonu \\
&& +\frac{1}{2 N (k+N+2)^2} \times
(-24 N^2-48 N) +\frac{1}{(k+N+2)^2} \times (-24 N)
-\frac{4 k}{(k+N+2)^2} \times ({\bf -2 N}+2)
\nonu \\
&& -\frac{2 (2 k+N)}{3 (k+N+2)^2} \times ({\bf 8 N}-8)
\nonu \\
&& =
-\frac{4 ({\bf 12 k^2 N^2}-2 k^2 N-10 k^2 {\bf +24 k N^3}+
  192 k N^2+317 k N-38 k-4 N^3+35 N^2+302 N+72)}{3 (k+N+2)^2 (6 k N+5 k+5 N+4)}
\nonu \\
&& \longrightarrow -\frac{8}{3}  \, \lambda  \, (\lambda +1),
\label{Eigensymmsymm31}
\eea}
where the limit in (\ref{largenk}) is taken at the final stage.
It is obvious that
the leading contributions in this limit come from the 
fifth and the sixth terms of (\ref{Eigensymmsymm31}).
The eigenvalue in this limit is the twice of   the one in
(\ref{eigenfzero}).
The boldface parts are increased by two.

By realizing that the eigenvalue of the zeromode
of $\Phi_0^{(1)} \, T$ is given by 
{\footnotesize \bea
\Bigg[ \frac{1}{2 N (k+N+2)^2} \times
  (8 N^2+16 N-24) +\frac{1}{(k+N+2)^2} \times (128 N^2-128 N^3)
    \Bigg]=
-\frac{2 (N-1) (N+5)}{(k+N+2)^2},
\label{eigensymmsymm31}
\eea}
from the first two terms of (\ref{Eigensymmsymm31}),
one can decompose this eigenvalue (\ref{eigensymmsymm31})
as the following form  
\bea
\phi_0^{(1)} (\tiny\yng(3);\tiny\yng(1))\,
h (\tiny\yng(3);\tiny\yng(1)) = \Bigg[ \frac{2(1-N)}{(k+N+2)} \Bigg]
  \times
\Bigg[ \frac{(N+5)}{(k+N+2)} \Bigg],
\label{OneTwoproduct}
\eea
where the conformal dimension for this representation
appearing in the subsection $3.3.3$ of \cite{Ahn1711} is substituted
into (\ref{OneTwoproduct}).
Then one sees that
the eigenvalue for the zeromode of the higher spin-$1$ current
in this representation can be read off and it is given by
\bea
\phi_0^{(1)} (\tiny\yng(3);\tiny\yng(1))
& = & \frac{2(1-N)}{(k+N+2)}
 \longrightarrow   -2 \la,
\label{1Symmf}
\eea
under the large $(N,k)$ 't Hooft like limit.
On the other hand,
the other eigenvalue of the zeromode for the higher spin current
appearing in the first term of (\ref{spinone}) in
this representation is given by
$(-\frac{4}{(k+5)},-\frac{8}{(k+7)},-\frac{12}{(k+9)})$
for $N=3,5,7$ respectively. 
The three-point function is given by
{\small
\bea
&& <\overline{\cal O}
(\overline{\tiny\yng(3)};\overline{\tiny\yng(1)})
  \,  {\cal O}
    (\tiny\yng(3);\tiny\yng(1)) \, \Phi_0^{(1)} >  =  \Bigg[
\frac{2 (1-N) }{(k+N+2)}
      \Bigg]
     <\overline{\cal O} (\overline{\tiny\yng(3)};
      \overline{\tiny\yng(1)})
      \, {\cal O} (\tiny\yng(3);\tiny\yng(1)) >
      \nonu \\
      && \longrightarrow
      -2\, \lambda \,
<\overline{\cal O}
(\overline{\tiny\yng(3)};\overline{\tiny\yng(1)})
   \, {\cal O} (\tiny\yng(3);\tiny\yng(1)) >,
\label{spinonethreepoint}
\eea}
where the limit in (\ref{largenk}) is taken at the final stage.

Note that one observes
the relation between the eigenvalues
\bea
\phi_2^{(1)} (\tiny\yng(3);\tiny\yng(1)) =   2 \phi_2^{(1)}
(\tiny\yng(1);0) = 2 \times
\Bigg[ -\frac{4}{3}  \, \lambda  \, (\lambda +1)
  \Bigg],
\label{12Product}
\eea
under the large $(N,k)$ 't Hooft like limit, as anticipated in
\cite{Ahn1711}. In (\ref{12Product}), one uses (\ref{eigenfzero}).

The other eigenvalue of the zeromode for the higher spin current
appearing in the third term of (\ref{vspin3}) in
this representation is given by
$(-240,-560,-1008)$ for $N=3,5,7$ respectively.
Note that there are triplet states in this representation
from (\ref{transtranstrans}).
Then the general $N$ dependence can be read off and it is
given by $(-16 N^2-32N)$.
From the other eigenvalues $(-52,-92,-132)$
of the zeromode for the higher spin current
appearing in the fourth term of (\ref{vspin3})
for $N=3,5,7$,  the general $N$ behavior
is given by $(-20N+8)$.
It turns out that
{\footnotesize
\bea
&& \phi_2^{(1)}(\tiny\yng(3);\tiny\yng(1)) =
\frac{16 (k-N)}{(6 k N+5 k+5 N+4)} \Bigg[
  -\frac{1}{4 (k+N+2)^2} \times (8 N^2+16 N-24)
  \nonu \\
  && + \frac{1}{32 N^2 (k+N+2)^2} \times (128 N^2-128 N^3)
  \Bigg]
\nonu \\
&& +\frac{1}{2 N (k+N+2)^2} \times
(-16 N^2-32 N) +\frac{1}{(k+N+2)^2} \times (-20 N+8)
-\frac{4 k}{(k+N+2)^2} \times (2-2 N)
\nonu \\
&& -\frac{2 (2 k+N)}{3 (k+N+2)^2} \times (8 N-8)
\nonu \\
&& =
-\frac{4 ({\bf 12 k^2 N^2}-2 k^2 N-10 k^2 {\bf +24 k N^3}+
  156 k N^2+215 k N-98 k-4 N^3+5 N^2+218 N+24)}{3 (k+N+2)^2 (6 k N+5 k+5 N+4)}
\nonu \\
&& \longrightarrow -\frac{8}{3}  \, \lambda  \, (\lambda +1).
\label{eigenn}
\eea}
Although the eigenvalue (\ref{eigenn}) is different from the one in
(\ref{Eigensymmsymm31}) at finite $(N,k)$, the large $(N,k)$ 't Hooft
like limit is the same. 

 It is straightforward to write down the three-point function
of the higher spin-$3$ current 
with the scalar operator corresponding to
 $(\tiny\yng(3);\tiny\yng(1))$
 and the scalar operator corresponding to
$ (\overline{\tiny\yng(3)};\overline{\tiny\yng(1)})$
 from (\ref{Eigensymmsymm31})
{\footnotesize
\bea
&& <\overline{\cal O} (\overline{\tiny\yng(3)};\overline{\tiny\yng(1)})
  \,  {\cal O} (\tiny\yng(3);\tiny\yng(1)) \, \Phi_2^{(1)} > \nonu \\
    && =  \Bigg[
-\frac{4 ({\bf 12 k^2 N^2}-2 k^2 N-10 k^2 {\bf +24 k N^3}+
  192 k N^2+317 k N-38 k-4 N^3+35 N^2+302 N+72)}{3 (k+N+2)^2 (6 k N+5 k+5 N+4)}
      \Bigg]
    \nonu \\
    && \times <\overline{\cal O} (\overline{\tiny\yng(3)};\overline{\tiny\yng(1)}) \,
    {\cal O} (\tiny\yng(3);\tiny\yng(1))>
     \rightarrow
 -\frac{8}{3}  \, \lambda  \, (\lambda +1)
    <\overline{\cal O} (\overline{\tiny\yng(3)};\overline{\tiny\yng(1)})
    \, {\cal O} (\tiny\yng(3);\tiny\yng(1))>
    \nonu \\
    && =
    \frac{4}{3} \, (\la+1) \,
 <\overline{\cal O}
(\overline{\tiny\yng(3)};\overline{\tiny\yng(1)})
  \,  {\cal O}
  (\tiny\yng(3);\tiny\yng(1)) \, \Phi_0^{(1)} >,
\label{TTToday}
\eea}
where the limit in (\ref{largenk}) is taken at the final stage
and one uses (\ref{spinonethreepoint}).
In (\ref{TTToday}), the three-point function is a multiple of the one in
the subsection $4.3.1$.

\subsubsection{ The eigenvalue in (\mbox{antisymm}; \mbox{antisymm})
representation with  three, and three boxes}

The eigenvalue of the zeromode for the higher spin current
appearing in the first term of (\ref{vspin3}) in
this representation is given by
$(-72,-72,-72)$ for $N=3,5,7$ respectively \footnote{
Similarly,
the eigenvalues appearing in the second term are $(0,0,0)$
for $N=3,5,7$.
The eigenvalue 
appearing in the third term  in
this representation is given by
$(0,0,0)$ for $N=3,5,7$ respectively. 
From the eigenvalues $(-24,-72,-120)$
appearing in the fourth term 
for $N=3,5,7$,  the general $N$ behavior
is given by $(-24N+48)$.
The eigenvalue 
appearing in the fifth term  in
this representation is given by
$(6,6,6)$
for $N=3,5,7$ respectively. 
Finally, from the eigenvalues $(-24,-24,-24)$
appearing in the last term 
for $N=3,5,7$,  the general $N$ behavior
can be seen.}.
One obtains the final eigenvalue as follows:
\bea
&& \phi_2^{(1)}(\tiny\yng(1,1,1);\tiny\yng(1,1,1))  = 
 \frac{16 (k-N)}{(6 k N+5 k+5 N+4)} \Bigg[
  -\frac{1}{4 (k+N+2)^2} \times (-72)
  + \frac{1}{32 N^2 (k+N+2)^2} \times 0
  \Bigg]
\nonu \\
& & +  \frac{1}{2 N (k+N+2)^2} \times
0 +\frac{1}{(k+N+2)^2} \times ({\bf -24 N}+48)
-\frac{4 k}{(k+N+2)^2} \times {\bf 6}
\nonu \\
&& -  \frac{2 (2 k+N)}{3 (k+N+2)^2} \times ({\bf -24})
 = 
\frac{8 ({\bf 6 k^2 N}+5 k^2 {\bf -6 k N^2}+36 k N+70 k-5 N^2-10 N+24)}{
  (k+N+2)^2 (6 k N+5 k+5 N+4)}
\nonu \\
& & \longrightarrow  -\frac{8}{N}  \, \lambda  \, (2\lambda -1),
\label{eigenantisymmantisymm33}
\eea
where the limit in (\ref{largenk}) is taken at the final stage.
It is obvious that
the leading contributions in this limit come from the 
fifth and the sixth terms of (\ref{eigenantisymmantisymm33}).
The eigenvalue in this limit is the third times of  the one in
(\ref{eigenff}).
The boldface parts are increased by three.

By realizing that the eigenvalue of the zeromode
of $\Phi_0^{(1)} \, T$ is given by 
\bea
\Bigg[ -\frac{1}{4 (k+N+2)^2} \times (-72)
  + \frac{1}{32 N^2 (k+N+2)^2} \times 0 \Bigg]
= \frac{18}{(k+N+2)^2},
\label{This1}
\eea
from the first two terms of (\ref{eigenantisymmantisymm33}),
one can decompose this eigenvalue (\ref{This1}) as the following form  
\bea
\phi_0^{(1)} (\tiny\yng(1,1,1);\tiny\yng(1,1,1))\,
h (\tiny\yng(1,1,1);\tiny\yng(1,1,1)) = \Bigg[ \frac{6}{(k+N+2)} \Bigg]
  \times
\Bigg[ \frac{3}{(k+N+2)} \Bigg],
\label{onetwopro}
\eea
where the conformal dimension for this representation
appearing in the subsection $3.4.1$ of \cite{Ahn1711} is substituted
into (\ref{onetwopro}).
Then one sees that
the eigenvalue for the zeromode of the higher spin-$1$ current
in this representation can be read off and it is given by,
together with (\ref{12ff}),
\bea
\phi_0^{(1)} (\tiny\yng(1,1,1);\tiny\yng(1,1,1))
 = \frac{6}{(k+N+2)} = 3 \phi_0^{(1)} (\tiny\yng(1);\tiny\yng(1)),
\label{1AntiAnti}
\eea
which
implies that
the multiplicity $3$ on the eigenvalue
$\phi_0^{(1)} (\tiny\yng(1);\tiny\yng(1))$ is valid at finite $(N,k)$
(as well as under the large $(N,k)$ 't Hooft like limit).
On the other hand,
the other eigenvalue of the zeromode for the higher spin-$1$ current
appearing in the first term of (\ref{spinone}) in
this representation is given by
$(\frac{6}{(k+5)},\frac{6}{(k+7)},\frac{6}{(k+9)})$
for $N=3,5,7$ respectively. 
The three-point function is given by
{\footnotesize
\bea
 <\overline{\cal O}
(\overline{\tiny\yng(1,1,1)};\overline{\tiny\yng(1,1,1)})
  \,  {\cal O}
  (\tiny\yng(1,1,1);\tiny\yng(1,1,1)) \, \Phi_0^{(1)} >   =  \Bigg[
\frac{6}{(k+N+2)}
      \Bigg]
     <\overline{\cal O} (\overline{\tiny\yng(1,1,1)};
      \overline{\tiny\yng(1,1,1)})
      \, {\cal O} (\tiny\yng(1,1,1);\tiny\yng(1,1,1)) > \rightarrow
      \frac{6 \lambda}{N} \,
<\overline{\cal O}
(\overline{\tiny\yng(1,1,1)};\overline{\tiny\yng(1,1,1)})
   \, {\cal O} (\tiny\yng(1,1,1);\tiny\yng(1,1,1)) >.
\label{threepointSpinone}
\eea}
where the limit in (\ref{largenk}) is taken at the final stage.

Note that one observes
the eigenvalue
\bea
\phi_2^{(1)} (\tiny\yng(1,1,1);\tiny\yng(1,1,1)) = 3 \phi_2^{(1)}
(\tiny\yng(1);\tiny\yng(1))
= 3 \times \Bigg[ -\frac{8}{3 N} \, \lambda  \, (2 \lambda -1) \Bigg],
\label{day}
\eea
under the large $(N,k)$ 't Hooft like limit, as anticipated in
\cite{Ahn1711}. In (\ref{day}), the previous result (\ref{eigenff})
is used.

 It is straightforward to write down the three-point function
of the higher spin-$3$ current 
with the scalar operator corresponding to
 $(\tiny\yng(1,1,1);\tiny\yng(1,1,1))$
 and the scalar operator corresponding to
$ (\overline{\tiny\yng(1,1,1)};\overline{\tiny\yng(1,1,1)})$
 from (\ref{eigenantisymmantisymm33})
{\small
\bea
&& <\overline{\cal O} (\overline{\tiny\yng(1,1,1)};
\overline{\tiny\yng(1,1,1)})
\,    {\cal O} (\tiny\yng(1,1,1);\tiny\yng(1,1,1)) \,
\Phi_2^{(1)} > \nonu \\
    && =  \Bigg[
\frac{8 ({\bf 6 k^2 N}+5 k^2 {\bf -6 k N^2}+36 k N+70 k-5 N^2-10 N+24)}{
  (k+N+2)^2 (6 k N+5 k+5 N+4)}
      \Bigg]
     <\overline{\cal O} (\overline{\tiny\yng(1,1,1)};
      \overline{\tiny\yng(1,1,1)})
\,    {\cal O} (\tiny\yng(1,1,1);\tiny\yng(1,1,1))>
\nonu \\
    && \longrightarrow
  -\frac{8}{N}  \, \lambda  \, (2\lambda -1)
  <\overline{\cal O} (\overline{\tiny\yng(1,1,1)};
  \overline{\tiny\yng(1,1,1)})
  \,  {\cal O} (\tiny\yng(1,1,1);\tiny\yng(1,1,1))>
  =
-\frac{4}{3} \, (2\la-1) \,
 <\overline{\cal O}
(\overline{\tiny\yng(1,1,1)};\overline{\tiny\yng(1,1,1)})
  \,  {\cal O}
    (\tiny\yng(1,1,1);\tiny\yng(1,1,1)) \, \Phi_0^{(1)} >, 
\label{day1}
\eea}
where the limit in (\ref{largenk}) is taken at the final stage
and the relation (\ref{threepointSpinone}) is used.
In (\ref{day1}), the three-point function is a multiple of the one in
the subsection $4.3.2$.

\subsubsection{ The eigenvalue in (\mbox{antisymm}; \mbox{antisymm})
representation with  three, and two boxes}

The eigenvalue of the zeromode for the higher spin current
appearing in the first term of (\ref{vspin3}) in
this representation is given by
$(-6,10,42)$ for $N=3,5,7$ respectively. 
Then the general $N$ dependence can be read off and it is
given by $(2N^2-8N)$ \footnote{
Similarly,
from the eigenvalues $(216,-600,-3528,-9720)$
appearing in the second term 
for $N=3,5,7,9$,  the general $N$ behavior
is given by $(-24N^3+96N^2)$.
The eigenvalue 
appearing in the third term  in
this representation is given by
$(24,-40,-168)$ for $N=3,5,7$ respectively. 
Then the general $N$ dependence can be read off and it is
given by $(32N-8N^2)$.
There are  eigenvalues $(0,0,0)$
appearing in the fourth term 
for $N=3,5,7$.
From the eigenvalues $(1,-1,-3)$
appearing in the fifth term 
for $N=3,5,7$,  the general $N$ behavior
is given by $(-N+4)$.
From the eigenvalues $(-4,4,12)$
appearing in the last term 
for $N=3,5,7$,  the general $N$ behavior
is given by $(4N-16)$.}.
By collecting all the contributions with correct
coefficients, one determines the final eigenvalue as follows:
{\footnotesize
  \bea
&& \phi_2^{(1)}(\tiny\yng(1,1,1);\tiny\yng(1,1)) =
\nonu \\
&& \frac{16 (k-N)}{(6 k N+5 k+5 N+4)} \Bigg[
  -\frac{1}{4 (k+N+2)^2} \times (2 N^2-8 N)
  + \frac{1}{32 N^2 (k+N+2)^2} \times (96 N^2-24 N^3)
  \Bigg]
\nonu \\
&& +\frac{1}{2 N (k+N+2)^2} \times
(32 N-8 N^2) +\frac{1}{(k+N+2)^2} \times 0
-\frac{4 k}{(k+N+2)^2} \times ({\bf -N}+4)
\nonu \\
&& -\frac{2 (2 k+N)}{3 (k+N+2)^2} \times ({\bf 4 N}-16)
 =
-\frac{4 (N-4) ({\bf 6 k^2 N}+5 k^2 {\bf +12 k N^2}+
  39 k N+28 k+4 N^2+14 N+12
)}{3 (k+N+2)^2 (6 k N+5 k+5 N+4)}
\nonu \\
&& \longrightarrow -\frac{4}{3}  \, \lambda  \, (\lambda +1),
\label{eigenantisymmantisymm32}
\eea}
where the limit in (\ref{largenk}) is taken at the final stage.
It is obvious that
the leading contributions in this limit come from the 
fifth and the sixth terms of (\ref{eigenantisymmantisymm32}).
The eigenvalue in this limit is the same as  the one in
(\ref{eigenfzero}).
The boldface parts are the same as each other. 

By realizing that the eigenvalue of the zeromode
of $\Phi_0^{(1)} \, T$ is given by 
{\footnotesize
  \bea
\Bigg[ -\frac{1}{4 (k+N+2)^2} \times (2 N^2-8 N)
 + \frac{1}{32 N^2 (k+N+2)^2} \times (96 N^2-24 N^3) \Bigg]=
 -\frac{(N-4) (2 N+3)}{4 (k+N+2)^2},
\label{dayday}
\eea}
from the first two terms of (\ref{eigenantisymmantisymm32}),
one can decompose this eigenvalue (\ref{dayday}) as the following form  
\bea
\phi_0^{(1)} (\tiny\yng(1,1,1);\tiny\yng(1,1))\,
h (\tiny\yng(1,1,1);\tiny\yng(1,1)) = \Bigg[ \frac{(4-N)}{(k+N+2)}
  \Bigg] \times
\Bigg[ \frac{(2N+3)}{4(k+N+2)} \Bigg],
\label{dayday3}
\eea
where the conformal dimension for this representation
appearing in the subsection $3.4.2$ of \cite{Ahn1711}
is substituted into (\ref{dayday3}).
Then one sees that
the eigenvalue for the zeromode of the higher spin-$1$ current
in this representation can be read off and it is given by
\bea
\phi_0^{(1)} (\tiny\yng(1,1,1);\tiny\yng(1,1))
 & = &  \frac{(4-N)}{(k+N+2)}
  \longrightarrow  -\la = \phi_0^{(1)} (\tiny\yng(1);0),
\label{1Antianti}
 \eea
under the large $(N,k)$ 't Hooft like limit.
On the other hand,
the other eigenvalue of the zeromode for the higher spin-$1$ current
appearing in the first term of (\ref{spinone}) in
this representation is given by
$(\frac{1}{(k+5)},-\frac{1}{(k+7)},-\frac{3}{(k+9)})$
for $N=3,5,7$ respectively. 

The three-point function of the higher spin-$1$ current
with two scalars is given by
{\footnotesize
\bea
 <\overline{\cal O}
(\overline{\tiny\yng(1,1,1)};\overline{\tiny\yng(1,1)})
  \,  {\cal O}
  (\tiny\yng(1,1,1);\tiny\yng(1,1)) \, \Phi_0^{(1)} >
  =  \Bigg[
\frac{(4-N)}{(k+N+2)}
      \Bigg]
     <\overline{\cal O} (\overline{\tiny\yng(1,1,1)};
      \overline{\tiny\yng(1,1)})
    \, {\cal O} (\tiny\yng(1,1,1);\tiny\yng(1,1)) >
\rightarrow  - \lambda
<\overline{\cal O}
(\overline{\tiny\yng(1,1,1)};\overline{\tiny\yng(1,1)})
   \, {\cal O} (\tiny\yng(1,1,1);\tiny\yng(1,1)) >,
\label{Threepointspinone}
\eea}
where the limit in (\ref{largenk}) is taken at the final stage.
Note that one observes
the eigenvalue
\bea
\phi_2^{(1)} (\tiny\yng(1,1,1);\tiny\yng(1,1)) =  \phi_2^{(1)}
(\tiny\yng(1);0),
\label{dayto}
\eea
under the large $(N,k)$ 't Hooft like limit, as anticipated in
\cite{Ahn1711}. Again the relation (\ref{eigenfzero})
is used in (\ref{dayto}).

 It is straightforward to write down the three-point function
of the higher spin-$3$ current 
with the scalar operator corresponding to
 $(\tiny\yng(1,1,1);\tiny\yng(1,1))$
 and the scalar operator corresponding to
$ (\overline{\tiny\yng(1,1,1)};\overline{\tiny\yng(1,1)})$
 from (\ref{eigenantisymmantisymm32})
{\footnotesize
\bea
&& <\overline{\cal O}
(\overline{\tiny\yng(1,1,1)};\overline{\tiny\yng(1,1)})
\,  {\cal O} (\tiny\yng(1,1,1);\tiny\yng(1,1)) \,
\Phi_2^{(1)} > \nonu \\
    && =  \Bigg[
-\frac{4 (N-4) ({\bf 6 k^2 N}+5 k^2 {\bf +12 k N^2}+
  39 k N+28 k+4 N^2+14 N+12
)}{3 (k+N+2)^2 (6 k N+5 k+5 N+4)}
      \Bigg]
     <\overline{\cal O} (\overline{\tiny\yng(1,1,1)};
      \overline{\tiny\yng(1,1)})
 \,   {\cal O} (\tiny\yng(1,1,1);\tiny\yng(1,1)) >
\nonu \\
&& \longrightarrow
 -\frac{4}{3}  \, \lambda  \, (\lambda +1)
 <\overline{\cal O}
 (\overline{\tiny\yng(1,1,1)};\overline{\tiny\yng(1,1)})
 \,  {\cal O} (\tiny\yng(1,1,1);\tiny\yng(1,1)) >=
\frac{4}{3} \, (\la+1) \,
 <\overline{\cal O}
(\overline{\tiny\yng(1,1,1)};\overline{\tiny\yng(1,1)})
  \,  {\cal O}
    (\tiny\yng(1,1,1);\tiny\yng(1,1)) \, \Phi_0^{(1)} >, 
\label{532018}
\eea}
where the limit in (\ref{largenk}) is taken at the final stage
and the relation (\ref{Threepointspinone}) is used.
In (\ref{532018}), the three-point function is the same as the one in
the subsection $4.3.1$ under the limit.

\subsubsection{ The eigenvalue in (\mbox{antisymm}; f)
representation with  three boxes}

The eigenvalue of the zeromode for the higher spin current
appearing in the first term of (\ref{vspin3}) in
this representation is given by
$(32,128,288)$ for $N=3,5,7$ respectively. 
Then the general $N$ dependence can be read off and it is
given by $(8N^2-16N+8)$ \footnote{
Similarly,
there are eigenvalues $(0,0,0,0)$
appearing in the second or third term 
for $N=3,5,7,9$.
From the eigenvalues $(32,64,96)$
appearing in the fourth term
for $N=3,5,7$,  the general $N$ behavior
is given by $(16N-16)$.
From the eigenvalues $(-4,-8,-12)$
appearing in the fifth term
for $N=3,5,7$,  the general $N$ behavior
is given by $(-2N+2)$.
From the eigenvalues $(16,32,48)$
appearing in the last term 
for $N=3,5,7$,  the general $N$ behavior
is given by $(8N-8)$.}.
By collecting all the contributions with correct
coefficients, one determines the final eigenvalue as follows:
\bea
&& \frac{16 (k-N)}{(6 k N+5 k+5 N+4)} \Bigg[
  -\frac{1}{4 (k+N+2)^2} \times (8 N^2-16 N+8)
  + \frac{1}{32 N^2 (k+N+2)^2} \times 0
  \Bigg]
\nonu \\
&& +\frac{1}{2 N (k+N+2)^2} \times
0 +\frac{1}{(k+N+2)^2} \times (16 N-16)
-\frac{4 k}{(k+N+2)^2} \times ({\bf -2 N}+2)
\nonu \\
&& -\frac{2 (2 k+N)}{3 (k+N+2)^2} \times ({ \bf 8 N}-8)
\label{eigenantisymmantisymm31}
\\
&& =
-\frac{8 (N-1) ({\bf 6 k^2 N}+5 k^2 {\bf +12 k N^2}-
  9 k N-38 k-2 N^2-10 N-24)}{
  3 (k+N+2)^2 (6 k N+5 k+5 N+4)}
 \longrightarrow -\frac{8}{3}  \, \lambda  \, (\lambda +1),
\nonu
 \eea
where the limit in (\ref{largenk}) is taken at the final stage.
It is obvious that
the leading contributions in this limit come from the 
fifth and the sixth terms of (\ref{eigenantisymmantisymm31}).
The eigenvalue in this limit is the twice of  the one in
(\ref{eigenfzero}).
The boldface parts are increased by two.

By realizing that the eigenvalue of the zeromode
of $\Phi_0^{(1)} \, T$ is given by 
{\small
  \bea
\Bigg[  -\frac{1}{4 (k+N+2)^2} \times (8 N^2-16 N+8)
  + \frac{1}{32 N^2 (k+N+2)^2} \times 0 \Bigg]  =
  -\frac{2 (N-1)^2}{(k+N+2)^2},
\label{that1}
\eea}
from the first two terms of (\ref{eigenantisymmantisymm31}),
one can decompose this eigenvalue (\ref{that1}) as the following form
\bea
\phi_0^{(1)} (\tiny\yng(1,1,1);\tiny\yng(1))\,
h (\tiny\yng(1,1,1);\tiny\yng(1)) = \Bigg[ \frac{-2(N-1)}{(k+N+2)}
  \Bigg] \times
\Bigg[ \frac{(N-1)}{(k+N+2)} \Bigg],
\label{pro12}
\eea
where the conformal dimension for this representation is substituted
into (\ref{pro12}).
Then one sees that
the eigenvalue for the zeromode of the higher spin-$1$ current
in this representation can be read off and it is given by
\bea
\phi_0^{(1)} (\tiny\yng(1,1,1);\tiny\yng(1))
 & = &  \frac{-2(N-1)}{(k+N+2)}
\longrightarrow  -2 \la,
\label{1Antif}
\eea
under the large $(N,k)$ 't Hooft like limit.
On the other hand,
the other eigenvalue of the zeromode for the higher spin-$1$ current
appearing in the first term of (\ref{spinone}) in
this representation is given by
$(-\frac{4}{(k+5)},-\frac{8}{(k+7)},-\frac{12}{(k+9)})$
for $N=3,5,7$ respectively. 

The three-point function of the higher spin-$1$ current
with two scalars is given by
{\footnotesize
\bea
 <\overline{\cal O}
(\overline{\tiny\yng(1,1,1)};\overline{\tiny\yng(1)})
  \,  {\cal O}
  (\tiny\yng(1,1,1);\tiny\yng(1)) \, \Phi_0^{(1)} >
   =  \Bigg[
 \frac{-2(N-1)}{(k+N+2)}
      \Bigg]
     <\overline{\cal O} (\overline{\tiny\yng(1,1,1)};
      \overline{\tiny\yng(1)})
    \, {\cal O} (\tiny\yng(1,1,1);\tiny\yng(1)) >
\rightarrow  -2  \, \lambda
<\overline{\cal O}
(\overline{\tiny\yng(1,1,1)};\overline{\tiny\yng(1)})
   \, {\cal O} (\tiny\yng(1,1,1);\tiny\yng(1)) >,
\label{Threespinone}
\eea}
where the limit in (\ref{largenk}) is taken at the final stage.
Note that one observes
the eigenvalue
\bea
\phi_2^{(1)} (\tiny\yng(1,1,1);\tiny\yng(1)) = 2 \phi_2^{(1)}
(\tiny\yng(1);0) = 2 \times \Bigg[ -\frac{4}{3}  \,
  \lambda  \, (\lambda +1)\Bigg],
\label{ProPro}
\eea
under the large $(N,k)$ 't Hooft like limit, as anticipated in
\cite{Ahn1711}. In (\ref{ProPro}),
one uses the previous relation (\ref{eigenfzero}).

It is straightforward to write down the three-point function
of the higher spin-$3$ current 
with the scalar operator corresponding to
 $(\tiny\yng(1,1,1);\tiny\yng(1))$
 and the scalar operator corresponding to
$ (\overline{\tiny\yng(1,1,1)};\overline{\tiny\yng(1)})$
 from (\ref{eigenantisymmantisymm31})
{\small
\bea
&& <\overline{\cal O}
(\overline{\tiny\yng(1,1,1)};\overline{\tiny\yng(1)})
  \,  {\cal O}
    (\tiny\yng(1,1,1);\tiny\yng(1)) \, \Phi_2^{(1)} > \nonu \\
    && =  \Bigg[
-\frac{8 (N-1) ({\bf 6 k^2 N}+5 k^2 {\bf +12 k N^2}-
  9 k N-38 k-2 N^2-10 N-24)}{
  3 (k+N+2)^2 (6 k N+5 k+5 N+4)}
      \Bigg]
     <\overline{\cal O} (\overline{\tiny\yng(1,1,1)};
      \overline{\tiny\yng(1)})
    \, {\cal O} (\tiny\yng(1,1,1);\tiny\yng(1)) >
\nonu \\
&& \longrightarrow  -\frac{8}{3}  \, \lambda  \, (\lambda +1)
<\overline{\cal O}
(\overline{\tiny\yng(1,1,1)};\overline{\tiny\yng(1)})
\, {\cal O} (\tiny\yng(1,1,1);\tiny\yng(1)) > =
\frac{4}{3} \, (\la+1) \,
 <\overline{\cal O}
(\overline{\tiny\yng(1,1,1)};\overline{\tiny\yng(1)})
  \,  {\cal O}
    (\tiny\yng(1,1,1);\tiny\yng(1)) \, \Phi_0^{(1)} >,
\label{threetoday}
\eea}
where the limit in (\ref{largenk}) is taken at the final stage
and the relation (\ref{Threespinone}) is used.
There is a simple relation the three-point function in (\ref{threetoday})
and the one in the subsection $4.3.1$.

\subsubsection{ The eigenvalue in $(\La_{+}; \La_{-})$
representation with  more than four boxes}

One can obtain the following branching rules
for the symmetric and antisymmetric representations with four boxes
\cite{Slansky,FK}
\bea
        {\tiny\yng(4)} & \rightarrow &
          ({\tiny\yng(4)},{\bf 1})_4+
      ({\tiny\yng(3)},{\bf 2})_{3-\frac{N}{2}} +
      ({\tiny\yng(2)},{\bf 3})_{2-N}
        +  ({\tiny\yng(1)},{\bf 4})_{1-\frac{3N}{2}} +
          +  ({\bf 1},{\bf 5})_{-2N},
        \nonu \\
              {\tiny\yng(1,1,1,1)}
              & \rightarrow &
({\tiny\yng(1,1,1,1)},{\bf 1})_4+
      ({\tiny\yng(1,1,1)},{\bf 2})_{3-\frac{N}{2}} +
      ({\tiny\yng(1,1)},{\bf 1})_{2-N}.              
\nonu
    \eea
    Then four-index symmetric (or antisymmetric) parts
    of the $SU(N+2)$ representation
can be obtained from the generators of the fundamental representation
of $SU(N+2)$ by using the projection operator.
By acting on the space
\bea
&& T_a \otimes {\bf 1} \otimes {\bf 1}
\otimes {\bf 1} +
        {\bf 1} \otimes T_a \otimes {\bf 1}
 \otimes {\bf 1} +{\bf 1} \otimes {\bf 1}
 \otimes T_a  \otimes {\bf 1} +
 {\bf 1} \otimes {\bf 1}
\otimes {\bf 1}
 \otimes T_a,
\nonu
\eea
where ${\bf 1}$ is $(N+2) \times (N+2)$ unit matrix,
the generators for the symmetric (or antisymmetric)
representation for the $SU(N+2)$ can be determined.

For more than five boxes, one can analyze similarly.
One observes that for the higher representations
$(\La_{+}, \La_{+})$, the eigenvalues (corresponding to the
higher spin-$3$ current) of the fourth, fifth and sixth
terms of (\ref{vspin3}) (under the large $(N,k)$ 't Hooft limit)
behave as $(-8N,2,-8)$ for one box, $(-16N, 4, -16)$
for two boxes and $(-24N, 6, -24)$ for three boxes.
This implies that the basic quantity is the one for one box.
The remaining one is a multiple of this quantity.

For the higher representations
$(\La_{+}, \La_{-})$, the eigenvalues of the fifth and sixth
terms of (\ref{vspin3}) (under the large $(N,k)$ 't Hooft limit)
behave as $(-N,4N)$ for one box which is equal to $(|\La_{+}|- |\La_{-}|)$,
$(-2N, 8N)$
for two boxes and $(-3N, 12N)$ for three boxes.
In this case also the basic quantity is the one for one box.
The remaining one is a multiple of this quantity.

Similarly,
 for the higher representations
$(\La_{+}, \La_{+})$, the eigenvalues (corresponding to the
higher spin-$1$ current) under the large $(N,k)$ 't Hooft limit
behave as $\frac{2}{(k+N+2)}$ for one box, $\frac{4}{(k+N+2)}$
for two boxes and $\frac{6}{(k+N+2)}$ for three boxes.
This implies that the basic quantity is the one for one box.
See also (\ref{12ff}), (\ref{12symmsymm}), (\ref{12antianti}),
(\ref{12SymmSymm}), and (\ref{1AntiAnti}).
The remaining one is a multiple of this quantity.

For the higher representations
$(\La_{+}, \La_{-})$, the eigenvalues  (of the higher spin-$1$ current)
under the large $(N,k)$ 't Hooft limit
behave as $-\frac{N}{(k+N+2)}$
for one box which is equal to $(|\La_{+}|- |\La_{-}|)$,
$-\frac{2N}{(k+N+2)}$
for two boxes and $-\frac{3N}{(k+N+2)}$ for three boxes.
In this case also the basic quantity is the one for one box.
The remaining one is a multiple of this quantity.
See also the equations (\ref{12f0}), (\ref{12symm0}), (\ref{12symmf}),
(\ref{12anti0}), (\ref{12antif}), (\ref{12Symm0}), (\ref{1Symmsymm}),
(\ref{1Symmf}), (\ref{1Antianti}) and (\ref{1Antif}).

For the three-point functions of the higher spin current
with two scalars, one expects that the following
relations satisfy
\bea
<\overline{\cal O}_{(\overline{\La}_{+};\overline{\La}_{+})}
  \,  {\cal O}_{
  (\La_{+};\La_{+})} \, \Phi_2^{(1)} > & \longrightarrow  & 
  -\frac{8 |\La_{+}|}{3 N}  \, \lambda  \, (2\lambda -1)
  <\overline{\cal O}_{(\overline{\La}_{+};\overline{\La}_{+})}
  \,  {\cal O}_{(\La_{+};\La_{+})}>
  \nonu \\
  & = &  -\frac{4}{3} \, (2\la -1) \,
  <\overline{\cal O}_{(\overline{\La}_{+};\overline{\La}_{+})}
  \,  {\cal O}_{(\La_{+};\La_{+})} \, \Phi_0^{(1)} >, \nonu \\
  <\overline{\cal O}_{
(\overline{\La}_{+};\overline{\La}_{-})}
  \,  {\cal O}_{(\La_{+};\La_{-})} \, \Phi_2^{(1)} >  & \longrightarrow  & 
  -\frac{4 (|\La_{+}| -|\La_{-}|)}{3}
  \, \lambda  \, (\lambda +1)
  <\overline{\cal O}_{(\overline{\La}_{+};\overline{\La}_{-})}
  \,  {\cal O}_{(\La_{+};\La_{-})}>
  \nonu \\
  & = & \frac{4}{3} \, (\la +1) \,
  <\overline{\cal O}_{(\overline{\La}_{+};\overline{\La}_{-})}
  \,  {\cal O}_{(\La_{+};\La_{-})} \, \Phi_0^{(1)} >.
\label{threegeneral}
  \eea
  In the first case of (\ref{threegeneral}),
  the case of $\La_-=\La_+$ is considered and the leading behavior
  is given by $\frac{1}{N}$ for finite $|\La_+|$.
  The coefficient of the two-point function is a multiple of
  the quantity appearing in (\ref{eigenff}).
  The three-point function of higher spin-$3$
  current is written in terms of the one of higher spin-$1$ current.
  One observes the simple factor $(2\la -1)$ appears in their ratios.
In the second case of (\ref{threegeneral}) where $|\La_-| \leq |\La_{+}|$,
 the coefficient of the two-point function is a multiple of
  the quantity appearing in (\ref{eigenfzero}).
  The simple factor $(\la +1)$ appears in the ratio of the three-point
  functions of higher spin-$3, 1$ currents.
  
\section{Conclusions and outlook }

We have found the $16$ higher spin currents where
the higher spin-$2$ currents  are given by (\ref{spin2final}),
the higher spin-$\frac{5}{2}$ currents are given by 
(\ref{spin5halfone}), (\ref{spin5halftwo}), (\ref{spin5halfthree}),
and (\ref{spin5halffour}),
the higher spin-$3$ current is given by
(\ref{spin3expression}) with (\ref{qspin3}),
(\ref{vspin3}) and (\ref{qvspin3}), together with the known higher spin-$1$
current (\ref{spinone}) and
the known higher spin-$\frac{3}{2}$ currents (\ref{spin3half}). 
Based on the explicit forms in (\ref{qspin3}) and (\ref{vspin3})
for the higher spin-$3$ current,  
the eigenvalues and the three-point functions are analyzed for various
higher representations. Under the large $(N,k)$ 't Hooft limit,
one has very simple expressions for the three-point functions of
the higher spin-$1,3$ currents
with two scalar operators in (\ref{threepointforq}) and (\ref{threegeneral}).

Let us present some open problems in some related directions of this paper.

$\bullet$
Checking of the remaining OPEs of Appendix (\ref{1116})

So far, some of the OPEs in Appenidx (\ref{1116})
using the adjoint spin-$1,\frac{1}{2}$ currents
has been not analyzed fully.
It would be interesting to observe whether there exist any nontrivial
identities between the various tensors.

$\bullet$ The eigenvalues and three-point functions
in other higher representations

We did not consider the eigenvalues for the higher representations
with mixed three boxes for the $\La_+$. One should obtain the corresponding
$SU(N+2)$ generators for several $N$ values. For $N=3$, they
were given in \cite{Ahn1711}.

$\bullet$ The next lowest $16$ higher spin currents in terms of
adjoint spin-$\frac{1}{2}, 1$ currents

We have considered the lowest $16$ higher spin currents. It would be
interesting to observe the three-point functions for the higher spin-$2,4$
current (in higher representations)
living in the next ${\cal N}=4$ multiplet.

$\bullet$ The lowest $16$ higher spin currents in terms of
adjoint spin-$\frac{1}{2}, 1$ currents in orthogonal Wolf space coset

The defining OPE relations between the $11$ currents and the
lowest $16$ higher spin currents ($s=2$) in \cite{AKK1703}
are valid in this orthogonal Wolf space coset.
The lowest higher spin-$2$ current has an explicit form in \cite{AKP1510}.
It is straightforward to continue to the present calculation and
obtain these $16$ higher spin currents using the adjoint
spin-$1,\frac{1}{2}$ currents.

$\bullet$ The bulk theory

It would be interesting to construct the bulk dual theory and
observe whether one can see the corresponding three-point functions
which should be equal to the ones in this paper. 

\vspace{.7cm}

\centerline{\bf Acknowledgments}

CA acknowledges warm hospitality from
C.N. Yang Institute for Theoretical Physics,
Stony Brook University.
CA would like to thank M.H. Kim for discussions.
This research was supported by Basic Science Research Program through
the National Research Foundation of Korea  
funded by the Ministry of Education  
(No. 2017R1D1A1A09079512).

\newpage

\appendix

\renewcommand{\theequation}{\Alph{section}\mbox{.}\arabic{equation}}

\section{The OPEs between the $11$ currents and the lowest
$16$ higher spin currents}

For convenience,
the explicit OPEs \cite{AKK1703} (by putting $s=1$)
between the $11$ currents and the $16$ higher spin
currents are given by
{\footnotesize
\bea
&& T(z)\,
\left(
\begin{array}{c}
 \Phi_0^{(1)} \\
\Phi_{\frac{1}{2}}^{(1),\mu}  \\
\Phi_1^{(1),\mu\nu}  \\
\Phi_{\frac{3}{2}}^{(1),\mu}  \\
\Phi_{2}^{(1)} \\
\end{array}
\right)
(w)  = 
\frac{1}{(z-w)^{2}}\, \left(
\begin{array}{c}
  \,\Phi_0^{(1)} \\
\frac{3}{2}\, \Phi_{\frac{1}{2}}^{(1),\mu}  \\
2 \, \Phi_1^{(1),\mu\nu}  \\
\frac{5}{2} \, \Phi_{\frac{3}{2}}^{(1),\mu}  \\
3 \, \Phi_{2}^{(1)} \\
\end{array}
\right)(w)+\frac{1}{(z-w)}\,
\left(
\begin{array}{c}
 \pa \, \Phi_0^{(1)} \\
\pa \, \Phi_{\frac{1}{2}}^{(1),\mu}  \\
\pa \, \Phi_1^{(1),\mu\nu}  \\
\pa \, \Phi_{\frac{3}{2}}^{(1),\mu}  \\
\pa \, \Phi_{2}^{(1)} \\
\end{array}
\right)
(w)+\cdots,
\nonu \\
\nonu \\  
&& G^{\mu}(z)\,\Phi_{2}^{(1)}(w)  =  
\frac{1}{(z-w)^{3}}\, \Bigg[
-\frac{8 (k-N) (12 k N+19 k+19 N+26)}{3 (k+N+2) (6 k N+5 k+5 N+4)}
  \Bigg] \,\Phi_{\frac{1}{2}}^{(1),\mu}(w)
\nonu \\ 
&& +  
\frac{1}{(z-w)^{2}}\Bigg[-5 \Phi_{\frac{3}{2}}^{(1),\mu}
+ \frac{24 (k-N)}{ (6 k N+5 k+5 N+4)} ( G^{\mu}  \Phi_{0}^{(1)}
+\frac{2}{3}\partial\Phi_{\frac{1}{2}}^{(1),\mu})
 -  \frac{8}{(2+k+N)}\varepsilon^{\mu \nu \rho \si} T^{\nu \rho}
 \Phi_{\frac{1}{2}}^{(1),\si}\Bigg]
\nonu \\ 
&& + \frac{1}{(z-w)} \Bigg[
-\partial\Phi_{\frac{3}{2}}^{(1),\mu}
-\frac{4\, i}{(2+k+N)} \varepsilon^{\mu \nu \rho \si}
\partial T^{\nu \rho}  \Phi_{\frac{1}{2}}^{(1),\si}
+   \nonu \\
&& \frac{8 (k-N)}{ (6 k N+5 k+5 N+4)}  
( \partial^{2}\Phi_{\frac{1}{2}}^{(1),\mu}-
2 T  \Phi_{\frac{1}{2}}^{(1),\mu}+\partial
G^{\mu}  \Phi_{0}^{(1)})\Bigg] + \cdots,
\nonu \\
&& G^{\mu}(z)\,\Phi_{\frac{3}{2}}^{(1),\nu}(w)  =  
\frac{1}{(z-w)^{3}}\Bigg[\frac{16(k-N)}{3(2+k+N)}
  \Bigg] \delta^{\mu \nu} \, \Phi_{0}^{(1)}(w)
 +  \frac{1}{(z-w)^{2}}\,\Bigg[-\,4\,\Phi_{1}^{(1),\mu \nu}
   \nonu \\
   && -\frac{2(k-N)}{3(2+k+N)}\, \varepsilon^{\mu \nu \rho \si}
  \, \Phi_{1}^{(1),\rho \si}\,\Bigg](w)\nonu
 +  \frac{1}{(z-w)}\, [\,\delta^{\mu \nu}\,(-\Phi_{2}^{(1)}
+\frac{16 (k-N)}{ (6 k N+5 k+5 N+4)}\, \Phi_{0}^{(1)} \, T
\,)
\nonu \\ 
&& - 
\partial\Phi_{1}^{(1),\mu \nu}
-\frac{(k-N)}{6(2+k+N)}\, \varepsilon^{\mu \nu \rho \si} \,
\partial\Phi_{1}^{(1),\rho \si}
+\frac{2\, i}{(2+k+N)}\,\varepsilon^{\mu \nu \rho \si} \, \partial {T}^{\rho \si}
\, \Phi_{0}^{(1)}
\nonu \\ 
&& -
\frac{2\, i}{(2+k+N)} \varepsilon^{\mu \nu \rho \si}  {T}^{\rho \si} 
\partial\Phi_{0}^{(1)}
+ \frac{2\, i}{(2+k+N)}
( T^{\mu \rho}  \Phi_{1}^{(1),\nu \rho}-
T^{\nu \rho}  \Phi_{1}^{(1),\mu \rho} )
 - \frac{2 }{(2+k+N)} \varepsilon^{\mu \nu \rho \si}  G^{\rho}
 \Phi_{\frac{1}{2}}^{(1),\si} ](w),
\nonu \\
&& {G}^{\mu}(z) \, \Phi_1^{(1),\nu\rho}(w)   = 
\frac{1}{(z-w)^2} \,  \Bigg[
  \frac{(N-k)}{(k+N+2)} \,(
  \delta^{\mu \nu} \, \Phi_{\frac{1}{2}}^{(1),\rho} -
\delta^{\mu \rho} \, \Phi_{\frac{1}{2}}^{(1),\nu})
  + \frac{(2+3k+3N)}{(k+N+2)} \, \varepsilon^{\mu \nu \rho \si} \,
  \Phi_{\frac{1}{2}}^{(1),\si}
\Bigg](w)
  \nonu \\
  && +  \frac{1}{(z-w)} \, \Bigg[
  \frac{i}{(k+N+2)} \, (\varepsilon^{\si \alpha \mu \rho} \,
  T^{\si \alpha} \, \Phi_{\frac{1}{2}}^{(1),\nu} -
\varepsilon^{\si \alpha \mu \nu} \,
  T^{\si \alpha} \, \Phi_{\frac{1}{2}}^{(1),\rho}
  ) + \varepsilon^{\mu \nu \rho \si } \, \pa \,
  \Phi_{\frac{1}{2}}^{(1),\si} 
  \nonu \\
  && -  \delta^{\mu \nu} \, ( \Phi_{\frac{3}{2}}^{(1),\rho}
  + \frac{(k-N)}{3(k+N+2)} \, \pa \, \Phi_{\frac{1}{2}}^{(1),\rho}
  + \frac{i}{(k+N+2)} \,  \varepsilon^{\alpha \beta \rho \si} \,
  T^{\alpha \beta} \, \Phi_{\frac{1}{2}}^{(1),\si}) \nonu \\
  && + 
\delta^{\mu \rho} \, ( \Phi_{\frac{3}{2}}^{(1),\nu}
  + \frac{(k-N)}{3(k+N+2)} \, \pa \, \Phi_{\frac{1}{2}}^{(1),\nu}
  + \frac{i}{(k+N+2)} \,  \varepsilon^{\alpha \beta \nu \si} \,
  T^{\alpha \beta} \, \Phi_{\frac{1}{2}}^{(1),\si}) 
  \Bigg](w) +\cdots,
  \nonu \\
&& G^{\mu}(z)\,\Phi_{\frac{1}{2}}^{(1),\nu}(w)  = 
-\frac{1}{(z-w)^{2}}\,2\, \delta^{\mu\nu}\,\Phi_{0}^{(1)}(w)
 +  
\frac{1}{(z-w)}\,\Bigg[-\delta^{\mu \nu}\,\partial\Phi_{0}^{(1)}+
  \frac{1}{2}\, \varepsilon^{\mu \nu \rho \si} \, \Phi_{1}^{(1),\rho \si}\,
  \Bigg](w)+\cdots,
\nonu \\
&& G^{\mu}(z) \, \Phi_{0}^{(1)}(w)  =  
-\frac{1}{(z-w)} \, \Phi_{\frac{1}{2}}^{(1),\mu}(w)
+\cdots,
\nonu \\
\nonu \\ 
&& T^{\mu \nu}(z)\,\Phi_{2}^{(1)}(w)  =  
\frac{1}{(z-w)^{2}}\,\Bigg[\,4\,i\,\Phi_{1}^{(1),\mu \nu}+
  \frac{16 (k-N)}{ (6 k N+5 k+5 N+4)}\, T^{\mu \nu}
  \, \Phi_{0}^{(1)}\,\Bigg](w)+\cdots,
\nonu \\
&& T^{\mu \nu}(z)\,\Phi_{\frac{3}{2}}^{(1),\rho}(w)  = 
\frac{1}{(z-w)^{2}}\Bigg[\,\frac{4i(k-N)}{3(2+k+N)}\,
  (\delta^{\mu \rho}\,\Phi_{\frac{1}{2}}^{(1),\nu}-\delta^{\nu \rho}\,
  \Phi_{\frac{1}{2}}^{(1),\mu})
 - 
  4i\, \varepsilon^{\mu \nu \rho \si}\, \Phi_{\frac{1}{2}}^{(1),\si}
\,\Bigg](w)
\nonu \\
&& -\frac{1}{(z-w)}\,i\, \Bigg[\delta^{\mu \rho}\,\Phi_{\frac{3}{2}}^{(1),\nu}
-\delta^{\nu \rho}\,\Phi_{\frac{3}{2}}^{(1),\mu} \Bigg]
(w) +  \cdots,
\qquad
 T^{\mu \nu}(z)\,\Phi_{1}^{(1), \rho \si}(w)
 =  
\frac{1}{(z-w)^{2}} 2 i \varepsilon^{\mu \nu \rho \si} \Phi_{0}^{(1)}(w)
\nonu \\
&& - 
\frac{1}{(z-w)} i \Bigg[\delta^{\mu \rho} \Phi_{1}^{(1),\nu \si}
  -\delta^{\mu \si} \Phi_{1}^{(1),\nu \rho}-\delta^{\nu \rho} \Phi_{1}^{(1),\mu \si}
  +\delta^{\nu \si} \Phi_{1}^{(1),\mu \rho}\,\Bigg](w),
\nonu \\
&& T^{\mu \nu}(z)\;\Phi_{\frac{1}{2}}^{(1),\rho}(w)
 =  
-\frac{1}{(z-w)}\,i\, \Bigg[\,\delta^{\mu \rho}\,\Phi_{\frac{1}{2}}^{(1),\nu}-
\delta^{\nu \rho}\,\Phi_{\frac{1}{2}}^{(1),\mu}\, \Bigg](w)+\cdots,
\qquad  T^{\mu\nu}(z)\;\Phi_{0}^{(1)}(w)  =  + \cdots.
\label{1116}
\eea}
These OPEs are obtained from the ${\cal N}=4$ primary condition
in the ${\cal N}=4$ superspace 
in the linear version \cite{AK1509}.

\section{ The OPEs between the spin-$\frac{3}{2}$ composite fields
and the spin-$2, \frac{5}{2}$ composite fields}

Let us present the OPEs between the spin-$\frac{3}{2}$ composite fields
appearing in the spin-$\frac{3}{2}$ currents (\ref{11currents}) and the
spin-$2$ composite fields appearing in the higher spin-$2$ currents
(\ref{spin2final}) as follows:
{\footnotesize
\bea
&& Q^{\bar{a}} \, V^{\bar{b} }(z) \, V^{\bar{c} } \, V^{\bar{d}} (w)
=
\frac{1}{(z-w)^2} \Bigg[
k \, g^{\bar{c} \bar{b}} \, Q^{\bar{a}} \, V^{\bar{d}} -
f^{\bar{c} \bar{b}}_{\,\,\,\,\,\, e} \, f^{\bar{d} e}_{\,\,\,\,\,\, g} \,
Q^{\bar{a}} \, V^{g} +
k \, g^{\bar{d} \bar{b}} \, Q^{\bar{a}} \, V^{\bar{c}}
  \Bigg](w)
\nonu \\
&& + 
\frac{1}{(z-w)} \, \Bigg[ k \,
  ( g^{\bar{c} \bar{b} } \, \pa \, Q^{\bar{a}}  \, V^{\bar{d}} 
+ g^{\bar{d} \bar{b} } \, \pa \, Q^{\bar{a}}  \, V^{\bar{c}}  ) 
+f^{\bar{c} \bar{b} }_{\,\,\,\,\,\, e} \, Q^{\bar{a}} \, V^{\bar{d}} \, V^{e}  
+f^{\bar{d} \bar{b} }_{\,\,\,\,\,\, e} \, Q^{\bar{a}} \, V^{\bar{c}} \, V^{e}   
+ f^{\bar{c} \bar{b} }_{\,\,\,\,\,\, e} \, f^{\bar{d} e }_{\,\,\,\,\,\, \bar{f}}
\, \pa \, ( Q^{\bar{a}} \, V^{\bar{f}} ) \Bigg](w),
\nonu
 \\
&& Q^{\bar{a}} \, V^{\bar{b} }(z) \,  Q^{\bar{c}} \, \pa \, Q^{\bar{d}}  (w)
=
\frac{1}{(z-w)^2} 
(k+N+2)   \Bigg[ g^{\bar{d} \bar{a}}  Q^{\bar{c}}  V^{\bar{b}}
  \Bigg](w)
 + 
\frac{ (k+N+2)}{(z-w)}  \Bigg[ 
g^{\bar{d} \bar{a} }   Q^{\bar{c}}  \pa   V^{\bar{b}}  -
g^{\bar{c} \bar{a} }   \pa  Q^{\bar{d}}   V^{\bar{b}}  \Bigg],
\nonu
 \\
&& Q^{\bar{a}} \, V^{\bar{b} }(z) \,  Q^{\bar{c}}  \, Q^{\bar{d}} \, 
V^{e} (w)
 =
\frac{1}{(z-w)^2} 
(k+N+2) \Bigg[ g^{\bar{c} \bar{a}} \,  f^{\bar{b} e}_{\,\,\,\,\,\, h} \, Q^{\bar{d}} \,
V^{h} - g^{\bar{d} \bar{a}} \,  f^{\bar{b} e}_{\,\,\,\,\,\, h} \, Q^{\bar{c}} \,
V^{h}
  \Bigg](w)
\label{firstfirst-1}
\\
&& + 
\frac{1}{(z-w)} \Bigg[ (k+N+2) \, ( 
g^{\bar{d} \bar{a} } \, Q^{\bar{c}}  \, V^{\bar{b}} \, V^{e} -
g^{\bar{c} \bar{a} } \, Q^{\bar{d}} \, V^{\bar{b}} \, V^{e}
)
+k \, g^{e \bar{b} } \, Q^{\bar{c}} \, Q^{\bar{d}} \, \pa \, Q^{\bar{a}} 
 +  f^{e \bar{b} }_{\,\,\,\,\,\, f} \, Q^{\bar{c}} \,Q^{\bar{d}} \,
Q^{\bar{a}} \, V^{f}
\Bigg] (w) + \cdots.
\nonu
\eea}
In particular, the explicit form for the first order pole
is very crucial to determine the higher spin-$\frac{5}{2}$ current.

The OPEs between the spin-$\frac{3}{2}$ composite fields
appearing in the spin-$\frac{3}{2}$ currents (\ref{11currents}) and the
spin-$\frac{5}{2}$
composite fields appearing in the higher spin-$\frac{5}{2}$
currents (\ref{spin5halfone})
are presented 
as follows:
{\footnotesize
\bea
&& Q^{\bar{a}} \, V^{\bar{b} }(z) \, 
\pa Q^{\bar{c}}  \, V^{\bar{d}} (w)
= \frac{1}{(z-w)^3} (k+N+2) \Bigg[  g^{\bar{c} \bar{a}} \,
 f^{\bar{b}  \bar{d} }_{\,\,\,\,\,\, e} \, V^e
  \Bigg](w)
\nonu \\
&& + \frac{1}{(z-w)^2} \Bigg[
-(k+N+2) \,  g^{\bar{c} \bar{a}} \, V^{\bar{b}} \, V^{\bar{d}}
- k \,  g^{\bar{d} \bar{b}} \, \pa \, Q^{\bar{c}} \, Q^{\bar{a}}
\Bigg](w)
\nonu \\
&& +  \frac{1}{(z-w)} \Bigg[ (k+N+2) \, (
-g^{\bar{c} \bar{a}} \, V^{\bar{d}} \, \pa \, V^{\bar{b}}
+\frac{1}{2} \, g^{\bar{c} \bar{a}} \, f^{\bar{b}  \bar{d} }_{\,\,\,\,\,\, e}
\, \pa^2 \, V^e )
 -   
k \, g^{\bar{d} \bar{b}} \, \pa \, Q^{\bar{c}} \, \pa \, Q^{\bar{a}} 
- f^{\bar{b}  \bar{d} }_{\,\,\,\,\,\, e}  \, Q^{\bar{a}} \, \pa \, Q^{\bar{c}} 
\, V^e \Bigg] (w) + \cdots,
\nonu \\
&& Q^{\bar{a}} \, V^{\bar{b} }(z) \, 
 Q^{\bar{c}} \, \pa \, V^{\bar{d}} (w)
 =
 \frac{1}{(z-w)^3} \Bigg[
(k+N+2) \, g^{\bar{c} \bar{a}} \,
 f^{\bar{b}  \bar{d} }_{\,\,\,\,\,\, e} \, V^e
   -2 k \, g^{\bar{d} \bar{b}} \,  Q^{\bar{c}} \, Q^{\bar{a}} \Bigg](w)
\nonu \\
&& + \frac{1}{(z-w)^2} \Bigg[
(k+N+2) \, g^{\bar{c} \bar{a}} \,
 f^{\bar{b}  \bar{d} }_{\,\,\,\,\,\, e} \, \pa \, V^e
- 2k\, g^{\bar{d} \bar{b}} \,
Q^{\bar{c}} \, \pa \, Q^{\bar{a}}  -   f^{\bar{d}  \bar{b} }_{\,\,\,\,\,\, e}
\,  Q^{\bar{c}} \, Q^{\bar{a}} \, V^e
\Bigg](w)
\nonu \\
&& + 
 \frac{1}{(z-w)} \Bigg[
   -(k+N+2) \, g^{\bar{c} \bar{a}}  \, V^{\bar{b}} \, \pa \, V^{\bar{d}}
-k \, g^{\bar{d} \bar{b}} \, Q^{\bar{c}} \, \pa^2 \, Q^{\bar{a}}
 -   f^{\bar{d}  \bar{b} }_{\,\,\,\,\,\, e}
 \, Q^{\bar{c}} \, \pa \, (  Q^{\bar{a}} \, V^e )  \Bigg](w) + \cdots,
 \nonu \\
&&  Q^{\bar{a}} \, V^{\bar{b} }(z) \, 
 Q^{\bar{c}}  \, V^{\bar{d}} \, V^e (w)
 =
 \frac{1}{(z-w)^3} \Bigg[ -(k+N+2) g^{\bar{c} \bar{a}} (k \,
   g^{\bar{b} \bar{d}} \, V^e +  f^{\bar{b}  \bar{d} }_{\,\,\,\,\,\, f}\,
   f^{f  e }_{\,\,\,\,\,\, g} \, V^g +k \, g^{\bar{b} e} \, V^{\bar{d}})
   \nonu \\
   && -  
   k \,  f^{\bar{d}  \bar{b} }_{\,\,\,\,\,\, f} \,  g^{e f}  \, Q^{\bar{c}} \,
   Q^{\bar{a}} \Bigg](w)
 + \frac{1}{(z-w)^2} \Bigg[ (k+N+2) \, g^{\bar{c} \bar{a}} \,
  f^{\bar{b}  \bar{d} }_{\,\,\,\,\,\, f} \, V^f
  \, V^e +
 (k+N+2) \, g^{\bar{c} \bar{a}} \,
  f^{\bar{b}  \bar{e} }_{\,\,\,\,\,\, f} \, V^{\bar{d}}
  \, V^f
  \nonu \\
  && -  k\, g^{\bar{d} \bar{b}} \, Q^{\bar{c}} \, Q^{\bar{a}} \, V^e
  - f^{\bar{d}  \bar{b} }_{\,\,\,\,\,\, f} \, ( k\, g^{ef} \, Q^{\bar{c}} \, \pa \,
  Q^{\bar{a}} +  f^{e f   }_{\,\,\,\,\,\, g} \, Q^{\bar{c}} \, Q^{\bar{a}} \, V^g )
  -   k\, g^{e \bar{b}} \, Q^{\bar{c}} \, Q^{\bar{a}} \, V^{\bar{d}}
  \Bigg](w)
\nonu \\
 && + \frac{1}{(z-w)} \Bigg[
   -(k+N+2) \, g^{\bar{c} \bar{a}} \, V^{\bar{b}}  \, V^{\bar{d}}  \, V^{e} 
 -k \, g^{\bar{d} \bar{b}} \, Q^{\bar{c}}  \, \pa \, Q^{\bar{a}} \, V^e
  -  f^{\bar{d}  \bar{b} }_{\,\,\,\,\,\, g}  
 \, Q^{\bar{c}} \, Q^{\bar{a}} \, V^e \, V^g 
 -  f^{\bar{d}  \bar{b} }_{\,\,\,\,\,\, g} \, f^{e g }_{\,\,\,\,\,\, h}
 \, Q^{\bar{c}} \, \pa \, (Q^{\bar{a}} \, V^h)
\nonu \\
&& -f^{e  \bar{b} }_{\,\,\,\,\,\, g} 
  \, Q^{\bar{c}} \, Q^{\bar{a}} \, V^{\bar{d}} \, V^{g}
 -   \frac{1}{2} \, k \, f^{\bar{d}  \bar{b} }_{\,\,\,\,\,\, g}  
  \, g^{e g} \, Q^{\bar{c}} \, \pa^2 \, Q^{\bar{a}} 
 - k \, g^{e \bar{b} } \, Q^{\bar{c}} \, \pa \, Q^{\bar{a}} \, 
V^{\bar{d}} 
  \Bigg] (w) + \cdots,
  \nonu \\
&&   Q^{\bar{a}} \, V^{\bar{b} }(z) \, 
Q^{\bar{c}}  \, Q^{\bar{d}} \, Q^{\bar{e}} \, V^{f} (w)
=
\frac{1}{(z-w)^3} (k+N+2) k \, g^{\bar{b} f} \, \Bigg[ - g^{\bar{c} \bar{a}} \, 
  \, Q^{\bar{d}} \, Q^{\bar{e}} +g^{\bar{d} \bar{a}}
  \, Q^{\bar{c}} \, Q^{\bar{e}}
-g^{\bar{e} \bar{a}} 
  \, Q^{\bar{c}} \, Q^{\bar{d}}
  \Bigg](w)
\nonu \\
&& + \frac{1}{(z-w)^2} \Bigg[
  (k+N+2) \,  g^{\bar{c} \bar{a}} \,  f^{\bar{b}  \bar{f} }_{\,\,\,\,\,\, g}
  \, Q^{\bar{d}} \, Q^{\bar{e}} \, V^g
  \nonu \\
  && -   (k+N+2) \,  g^{\bar{d} \bar{a}} \,  f^{\bar{b}  \bar{f} }_{\,\,\,\,\,\, g}
  \, Q^{\bar{c}} \, Q^{\bar{e}} \, V^g
+  (k+N+2) \,  g^{\bar{e} \bar{a}} \,  f^{\bar{b}  \bar{f} }_{\,\,\,\,\,\, g}
  \, Q^{\bar{c}} \, Q^{\bar{d}} \, V^g
   -  k \, g^{\bar{f} \bar{b}} \,  Q^{\bar{c}} \, Q^{\bar{d}} \,
   Q^{\bar{e}} \, Q^{\bar{a}} 
  \Bigg](w)
\nonu \\
&& + 
\frac{1}{(z-w)} \Bigg[ (k+N+2) \, ( 
g^{\bar{e} \bar{a}  } \,  Q^{\bar{d}} \,Q^{\bar{c}} \,V^{\bar{b}}\,  V^{f} -
g^{\bar{d} \bar{a}  } \,  Q^{\bar{e}} \,Q^{\bar{c}}\, V^{\bar{b}} \, V^{f} 
  +  
g^{\bar{c} \bar{a}  } \,  Q^{\bar{e}} \,Q^{\bar{d}} \,V^{\bar{b}} \, V^{f} ) 
+  f^{f \bar{b} }_{\,\,\,\,\,\, g}
 \,Q^{\bar{e}} \,Q^{\bar{d}} \,Q^{\bar{c}} \,Q^{\bar{a}}\,V^{g}
\nonu \\ 
&& +  
k \,g^{f \bar{b}}  \,Q^{\bar{e}} \,Q^{\bar{d}} \,Q^{\bar{c}} \,\pa \,
Q^{\bar{a}}
 \Bigg] (w) + \cdots.
\label{finalfirst}
\eea}
Some of these expressions are given in \cite{AK1411}.
In particular, the explicit form for the first order pole
is very crucial to determine the higher spin-$3$ current.

\section{The higher spin-$\frac{5}{2}$ currents using the different
tensor components}

When one looks at the equation (\ref{3halfandtwo}),
there are three cases of the higher spin-$\frac{5}{2}$
current for each $\rho$ index for given
indices $\mu=\nu$ (or for each $\nu$ index for given indices $\mu=\rho$).
Note that for nontrivial case, the index $\rho$
is not equal to $\mu=\nu$ (or the index $\nu$ is not equal to $\mu=\rho$).
For three different $\rho, \mu$ and $\nu$ cases,
there is no higher spin-$\frac{5}{2}$ current in the first order pole.

\subsection{ The higher spin-$\frac{5}{2}$ current
  $\Phi_{\frac{3}{2}}^{(1),1}$}
In (\ref{spin5halfthree}), one of the
expression for this higher spin-$\frac{5}{2}$ current
is described. Let us consider two other
expressions for the same higher spin-$\frac{5}{2}$ current as follows.
For example, one has the following OPE from Appendix $A$,
with $\mu=\rho=3$ and $\nu=1$ in $SO(4)$ basis,
{\footnotesize
  \bea
G^3(z) \, \Phi_{1}^{(1),13}(w) &=&
\frac{1}{(z-w)^2} \, \frac{(k-N)}{(k+N+2)} \,\Phi_{\frac{1}{2}}^{(1),1}(w)
+ \frac{1}{(z-w)} \Bigg[
\Phi_{\frac{3}{2}}^{(1),1} 
+\frac{2 i}{(k+N+2)} \, T^{23} \, \Phi_{\frac{1}{2}}^{(1),4}
\nonu \\
& + & \frac{2 i}{(k+N+2)}  T^{34}  \Phi_{\frac{1}{2}}^{(1),2}
-\frac{4 i}{(k+N+2)}  T^{24}  \Phi_{\frac{1}{2}}^{(1),3}
-  \frac{(N-k)}{3(k+N+2)}  \pa  \Phi_{\frac{1}{2}}^{(1),1}
\Bigg](w).
\label{313}
\eea}
From this expression of the first order pole,
one can write down the higher spin-$\frac{5}{2}$
current with the help of (\ref{TPhirel2}) as follows:
{\footnotesize
\bea
&& \Phi_{\frac{3}{2}}^{(1),1}(z)  = 
\frac{i}{(k+N+2)^2} \, \Bigg[ -\frac{1}{2N} \, h^1_{\bar{a} \bar{b}}\,
  d^2_{\bar{d} \bar{e}} +
\frac{1}{2N} \, h^3_{\bar{a} \bar{b}}\,
  d^0_{\bar{d} \bar{e}}
+ \frac{1}{N} \, h^2_{\bar{a} \bar{b}}\,
  d^1_{\bar{d} \bar{e}}
  \nonu \\
  &&  +  2  \, h^1_{\bar{b} \bar{d}}\,
  d^2_{\bar{a} \bar{e}} -  h^3_{\bar{b} \bar{e}}\,
  d^0_{\bar{a} \bar{d}} -  h^2_{\bar{b} \bar{d}}\,
  d^1_{\bar{a} \bar{e}} \Bigg] f^{\bar{a} \bar{b}}_{\,\,\,\,\,\,c}
\, Q^{\bar{d}} \, V^c \, V^{\bar{e}}
-\frac{4i(3+2k+N)}{3(k+N+2)^2} \, d^{3}_{\bar{a} \bar{b}} \, \pa
Q^{\bar{a}} \, V^{\bar{b}}(z)
\label{spin5halfthree-1}
 \\
& & + \frac{i}{(k+N+2)^3} \, \Bigg[ -d^0_{\bar{a} \bar{b}} \, h^3_{\bar{c}
    \bar{d}} - d^1_{\bar{c} \bar{d}} \, h^2_{\bar{a} \bar{b}} +
  d^2_{\bar{c} \bar{d}} \, h^1_{\bar{a} \bar{b}} + d^0_{\bar{c} \bar{d}}
  \, h^3_{\bar{a} \bar{b}}\Bigg] Q^{\bar{a}} \, Q^{\bar{b}} \, Q^{\bar{c}} \,
V^{\bar{d}}(z) 
 + \frac{4i(k+2N)}{3(k+N+2)^2} \, d^{3}_{\bar{a} \bar{b}} \, 
Q^{\bar{a}} \, \pa \, V^{\bar{b}}(z).
\nonu
\eea}
Compared to the previous result in (\ref{spin5halfthree}),
the difference arises in the cubic terms.

Furthermore, one can also consider the following OPE from Appendix $A$
with $\mu=\rho=4$ and $\nu=1$ in $SO(4)$ basis 
{\footnotesize
  \bea
G^4(z) \, \Phi_{1}^{(1),14}(w) &=&
\frac{1}{(z-w)^2} \, \frac{(k-N)}{(k+N+2)} \,\Phi_{\frac{1}{2}}^{(1),1}(w)
+ \frac{1}{(z-w)} \Bigg[
\Phi_{\frac{3}{2}}^{(1),1} 
-\frac{2 i}{(k+N+2)} \, T^{24} \, \Phi_{\frac{1}{2}}^{(1),3}
\nonu \\
& + & \frac{2 i}{(k+N+2)}  T^{34}  \Phi_{\frac{1}{2}}^{(1),2}
+\frac{4 i}{(k+N+2)}  T^{23}  \Phi_{\frac{1}{2}}^{(1),4}
-  \frac{(N-k)}{3(k+N+2)}  \pa  \Phi_{\frac{1}{2}}^{(1),1}
\Bigg](w).
\label{414}
\eea}
Similarly, 
the same higher spin-$\frac{5}{2}$
current with the help of (\ref{TPhirel2}) 
can be written as
{\footnotesize
\bea
&& \Phi_{\frac{3}{2}}^{(1),1}(z)  = 
\frac{i}{(k+N+2)^2} \, \Bigg[ \frac{1}{2N} \, h^2_{\bar{a} \bar{b}}\,
  d^1_{\bar{d} \bar{e}} +
\frac{1}{2N} \, h^3_{\bar{a} \bar{b}}\,
  d^0_{\bar{d} \bar{e}}
- \frac{1}{N} \, h^1_{\bar{a} \bar{b}}\,
  d^2_{\bar{d} \bar{e}}
  \nonu \\
  & & -  2  \, h^2_{\bar{b} \bar{d}}\,
  d^1_{\bar{a} \bar{e}} -  h^3_{\bar{b} \bar{e}}\,
  d^0_{\bar{a} \bar{d}} +  h^1_{\bar{b} \bar{d}}\,
  d^2_{\bar{a} \bar{e}} \Bigg] f^{\bar{a} \bar{b}}_{\,\,\,\,\,\,c}
\, Q^{\bar{d}} \, V^c \, V^{\bar{e}}
-\frac{4i(3+2k+N)}{3(k+N+2)^2} \, d^{3}_{\bar{a} \bar{b}} \, \pa
Q^{\bar{a}} \, V^{\bar{b}}(z)
\label{spin5halfthree-2}
\\
& & + \frac{i}{(k+N+2)^3} \, \Bigg[ -d^0_{\bar{a} \bar{b}} \, h^3_{\bar{c}
    \bar{d}} - d^1_{\bar{c} \bar{d}} \, h^2_{\bar{a} \bar{b}} +
  d^2_{\bar{c} \bar{d}} \, h^1_{\bar{a} \bar{b}} + d^0_{\bar{c} \bar{d}}
  \, h^3_{\bar{a} \bar{b}}\Bigg] Q^{\bar{a}} \, Q^{\bar{b}} \, Q^{\bar{c}} \,
V^{\bar{d}}(z) 
 + \frac{4i(k+2N)}{3(k+N+2)^2} \, d^{3}_{\bar{a} \bar{b}} \, 
Q^{\bar{a}} \, \pa \, V^{\bar{b}}(z).
\nonu
\eea}
Compared to the previous results in (\ref{spin5halfthree}) and
Appendix (\ref{spin5halfthree-1}),
the difference arises in the cubic terms.

Therefore, one has two relations between three identical results.
From (\ref{spin5halfthree}) and
Appendix (\ref{spin5halfthree-1}), one has
\bea
&& \Bigg[ -\frac{1}{2N} \, h^1_{\bar{a} \bar{b}}\,
  d^2_{\bar{d} \bar{e}} +
\frac{1}{2N} \, h^2_{\bar{a} \bar{b}}\,
  d^1_{\bar{d} \bar{e}}
+ \frac{1}{N} \, h^3_{\bar{a} \bar{b}}\,
  d^0_{\bar{d} \bar{e}}
 -  2  \, h^0_{\bar{b} \bar{d}}\,
  d^3_{\bar{a} \bar{e}} +  h^2_{\bar{b} \bar{e}}\,
  d^1_{\bar{a} \bar{d}} -  h^2_{\bar{b} \bar{d}}\,
  d^1_{\bar{a} \bar{e}} \Bigg] f^{\bar{a} \bar{b}}_{\,\,\,\,\,\,c}
=
\nonu \\
&& \Bigg[ -\frac{1}{2N} \, h^1_{\bar{a} \bar{b}}\,
  d^2_{\bar{d} \bar{e}} +
\frac{1}{2N} \, h^3_{\bar{a} \bar{b}}\,
  d^0_{\bar{d} \bar{e}}
+ \frac{1}{N} \, h^2_{\bar{a} \bar{b}}\,
  d^1_{\bar{d} \bar{e}}
 +  2  \, h^1_{\bar{b} \bar{d}}\,
  d^2_{\bar{a} \bar{e}} -  h^3_{\bar{b} \bar{e}}\,
  d^0_{\bar{a} \bar{d}} -  h^2_{\bar{b} \bar{d}}\,
  d^1_{\bar{a} \bar{e}} \Bigg] f^{\bar{a} \bar{b}}_{\,\,\,\,\,\,c}.
\nonu
\eea
This can be simplified further as
\bea
\Bigg[  -
\frac{1}{2N} \, h^2_{\bar{a} \bar{b}}\,
  d^1_{\bar{d} \bar{e}}
+ \frac{1}{2N} \, h^3_{\bar{a} \bar{b}}\,
  d^0_{\bar{d} \bar{e}}
 -   h^0_{\bar{b} \bar{d}}\,
  d^3_{\bar{a} \bar{e}}  -   h^1_{\bar{b} \bar{d}}\,
  d^2_{\bar{a} \bar{e}}  \Bigg] f^{\bar{a} \bar{b}}_{\,\,\,\,\,\,c}
=0,
\label{Iden1}
\eea
by using the identity
\bea
\Bigg[ h^{\mu}_{\bar{b} \bar{d}}\,
d^{\nu}_{\bar{a} \bar{e}} -
h^{\nu}_{\bar{b} \bar{e}}\,
d^{\mu}_{\bar{a} \bar{d}} \Bigg] f^{\bar{a} \bar{b}}_{\,\,\,\,\,\,c} =0.
\label{simpleidentity}
\eea
This can be checked for several $N$ values.

From (\ref{spin5halfthree}) and Appendix
(\ref{spin5halfthree-2}), one has
\bea
&& \Bigg[ -\frac{1}{2N} \, h^1_{\bar{a} \bar{b}}\,
  d^2_{\bar{d} \bar{e}} +
\frac{1}{2N} \, h^2_{\bar{a} \bar{b}}\,
  d^1_{\bar{d} \bar{e}}
+ \frac{1}{N} \, h^3_{\bar{a} \bar{b}}\,
  d^0_{\bar{d} \bar{e}}
 -  2  \, h^0_{\bar{b} \bar{d}}\,
  d^3_{\bar{a} \bar{e}} +  h^2_{\bar{b} \bar{e}}\,
  d^1_{\bar{a} \bar{d}} -  h^2_{\bar{b} \bar{d}}\,
  d^1_{\bar{a} \bar{e}} \Bigg] f^{\bar{a} \bar{b}}_{\,\,\,\,\,\,c}
=
\nonu \\
&& \Bigg[ \frac{1}{2N} \, h^2_{\bar{a} \bar{b}}\,
  d^1_{\bar{d} \bar{e}} +
\frac{1}{2N} \, h^3_{\bar{a} \bar{b}}\,
  d^0_{\bar{d} \bar{e}}
- \frac{1}{N} \, h^1_{\bar{a} \bar{b}}\,
  d^2_{\bar{d} \bar{e}}
   -  2  \, h^2_{\bar{b} \bar{d}}\,
  d^1_{\bar{a} \bar{e}} -  h^3_{\bar{b} \bar{e}}\,
  d^0_{\bar{a} \bar{d}} +  h^1_{\bar{b} \bar{d}}\,
  d^2_{\bar{a} \bar{e}} \Bigg] f^{\bar{a} \bar{b}}_{\,\,\,\,\,\,c}.
\nonu
\eea
Moreover, one obtains with the help of Appendix (\ref{simpleidentity})
\bea
\Bigg[ \frac{1}{2N} \, h^1_{\bar{a} \bar{b}}\,
  d^2_{\bar{d} \bar{e}} 
+ \frac{1}{2N} \, h^3_{\bar{a} \bar{b}}\,
  d^0_{\bar{d} \bar{e}}
 -   h^0_{\bar{b} \bar{d}}\,
  d^3_{\bar{a} \bar{e}}  +  h^2_{\bar{b} \bar{d}}\,
  d^1_{\bar{a} \bar{e}}
  \Bigg] f^{\bar{a} \bar{b}}_{\,\,\,\,\,\,c}
=0.
\label{Iden2}
\eea
The identities Appendices (\ref{Iden1}) and (\ref{Iden2})
are nontrivial relations which will be useful to check other relations
appearing in Appendix $A$. 
It would be interesting to prove these identities in general without using
the above defining relations. 

\subsection{ The higher spin-$\frac{5}{2}$ current
  $\Phi_{\frac{3}{2}}^{(1),2}$}

In (\ref{spin5halffour}), one of the
expression for this higher spin-$\frac{5}{2}$ current
is found. Let us consider two other
expressions for the same higher spin-$\frac{5}{2}$ current as follows.
Let us consider $\mu=4=\rho$ and $\nu=2$ in $SO(4)$ basis
and the OPE from Appendix $A$ can be written as
{\footnotesize
  \bea
{G}^4(z) \, \Phi_{1}^{(1),24}(w) &=&
\frac{1}{(z-w)^2} \, \frac{(k-N)}{(k+N+2)} \,\Phi_{\frac{1}{2}}^{(1),2}(w)
+ \frac{1}{(z-w)} \Bigg[
\Phi_{\frac{3}{2}}^{(1),2} 
+\frac{2 i}{(k+N+2)} \, T^{14} \, \Phi_{\frac{1}{2}}^{(1),3}
\nonu \\
& - & \frac{2 i}{(k+N+2)}  T^{34}  \Phi_{\frac{1}{2}}^{(1),1}
-\frac{4 i}{(k+N+2)}  T^{13}  \Phi_{\frac{1}{2}}^{(1),4}
-  \frac{(N-k)}{3(k+N+2)}  \pa  \Phi_{\frac{1}{2}}^{(1),2}
\Bigg](w).
\label{424}
\eea}
From the first order pole,
after substituting the relations in (\ref{TPhirel3}), the
higher spin-$\frac{5}{2}$ current can be described as
{\footnotesize
  \bea
&& \Phi_{\frac{3}{2}}^{(1),2}(z)  = 
\frac{i}{(k+N+2)^2} \, \Bigg[ -\frac{1}{2N} \, h^1_{\bar{a} \bar{b}}\,
  d^1_{\bar{d} \bar{e}} -
\frac{1}{2N} \, h^3_{\bar{a} \bar{b}}\,
  d^3_{\bar{d} \bar{e}}
- \frac{1}{N} \, h^2_{\bar{a} \bar{b}}\,
  d^2_{\bar{d} \bar{e}}
  \nonu \\
  && -  2  \, h^2_{\bar{b} \bar{d}}\,
  d^2_{\bar{a} \bar{e}} -  h^3_{\bar{b} \bar{d}}\,
  d^3_{\bar{a} \bar{e}} -  h^1_{\bar{b} \bar{d}}\,
  d^1_{\bar{a} \bar{e}} \Bigg] f^{\bar{a} \bar{b}}_{\,\,\,\,\,\,c}
\, Q^{\bar{d}} \, V^c \, V^{\bar{e}}
-\frac{4i(3+2k+N)}{3(k+N+2)^2} \, d^{0}_{\bar{a} \bar{b}} \, \pa
Q^{\bar{a}} \, V^{\bar{b}}(z)
\label{spin5halffour-1}
\\
&& + \frac{i}{(k+N+2)^3} \, \Bigg[ -d^3_{\bar{c} \bar{d}} \, h^3_{\bar{a}
    \bar{b}} - d^2_{\bar{c} \bar{d}} \, h^2_{\bar{a} \bar{b}} -
  d^0_{\bar{a} \bar{b}} \, h^0_{\bar{c} \bar{d}} - 
  \, d^1_{\bar{c} \bar{d}} h^1_{\bar{a} \bar{b}}
  \Bigg] Q^{\bar{a}} \, Q^{\bar{b}} \, Q^{\bar{c}} \,
V^{\bar{d}}(z) 
 + \frac{4i(k+2N)}{3(k+N+2)^2} \, d^{0}_{\bar{a} \bar{b}} \, 
Q^{\bar{a}} \, \pa \, V^{\bar{b}}(z).
\nonu
\eea}
Compared to the previous results in (\ref{spin5halffour}) and
Appendix (\ref{spin5halffour-1}),
the difference arises in the cubic terms as before.
Therefore, one has a relation between two identical results.
From (\ref{spin5halffour}) and Appendix
(\ref{spin5halffour-1}), one has
\bea
&& \Bigg[ -\frac{1}{2N} \, h^2_{\bar{a} \bar{b}}\,
  d^2_{\bar{d} \bar{e}} -
\frac{1}{2N} \, h^3_{\bar{a} \bar{b}}\,
  d^3_{\bar{d} \bar{e}}
- \frac{1}{N} \, h^1_{\bar{a} \bar{b}}\,
  d^1_{\bar{d} \bar{e}}
-  2  \, h^1_{\bar{b} \bar{d}}\,
  d^1_{\bar{a} \bar{e}} -  h^3_{\bar{b} \bar{d}}\,
  d^3_{\bar{a} \bar{e}} -  h^2_{\bar{b} \bar{d}}\,
  d^2_{\bar{a} \bar{e}} \Bigg] f^{\bar{a} \bar{b}}_{\,\,\,\,\,\,c}=
\nonu \\
&& \Bigg[ -\frac{1}{2N} \, h^1_{\bar{a} \bar{b}}\,
  d^1_{\bar{d} \bar{e}} -
\frac{1}{2N} \, h^3_{\bar{a} \bar{b}}\,
  d^3_{\bar{d} \bar{e}}
- \frac{1}{N} \, h^2_{\bar{a} \bar{b}}\,
  d^2_{\bar{d} \bar{e}}
  -  2  \, h^2_{\bar{b} \bar{d}}\,
  d^2_{\bar{a} \bar{e}} -  h^3_{\bar{b} \bar{d}}\,
  d^3_{\bar{a} \bar{e}} -  h^1_{\bar{b} \bar{d}}\,
  d^1_{\bar{a} \bar{e}} \Bigg] f^{\bar{a} \bar{b}}_{\,\,\,\,\,\,c}.
\nonu
\eea
This leads to
\bea
\Bigg[ \frac{1}{2N} \, h^2_{\bar{a} \bar{b}}\,
  d^2_{\bar{d} \bar{e}} 
- \frac{1}{2N} \, h^1_{\bar{a} \bar{b}}\,
d^1_{\bar{d} \bar{e}}
-     h^1_{\bar{b} \bar{d}}\,
  d^1_{\bar{a} \bar{e}}  +  h^2_{\bar{b} \bar{d}}\,
  d^2_{\bar{a} \bar{e}} 
\Bigg] f^{\bar{a} \bar{b}}_{\,\,\,\,\,\,c} =0,
\label{Iiden1}
\eea
which was observed in (\ref{1122}) in different context.

Furthermore, one can also consider the following OPE
with $\mu=\nu=1$ and $\rho=2$ in $SO(4)$ basis
{\footnotesize
  \bea
{G}^1(z) \, \Phi_{1}^{(1),12}(w) &=&
-\frac{1}{(z-w)^2} \, \frac{(k-N)}{(k+N+2)} \,\Phi_{\frac{1}{2}}^{(1),2}(w)
+ \frac{1}{(z-w)} \Bigg[
-\Phi_{\frac{3}{2}}^{(1),2} 
+\frac{2 i}{(k+N+2)} \, T^{13} \, \Phi_{\frac{1}{2}}^{(1),4}
\nonu \\
& - & \frac{2 i}{(k+N+2)}  T^{14}  \Phi_{\frac{1}{2}}^{(1),3}
+\frac{4 i}{(k+N+2)}  T^{34}  \Phi_{\frac{1}{2}}^{(1),1}
+  \frac{(N-k)}{3(k+N+2)}  \pa  \Phi_{\frac{1}{2}}^{(1),2}
\Bigg](w).
\label{112}
\eea}
Similarly, from the first order pole,
the higher spin-$\frac{5}{2}$
current with the help of (\ref{TPhirel3}) 
can be written as
{\footnotesize
\bea
&& \Phi_{\frac{3}{2}}^{(1),2}(z)  = 
\frac{i}{(k+N+2)^2} \, \Bigg[ -\frac{1}{2N} \, h^2_{\bar{a} \bar{b}}\,
  d^2_{\bar{d} \bar{e}} -
\frac{1}{2N} \, h^1_{\bar{a} \bar{b}}\,
  d^1_{\bar{d} \bar{e}}
- \frac{1}{N} \, h^3_{\bar{a} \bar{b}}\,
  d^3_{\bar{d} \bar{e}}
  \nonu \\
  && -  2  \, h^3_{\bar{b} \bar{d}}\,
  d^3_{\bar{a} \bar{e}} -  h^2_{\bar{b} \bar{d}}\,
  d^2_{\bar{a} \bar{e}} -  h^1_{\bar{b} \bar{d}}\,
  d^1_{\bar{a} \bar{e}} \Bigg] f^{\bar{a} \bar{b}}_{\,\,\,\,\,\,c}
\, Q^{\bar{d}} \, V^c \, V^{\bar{e}}
-\frac{4i(3+2k+N)}{3(k+N+2)^2} \, d^{0}_{\bar{a} \bar{b}} \, \pa
Q^{\bar{a}} \, V^{\bar{b}}(z)
\label{spin5halffour-2}
\\
&& + \frac{i}{(k+N+2)^3} \, \Bigg[ -d^3_{\bar{c} \bar{d}} \, h^3_{\bar{a}
    \bar{b}} - d^2_{\bar{c} \bar{d}} \, h^2_{\bar{a} \bar{b}} -
  d^0_{\bar{a} \bar{b}} \, h^0_{\bar{c} \bar{d}} - 
  \, d^1_{\bar{c} \bar{d}} h^1_{\bar{a} \bar{b}}
  \Bigg] Q^{\bar{a}} \, Q^{\bar{b}} \, Q^{\bar{c}} \,
V^{\bar{d}}(z) 
+ \frac{4i(k+2N)}{3(k+N+2)^2} \, d^{0}_{\bar{a} \bar{b}} \, 
Q^{\bar{a}} \, \pa \, V^{\bar{b}}(z).
\nonu
\eea}
Compared to the previous results in (\ref{spin5halffour}) and
Appendix (\ref{spin5halffour-2}),
the difference arises in the cubic terms.
From (\ref{spin5halffour}) and Appendix
(\ref{spin5halffour-2}), one has
\bea
&& \Bigg[ -\frac{1}{2N} \, h^2_{\bar{a} \bar{b}}\,
  d^2_{\bar{d} \bar{e}} -
\frac{1}{2N} \, h^3_{\bar{a} \bar{b}}\,
  d^3_{\bar{d} \bar{e}}
- \frac{1}{N} \, h^1_{\bar{a} \bar{b}}\,
  d^1_{\bar{d} \bar{e}}
-  2  \, h^1_{\bar{b} \bar{d}}\,
  d^1_{\bar{a} \bar{e}} -  h^3_{\bar{b} \bar{d}}\,
  d^3_{\bar{a} \bar{e}} -  h^2_{\bar{b} \bar{d}}\,
  d^2_{\bar{a} \bar{e}} \Bigg] f^{\bar{a} \bar{b}}_{\,\,\,\,\,\,c}=
\nonu \\
&& \Bigg[ -\frac{1}{2N} \, h^2_{\bar{a} \bar{b}}\,
  d^2_{\bar{d} \bar{e}} -
\frac{1}{2N} \, h^1_{\bar{a} \bar{b}}\,
  d^1_{\bar{d} \bar{e}}
- \frac{1}{N} \, h^3_{\bar{a} \bar{b}}\,
  d^3_{\bar{d} \bar{e}}
 -  2  \, h^3_{\bar{b} \bar{d}}\,
  d^3_{\bar{a} \bar{e}} -  h^2_{\bar{b} \bar{d}}\,
  d^2_{\bar{a} \bar{e}} -  h^1_{\bar{b} \bar{d}}\,
  d^1_{\bar{a} \bar{e}} \Bigg] f^{\bar{a} \bar{b}}_{\,\,\,\,\,\,c}.
\nonu
\eea
Then one has
\bea
\Bigg[ 
\frac{1}{2N} \, h^3_{\bar{a} \bar{b}}\,
  d^3_{\bar{d} \bar{e}}
- \frac{1}{2N} \, h^1_{\bar{a} \bar{b}}\,
  d^1_{\bar{d} \bar{e}}
-   h^1_{\bar{b} \bar{d}}\,
  d^1_{\bar{a} \bar{e}} +  h^3_{\bar{b} \bar{d}}\,
  d^3_{\bar{a} \bar{e}}  \Bigg] f^{\bar{a} \bar{b}}_{\,\,\,\,\,\,c}=0,
\label{Iiden2}
\eea
which was observed in (\ref{1133}) in section $3$.
It would be interesting to prove these identities
Appendices
(\ref{Iiden1}) and (\ref{Iiden2}) from the group theoretical context.

\subsection{ The higher spin-$\frac{5}{2}$ current
  $\Phi_{\frac{3}{2}}^{(1),3}$}

In (\ref{spin5halfone}), one of the
expression for this higher spin-$\frac{5}{2}$ current
is obtained. Let us consider two other
expressions for the same higher spin-$\frac{5}{2}$ current as follows.
Let us consider $\mu=4=\rho$ and $\nu=3$ in $SO(4)$ basis
and the OPE from Appendix $A$ can be written as
{\footnotesize
  \bea
&& {G}^4(z) \, \Phi_{1}^{(1),34}(w) =
\frac{1}{(z-w)^2} \, \frac{(k-N)}{(k+N+2)} \,\Phi_{\frac{1}{2}}^{(1),3}(w)
+ \frac{1}{(z-w)} \Bigg[
\Phi_{\frac{3}{2}}^{(1),3} 
-\frac{2 i}{(k+N+2)} \, T^{14} \, \Phi_{\frac{1}{2}}^{(1),2}
\nonu \\
&& +  \frac{2 i}{(k+N+2)}  T^{24}  \Phi_{\frac{1}{2}}^{(1),1}
+\frac{4 i}{(k+N+2)}  T^{12}  \Phi_{\frac{1}{2}}^{(1),4}
-  \frac{(N-k)}{3(k+N+2)}  \pa  \Phi_{\frac{1}{2}}^{(1),3}
\Bigg](w).
\label{434}
\eea}
After substituting the relations in (\ref{TPhirel}), the
higher spin-$\frac{5}{2}$ current can be described as
{\footnotesize
\bea
&& \Phi_{\frac{3}{2}}^{(1),3}(z)  = 
\frac{i}{(k+N+2)^2} \, \Bigg[ -\frac{1}{2N} \, h^2_{\bar{a} \bar{b}}\,
  d^3_{\bar{d} \bar{e}} +
\frac{1}{2N} \, h^1_{\bar{a} \bar{b}}\,
  d^0_{\bar{d} \bar{e}}
+ \frac{1}{N} \, h^3_{\bar{a} \bar{b}}\,
  d^2_{\bar{d} \bar{e}}
  \nonu \\
  && +  2  \, h^2_{\bar{b} \bar{d}}\,
  d^3_{\bar{a} \bar{e}} -  h^2_{\bar{b} \bar{e}}\,
  d^3_{\bar{a} \bar{d}} -  h^0_{\bar{b} \bar{d}}\,
  d^1_{\bar{a} \bar{e}} \Bigg] f^{\bar{a} \bar{b}}_{\,\,\,\,\,\,c}
\, Q^{\bar{d}} \, V^c \, V^{\bar{e}}
-\frac{4i(3+2k+N)}{3(k+N+2)^2} \, d^{1}_{\bar{a} \bar{b}} \, \pa
Q^{\bar{a}} \, V^{\bar{b}}(z)
\label{spin5halfone-1}
\\
&& + \frac{i}{(k+N+2)^3} \, \Bigg[ d^2_{\bar{b} \bar{d}} \, h^3_{\bar{a}
    \bar{c}} + d^3_{\bar{a} \bar{d}} \, h^2_{\bar{b} \bar{c}} -
  d^0_{\bar{a} \bar{b}} \, h^1_{\bar{c} \bar{d}} + d^0_{\bar{c} \bar{d}}
  \, h^1_{\bar{a} \bar{b}}\Bigg] Q^{\bar{a}} \, Q^{\bar{b}} \, Q^{\bar{c}} \,
V^{\bar{d}}(z) 
+ \frac{4i(k+2N)}{3(k+N+2)^2} \, d^{1}_{\bar{a} \bar{b}} \, 
Q^{\bar{a}} \, \pa \, V^{\bar{b}}(z).
\nonu
\eea}
Compared to the previous results in (\ref{spin5halfone}) and
Appendix (\ref{spin5halfone-1}),
the difference arises in the cubic terms as before.
From (\ref{spin5halfone}) and Appendix
(\ref{spin5halfone-1}), one has
\bea
\Bigg[ \frac{1}{2N} \, h^3_{\bar{a} \bar{b}}\,
  d^2_{\bar{d} \bar{e}} -
\frac{1}{2N} \, h^2_{\bar{a} \bar{b}}\,
  d^3_{\bar{d} \bar{e}}
+ \frac{1}{N} \, h^1_{\bar{a} \bar{b}}\,
  d^0_{\bar{d} \bar{e}}
  -  2  \, h^0_{\bar{b} \bar{d}}\,
  d^1_{\bar{a} \bar{e}} -  h^2_{\bar{b} \bar{e}}\,
  d^3_{\bar{a} \bar{d}} +  h^2_{\bar{b} \bar{d}}\,
  d^3_{\bar{a} \bar{e}} \Bigg] f^{\bar{a} \bar{b}}_{\,\,\,\,\,\,c}=
\nonu \\
\Bigg[ -\frac{1}{2N} \, h^2_{\bar{a} \bar{b}}\,
  d^3_{\bar{d} \bar{e}} +
\frac{1}{2N} \, h^1_{\bar{a} \bar{b}}\,
  d^0_{\bar{d} \bar{e}}
+ \frac{1}{N} \, h^3_{\bar{a} \bar{b}}\,
  d^2_{\bar{d} \bar{e}}
 +  2  \, h^2_{\bar{b} \bar{d}}\,
  d^3_{\bar{a} \bar{e}} -  h^2_{\bar{b} \bar{e}}\,
  d^3_{\bar{a} \bar{d}} -  h^0_{\bar{b} \bar{d}}\,
  d^1_{\bar{a} \bar{e}} \Bigg] f^{\bar{a} \bar{b}}_{\,\,\,\,\,\,c}.
\nonu
\eea
Then it is easy to see that
\bea
\Bigg[ -\frac{1}{2N} \, h^3_{\bar{a} \bar{b}}\,
  d^2_{\bar{d} \bar{e}} 
+ \frac{1}{2N} \, h^1_{\bar{a} \bar{b}}\,
  d^0_{\bar{d} \bar{e}}
  -   h^0_{\bar{b} \bar{d}}\,
  d^1_{\bar{a} \bar{e}}  -  h^2_{\bar{b} \bar{d}}\,
  d^3_{\bar{a} \bar{e}} \Bigg] f^{\bar{a} \bar{b}}_{\,\,\,\,\,\,c}=0.
\label{iden1}
\eea

Let us consider $\mu=1=\nu$ and $\rho=3$ in $SO(4)$ basis
and the OPE from Appendix $A$ can be written as
{\footnotesize
  \bea
&& {G}^1(z) \, \Phi_{1}^{(1),13}(w) =
\frac{1}{(z-w)^2} \, \frac{(k-N)}{(k+N+2)} \,\Phi_{\frac{1}{2}}^{(1),3}(w)
+ \frac{1}{(z-w)} \Bigg[
-\Phi_{\frac{3}{2}}^{(1),3} 
-\frac{2 i}{(k+N+2)} \, T^{12} \, \Phi_{\frac{1}{2}}^{(1),4}
\nonu \\
&& +  \frac{2 i}{(k+N+2)}  T^{14}  \Phi_{\frac{1}{2}}^{(1),2}
-\frac{4 i}{(k+N+2)}  T^{24}  \Phi_{\frac{1}{2}}^{(1),1}
+  \frac{(N-k)}{3(k+N+2)}  \pa  \Phi_{\frac{1}{2}}^{(1),3}
\Bigg](w).
\label{113}
\eea}
Again, after substituting the relations in (\ref{TPhirel}), the
higher spin-$\frac{5}{2}$ current can be described as
{\footnotesize
\bea
&& \Phi_{\frac{3}{2}}^{(1),3}(z)  = 
\frac{i}{(k+N+2)^2} \, \Bigg[ \frac{1}{2N} \, h^3_{\bar{a} \bar{b}}\,
  d^2_{\bar{d} \bar{e}} +
\frac{1}{2N} \, h^1_{\bar{a} \bar{b}}\,
  d^0_{\bar{d} \bar{e}}
- \frac{1}{N} \, h^2_{\bar{a} \bar{b}}\,
  d^3_{\bar{d} \bar{e}}
  \nonu \\
  && -  2  \, h^3_{\bar{b} \bar{d}}\,
  d^2_{\bar{a} \bar{e}} +  h^3_{\bar{b} \bar{e}}\,
  d^2_{\bar{a} \bar{d}} -  h^0_{\bar{b} \bar{d}}\,
  d^1_{\bar{a} \bar{e}} \Bigg] f^{\bar{a} \bar{b}}_{\,\,\,\,\,\,c}
\, Q^{\bar{d}} \, V^c \, V^{\bar{e}}
-\frac{4i(3+2k+N)}{3(k+N+2)^2} \, d^{1}_{\bar{a} \bar{b}} \, \pa
Q^{\bar{a}} \, V^{\bar{b}}(z)
\label{spin5halfone-2}
\\
&& + \frac{i}{(k+N+2)^3} \, \Bigg[ d^2_{\bar{b} \bar{d}} \, h^3_{\bar{a}
    \bar{c}} + d^3_{\bar{a} \bar{d}} \, h^2_{\bar{b} \bar{c}} -
  d^0_{\bar{a} \bar{b}} \, h^1_{\bar{c} \bar{d}} + d^0_{\bar{c} \bar{d}}
  \, h^1_{\bar{a} \bar{b}}\Bigg] Q^{\bar{a}} \, Q^{\bar{b}} \, Q^{\bar{c}} \,
V^{\bar{d}}(z) 
+ \frac{4i(k+2N)}{3(k+N+2)^2} \, d^{1}_{\bar{a} \bar{b}} \, 
Q^{\bar{a}} \, \pa \, V^{\bar{b}}(z).
\nonu
\eea}
Compared to the previous results in (\ref{spin5halfone}) and
Appendix (\ref{spin5halfone-2}),
the difference arises in the cubic terms.
From (\ref{spin5halfone}) and Appendix
(\ref{spin5halfone-2}), one has
\bea
&& \Bigg[ \frac{1}{2N} \, h^3_{\bar{a} \bar{b}}\,
  d^2_{\bar{d} \bar{e}} -
\frac{1}{2N} \, h^2_{\bar{a} \bar{b}}\,
  d^3_{\bar{d} \bar{e}}
+ \frac{1}{N} \, h^1_{\bar{a} \bar{b}}\,
  d^0_{\bar{d} \bar{e}}
  -  2  \, h^0_{\bar{b} \bar{d}}\,
  d^1_{\bar{a} \bar{e}} -  h^2_{\bar{b} \bar{e}}\,
  d^3_{\bar{a} \bar{d}} +  h^2_{\bar{b} \bar{d}}\,
  d^3_{\bar{a} \bar{e}} \Bigg] f^{\bar{a} \bar{b}}_{\,\,\,\,\,\,c}=
\nonu \\
&&
\Bigg[ \frac{1}{2N} \, h^3_{\bar{a} \bar{b}}\,
  d^2_{\bar{d} \bar{e}} +
\frac{1}{2N} \, h^1_{\bar{a} \bar{b}}\,
  d^0_{\bar{d} \bar{e}}
- \frac{1}{N} \, h^2_{\bar{a} \bar{b}}\,
  d^3_{\bar{d} \bar{e}}
  -  2  \, h^3_{\bar{b} \bar{d}}\,
  d^2_{\bar{a} \bar{e}} +  h^3_{\bar{b} \bar{e}}\,
  d^2_{\bar{a} \bar{d}} -  h^0_{\bar{b} \bar{d}}\,
  d^1_{\bar{a} \bar{e}} \Bigg] f^{\bar{a} \bar{b}}_{\,\,\,\,\,\,c}.
\nonu
\eea
Moreover, one has from Appendix (\ref{simpleidentity})
\bea
\Bigg[ 
\frac{1}{2N} \, h^2_{\bar{a} \bar{b}}\,
  d^3_{\bar{d} \bar{e}}
+ \frac{1}{2N} \, h^1_{\bar{a} \bar{b}}\,
  d^0_{\bar{d} \bar{e}}
  -   h^0_{\bar{b} \bar{d}}\,
  d^1_{\bar{a} \bar{e}}
   +   h^3_{\bar{b} \bar{d}}\,
  d^2_{\bar{a} \bar{e}}
  \Bigg] f^{\bar{a} \bar{b}}_{\,\,\,\,\,\,c}=0.
\label{iden2}
\eea
It would be interesting to see where one can use these two identities
Appendices (\ref{iden1}) and (\ref{iden2}) in the various OPEs.

\subsection{ The higher spin-$\frac{5}{2}$ current
  $\Phi_{\frac{3}{2}}^{(1),4}$}

In (\ref{spin5halftwo}), one of the
expression for this higher spin-$\frac{5}{2}$ current
is determined. Let us consider two other
expressions for the same higher spin-$\frac{5}{2}$ current as follows.
Let us consider $\mu=1=\nu$ and $\rho=4$ in $SO(4)$ basis
and the OPE from Appendix $A$ can be written as
{\footnotesize
  \bea
{G}^1(z) \, \Phi_{1}^{(1),14}(w)  & = &
-\frac{1}{(z-w)^2} \, \frac{(k-N)}{(k+N+2)} \,\Phi_{\frac{1}{2}}^{(1),4}(w)
+ \frac{1}{(z-w)} \Bigg[
-\Phi_{\frac{3}{2}}^{(1),4} 
+\frac{2 i}{(k+N+2)} \, T^{12} \, \Phi_{\frac{1}{2}}^{(1),3}
\nonu \\
& - &  \frac{2 i}{(k+N+2)}  T^{13}  \Phi_{\frac{1}{2}}^{(1),2}
+\frac{4 i}{(k+N+2)}  T^{23}  \Phi_{\frac{1}{2}}^{(1),1}
+  \frac{(N-k)}{3(k+N+2)}  \pa  \Phi_{\frac{1}{2}}^{(1),4}
\Bigg](w).
\label{114}
\eea}
Again, after substituting the relations in (\ref{TPhirel1}), the
higher spin-$\frac{5}{2}$ current can be described as
{\footnotesize
\bea
&& \Phi_{\frac{3}{2}}^{(1),4}(z)  = 
\frac{i}{(k+N+2)^2} \, \Bigg[ \frac{1}{2N} \, h^3_{\bar{a} \bar{b}}\,
  d^1_{\bar{d} \bar{e}} -
\frac{1}{2N} \, h^2_{\bar{a} \bar{b}}\,
  d^0_{\bar{d} \bar{e}}
- \frac{1}{N} \, h^1_{\bar{a} \bar{b}}\,
  d^3_{\bar{d} \bar{e}}
  \nonu \\
  & & -  2  \, h^3_{\bar{b} \bar{d}}\,
  d^1_{\bar{a} \bar{e}} +  h^2_{\bar{b} \bar{e}}\,
  d^0_{\bar{a} \bar{d}} +  h^1_{\bar{b} \bar{d}}\,
  d^3_{\bar{a} \bar{e}} \Bigg] f^{\bar{a} \bar{b}}_{\,\,\,\,\,\,c}
\, Q^{\bar{d}} \, V^c \, V^{\bar{e}}
+\frac{4i(3+2k+N)}{3(k+N+2)^2} \, d^{2}_{\bar{a} \bar{b}} \, \pa
Q^{\bar{a}} \, V^{\bar{b}}(z)
\label{spin5halftwo-1}
\\
& & + \frac{i}{(k+N+2)^3} \, \Bigg[ -d^1_{\bar{c} \bar{d}} \, h^3_{\bar{a}
    \bar{b}} + d^0_{\bar{a} \bar{b}} \, h^2_{\bar{c} \bar{d}} -
  d^0_{\bar{a} \bar{d}} \, h^2_{\bar{b} \bar{c}} + d^3_{\bar{c} \bar{d}}
  \, h^1_{\bar{a} \bar{b}}\Bigg] Q^{\bar{a}} \, Q^{\bar{b}} \, Q^{\bar{c}} \,
V^{\bar{d}}(z) 
- \frac{4i(k+2N)}{3(k+N+2)^2} \, d^{2}_{\bar{a} \bar{b}} \, 
Q^{\bar{a}} \, \pa \, V^{\bar{b}}(z).
\nonu
\eea}
Compared to the previous results in (\ref{spin5halftwo}) and
Appendix (\ref{spin5halftwo-1}),
the difference arises in the cubic terms as before.
From (\ref{spin5halftwo}) and
Appendix 
(\ref{spin5halftwo-1}), one has
\bea
&& \Bigg[ \frac{1}{2N} \, h^3_{\bar{a} \bar{b}}\,
  d^1_{\bar{d} \bar{e}} -
\frac{1}{2N} \, h^1_{\bar{a} \bar{b}}\,
  d^3_{\bar{d} \bar{e}}
- \frac{1}{N} \, h^2_{\bar{a} \bar{b}}\,
  d^0_{\bar{d} \bar{e}}
  +  2  \, h^0_{\bar{b} \bar{d}}\,
  d^2_{\bar{a} \bar{e}} -  h^1_{\bar{b} \bar{e}}\,
  d^3_{\bar{a} \bar{d}} +  h^1_{\bar{b} \bar{d}}\,
  d^3_{\bar{a} \bar{e}} \Bigg] f^{\bar{a} \bar{b}}_{\,\,\,\,\,\,c}=
\nonu \\
&& \Bigg[ \frac{1}{2N} \, h^3_{\bar{a} \bar{b}}\,
  d^1_{\bar{d} \bar{e}} -
\frac{1}{2N} \, h^2_{\bar{a} \bar{b}}\,
  d^0_{\bar{d} \bar{e}}
- \frac{1}{N} \, h^1_{\bar{a} \bar{b}}\,
  d^3_{\bar{d} \bar{e}}
  -  2  \, h^3_{\bar{b} \bar{d}}\,
  d^1_{\bar{a} \bar{e}} +  h^2_{\bar{b} \bar{e}}\,
  d^0_{\bar{a} \bar{d}} +  h^1_{\bar{b} \bar{d}}\,
  d^3_{\bar{a} \bar{e}} \Bigg] f^{\bar{a} \bar{b}}_{\,\,\,\,\,\,c}.
\nonu
\eea
Furthermore, one has by using Appendix (\ref{simpleidentity})
\bea
\Bigg[ 
\frac{1}{2N} \, h^1_{\bar{a} \bar{b}}\,
  d^3_{\bar{d} \bar{e}}
- \frac{1}{2N} \, h^2_{\bar{a} \bar{b}}\,
  d^0_{\bar{d} \bar{e}}
  +  h^0_{\bar{b} \bar{d}}\,
  d^2_{\bar{a} \bar{e}}
 +   h^3_{\bar{b} \bar{d}}\,
  d^1_{\bar{a} \bar{e}} 
  \Bigg] f^{\bar{a} \bar{b}}_{\,\,\,\,\,\,c}=0.
\label{Ide1}
\eea

Let us consider $\mu=3=\nu$ and $\rho=4$ in $SO(4)$ basis
and the OPE from Appendix $A$ can be written as
{\footnotesize
  \bea
{G}^3(z) \, \Phi_{1}^{(1),34}(w) &=&
-\frac{1}{(z-w)^2} \, \frac{(k-N)}{(k+N+2)} \,\Phi_{\frac{1}{2}}^{(1),4}(w)
+ \frac{1}{(z-w)} \Bigg[
-\Phi_{\frac{3}{2}}^{(1),4} 
-\frac{2 i}{(k+N+2)} \, T^{13} \, \Phi_{\frac{1}{2}}^{(1),2}
\nonu \\
& + & \frac{2 i}{(k+N+2)}  T^{23}  \Phi_{\frac{1}{2}}^{(1),1}
+\frac{4 i}{(k+N+2)}  T^{12}  \Phi_{\frac{1}{2}}^{(1),3}
+  \frac{(N-k)}{3(k+N+2)}  \pa  \Phi_{\frac{1}{2}}^{(1),4}
\Bigg](w).
\label{334}
\eea}
Again, after substituting the relations in (\ref{TPhirel1}), the
higher spin-$\frac{5}{2}$ current can be described as
{\footnotesize
\bea
&& \Phi_{\frac{3}{2}}^{(1),4}(z)  = 
\frac{i}{(k+N+2)^2} \, \Bigg[ -\frac{1}{2N} \, h^2_{\bar{a} \bar{b}}\,
  d^0_{\bar{d} \bar{e}} -
\frac{1}{2N} \, h^1_{\bar{a} \bar{b}}\,
  d^3_{\bar{d} \bar{e}}
+ \frac{1}{N} \, h^3_{\bar{a} \bar{b}}\,
  d^1_{\bar{d} \bar{e}}
  \nonu \\
  && +  2  \, h^1_{\bar{b} \bar{d}}\,
  d^3_{\bar{a} \bar{e}} +  h^2_{\bar{b} \bar{e}}\,
  d^0_{\bar{a} \bar{d}} -  h^3_{\bar{b} \bar{d}}\,
  d^1_{\bar{a} \bar{e}} \Bigg] f^{\bar{a} \bar{b}}_{\,\,\,\,\,\,c}
\, Q^{\bar{d}} \, V^c \, V^{\bar{e}}
+\frac{4i(3+2k+N)}{3(k+N+2)^2} \, d^{2}_{\bar{a} \bar{b}} \, \pa
Q^{\bar{a}} \, V^{\bar{b}}(z)
\label{spin5halftwo-2}
\\
&& + \frac{i}{(k+N+2)^3} \, \Bigg[ -d^1_{\bar{c} \bar{d}} \, h^3_{\bar{a}
    \bar{b}} + d^0_{\bar{a} \bar{b}} \, h^2_{\bar{c} \bar{d}} -
  d^0_{\bar{a} \bar{d}} \, h^2_{\bar{b} \bar{c}} + d^3_{\bar{c} \bar{d}}
  \, h^1_{\bar{a} \bar{b}}\Bigg] Q^{\bar{a}} \, Q^{\bar{b}} \, Q^{\bar{c}} \,
V^{\bar{d}}(z) 
- \frac{4i(k+2N)}{3(k+N+2)^2} \, d^{2}_{\bar{a} \bar{b}} \, 
Q^{\bar{a}} \, \pa \, V^{\bar{b}}(z).
\nonu
\eea}
Compared to the previous results in (\ref{spin5halftwo}) and
Appendix (\ref{spin5halftwo-2}),
the difference arises in the cubic terms.
From (\ref{spin5halftwo}) and Appendix
(\ref{spin5halftwo-2}), one has
\bea
&& \Bigg[ \frac{1}{2N} \, h^3_{\bar{a} \bar{b}}\,
  d^1_{\bar{d} \bar{e}} -
\frac{1}{2N} \, h^1_{\bar{a} \bar{b}}\,
  d^3_{\bar{d} \bar{e}}
- \frac{1}{N} \, h^2_{\bar{a} \bar{b}}\,
  d^0_{\bar{d} \bar{e}}
  +  2  \, h^0_{\bar{b} \bar{d}}\,
  d^2_{\bar{a} \bar{e}} -  h^1_{\bar{b} \bar{e}}\,
  d^3_{\bar{a} \bar{d}} +  h^1_{\bar{b} \bar{d}}\,
  d^3_{\bar{a} \bar{e}} \Bigg] f^{\bar{a} \bar{b}}_{\,\,\,\,\,\,c}=
\nonu \\
&& \Bigg[ -\frac{1}{2N} \, h^2_{\bar{a} \bar{b}}\,
  d^0_{\bar{d} \bar{e}} -
\frac{1}{2N} \, h^1_{\bar{a} \bar{b}}\,
  d^3_{\bar{d} \bar{e}}
+ \frac{1}{N} \, h^3_{\bar{a} \bar{b}}\,
  d^1_{\bar{d} \bar{e}}
  +  2  \, h^1_{\bar{b} \bar{d}}\,
  d^3_{\bar{a} \bar{e}} +  h^2_{\bar{b} \bar{e}}\,
  d^0_{\bar{a} \bar{d}} -  h^3_{\bar{b} \bar{d}}\,
  d^1_{\bar{a} \bar{e}} \Bigg] f^{\bar{a} \bar{b}}_{\,\,\,\,\,\,c}.
\nonu
\eea
Then by using the identity Appendix (\ref{simpleidentity}), one has
\bea
\Bigg[ -\frac{1}{2N} \, h^3_{\bar{a} \bar{b}}\,
  d^1_{\bar{d} \bar{e}} 
- \frac{1}{2N} \, h^2_{\bar{a} \bar{b}}\,
  d^0_{\bar{d} \bar{e}}
  +   h^0_{\bar{b} \bar{d}}\,
  d^2_{\bar{a} \bar{e}} -  h^1_{\bar{b} \bar{d}}\,
  d^3_{\bar{a} \bar{e}}
  \Bigg] f^{\bar{a} \bar{b}}_{\,\,\,\,\,\,c}=0.
\label{Ide2}
\eea
As before,
it is an open problems to observe the identities Appendices
(\ref{Ide1}) and (\ref{Ide2}) in the general context.

\section{Different routes for obtaining the higher spin-$3$ current }

Recall that the higher spin-$3$ current arises
in the OPE between $G^{\mu}(z)$ and
the higher spin-$\frac{5}{2}$ current $\Phi_{\frac{3}{2}}^{(1),\nu}(w)$
appearing in Appendix (\ref{1116}) with the condition $\mu=\nu$.
In section $3$, the $\mu=\nu=3$ case in $SO(4)$ basis is considered.
In this Appendix, we describe the other cases, $\mu=\nu=1,2$ and $4$.

\subsection{ From the second higher spin-$\frac{5}{2}$ current
$\Phi_{\frac{3}{2}}^{(1),4}$}

Let us introduce some part of the first order pole in (\ref{224})
as follows:
{\footnotesize
\bea
\tilde{\Phi}_{\frac{3}{2}}^{(1),4}
 \equiv 
\frac{2 i}{(k+N+2)}  T^{12}  \Phi_{\frac{1}{2}}^{(1),3}
+\frac{2 i}{(k+N+2)}  T^{23}  \Phi_{\frac{1}{2}}^{(1),1}
-\frac{4 i}{(k+N+2)} \, T^{13}  \Phi_{\frac{1}{2}}^{(1),2}
+  \frac{(N-k)}{3(k+N+2)}  \pa  \Phi_{\frac{1}{2}}^{(1),4}.
\nonu
\eea}
In order to obtain the higher spin-$3$ current
using the higher spin-$\frac{5}{2}$ current (\ref{spin5halftwo}),
one should also calculate the OPE
between $G^4(z)$ and $\tilde{\Phi}_{\frac{3}{2}}^{(1),4}(w)$ ($\mu=\nu=4$)
in $SO(4)$ basis.
From the defining OPEs in Appendix (\ref{1116}), one can calculate
this OPE as follows:
{\footnotesize
\bea
{G}^4(z) \,
\tilde{\Phi}_{\frac{3}{2}}^{(1),4}(w) & = &
\frac{1}{(z-w)^3} \, \Bigg[ \frac{4(k-N)}{3(k+N+2)}
  \Bigg] \, \Phi_0^{(1)}(w) 
+ \frac{1}{(z-w)^2} \, \Bigg[ \frac{(k-N)}{(k+N+2)}
  \Bigg] \, \pa \, \Phi_0^{(1)}(w)
\nonu \\
&+& \frac{1}{(z-w)} \Bigg[
-\frac{2 i}{(k+N+2)} \, T^{12} \, \Phi_{1}^{(1),12}
-\frac{2 i}{(k+N+2)} \, T^{23} \, \Phi_{1}^{(1),23}
\nonu \\
& - & \frac{4 i}{(k+N+2)} \, T^{13} \, \Phi_{1}^{(1),13}
+  \frac{(k-N)}{3(k+N+2)} \, \pa^2 \, \Phi_{0}^{(1)}
  \Bigg](w) 
+\cdots.
\label{44other}
\eea}
Then one can write down the higher spin-$3$ current 
as in (\ref{simplespin3}).
In other words, one has $1)$ the first order pole in (\ref{poleoneone})
with $\mu=\nu=4$ in $SO(4)$ basis, which comes from the OPE between
$G^4(z)$ and the first order pole of (\ref{5halfcombination}),
$2)$ the second term of (\ref{simplespin3}) and 
$3)$ the first order pole of Appendix (\ref{44other}).
It is an open problem to observe any nontrivial identities
between the various tensors from the two expressions
for the same higher spin-$3$ current.

There exists other expression for the same higher spin-$\frac{5}{2}$
current in Appendix (\ref{spin5halftwo-1}). 
From its defining equation in Appendix
(\ref{114}), one introduces the following
quantity
{\footnotesize
  \bea
\check{\Phi}_{\frac{3}{2}}^{(1),4}
 \equiv 
\frac{2 i}{(k+N+2)}  T^{12}  \Phi_{\frac{1}{2}}^{(1),3}
-  \frac{2 i}{(k+N+2)}  T^{13}  \Phi_{\frac{1}{2}}^{(1),2}
+\frac{4 i}{(k+N+2)} \, T^{23}  \Phi_{\frac{1}{2}}^{(1),1}
+   \frac{(N-k)}{3(k+N+2)}  \pa  \Phi_{\frac{1}{2}}^{(1),4}.
\nonu
\eea}
The following OPE can be calculated from Appendix (\ref{1116}) 
{\footnotesize
\bea
{G}^4(z) \,
\check{\Phi}_{\frac{3}{2}}^{(1),4}(w) & = &
\frac{1}{(z-w)^3} \, \Bigg[ \frac{4(k-N)}{3(k+N+2)}
  \Bigg] \, \Phi_0^{(1)}(w) 
+ \frac{1}{(z-w)^2} \, \Bigg[ \frac{(k-N)}{(k+N+2)}
  \Bigg] \, \pa \, \Phi_0^{(1)}(w)
\nonu \\
&+& \frac{1}{(z-w)} \Bigg[
-\frac{2 i}{(k+N+2)} \, T^{12} \, \Phi_{1}^{(1),12}
-\frac{2 i}{(k+N+2)} \, T^{13} \, \Phi_{1}^{(1),13}
\nonu \\
& - & \frac{4 i}{(k+N+2)} \, T^{23} \, \Phi_{1}^{(1),23}
+  \frac{(k-N)}{3(k+N+2)} \, \pa^2 \, \Phi_{0}^{(1)}
  \Bigg](w) 
+\cdots.
\label{44other-1}
\eea}
Then as before, one can write down the higher spin-$3$ current 
as in (\ref{simplespin3}).
In other words, one has $1)$ the first order pole in (\ref{poleoneone})
with $\mu=\nu=4$ in $SO(4)$ basis,  which comes from the OPE between
$G^4(z)$ and the first order pole of  (\ref{5halfcombination}),
$2)$ the second term of (\ref{simplespin3}) and 
$3)$ the first order pole of Appendix (\ref{44other-1}).
One expects that there will be nontrivial identities
by identifying two expressions for the same higher spin-$3$ current.

There exists other expression for the same higher spin-$\frac{5}{2}$
current in Appendix (\ref{spin5halftwo-2}). 
From its defining equation in Appendix
(\ref{114}), one introduces the following
quantity
{\footnotesize
  \bea
\breve{\Phi}_{\frac{3}{2}}^{(1),4}
 \equiv 
-\frac{2 i}{(k+N+2)}  T^{13}  \Phi_{\frac{1}{2}}^{(1),2}
+  \frac{2 i}{(k+N+2)}  T^{23}  \Phi_{\frac{1}{2}}^{(1),1}
+\frac{4 i}{(k+N+2)}  T^{12}  \Phi_{\frac{1}{2}}^{(1),3}
+   \frac{(N-k)}{3(k+N+2)}  \pa  \Phi_{\frac{1}{2}}^{(1),4}.
\nonu
\eea}
The following OPE can be calculated from Appendix (\ref{1116}) 
\bea
{G}^4(z) \,
\breve{\Phi}_{\frac{3}{2}}^{(1),4}(w) & = &
\frac{1}{(z-w)^3} \, \Bigg[ \frac{4(k-N)}{3(k+N+2)}
  \Bigg] \, \Phi_0^{(1)}(w) 
+ \frac{1}{(z-w)^2} \, \Bigg[ \frac{(k-N)}{(k+N+2)}
  \Bigg] \, \pa \, \Phi_0^{(1)}(w)
\nonu \\
&+& \frac{1}{(z-w)} \Bigg[
-\frac{2 i}{(k+N+2)} \, T^{13} \, \Phi_{1}^{(1),13}
-\frac{2 i}{(k+N+2)} \, T^{23} \, \Phi_{1}^{(1),23}
\nonu \\
& - & \frac{4 i}{(k+N+2)} \, T^{12} \, \Phi_{1}^{(1),12}
+  \frac{(k-N)}{3(k+N+2)} \, \pa^2 \, \Phi_{0}^{(1)}
  \Bigg](w) 
+\cdots.
\label{44other-2}
\eea
The higher spin-$3$ current can be determined 
as in (\ref{simplespin3}).
In other words, one has $1)$ the first order pole in (\ref{poleoneone})
with $\mu=\nu=4$,  which comes from the OPE between
$G^4(z)$ and the first order pole of (\ref{5halfcombination}),
$2)$ the second term of (\ref{simplespin3}) and 
$3)$ the first order pole of Appendix (\ref{44other-2}).
There should be nontrivial identities
by identifying two expressions for the same higher spin-$3$ current.

\subsection{ From the third higher spin-$\frac{5}{2}$ current
$\Phi_{\frac{3}{2}}^{(1),1}$}

Let us introduce some part of the first order pole in (\ref{212})
as follows:
{\footnotesize
  \bea
\tilde{\Phi}_{\frac{3}{2}}^{(1),1}
 \equiv 
\frac{2 i}{(k+N+2)}  T^{23}  \Phi_{\frac{1}{2}}^{(1),4}
-\frac{2 i}{(k+N+2)}  T^{24}  \Phi_{\frac{1}{2}}^{(1),3}
+\frac{4 i}{(k+N+2)}  T^{34}  \Phi_{\frac{1}{2}}^{(1),2}
-  \frac{(N-k)}{3(k+N+2)}  \pa  \Phi_{\frac{1}{2}}^{(1),1}.
\nonu
\eea}
One should also calculate the OPE
between $G^1(z)$ and $\tilde{\Phi}_{\frac{3}{2}}^{(1),1}(w)$ ($\mu=\nu=1$).
From the defining OPEs in Appendix (\ref{1116}), one can calculate
this OPE as follows:
\bea
{G}^1(z) \,
\tilde{\Phi}_{\frac{3}{2}}^{(1),1}(w) & = &
\frac{1}{(z-w)^3} \, \Bigg[ \frac{-4(k-N)}{3(k+N+2)}
  \Bigg] \, \Phi_0^{(1)}(w) 
+ \frac{1}{(z-w)^2} \, \Bigg[ \frac{-(k-N)}{(k+N+2)}
  \Bigg] \, \pa \, \Phi_0^{(1)}(w)
\nonu \\
&+& \frac{1}{(z-w)} \Bigg[
\frac{2 i}{(k+N+2)} \, T^{23} \, \Phi_{1}^{(1),23}
+\frac{2 i}{(k+N+2)} \, T^{24} \, \Phi_{1}^{(1),24}
\nonu \\
& + & \frac{4 i}{(k+N+2)} \, T^{34} \, \Phi_{1}^{(1),34}
-  \frac{(k-N)}{3(k+N+2)} \, \pa^2 \, \Phi_{0}^{(1)}
  \Bigg](w) 
+\cdots.
\label{11other}
\eea
Then one can write down the higher spin-$3$ current 
as in (\ref{simplespin3}).
In other words, one has $1)$ the first order pole in (\ref{poleoneone})
with $\mu=\nu=4$ in $SO(4)$ basis, which comes from the OPE between
$G^4(z)$ and the first order pole of (\ref{5halfcombination}),
$2)$ the second term of (\ref{simplespin3}) and 
$3)$ the first order pole of Appendix (\ref{11other}).

There exists other expression for the same higher spin-$\frac{5}{2}$
current in Appendix (\ref{spin5halfthree-1}). 
From its defining equation in Appendix
(\ref{313}), one introduces the following
quantity
{\footnotesize
\bea
\check{\Phi}_{\frac{3}{2}}^{(1),1}
 \equiv 
\frac{2 i}{(k+N+2)}  T^{23}  \Phi_{\frac{1}{2}}^{(1),4}
+  \frac{2 i}{(k+N+2)}  T^{34}  \Phi_{\frac{1}{2}}^{(1),2}
-\frac{4 i}{(k+N+2)}  T^{24}  \Phi_{\frac{1}{2}}^{(1),3}
-  \frac{(N-k)}{3(k+N+2)}  \pa  \Phi_{\frac{1}{2}}^{(1),1}.
\nonu
\eea}
The following OPE can be calculated from Appendix (\ref{1116}) 
\bea
{G}^1(z) \,
\check{\Phi}_{\frac{3}{2}}^{(1),1}(w) & = &
\frac{1}{(z-w)^3} \, \Bigg[ \frac{-4(k-N)}{3(k+N+2)}
  \Bigg] \, \Phi_0^{(1)}(w) 
+ \frac{1}{(z-w)^2} \, \Bigg[ \frac{-(k-N)}{(k+N+2)}
  \Bigg] \, \pa \, \Phi_0^{(1)}(w)
\nonu \\
&+& \frac{1}{(z-w)} \Bigg[
\frac{2 i}{(k+N+2)} \, T^{23} \, \Phi_{1}^{(1),23}
+\frac{2 i}{(k+N+2)} \, T^{34} \, \Phi_{1}^{(1),34}
\nonu \\
& + & \frac{4 i}{(k+N+2)} \, T^{24} \, \Phi_{1}^{(1),24}
-  \frac{(k-N)}{3(k+N+2)} \, \pa^2 \, \Phi_{0}^{(1)}
  \Bigg](w) 
+\cdots.
\label{11other-1}
\eea
The higher spin-$3$ current can be determined 
as in (\ref{simplespin3}).
In other words, one has the first order pole in (\ref{poleoneone})
with $\mu=\nu=1$ in $SO(4)$ basis,  which comes from the OPE between
$G^1(z)$ and the first order pole of (\ref{5halfcombination}),
the second term of (\ref{simplespin3}) and 
the first order pole of Appendix (\ref{11other-1}).

There exists other expression for the same higher spin-$\frac{5}{2}$
current in Appendix (\ref{spin5halfthree-2}). 
From its defining equation in Appendix
(\ref{414}), one introduces the following
quantity
{\footnotesize
\bea
\breve{\Phi}_{\frac{3}{2}}^{(1),1}
 \equiv 
-\frac{2 i}{(k+N+2)}  T^{24}  \Phi_{\frac{1}{2}}^{(1),3}
+  \frac{2 i}{(k+N+2)}  T^{34}  \Phi_{\frac{1}{2}}^{(1),2}
+\frac{4 i}{(k+N+2)}  T^{23}  \Phi_{\frac{1}{2}}^{(1),4}
-   \frac{(N-k)}{3(k+N+2)}  \pa  \Phi_{\frac{1}{2}}^{(1),1}.
\nonu
\eea}
The following OPE can be calculated from Appendix (\ref{1116}) 
\bea
{G}^1(z) \,
\breve{\Phi}_{\frac{3}{2}}^{(1),1}(w) & = &
\frac{1}{(z-w)^3} \, \Bigg[ \frac{-4(k-N)}{3(k+N+2)}
  \Bigg] \, \Phi_0^{(1)}(w) 
+ \frac{1}{(z-w)^2} \, \Bigg[ \frac{-(k-N)}{(k+N+2)}
  \Bigg] \, \pa \, \Phi_0^{(1)}(w)
\nonu \\
&+& \frac{1}{(z-w)} \Bigg[
\frac{2 i}{(k+N+2)} \, T^{24} \, \Phi_{1}^{(1),24}
+\frac{2 i}{(k+N+2)} \, T^{34} \, \Phi_{1}^{(1),34}
\nonu \\
& + & \frac{4 i}{(k+N+2)} \, T^{23} \, \Phi_{1}^{(1),23}
-  \frac{(k-N)}{3(k+N+2)} \, \pa^2 \, \Phi_{0}^{(1)}
  \Bigg](w) 
+\cdots.
\label{11other-2}
\eea
Then one can write down the higher spin-$3$ current 
as in (\ref{simplespin3}).
In other words, one has $1)$ the first order pole in (\ref{poleoneone})
with $\mu=\nu=4$ in $SO(4)$ basis, which comes from the OPE between
$G^4(z)$ and the first order pole of (\ref{5halfcombination}),
$2)$ the second term of (\ref{simplespin3}) and 
$3)$ the first order pole of Appendix (\ref{11other-2}).

\subsection{ From the fourth higher spin-$\frac{5}{2}$ current
$\Phi_{\frac{3}{2}}^{(1),2}$}

Let us introduce some part of the first order pole in (\ref{323})
as follows:
{\footnotesize
  \bea
\tilde{\Phi}_{\frac{3}{2}}^{(1),2}
 \equiv 
-\frac{2 i}{(k+N+2)}  T^{13}  \Phi_{\frac{1}{2}}^{(1),4}
-\frac{2 i}{(k+N+2)}  T^{34}  \Phi_{\frac{1}{2}}^{(1),1}
+\frac{4 i}{(k+N+2)}  T^{14}  \Phi_{\frac{1}{2}}^{(1),3}
-  \frac{(N-k)}{3(k+N+2)}  \pa  \Phi_{\frac{1}{2}}^{(1),2}.
\nonu
\eea}
One should also calculate the OPE
between $G^2(z)$ and $\tilde{\Phi}_{\frac{3}{2}}^{(1),2}(w)$ ($\mu=\nu=2$).
From the defining OPEs in Appendix (\ref{1116}), one can calculate
this OPE as follows:
\bea
{G}^2(z) \,
\tilde{\Phi}_{\frac{3}{2}}^{(1),2}(w) & = &
\frac{1}{(z-w)^3} \, \Bigg[ \frac{-4(k-N)}{3(k+N+2)}
  \Bigg] \, \Phi_0^{(1)}(w) 
+ \frac{1}{(z-w)^2} \, \Bigg[ \frac{-(k-N)}{(k+N+2)}
  \Bigg] \, \pa \, \Phi_0^{(1)}(w)
\nonu \\
&+& \frac{1}{(z-w)} \Bigg[
\frac{2 i}{(k+N+2)} \, T^{13} \, \Phi_{1}^{(1),13}
+\frac{2 i}{(k+N+2)} \, T^{34} \, \Phi_{1}^{(1),34}
\nonu \\
& - & \frac{4 i}{(k+N+2)} \, T^{14} \, \Phi_{1}^{(1),14}
-  \frac{(k-N)}{3(k+N+2)} \, \pa^2 \, \Phi_{0}^{(1)}
  \Bigg](w) 
+\cdots.
\label{22other}
\eea
Then one can write down the higher spin-$3$ current 
as in (\ref{simplespin3}).
In other words, one has $1)$ the first order pole in (\ref{poleoneone})
with $\mu=\nu=2$ in $SO(4)$ basis, which comes from the OPE between
$G^4(z)$ and the first order pole of  (\ref{5halfcombination}),
$2)$ the second term of (\ref{simplespin3}) and 
$3)$ the first order pole of Appendix (\ref{22other}).

There exists other expression for the same higher spin-$\frac{5}{2}$
current in Appendix (\ref{spin5halffour-1}). 
From its defining equation in Appendix
(\ref{424}), one introduces the following
quantity
{\footnotesize
  \bea
\check{\Phi}_{\frac{3}{2}}^{(1),2}
 \equiv 
+\frac{2 i}{(k+N+2)}  T^{14}  \Phi_{\frac{1}{2}}^{(1),3}
-  \frac{2 i}{(k+N+2)}  T^{34}  \Phi_{\frac{1}{2}}^{(1),1}
-\frac{4 i}{(k+N+2)}  T^{13}  \Phi_{\frac{1}{2}}^{(1),4}
-   \frac{(N-k)}{3(k+N+2)}  \pa  \Phi_{\frac{1}{2}}^{(1),2}.
\nonu
\eea}
The following OPE can be calculated from Appendix (\ref{1116}) 
\bea
{G}^2(z) \,
\check{\Phi}_{\frac{3}{2}}^{(1),2}(w) & = &
\frac{1}{(z-w)^3} \, \Bigg[ \frac{-4(k-N)}{3(k+N+2)}
  \Bigg] \, \Phi_0^{(1)}(w) 
+ \frac{1}{(z-w)^2} \, \Bigg[ \frac{-(k-N)}{(k+N+2)}
  \Bigg] \, \pa \, \Phi_0^{(1)}(w)
\nonu \\
&+& \frac{1}{(z-w)} \Bigg[
\frac{2 i}{(k+N+2)} \, T^{14} \, \Phi_{1}^{(1),14}
+\frac{2 i}{(k+N+2)} \, T^{34} \, \Phi_{1}^{(1),34}
\nonu \\
& + & \frac{4 i}{(k+N+2)} \, T^{13} \, \Phi_{1}^{(1),13}
-  \frac{(k-N)}{3(k+N+2)} \, \pa^2 \, \Phi_{0}^{(1)}
  \Bigg](w) 
+\cdots.
\label{22other-1}
\eea
Then one can write down the higher spin-$3$ current 
as in (\ref{simplespin3}).
In other words, one has the first order pole in (\ref{poleoneone})
with $\mu=\nu=2$ in $SO(4)$ basis, which comes from the OPE between
$G^2(z)$ and the first order pole of (\ref{5halfcombination}),
the second term of (\ref{simplespin3}) and 
the first order pole of Appendix (\ref{22other-1}).

There exists other expression for the same higher spin-$\frac{5}{2}$
current in Appendix (\ref{spin5halffour-2}). 
From its defining equation in Appendix
(\ref{112}), one introduces the following
quantity
{\footnotesize
  \bea
\breve{\Phi}_{\frac{3}{2}}^{(1),2}
 \equiv 
-\frac{2 i}{(k+N+2)}  T^{13}  \Phi_{\frac{1}{2}}^{(1),4}
+ \frac{2 i}{(k+N+2)}  T^{14}  \Phi_{\frac{1}{2}}^{(1),3}
-\frac{4 i}{(k+N+2)}  T^{34}  \Phi_{\frac{1}{2}}^{(1),1}
-   \frac{(N-k)}{3(k+N+2)}  \pa  \Phi_{\frac{1}{2}}^{(1),2}.
\nonu
\eea}
The following OPE can be calculated from Appendix (\ref{1116}) 
\bea
{G}^2(z) \,
\breve{\Phi}_{\frac{3}{2}}^{(1),2}(w) & = &
\frac{1}{(z-w)^3} \, \Bigg[ \frac{-4(k-N)}{3(k+N+2)}
  \Bigg] \, \Phi_0^{(1)}(w) 
+ \frac{1}{(z-w)^2} \, \Bigg[ \frac{-(k-N)}{(k+N+2)}
  \Bigg] \, \pa \, \Phi_0^{(1)}(w)
\nonu \\
&+& \frac{1}{(z-w)} \Bigg[
\frac{2 i}{(k+N+2)} \, T^{13} \, \Phi_{1}^{(1),13}
+\frac{2 i}{(k+N+2)} \, T^{14} \, \Phi_{1}^{(1),14}
\nonu \\
& + & \frac{4 i}{(k+N+2)} \, T^{34} \, \Phi_{1}^{(1),34}
-  \frac{(k-N)}{3(k+N+2)} \, \pa^2 \, \Phi_{0}^{(1)}
  \Bigg](w) 
+\cdots.
\label{22other-2}
\eea
Then one can write down the higher spin-$3$ current 
as in (\ref{simplespin3}).
In other words, one has the first order pole in (\ref{poleoneone})
with $\mu=\nu=4$ in $SO(4)$ basis, which comes from the OPE between
$G^4(z)$ and the first order pole of (\ref{5halfcombination}),
the second term of (\ref{simplespin3}) and 
the first order pole of Appendix (\ref{22other-2}).


\end{document}